\documentclass[a4paper,11pt]{article}
\pdfoutput=1 

\usepackage{amssymb}
\usepackage{dsfont}
\usepackage[T1]{fontenc} 
\usepackage{physics}
\usepackage{comment}
\usepackage{amsmath}
\usepackage{graphicx}
\usepackage{array, makecell}
\usepackage{float}
\setcellgapes{3pt}
\usepackage{xcolor}
\usepackage{amssymb}
\usepackage[normalem]{ulem}
\usepackage[numbers,sort&compress]{natbib}
\usepackage{geometry}
\usepackage{pdflscape}

\definecolor{azure}{rgb}{0.0, 0.5, 1.0}
\definecolor{darkblue}{rgb}{0.15,0.35,0.7}
\definecolor{reddish}{rgb}{0.65, 0.2, 0.2}
\definecolor{brandeisblue}{rgb}{0.0, 0.44, 1.0}
\definecolor{ceruleanblue}{rgb}{0.16, 0.32, 0.75}
\definecolor{indigo(dye)}{rgb}{0.0, 0.25, 0.42}
\definecolor{dgrey}{rgb}{0.3,0.3,0.3}
\definecolor{grey}{rgb}{0.9,0.9,0.9}

\usepackage[linktocpage=true]{hyperref}
\hypersetup{
colorlinks=true,
citecolor=ceruleanblue,
linkcolor=ceruleanblue,
urlcolor=ceruleanblue,
pdfauthor={},
pdftitle={},
pdfsubject={}
}

\usepackage{cleveref}
\usepackage{bm}
\crefname{lem}{lemma}{lemmas}
\crefname{thm}{theorem}{theorems}
\crefname{cor}{corollary}{corollaries}
\crefname{rem}{remark}{remarks}
\crefname{prop}{proposition}{propositions}

\usepackage{colortbl}
\definecolor{dgreen}{rgb}{0, 0.55, 0}
\definecolor{llightyellow}{rgb}{1.0, 0.95, 0.7}
\definecolor{llightblue}{rgb}{0.7, 0.9, 1.0}
\definecolor{llightpink}{rgb}{1.0, 0.85, 0.95}
\definecolor{llightgreen}{rgb}{0.7, 1.0, 0.4}
\colorlet{lightyellow}{llightyellow!50!white}
\colorlet{lightblue}{llightblue!50!white}
\colorlet{lightgreen}{llightgreen!50!white}
\colorlet{lightpink}{llightpink!50!white}

\usepackage[framemethod=TikZ]{mdframed} 
\usepackage{tikz-cd} 
\usetikzlibrary{arrows,snakes,shapes.arrows,decorations.markings}
     \tikzset{>=triangle 90}
     \tikzstyle{bbc}=[draw,circle,fill=black,scale=.75]
     \tikzstyle{rc}=[circle,fill=red,scale=.6]
     \tikzstyle{wc}=[draw,circle,scale=.75]

\usetikzlibrary{cd}

\tikzset{snake it/.style={decorate, decoration=snake}}

\tikzset{
	on each segment/.style={
		decorate,
		decoration={
			show path construction,
			moveto code={},
			lineto code={
				\path [#1]
				(\tikzinputsegmentfirst) -- (\tikzinputsegmentlast);
			},
			curveto code={
				\path [#1] (\tikzinputsegmentfirst)
				.. controls
				(\tikzinputsegmentsupporta) and (\tikzinputsegmentsupportb)
				..
				(\tikzinputsegmentlast);
			},
			closepath code={
				\path [#1]
				(\tikzinputsegmentfirst) -- (\tikzinputsegmentlast);
			},
		},
	},
	mid arrow/.style={postaction={decorate,decoration={
				markings,
				mark=at position .5 with {\arrow[#1]{stealth}}
	}}},
}

\usetikzlibrary{decorations.markings}

\tikzset{line/.style={line width=0.25mm},
curve/.style={line,smooth,tension=1},
->-/.style={decoration={
  markings,
  mark=at position #1 with {\arrow[>=stealth]{>}}},postaction={decorate}},
-<-/.style={decoration={
  markings,
  mark=at position #1 with {\arrow[>=stealth]{<}}},postaction={decorate}},
}

\tikzset{bg/.style={opacity=.5}}

\tikzset{
    partial ellipse/.style args={#1:#2:#3}{
        insert path={+ (#1:#3) arc (#1:#2:#3)}
    }
}

\makeatletter
\renewcommand\section{\@startsection {section}{1}{\z@}%
                               {-3.5ex \@plus -1ex \@minus -.2ex}
                               {2.3ex \@plus.2ex}%
                               {\normalfont\large\bfseries}}
\renewcommand\subsection{\@startsection{subsection}{2}{\z@}%
                                 {-3.25ex\@plus -1ex \@minus -.2ex}%
                                 {1.5ex \@plus .2ex}%
                                 {\normalfont\bfseries}}
\makeatother

\let\non\nonumber

\newfont{\goth}{ygoth.tfm scaled 1200}                   

\numberwithin{equation}{section}

\newcommand\ra{\mathrm{a}}
\newcommand\rb{\mathrm{b}}
\newcommand\rc{\mathrm{c}}
\newcommand\rd{\mathrm{d}}
\newcommand\rt{\mathrm{t}}
\newcommand\rg{\mathrm{g}}

\newcommand{\be}{\begin{equation}}
\newcommand{\ee}{\end{equation}}
\newcommand{\bee}{\begin{equation} \begin{aligned}}
\newcommand{\eee}{\end{aligned} \end{equation}}

\newcommand{\CC}{\mathcal{C}}

\newcommand{\CE}{\mathcal{E}}

\newcommand{\CH}{\mathcal{H}}
\newcommand{\CL}{\mathcal{L}}
\newcommand{\CN}{\mathcal{N}}

\newcommand{\CO}{\mathcal{O}}

\newcommand\doubleA{\mathbb{A}}
\newcommand\doubleB{\mathbb{B}}
\newcommand\doubleC{\mathbb{C}}

\newcommand\doubleG{\mathbb{G}}
\newcommand\doubleH{\mathbb{H}}

\newcommand\doubleK{\mathbb{K}}

\newcommand\doubleZ{\mathbb{Z}}

\newcommand\scriptA{\mathcal{A}}

\newcommand\scriptC{\mathcal{C}}
\newcommand\scriptD{\mathcal{D}}
\newcommand\scriptE{\mathcal{E}}

\newcommand\scriptH{\mathcal{H}}

\newcommand\scriptL{\mathcal{L}}
\newcommand\scriptM{\mathcal{M}}
\newcommand\scriptN{\mathcal{N}}
\newcommand\scriptO{\mathcal{O}}
\newcommand\scriptP{\mathcal{P}}

\newcommand\scriptS{\mathcal{S}}
\newcommand\scriptT{\mathcal{T}}

\newcommand\scriptX{\mathcal{X}}

\newcommand\scriptZ{\mathcal{Z}}

\newcommand{\VEC}{\operatorname{Vec}}

\newcommand\DW{\operatorname{DW}}
\newcommand\Spin{\operatorname{Spin}}
\newcommand\wo{\widehat{\omega}}

\newcommand{\dsi}{\mathds{1}}
\newcommand{\IZ}{\mathbb{Z}}

\newcommand{\ochi}{\overline{\chi}}

\newcommand{\ii}{\mathsf{i}}

\newcommand\Rep{\operatorname{Rep}}
\newcommand\TY{\operatorname{TY}}
\newcommand\Hom{\operatorname{Hom}}
\newcommand\He{\operatorname{He}}
\newcommand\p{\operatorname{p}}
\newcommand\EqBr{\operatorname{EqBr}}

\newcommand{\Aut}{\operatorname{Aut}}

\newcommand{\ha}{\hat{a}}

\begin{document} 

\begin{titlepage}
\begin{center}

\hfill         \phantom{xxx}  

\vskip 2 cm {\Large \bf Exploring $G$-ality defects in 2-dim QFTs} 

\vskip 1.25 cm {\bf Da-Chuan Lu${}^1$, Zhengdi Sun${}^{1,2}$, Zipei Zhang${}^1$}\non \\

\vskip 0.2 cm
 {\it ${}^{1}$ Department of Physics, University of California, San Diego, CA 92093, USA}

\vskip 0.2 cm

{\it ${}^{2}$ Mani L. Bhaumik Institute for Theoretical Physics, Department of Physics and Astronomy, University of California Los Angeles, CA 90095, USA
}
 
\end{center}
\vskip 1.5 cm

\begin{abstract}
\noindent 
The Tambara-Yamagami (TY) fusion category symmetry $\text{TY}(\mathbb{A},\chi,\epsilon)$ describes the enhanced non-invertible self-duality symmetry of a $2$-dim QFT under gauging a finite Abelian group $\mathbb{A}$. We generalize the enhanced non-invertible symmetries by considering twisted gauging which allows stacking $\mathbb{A}$-SPTs before and after the gauging. Such non-invertible symmetries can be obtained from invertible anyon permutation symmetries of the $3$-dim SymTFT. Consider a finite group $G$ formed by (un)twisted gaugings of $\mathbb{A}$, a $2$-dim QFT invariant under topological manipulations in $G$ admits non-invertible \textit{$G$-ality defects}. We study the classification and the physical implication of the $G$-ality defects using the SymTFT and the group-theoretical fusion categories, with three concrete examples. 1) Triality with $\mathbb{A} = \mathbb{Z}_N \times \mathbb{Z}_N$ where $N$ is coprime with $3$. The classification was previously determined by Jordan and Larson where the data is similar to the $\text{TY}$ fusion categories, and we determine the anomaly of these fusion categories. 2) $p$-ality with $\mathbb{A} = \mathbb{Z}_p \times \mathbb{Z}_p$ where $p$ is an odd prime. We consider two such categories $\mathcal{P}_{\pm,m}$ which are distinguished by different choices of the symmetry fractionalization, a new data that does not appear in the TY classification, and show that they have distinct anomaly structures and spin selection rules. 3) $S_3$-ality with $\mathbb{A} = \mathbb{Z}_N \times \mathbb{Z}_N$. We study their classification explicitly for $N < 20$ via SymTFT, and provide a group-theoretical construction for certain $N$. We find $N=5$ is the minimal $N$ to admit an $S_3$-ality and $N=11$ is the minimal $N$ to admit a group-theoretical $S_3$-ality.

\baselineskip=18pt

\end{abstract}
\end{titlepage}

\tableofcontents

\flushbottom

\newpage

\section{Introduction}
Generalized global symmetry provides powerful tools to reveal the universal behaviours of various theories. It helps build connections of phenomena in different phases and even seemingly unrelated models. For recent reviews on the developments of non-invertible symmetry in various fields, see \cite{gaiotto:generalizedsym,McGreevy:2022oyu,Cordova:2022ruw,daniel2023review,sakura2023review,lakshya2023review,shuheng2023review,runkel2023review,Costa:2024wks}. While ordinary global symmetry and higher-form symmetry can distinguish different phases of matter, certain fusion category symmetries can further constrain the relation between different phases and determine the phase diagram. 

A typical example is the $(1+1)$d Ising model, where the Kramers-Wannier duality maps the ordered and disordered phases to each other. When imposing the self-duality, the theory stays at the critical line between the two gapped phases; it could be first order or second order depending on the details of the model. The combination of self-duality and the original $\IZ_2$ symmetry gives the Tambara-Yamagami fusion category symmetry $\TY(\IZ_2,+1)$. For the ordinary transverse field Ising model, the critical point is the gapless Ising CFT with central charge $\frac{1}{2}$. Further analysis shows it is possible to drive the Ising CFT ($c=\frac{1}{2}$) to tricritical Ising CFT ($c=\frac{7}{10}$) by an irrelevant self-duality preserving operator with large strength \cite{affleck2015triising,fendley2018xzz}, and eventually entering the gapped phase with 3 ground state degeneracy which corresponds to the first order transition. The anomaly of the fusion category symmetry $\TY(\IZ_2,+1)$ obstructs a symmetric gapped phase with a unique ground state \cite{clay2023nianomaly,symtft2023kaidi2,Choi:2023xjw,shuheng2024maj,Chang:2018iay}.

A less trivial example is the triality (order-3) for the $\IZ_2\times \IZ_2$ symmetric Hamiltonians. The triality is generated by twisted gauging $\IZ_2\times \IZ_2$ \cite{Thorngren:2019iar} and it permutes $\IZ_2\times \IZ_2$ symmetry protected topological phase (SPT) to $\IZ_2\times \IZ_2$ spontaneously symmetry breaking (SSB) phase to $\IZ_2\times \IZ_2$ symmetric gapped phase (Sym) \cite{Thorngren:2019iar}. When imposing the self-triality, the theory stays at the multicritical point which is described by the $c=1$ compact boson theory at the Kosterlitz-Thouless (KT) point \cite{moradi2023topoholo}. As a further generalization, in \cite{Lu:2024ytl} some of the authors of this work study the triality and $p$-ality (order-$p$) transformations for the $\IZ_p\times \IZ_p$ symmetric Hamiltonians and their corresponding phases, where $p$ is a prime number. All these maps have the features that local $\mathbb{Z}_p\times \mathbb{Z}_p$-symmetric operators remain local under the map while local operators charged under $\mathbb{Z}_p\times\mathbb{Z}_p$ are mapped to non-local operators in the dual theory. More generally, such maps form a group, for instance, duality, triality and $p$-ality correspond to $\IZ_2$, $\IZ_3$ and $\IZ_p$ groups respectively. As a simple non-Abelian case, we study the $S_3$-ality in the main text. We will restrict the $G$-ality to the cases where all the group elements are (un)twisted gauging. For example, one may combine the order 3 twisted gauging of $\IZ_2\times \IZ_2$ and order-2 untwisted gauging of $\IZ_2\times \IZ_2$ to obtain a type of ``$S_3$-ality'' in $\IZ_2\times \IZ_2$ (see e.g. \cite{Ando:2024hun,Lu:2024ytl}). However, composing the order-3 twisted gauging with order-2 untwisted gauging leads to an order-2 transformation described by a unitary SPT entangler instead of some gauging operation. This means its corresponding line operator has quantum dimension 1; this is different from a line operator implementing gauging which will have quantum dimension $2$. So this is not the case we consider. As we will show later, $\IZ_5\times \IZ_5$ is the minimal symmetry to allow a $S_3$-ality enhancement with every element being (un)twisted gauging, and $\IZ_{11}\times \IZ_{11}$ is the minimal symmetry to have group-theoretical $S_3$-ality. The $G$-ality will constrain the theory at the multicritical point if it is invariant under the $G$-ality transformation and it admits corresponding non-invertible symmetry. For $G=\IZ_p$, the $p$-ality constrains the multicritical point in a similar way as in \cite{Lanzetta:2022lze}. Such $G$-ality non-invertible symmetries can be realized not only in continuum but also in quantum spin chain models via (un)twisted gauging and operator mappings \cite{quella2013kt,moradi2023topoholo,Garre-Rubio2023gauging,verstraete2023mpo,verstraete2024mpo,Sahand2024cluster,Lu:2024ytl,Ando:2024hun,linhao2023kt,yabo2024nispt,cao2024tydipole}. In particular, the triality and $p$-ality discussed in this paper are realized on lattice models in a companion paper \cite{Lu:2024ytl}.

For a 2d QFT $\mathcal{X}$ with a given fusion category symmetry $\mathcal{C}$, it is possible that $\mathcal{X}$ actually admits additional symmetries. For instance, one may consider applying invertible topological manipulation $\scriptT$ to the theory $\mathcal{X}$, and check if the theory $\scriptX$ is invariant under this topological manipulation. Such manipulations could be tensor autoequivalences of the symmetry $\mathcal{C}$ or discrete gauging some algebraic object $\mathcal{A}$ in $\scriptC$, or some combinations of both. If the theory $\scriptX$ is invariant under some of these invertible topological manipulations (which form a finite group $G$), then $\scriptX$ admits additional topological defect lines besides $\mathcal{C}$, constructed by performing the topological manipulation on half-space. Often, the new topological defects together with the known ones $\scriptC$ form a bigger fusion category $\mathcal{E}_{G}\mathcal{C}$, known as a $G$-extension of $\mathcal{C}$ \cite{Etingof:2009yvg,Gelaki:2009blp,meir2012module}. $\scriptE_G\scriptC$ naturally admits a $G$-grading, meaning there is a direct sum decomposition
\begin{equation}
    \scriptE_G\scriptC = \bigoplus_{g\in G} (\scriptE_G\scriptC)_g ~, \quad (\scriptE_G \scriptC)_\dsi = \scriptC ~.
\end{equation}
Each grading component $(\scriptE_G\scriptC)_g$ consists of simple objects generated by fusing simple TDLs in $\scriptC$ with the defect constructed from the half-space topological manipulation. 

A special case is when every non-trivial grading component of $\mathcal{E}_{G}\mathcal{C}$ contains a unique simple object, which we denote as $\underline{\scriptE}_{G}\mathcal{C}$. In this case, the additional topological defects are given by $\mathcal{N}_g$, where $g \in G-\{\dsi\}$, with the fusion rules
\begin{equation}\label{eq:G_ality_fusion_rule}
    \mathcal{N}_g \times \mathcal{L} = \mathcal{L} \times \mathcal{N}_g = \langle \mathcal{L} \rangle \mathcal{N}_g ~, \quad \mathcal{N}_g \times \mathcal{N}_{g'} = \begin{cases} & \displaystyle \sum_{\text{simple }\scriptL} \langle \mathcal{L} \rangle \mathcal{L} ~, \quad gg' = \dsi ~, \\ & \sqrt{|\mathcal{C}|} \scriptN_{gg'} ~, \quad \text{otherwise} ~, \end{cases} ~,
\end{equation}
where $\langle \mathcal{L}\rangle$ is the quantum dimension of the line $\mathcal{L}$ and $\sqrt{|\scriptC|}$ is the total quantum dimension of the fusion category $\scriptC$. Consistency of the fusion rules requires $\sqrt{|\mathcal{C}|}$ to be an integer. For this type of extension, since $\mathcal{N}_g$ absorbs every defect line in $\mathcal{C}$, the corresponding topological manipulation is the gauging of the entire $\mathcal{C}$ fusion category symmetry. We will name these $\mathcal{N}_g$'s as \textit{$G$-ality} defects, generalizing the well-known \textit{duality} defect for the case $G = \doubleZ_2$. 

There can be multiple fusion categories with fusion rule \eqref{eq:G_ality_fusion_rule}. Different fusion categories with the same fusion rule are distinguished by inequivalent $F$-symbols, and in principle, one can determine all the inequivalent $F$-symbols by solving the constraint equations known as the pentagon equations. However, due to the large number of equations as well as the large number of gauge redundancies in the $F$-symbols (for the classification purpose), this is generically hard in practice. On the other hand, the fusion category must be unambiguously specified before one can study any physical implications such as the anomaly and the spin selection rules, as the additional data will change the physical implications generically.

There are two alternative ways to determine unambiguously the fusion category. First, given any fusion category $\widetilde{\scriptC}$, gauging an algebraic object $\scriptA$ will generically lead to a different fusion category $\widetilde{\scriptC}' = {}_{\scriptA}\widetilde{\scriptC}_{\scriptA}$. A special case is to start with $\widetilde{\scriptC} = \VEC_{\mathbb{G}}^\omega$, the fusion category describing invertible $\mathbb{G}$ symmetry with 't Hooft anomaly $[\omega] \in H^3(\mathbb{G},U(1))$. Gauging an anomaly-free subgroup $\mathbb{H}$ of $\mathbb{G}$ with discrete torsion $[\psi] \in H^2(\mathbb{H},U(1))$ leads to the so-called \textit{group-theoretical fusion categories}, denoted as $\scriptC(\mathbb{G},\omega;\mathbb{H},\psi)$. In the case that the simple objects and the fusion rule of $\widetilde{\scriptC}'$ coincide with \eqref{eq:G_ality_fusion_rule}, $(\widetilde{\scriptC},\scriptA)$ does specify unambiguously a fusion category with the fusion rule \eqref{eq:G_ality_fusion_rule}. Furthermore, the physical implications can be relatively easily derived provided they are known for $\widetilde{\scriptC}$. 

On the other hand, the SymTFT $\mathcal{Z}(\scriptC)$ of the fusion category symmetry $\scriptC$ provides an alternative way to classify all possible $G$-extensions of a fusion category $\mathcal{C}$\cite{Etingof:2009yvg}. Generically, the symmetry and its properties (such as 't Hooft anomaly and topological sectors) of a $d$-dim QFT can be captured by the $(d+1)$-dim topological field theory known as the Symmetry TFT (SymTFT) \cite{Kitaev:2011dxc,symtft2019XGW,symtft2020XGW2,symtft2021XGW3,symtft2023XGW4,symtft2022XGW5,symtft2021Gaiotto,symtft2021Sakura,moradi2023topoholo,symtft2022Apruzzi,Freed:2022qnc,Kaidi:2022cpf,symtft2022Kulp,symtft2023kaidi2,Brennan:2024fgj,Bonetti:2024cjk,DelZotto:2024tae,Argurio:2024oym,Franco:2024mxa,Putrov:2024uor,Huang:2024ror,Freed:2022qnc,Lin:2022dhv,Bhardwaj:2024igy,Antinucci:2024ltv,Bhardwaj:2024qiv,Bhardwaj:2024qrf,Bhardwaj:2023bbf,Bhardwaj:2023ayw,Bhardwaj:2023wzd,Sun:2023xxv,Zhang:2023wlu,Copetti:2024onh,Antinucci:2023ezl,Choi:2024wfm,Choi:2024tri,Cordova:2023bja,Antinucci:2024zjp,Bhardwaj:2024ydc,Chen:2023qnv,Cui:2024cav}. For the generalized $0$-form fusion category symmetry $\scriptC$ in 2d QFT, its SymTFT in 3d is described by the specific Modular Tensor Category (MTC) $\scriptZ(\scriptC)$ known as the Drinfeld center $\mathcal{Z}(\scriptC)$ of $\scriptC$. To specify the classification data, one must first choose a homomorphism $\rho$ from $G$ to the symmetry group $\EqBr(\scriptZ(\scriptC))$ of the $\mathcal{Z}(\scriptC)$\footnote{For more details on $\EqBr(\mathcal{B})$ where $\mathcal{B}$ is a MTC, see  \cite{Etingof:2009yvg,2013arXiv1309.5026N,Fuchs:2014ema,Barkeshli:2014cna}.}. After such a choice, one must check the corresponding Postnikov class $\mathcal{O}^3_\rho$, which is valued in $H^3_\rho(G,\mathbf{A})$ where $\mathbf{A}$ is the set of Abelian anyons in $\scriptZ(\scriptC)$, vanishes. Then, one chooses a symmetry fractionalization class $[\nu] \in H^2_\rho(G,\mathbf{A})$, then checks the vanishing of the 't Hooft anomaly $\mathcal{O}^4(\rho,[\nu])$ of $G$ which is valued in $H^4(G,U(1))$. Notice that here the 't Hooft anomaly of $G$ may depend on the choice of $[\nu]$ \cite{Etingof:2009yvg,Barkeshli:2014cna,Delmastro:2022pfo,Brennan:2022tyl}. Finally, one chooses the discrete torsion for the $G$ symmetry, which is a class in $H^3(G,U(1))$, such a choice is known as the FS indicator, which we will denote as $\alpha$. Together, $(\rho,[\nu],\alpha)$ such that two obstruction classes $\mathcal{O}^3_\rho$ and $\mathcal{O}^4(\rho,[\nu])$ vanish specifies a $G$-extension. Furthermore, the two classifications are equivalent if they are related by the action of the group of the autoequivalence of $\mathcal{C}$. One may notice that this is the same data which is required for gauging the $G$-symmetry in the $\mathcal{Z}(\scriptC)$. Indeed, gauging $G$ leads to the SymTFT of the particular $G$-extension of $\scriptC$.

The most famous example is $G = \doubleZ_2$ and $\scriptC = \VEC_{\mathbb{A}}$. The corresponding fusion category $\underline{\scriptE}_{\doubleZ_2}\VEC_{\mathbb{A}}$ is known as the Tambara-Yamagami fusion category $\TY(\mathbb{A}, \chi, \epsilon)$, which describes the self-duality under gauging the Abelian group $\doubleA$. These fusion categories are first classified by solving pentagon equations explicitly \cite{tambara1998tensor}, where the classification data is given by a choice of symmetric, non-degenerate bicharacter $\chi$ of the Abelian group $\mathbb{A}$ together with a FS indicator $\epsilon = \pm 1$. Its classification in terms of the SymTFT is first understood in \cite{Etingof:2009yvg}. The choice of bicharacter $\chi$ corresponds to a choice of $\doubleZ_2$ symmetry which maps the electric lines labeled by $\mathbb{A}$ to magnetic lines labeled by $\widehat{\mathbb{A}}$, and it can be shown for these $\doubleZ_2$ symmetries, the twisted group cohomology $H^n_{\rho_\chi}(\doubleZ_2,\mathbb{A} \times \widehat{\mathbb{A}}) = \doubleZ_1$ for $n \geq 1$, therefore, the obstruction to symmetry fractionalization trivially vanishes and there is a unique choice of the symmetry fractionalization. Furthermore, $H^4(\doubleZ_2,U(1)) = \doubleZ_1$ which means there is no 't Hooft anomaly for $\doubleZ_2$ 0-form symmetry in 3d, and the only data left after fixing $\chi$ is a choice of $H^3(\doubleZ_2,U(1)) \simeq \doubleZ_2$ and this corresponds to the choice of FS indicator on the boundary.

Some generalizations have been considered. Fixing $G = \doubleZ_2$, it is possible to consider the fusion category $\scriptC$ containing non-invertible symmetries. In \cite{yichul2024selfdual,Diatlyk:2023fwf,Perez-Lona:2023djo}, one considers $\scriptC = \Rep(H_8)$ and gauging this $\Rep(H_8)$ symmetry is an order-$2$ topological manipulation, which will lead to the fusion category $\underline{\scriptE}_{\doubleZ_2}\Rep(H_8)$. Furthermore, by directly solving the pentagon equations, \cite{Choi:2023xjw} finds there are 8 inequivalent fusion categories with the same fusion rule. Alternatively, one may consider fixing $\scriptC = \VEC_{\mathbb{A}}$ and considering $G$ other than $\doubleZ_2$. In \cite{jordan2009classification,Thorngren:2019iar,Thorngren:2021yso,2023triality}, $G = \doubleZ_3$ is considered and the corresponding non-invertible symmetries are known as \textit{triality} defects. In the physics literature \cite{Thorngren:2019iar,Thorngren:2021yso,2023triality}, $\mathbb{A} = \doubleZ_2 \times \doubleZ_2$ is considered and a concrete example of the triality defect is discovered at the KT point of $c = 1$ compact boson \cite{Thorngren:2021yso}. Based on the classification result acquired in \cite{jordan2009classification}, which claims there are two inequivalent triality fusion categories under twisted gauging $\doubleZ_2 \times \doubleZ_2$ modulo the choice of $\doubleZ_3$ FS indicator, \cite{2023triality} identifies the triality defect at the KT point belongs to the group-theoretical one in the classification and computes the corresponding $F$-symbols. Anomaly of this particular class of fusion category is determined using group-theoretical techniques and the spin selection rules are also computed. Recently, a similar construction of finding non-invertible symmetries by checking if the theory is self-dual under (twisted) gauging has been generalized to QFT with dimension $d > 2$, see \cite{Kaidi:2021xfk,Kaidi:2022cpf,shuheng:noninvstd,Choi:2024rjm,Choi:2021kmx,Choi:2022zal,clay2023nianomaly,Cordova:2023bja,Cui:2024cav,Apruzzi:2024cty}.

In this work, we aim to generalize the previous study and consider the $G$-ality defects under twisted gauging of Abelian symmetry $\mathbb{A}$ in $2d$ QFTs. As we will see, although some simplifications appearing in the classification of TY fusion category remain, new phenomena still occur. Following the classification of $G$-extension derived in \cite{Etingof:2009yvg}, we must first choose a $G^{(0)}$\footnote{Here, we use $G^{(0)}$ to denote a specific symmetry group in the bulk, which is isomorphic to the abstract group $G$.} subgroup of the symmetry group $\EqBr(\scriptZ(\VEC_{\mathbb{A}}))$ of the SymTFT of $\mathbb{A}$. The latter group is known as the split orthogonal group $O(\mathbb{A} \oplus \widehat{\mathbb{A}})$, and has been studied in \cite{jordan2009classification,Fuchs:2014ema}. \cite{jordan2009classification} identifies the condition on $G^{(0)}$ that leads to $G$-ality defects on the boundary and also discusses the equivalence relations on the choice of $G^{(0)}$ in the classification. In this work, by establishing the connection between bulk symmetry and the corresponding topological manipulation, we provide a physical understanding of this result. As one would expect, only the $G^{(0)}$ whose every non-trivial element corresponds to twisted gauging the full $\mathbb{A}$ symmetry on the boundary will lead to $G$-ality defects. In the bulk, this condition is the same as some component, which we call $\beta$, of the bulk symmetry is invertible. We will denote these $G^{(0)}$'s as $G^{(0)}$ with invertible $\beta$ component, or even shortly $\beta$-invertible $G^{(0)}$\footnote{By that, we mean each non-trivial group element $U$ in $G^{(0)}$ has invertible $\beta$-component.}. Furthermore, there is a special subgroup $I_{\doubleA} \subset O(\mathbb{A}\oplus\widehat{\mathbb{A}})$ consisting of all the bulk symmetries which implement outer automorphisms of $\mathbb{A}$ and stacking $\mathbb{A}$-SPT on the boundary. Here, $I_{\doubleA}$ is the group of autoequivalences of the boundary fusion category $\VEC_{\doubleA}$. Conjugating $G^{(0)}$ by any element in $I_\mathbb{A}$ will lead to an equivalent fusion category. Given a generic theory $\scriptX$ with $\mathbb{A}$ symmetry, its partition function coupling to background field is generically ambiguous up to relabeling background fields and stacking SPT, and $I_{\mathbb{A}}$ exactly characterizes that. Notice that for the case of $G = \doubleZ_2$ which corresponds to the TY fusion category, such an issue exists as well. But because there exists a nice choice of the representative of the $\doubleZ_2$ symmetry such that they always correspond to untwisted gauging, the other possibilities are generically not discussed. We will comment on this more in Section \ref{sec:G_ext_general} and Section \ref{sec:trans_part}. For a generic $G$, however, there is really no nice representative of $G^{(0)}$ and such subtlety becomes more salient. Thus, our discussion shows that $G$-ality defects can be identified by testing partition function invariance under twisted gauging, which also determines the equivalence class of $[G^{(0)}]$ in the classification data. 

Once a specific $G^{(0)}$ with invertible $\beta$-component is chosen, one must then ask if there would be any obstruction to the symmetry fractionalization of $G^{(0)}$. Generically, this is hard to determine, but as shown in \cite{Benini:2018reh}, for Abelian TQFT, the obstruction class $\mathcal{O}^3_\rho \in H^3_{\rho}(G,\mathbf{A})$ is always trivial. Therefore, we can directly proceed to choose a symmetry fractionalization class in $H^2_{\rho}(G,\mathbf{A})$. Unlike the TY case, generally speaking, this data could be non-trivial, and will provide additional challenges for the study of the classification. However, in the special case where the order $|G|$ of the group $G$ is coprime with the order $|\doubleA|$ (therefore with $|\mathbf{A}| = |\doubleA \times \widehat{\mathbb{A}}|$), by Zassenhaus theorem $H^2_{\rho}(G,\mathbf{A}) = \doubleZ_1$ (see e.g. \cite{weibel1994introduction}), hence the choice of the symmetry fractionalization class is unique. 

Finally, once the symmetry fractionalization class is chosen, we must then check the 't Hooft anomaly. But for many simple $G$ such as $G = \doubleZ_N, S_3$ etc, $H^4(G,U(1))$ is trivial. Hence, we only need to pick a discrete torsion $H^3(G,U(1))$, which corresponds to a choice of the FS indicator for the $G$-ality defect.

To summarize, from the classification point of view, we expect the complication from inequivalent choices of $G^{(0)}$ with invertible $\beta$-component, the new phenomenon arising from the choice of symmetry fractionalization class $H^2(G,\mathbb{A}\times \widehat{\mathbb{A}})$, and the potential obstruction in $H^4(G,U(1))$ for some $G$ (such as $\doubleZ_2 \times \doubleZ_2$).

Before delving into the physical implications, the $G$-ality defects must be unambiguously specified. This can be done either by specifying a particular choice of the classification data, or by providing a group-theoretical construction $\scriptC(\mathbb{G},\omega;\mathbb{H},\psi)$. In this work, when treating the case where there are multiple choices of the symmetry fractionalization in the bulk, we will only focus on the case where group-theoretical constructions exist. This even simplifies the bulk SymTFTs as they are Dijkgraaf-Witten (DW) theories with suitable twists. We will leave the understanding of the effect of symmetry fractionalizations in a generic non-DW SymTFT for future studies.

In this work, we focus on the study of the anomaly of the $G$-ality defect. Here, we are interested in the notion of the anomaly that obstructs the existence of the trivially symmetric gapped phase. As pointed out in \cite{Choi:2023xjw,Bhardwaj:2017xup,Thorngren:2019iar,Zhang:2023wlu}, this is equivalent to the obstruction to gauging the full fusion category symmetry with the algebraic object of the form
\begin{equation}
    \scriptA = \bigoplus_{\text{simple} \, \mathcal{L}} \langle \mathcal{L} \rangle \mathcal{L} ~,
\end{equation}
where $\langle \mathcal{L} \rangle$ is the quantum dimension of $\mathcal{L}$. Then, it immediately follows that if the $G$-ality defect has non-integer quantum dimension, then it must be anomalous. While this does provide a useful constraint when $G = \doubleZ_2$, for any other $G$ this is trivially true following from the fusion rule in \eqref{eq:G_ality_fusion_rule}. Generally, the symmetric gapped phase of a fusion category symmetry $\scriptC$ in $2d$ is classified by module categories $\scriptM$ over $\scriptC$, where the number of ground states is the number of simple objects in $\scriptM$ \cite{Thorngren:2019iar}. Therefore, looking for a trivially symmetric gapped phase is the same as looking for $\scriptM$ with a unique simple object. \cite{meir2012module} provides a framework for classifying the module categories over any graded fusion category; therefore, it can be used to study the anomaly of the duality defects as demonstrated in \cite{Thorngren:2019iar} in a physics-friendly language. The same problem of the anomaly of the duality defects has recently been studied in \cite{Zhang:2023wlu,Antinucci:2023ezl,Cordova:2023bja}, which could be generalized to $G$-ality defects in general. 

In this work, however, we will take an alternative approach by combining a key insight in \cite{Thorngren:2021yso,Zhang:2023wlu} together with the approach used in \cite{2023triality}. From the boundary theory point of view, a trivially gapped phase of the $\underline{\scriptE}_G\VEC_{\mathbb{A}}$ must be a trivially gapped phase of the $\VEC_{\mathbb{A}}$ symmetry, which is an $\mathbb{A}$-SPT classified by $H^2(\mathbb{A},U(1))$. Furthermore, this $\mathbb{A}$-SPT must be invariant under the topological manipulation $G$, which leads to a necessary condition for $\underline{\scriptE}_G\VEC_{\mathbb{A}}$ to be anomaly-free, see \cite{Thorngren:2019iar} for more discussion. In the SymTFT, this condition is equivalent to checking if there exists a $G$-stable magnetic Lagrangian algebra in $\scriptZ(\VEC_{\mathbb{A}})$. For the $\underline{\scriptE}_G\VEC_{\mathbb{A}}$ which passes this check, an additional check in the SymTFT is needed to be done to determine if the $\underline{\scriptE}_G\VEC_{\mathbb{A}}$ is anomalous \cite{Antinucci:2023ezl,Cordova:2023bja}. In this work, however, we present an alternative approach. As pointed out in \cite{Gelaki:2009blp,Sun:2023xxv}, the existence of the $G$-stable Lagrangian algebra (not necessarily magnetic) guarantees that $\underline{\scriptE}_G\VEC_{\mathbb{A}}$ is group-theoretical. And the symmetric gapped phases for any group-theoretical fusion category have been classified in \cite{ostrik2002module}, which we will use to finish the rest of the analysis and determine the anomaly. It would be interesting to figure out how to analyze the anomaly purely in the SymTFT and we leave this for future study.

Notice that the group-theoretical $G$-ality defect, one can also conveniently derive the spin selection rules by matching the $\mathcal{N}_g$ defect Hilbert space with the topological sectors of the invertible symmetry $\mathbb{G}$. The spin selection rule of the latter can be straightforwardly computed \cite{Coste:2000tq}. Notice that it is also relatively easy to compute the $F$-symbol for the group-theoretical $G$-ality defects without solving pentagon equations as demonstrated in \cite{2023triality}. But notice that interesting physical consequences can be derived without referring to those $F$-symbols.

\subsection{Summary of the main result}
In this work, we explore three examples of $G$-ality defects. 

As a warm-up, we start with $G = \doubleZ_3$ and $\mathbb{A} = \doubleZ_N \times \doubleZ_N$ where $\gcd(N,3) = 1$. The choice of $N$ ensures the unique choice of the symmetry fractionalization class; therefore, its classification is similar to the case of TY fusion category and has been worked out in \cite{jordan2009classification}. We work out the details of the corresponding bulk data to demonstrate our approaches, and determine the anomaly of these triality fusion categories using the group-theoretical approach mentioned previously.

Our next example involves non-trivial choices of the symmetry fractionalization classes in the classification data. We provide two group-theoretical fusion categories realizing $G = \doubleZ_p$ and $\doubleA = \doubleZ_p \times \doubleZ_p$ where $p$ is an odd prime number. We find that while two constructions correspond to the same choice of the bulk symmetries, they correspond to different choices of the symmetry fractionalization classes and, therefore, are distinct fusion categories. As a result, they have different anomaly structures (when turning on non-trivial FS indicator) and distinct spin selection rules. 

Our final example is the case where $G = S_3$ and $\doubleA = \doubleZ_N \times \doubleZ_N$. In this case, $G$ contains two generators as well as being non-Abelian. One can think of this case as the triality defect and the duality defect would fuse in a non-commutative fashion\footnote{Notice that recently in \cite{Damia:2024xju}, an example of non-commutative fusions between three duality defects is also discovered. However, not all duality lines in \cite{Damia:2024xju} implement the gauging of the full $\doubleZ_2 \times \doubleZ_2$ symmetry.}. We discuss this example in order to address some subtleties on the gauge fixing on studying inequivalent $G$ when there are multiple generators in $G$.

Below, we summarize our results in the above three examples in more detail.

\subsubsection*{Example I: $\mathbb{A} = \doubleZ_N \times \doubleZ_N$ and $G = \doubleZ_3$ where $\gcd(N,3) = 1$:}

This case is considered in \cite{jordan2009classification} and $N$ is chosen such that $H^{*}_{[\rho]}(\doubleZ_3,\mathbf{A})$ is trivial, thus the fusion categories are classified by the choice of the $\beta$-invertible $\doubleZ_3$ symmetry (up to conjugation by a group element in $I_{\doubleZ_N \times \doubleZ_N}$) in the SymTFT and the discrete torsion (or the FS indicator from the boundary point of view) $H^3(\doubleZ_3, U(1)) = \doubleZ_3$, similar to the classification of the TY fusion categories. 

With a given $N$ where $\gcd(N,3) = 1$, consider its prime decomposition
\begin{equation}\label{eq:prime_decomp_N}
    N = \prod_{i = 1}^n p_i^{r_i} ~.
\end{equation}
There are $3 \cdot 2^{n}$ numbers of inequivalent fusion categories $\underline{\scriptE}_{\doubleZ_3}\VEC_{\doubleZ_N \times \doubleZ_N}$. Here, the overall factor $3$ denotes three different choices of the FS indicator in $H^3(\doubleZ_3, U(1)) \simeq \doubleZ_3$. The factor $2^n$ denotes inequivalent choices of the $\doubleZ_3$ symmetry in the bulk. More concretely, for each prime factor $p_i^{r_i}$, there are two inequivalent choices of $\doubleZ_3$ symmetries, and a generic $\doubleZ_3$ symmetry is obtained by specifying the choice for each prime factor.

We construct the concrete forms of the generators of the $\doubleZ_3$ symmetry for each prime factor $\doubleZ_{p^r} \times \doubleZ_{p^r}$ and denote them as $T_1$ and $T_2$. We also construct the corresponding condensation defects and derive the corresponding form of the twisted gauging of $\doubleZ_{p^r} \times \doubleZ_{p^r}$. This allows us to determine whether a concrete QFT admits which type of the triality defects (up to FS indicator). As an example, we confirm that the diagonal $\Spin(8)$ WZW model admits both types of the triality defect under twisted gauging $\doubleZ_2 \times \doubleZ_2$. 

For generic $\doubleZ_N \times \doubleZ_N$, with $\displaystyle N=\prod_{i=1}^n p_i^{r_i}$, the corresponding triality defect is specified by $(\mathbf{T}, \alpha)$. Here, $\displaystyle \mathbf{T} = \bigotimes_{i =1}^n T^{(i)}$ is the generator of the bulk $G = \doubleZ_3$-symmetry and is specified by a choice of $T^{(i)} = T_1 \, \text{or} \, T_2$ for each prime factor. Notice that $\mathbf{T}$ is the direct analog of the non-degenerate symmetric bicharacter $\chi$ which determines the bulk $\doubleZ_{2}^{em}$-symmetry in the classification of the duality defect. And $\alpha$ is the choice of the FS indicator, which now takes value in $H^3(\doubleZ_3, U(1)) = \doubleZ_3$ because we are studying the triality defects. The anomaly of these triality fusion categories is determined using the group-theoretical fusion category technique. For the triality fusion category $\underline{\scriptE}^{(\mathbf{T},\alpha)}_{\doubleZ_3} \VEC_{\doubleZ_N \times \doubleZ_N}$, it is anomaly-free if and only if
\begin{enumerate}
    \item In the prime decomposition \eqref{eq:prime_decomp_N}, every prime factor $p_i = 1 \mod 3$.
    \item The FS indicator $\alpha$ is trivial.
\end{enumerate}

\subsubsection*{Example II: $\mathbb{A} = \doubleZ_p \times \doubleZ_p$ and $G = \doubleZ_p$ where $p$ is an odd prime}

In this case, we construct two group-theoretical fusion categories realizing $G = \doubleZ_p$ and $\mathbb{A} = \doubleZ_p \times \doubleZ_p$. The two classes of $p$-ality defects correspond to the same bulk $\doubleZ_p$-symmetry but different choices of symmetry fractionalization. We see that they lead to distinct physical implications such as the anomaly and the spin selection rules.

We denote the two classes of $p$-ality fusion categories as $\scriptP_{\pm, m}$ where $m \in \doubleZ_p$ labels different choices of the FS indicator. Both of them admit several group-theoretical constructions, and in particular, $\scriptP_{m,\pm}$ can be constructed by discrete gauging $\doubleH = \doubleZ_p^{\rb} \times \doubleZ_p^{\rc}$ from the Abelian invertible symmetries $\doubleG = \doubleZ_p^{\ra} \times \doubleZ_p^{\rb} \times \doubleZ_p^{\rc}$ with 't Hooft anomaly $\omega_{\pm,m}$ given by
\begin{equation}
\begin{aligned}
    & \omega_{+,m}(\ra^{i_1}\rb^{j_1}\rc^{k_1},\ra^{i_2}\rb^{j_2}\rc^{k_2},\ra^{i_3}\rb^{j_3}\rc^{k_3}) = e^{\frac{2\pi i}{p} j_1 k_2 i_3 + \frac{2\pi i m }{p^2}i_1 (i_2 + i_3 - [i_2 + i_3]_p)} ~, \\
    & \omega_{-,m}(\ra^{i_1}\rb^{j_1}\rc^{k_1},\ra^{i_2}\rb^{j_2}\rc^{k_2},\ra^{i_3}\rb^{j_3}\rc^{k_3}) = e^{\frac{2\pi i}{p} j_1 k_2 i_3 + \frac{2\pi i}{p^2} j_1 (i_2 + i_3 - [i_2 + i_3]_p) + \frac{2\pi i m }{p^2}i_1 (i_2 + i_3 - [i_2 + i_3]_p)} ~.
\end{aligned}
\end{equation}
$\scriptP_{+,m}$ is anomaly-free when the FS indicator $m$ is trivial and the spin selection rule of the $p$-ality defect Hilbert space is given by
\begin{equation}
    s \in \frac{1}{p}\doubleZ + \frac{m}{p^2} ~, \quad m \in \doubleZ_p ~.
\end{equation}
For $\scriptP_{-,m}$, however, it is anomaly-free regardless of the FS indicators and the spin selection rule of the $p$-ality defect Hilbert space is given by
\begin{equation}
    s \in \frac{1}{p^2} \doubleZ ~,
\end{equation}
independent of the FS indicator as well. 

We then study the $p$-ality defects from the SymTFT. We start with the SymTFT $\scriptZ(\VEC_{\doubleZ_p\times \doubleZ_p})$ and identify the bulk data associated to $\scriptP_{\pm,m}$. The upshot is that they correspond to the same choices of the bulk symmetry group $\doubleZ_p$, therefore their distinction must result in different choices of the symmetry fractionalization classes in $H^2_\rho(\doubleZ_p, (\doubleZ_p)^4) = \doubleZ_p \times \doubleZ_p$. By explicitly establishing the isomorphism between the choice of symmetry fractionalization classes and the mixed 't Hooft anomaly between $\doubleZ_p^a$ and $\doubleZ_p^b \times \doubleZ_p^c$ classified by $H^2(\doubleZ_p^a, H^1(\doubleZ_p^b \times \doubleZ_p^c,U(1))) \simeq \doubleZ_p \times \doubleZ_p$, and we argue that the additional type II anomaly between $\omega_{\pm,m}$ should be viewed as different choices of the symmetry fractionalization class in the bulk.

In addition, the group-theoretical construction mentioned above also allows us to explicitly compute the data of the SymTFT for the entire $\scriptP_{\pm,m}$, which is given by the DW theory of the gauge group $\doubleZ_p^{\ra} \times \doubleZ_p^{\rb} \times \doubleZ_p^{\rc}$ with the twist $\omega_{\pm,m}$ respectively. We compute the anomaly and the topological sectors for $\scriptP_{\pm,m}$ from the bulk and find it matching the above result acquired from the boundary calculation.

\subsubsection*{Example III: $\mathbb{A} = \doubleZ_N \times \doubleZ_N$ and $G = S_3$}

Our final example is $\mathbb{A} = \doubleZ_N \times \doubleZ_N$ and $G = S_3$. Notice that the group cohomology $H^4(S_3,U(1))$ is trivial identically; therefore, there is no obstruction class to consider for this case. We consider this example as it contains some new subtleties in determining the choice of the $\beta$-invertible $G^{(0)}$ symmetries in the bulk.

For the $N$-ality defects, one can use the gauge transformation to make the generator of the bulk $\doubleZ_N$ symmetry take a certain form \eqref{eq:gf_generator}, thus simplifying the classification of the choice of the $\doubleZ_N$ subgroup of the anyon permutation symmetries. For $G = S_3$, however, one cannot assume both generators of the $S_3$ take the form \eqref{eq:gf_generator}, as the gauge transformation generally can only make one of them be of that form. 

Taking this subtlety into account, we numerically search for $\beta$-invertible $S_3$ symmetries in the SymTFT for $N < 20$. We find that for any $N$ containing a factor of $2$ or $3$, there are no $\beta$-invertible $S_3$ symmetries in the SymTFT, thus there cannot be $S_3$-ality defects for these $N$. We are able to prove this analytically when $N$ is even, and we conjecture this is true generally for $N$ being a multiple of $3$ based on our numerical evidence. 

Assuming this is the case, then $H^2(S_3,\mathbf{A})$ is trivial for $\scriptZ(\VEC_{\doubleZ_N \times \doubleZ_N})$ admitting a $\beta$-invertible $S_3$ symmetry. Thus, the $\underline{\scriptE}_{S_3}\VEC_{\doubleZ_N\times \doubleZ_N}$ is classified by a choice of inequivalent $\beta$-invertible $S_3$ symmetry as well as the FS indicator $\alpha \in H^3(S_3,U(1))$. We work out the classification of the $\underline{\scriptE}_{S_3}\VEC_{\doubleZ_N\times \doubleZ_N}$ explicitly for $N < 20$, and summarize the result in Table \ref{tab:S3_class} and Appendix \ref{app:S3_class}.

The minimal $N$ where $\underline{\scriptE}_{S_3}\VEC_{\doubleZ_N\times \doubleZ_N}$ exists is $N = 5$, while the minimal $N$ where a group-theoretical $\underline{\scriptE}_{S_3}\VEC_{\doubleZ_N\times \doubleZ_N}$ exists is $N = 11$. For $N < 20$, none of the $\underline{\scriptE}_{S_3}\VEC_{\doubleZ_N\times \doubleZ_N}$ admits a self-$S_3$-ality $\doubleZ_N \times \doubleZ_N$ SPT phase, therefore they are all anomalous.

We provide a group-theoretical construction of the $S_3$-ality defects which works for $N$ coprime with $2,3,5,7$, which takes the form $\scriptC((\doubleZ_{N}^\ra\times \doubleZ_{N}^\rb)\rtimes S_3, \omega_k, \doubleZ_{N}^{\ra\rb^3},1)$, where $\omega_k \in H^3(S_3, U(1))$ determines the choice of the FS indicator and the action of the automorphism of $S_3$ on $\doubleZ_N^\ra \times \doubleZ_N^\rb$ is given in \eqref{eq:S3_ext}. This group-theoretical construction allows one to easily find concrete 2d CFT examples admitting $S_3$-ality defects. For instance, one can consider embedding $(\doubleZ_{N}^\ra\times \doubleZ_{N}^\rb)\rtimes S_3$ into $(U(1) \times U(1))\rtimes S_3$ and find the 2d CFT with such symmetry. We provide a concrete example constructed out of two compact bosons.

The paper is organized as follows. In Section \ref{sec:review}, after reviewing some basic knowledge of the topological defect lines in 2d QFT, we outline the generic method of using SymTFT to study $G$-ality defects in $\underline{\scriptE}_G\VEC_{\doubleA}$. In Section \ref{sec:triality}, we consider the warm-up example where $\doubleA = \doubleZ_N \times \doubleZ_N$ and $G = \doubleZ_3$ with $\gcd(N,3) = 1$. The classification of this case has been studied in \cite{jordan2009classification}, and we use this as a warm-up example to demonstrate our approach and we derive the anomaly of these triality defects. In Section \ref{Sec:p_ality}, we consider the case where $\doubleA = \doubleZ_p\times \doubleZ_p$ and $G = \doubleZ_p$ where $p$ is an odd prime. We provide two classes of the $p$-ality defects which both admit group-theoretical constructions, and show these two classes of $p$-ality defects differ by a choice of symmetry fractionalization classes in the SymTFT from the classification point of view. We also derive the anomaly structure and the topological sectors for these two fusion categories. In Section \ref{sec:S3_ality}, we explore the case where $\doubleA = \doubleZ_N \times \doubleZ_N$ and $G = S_3$. We compute the classification explicitly for $N < 20$ and provide a group-theoretical construction of $S_3$-ality defects which works for $N$ coprime with $2,3,5,7$.

\subsection*{Summary of notations}
\begin{enumerate}
    \item $\doubleA,\doubleH,\doubleK,\doubleG$ denote the finite group symmetry on the 2d boundary.
    \item $\scriptC$ denotes the fusion category symmetry on the 2d boundary, and $\scriptZ(\scriptC)$ denotes its Drinfeld center.
    \item $\scriptA$ denotes a symmetric $\Delta$-separable Frobenius algebraic object (that is, a gaugable algebraic object) in a fusion category $\scriptC$, and $\scriptM$'s denote $\scriptA$-bimodule objects in $\scriptC$ which correspond to the dual symmetry after gauging $\scriptA$.
    \item $\mathbf{A}$ denotes the group of Abelian anyons in the SymTFT $\scriptZ(\scriptC)$.
    \item $G$ denotes the grading group of the boundary fusion category; and in the extension $\underline{\scriptE}_G\scriptC$, $G$-ality defects are labeled by $\scriptN_g$ where $g \in G/\{\dsi\}$. In the SymTFT $\scriptZ(\scriptC)$, $G$ becomes an invertible symmetry given by the finite group $G$.
    \item $\scriptL_a$ denotes a generic TDL on the 2d boundary; in the SymTFT bulk, $\scriptL$ ($\scriptL_{\operatorname{e}}$, $\scriptL_{\operatorname{m}}$) is also used to denote (electric, magnetic) Lagrangian algebras. The reader should be able to distinguish the two based on context.
    \item When parameterizing group elements of some specific group such as $\doubleZ_N$, $S_3$ etc, we use $\ra,\rb,\rc,\rt,\rd,\cdots$ to denote group elements with multiplicative notation; and we use $a,b,c,i,j,k,\cdots$ to denote group elements with additive notation.
\end{enumerate}

\section{Review}\label{sec:review}
In this section, we first briefly review some defining properties of the topological defect lines (TDLs) in 2d CFT \cite{Chang:2018iay,Bhardwaj:2017xup} in Section \ref{sec:review_TDL}. Mathematically, they are described by a \textit{fusion category} $\scriptC$. Equipping $\scriptC$ with $G$-ality defects is described by a certain type of $G$-extension $\underline{\scriptE}_G\scriptC$ of the fusion category $\scriptC$, and the classification of the latter together with its relation to half-space gauging is briefly reviewed in Section \ref{sec:review_G_ality}. Next, we review some basic properties of the 3d SymTFTs, which describe $0$-form global symmetries to set up the notations in Section \ref{sec:symtft_review}. Finally, we review the result in \cite{Etingof:2009yvg,jordan2009classification} on how to use the Drinfeld center $\mathcal{Z}(\VEC_{\doubleA})$ to study $G$-ality defects from $\VEC_{\doubleA}$ in the language of SymTFTs in Section \ref{sec:G_ext_general}.

\subsection{Review of TDLs in CFT and group-theoretical fusion categories}\label{sec:review_TDL}
The non-invertible $0$-form symmetries in the 2d conformal field theories (CFTs) are generated by topological defect lines (TDLs) \cite{Fuchs:2002cm,Bhardwaj:2017xup,Chang:2018iay,Thorngren:2019iar,Thorngren:2021yso}. These TDLs are line operators commuting with the energy-momentum tensor. We denote a generic TDL as $\scriptL_a$ where $a$ labels the line type, and the trivial TDL is denoted as $\dsi$. Given two TDLs $\scriptL_a$ and $\scriptL_b$, we can fuse them via putting them on top of each other to generate a new TDL, which then in general decomposes into a \textit{finite} sum of other TDLs
\begin{equation}\label{eq:fus_rule}
    \scriptL_a \times \scriptL_b = \bigoplus_{c} N_{ab}^c \scriptL_c ~, \quad  N_{ab}^c \in \doubleZ_{\geq 0} ~.
\end{equation}
This leads to the notion of \textit{simple} TDLs, which are the lines that cannot be written as a sum of two TDLs. 

When $N_{ab}^c > 0$, two TDLs $\scriptL_a$ and $\scriptL_b$ can join each other locally and become the $\scriptL_c$ at a topological trivalent junction. The set of topological junctions with a given $\scriptL_a,\scriptL_b,\scriptL_c$ forms a complex vector space $V_{ab}^c$ with dimension $N_{ab}^c$. We then fix a basis of $V_{ab}^c$ and use the Greek letters $\mu,\nu,\cdots = 1,2, \cdots, N_{ab}^c$ to label the basis vectors:
\begin{equation}
    \begin{tikzpicture}[scale=0.8,baseline={([yshift=-.5ex]current bounding box.center)},vertex/.style={anchor=base,
    circle,fill=black!25,minimum size=18pt,inner sep=2pt},scale=0.50]
    \draw[thick, black] (-2,-2) -- (0,0);
    \draw[thick, black] (+2,-2) -- (0,0);
    \draw[thick, black] (0,0) -- (0,2);
    \draw[thick, black, -stealth] (-2,-2) -- (-1,-1);
    \draw[thick, black, -stealth] (+2,-2) -- (1,-1);
    \draw[thick, black, -stealth] (0,0) -- (0,1);
    \filldraw[thick, black] (0,0) circle (3pt);
    \node[black, right] at (0,0) {\footnotesize $\mu$};
    \node[black, below] at (-2,-2) {$\mathcal{L}_a$};
    \node[black, below] at (2,-2) {$\mathcal{L}_b$};
    \node[black, above] at (0,2) {$\mathcal{L}_c$};
    
\end{tikzpicture}, \quad \mu = 1,2,\cdots, N^{c}_{ab} \,.
\end{equation}
The associativity only holds up to isomorphism, which is characterized by a collection of $\doubleC$-numbers known as the $F$-symbols,
\begin{equation}
\begin{tikzpicture}[baseline={([yshift=-1ex]current bounding box.center)},vertex/.style={anchor=base,
    circle,fill=black!25,minimum size=18pt,inner sep=2pt},scale=0.7]
	\draw[thick, black, -<-=0.5] (0,0) -- (-0.75,-0.75);
	\draw[thick, black, -<-=0.5] (0,0) -- (+0.75,-0.75);
	\draw[thick, black, ->-=0.5] (0,0) -- (+0.75,+0.75);
	\draw[thick, black, ->-=0.5] (+0.75,+0.75) -- (1.5,1.5);
	\draw[thick, black, -<-=0.5] (0.75,0.75) -- (2.25,-0.75);
	
	\node[below, black] at (-0.75,-0.75) {\scriptsize $\mathcal{L}_a$};
	\node[below, black] at (+0.75,-0.75) {\scriptsize $\mathcal{L}_b$};
	\node[below, black] at (+2.25,-0.75) {\scriptsize $\mathcal{L}_c$};
	\node[right, black] at (0.25,0.2) {\scriptsize $\mathcal{L}_e$};
 	\node[left, black] at (0,0) {\scriptsize $\mu$};
    \node[left, black] at (0.75,0.75) {\scriptsize $\nu$};
	\node[above, black] at (1.5,1.5) {\scriptsize $\mathcal{L}_d$};
	\filldraw[black] (0,0) circle (2pt);
	\filldraw[black] (0.75,0.75) circle (2pt);
	
	\node[black] at (5,0.25) {$\displaystyle = \sum\limits_{f,\rho,\sigma} \left[F^{abc}_d\right]_{(e,\mu,\nu),(f,\rho,\sigma)}$};
	
	\draw[thick, black,->-=0.5] (9-0.75,-0.75) -- (9+0.75,0.75);
	\draw[thick, black,->-=0.5] (9+0.75,+0.75) -- (9+1.5,1.5);
        \draw[thick, black,->-=0.5] (9+0.75,-0.75) -- (9+1.5,0.);
        \draw[thick, black,->-=0.5] (9+2.25,-0.75) -- (9+1.5,0.);
        \draw[thick, black,->-=0.5] (9+1.5,0.) -- (9+0.75,+0.75);
	
	\node[below, black] at (9-0.75,-0.75) {\scriptsize $\mathcal{L}_a$};
	\node[below, black] at (9+0.75,-0.75) {\scriptsize $\mathcal{L}_b$};
	\node[below, black] at (9+2.25,-0.75) {\scriptsize $\mathcal{L}_c$};
	\node[right, black] at (9+1.1,0.5) {\scriptsize $\mathcal{L}_f$};
	\node[above, black] at (9+1.5,1.5) {\scriptsize $\mathcal{L}_d$};

    \node[left, black] at (9+1.5,0) {\scriptsize $\rho$};
	\node[left, black] at (9+0.75,+0.75) {\scriptsize $\sigma$};
	\filldraw[black] (9+0.75,+0.75) circle (1.5pt);
	\filldraw[black] (9+1.5,0) circle (1.5pt);	
\end{tikzpicture} ~.
\end{equation}
Furthermore, the associativity maps or the $F$-symbols should satisfy the ``pentagon'' equations for consistency. Loosely peaking, the fusion and associativity maps of these TDLs form the mathematical structure known as the fusion category. See \cite{shuheng2023review,runkel2023review,Bhardwaj:2017xup} for more detailed reviews on categorical symmetry in 2d theories. Notice that generically the simple objects and their fusion rules \eqref{eq:fus_rule} is not enough to uniquely specify a fusion category, as there could be multiple inequivalent solutions to the pentagon equations.

The action of a TDL $\scriptL_a$ on a local operator $\CO$ is obtained by encircling the local operator with a closed loop of TDL labeled by $\CL_a$,
\begin{equation}
\begin{tikzpicture}[baseline={([yshift=+.5ex]current bounding box.center)},vertex/.style={anchor=base,
    circle,fill=black!25,minimum size=18pt,inner sep=2pt},scale=0.75]
    \draw[red, ->-=0.5, line width = 0.3mm] (0,0) circle (1);
    \filldraw[black] (0,0) circle (1.5pt);
    \node[black, below] at (0,0) {$\mathcal{O}$};
    \node[red, right] at (+1,0) {$\mathcal{L}_a$};
    \draw[thick, ->-=1] (2,0) -- (4,0);
    \filldraw[black] (5,0) circle (1.5pt);
    \node[black, below] at (5,0) {$\mathcal{L}_a\cdot\mathcal{O}$};
    \end{tikzpicture} ~.
\end{equation}
We denote such an action of a topological line $\CL_a$ on a local operator $\mathcal{O}$ as $\CL_a \cdot \mathcal{O}$. A topological defect line $\mathcal{L}_a$ may end on a (non-local) operator $\mathcal{O}$, and under the state-operator map, such a non-local operator $\mathcal{O}$ is mapped to a state in the Hilbert space $\mathcal{H}_{\mathcal{L}_a}$ quantized on $S^1$ but with the boundary condition twisted by $\mathcal{L}_a$,
\begin{equation}
\begin{tikzpicture}[scale=0.7]
        \draw [thick] (0,0) ellipse (1.25 and 0.5);
        \draw [thick] (-1.25,0) -- (-1.25,-3.5);
        \draw [thick] (-1.25,-3.5) arc (180:360:1.25 and 0.5);
        \draw [thick, dashed] (-1.25,-3.5) arc (180:360:1.25 and -0.5);
        \draw [thick] (1.25,-3.5) -- (1.25,0);
        \fill [gray,opacity=0.25] (-1.25,0) -- (-1.25,-3.5) arc (180:360:1.25 and 0.5) -- (1.25,0) arc (0:180:1.25 and -0.5);
        \draw [ultra thick, -<-=0.5] (0,-0.5) -- (0, -3.5-0.5);
        \node[black, below] at (0,-4) {$\ket{\CO}\in \CH_{\CL_a}$};
        \node[black, right] at (0,-2) {$\CL_a$};
        \node[black] at (3,-2) {$\longleftrightarrow$};
        \draw [ultra thick, -<-=0.5] (4.5,-0.5) -- (4.5, -3.5);
        \filldraw[black] (4.5,-3.5) circle (2pt);
        \node[black, below] at (4.5,-3.5) {$\CO(x)$};
    \end{tikzpicture}
\end{equation}
$\mathcal{H}_{\mathcal{L}_a}$ is known as the defect (or twisted) Hilbert space. The (twisted) torus partition function with simple TDLs is defined as,
\begin{equation}
Z[\mathcal{L}_1,\mathcal{L}_2,\mathcal{L}_3;\mu,\nu](\tau) = \Tr_{\mathcal{H}_{\mathcal{L}_1}}\left(\widehat{\mathcal{L}_2}_{(\mathcal{L}_3,\mu,\nu)} q^{L_0 - c/24} \overline{q}^{\overline{L}_0 - c/24}\right) 
\end{equation}
which corresponds to the torus partition function with the network of topological defect lines inserted,
\begin{equation}
   Z[\mathcal{L}_1,\mathcal{L}_2,\mathcal{L}_3;\mu,\nu](\tau) = \begin{tikzpicture}[baseline={([yshift=+.5ex]current bounding box.center)},vertex/.style={anchor=base,
    circle,fill=black!25,minimum size=18pt,inner sep=2pt},scale=0.5]
    \filldraw[grey] (-2,-2) rectangle ++(4,4);
    \draw[thick, dgrey] (-2,-2) -- (-2,+2);
    \draw[thick, dgrey] (+2,+2) -- (+2,-2);
    \draw[thick, dgrey] (-2,-2) -- (+2,-2);
    \draw[thick, dgrey] (+2,+2) -- (-2,+2);
    \draw[thick, black, -stealth] (0,-2) -- (0.354,-1.354);
    \draw[thick, black] (0,-2) -- (0.707,-0.707);
    \draw[thick, black, -stealth] (2,0) -- (1.354,-0.354);
    \draw[thick, black] (2,0) -- (0.707,-0.707);
    \draw[thick, black, -stealth] (-0.707,0.707) -- (-0.354,1.354);
    \draw[thick, black] (0,2) -- (-0.707,0.707);
    \draw[thick, black, -stealth] (-0.707,0.707) -- (-1.354,0.354);
    \draw[thick, black] (-2,0) -- (-0.707,0.707);
    \draw[thick, black, -stealth] (0.707,-0.707) -- (0,0);
    \draw[thick, black] (0.707,-0.707) -- (-0.707,0.707);
    \filldraw[black] (0.707,-0.707) circle (2pt);
    \filldraw[black] (-0.707,0.707) circle (2pt);
    \node[black, below] at (0,-2) {\scriptsize $\mathcal{L}_1$};
    \node[black, right] at (2,0) {\scriptsize $\mathcal{L}_2$};
    \node[black, above] at (0.2,0) {\scriptsize $\mathcal{L}_3$};
    \node[black, below] at (0.9,-0.607) {\scriptsize $\mu$};
    \node[black, below] at (-0.707,0.707) {\scriptsize $\nu$};
\end{tikzpicture} ~.
\end{equation}
For instance, the twisted partition function
\begin{equation}
    Z[\scriptL_1,\dsi,\scriptL_1] = \Tr_{\scriptH_{\scriptL_1}}q^{L_0 - c/24} \overline{q}^{\overline{L}_0 - c/24}
\end{equation}
captures the trace over defect Hilbert space $\scriptH_{\scriptL_1}$. The twisted torus partition function is covariant under the modular transformations. For example, as shown in \cite{Thorngren:2021yso}, under the modular $S$-transformation, the twisted torus partition function becomes
\begin{equation}    
    Z[\mathcal{L}_1,\mathcal{L}_2,\mathcal{L}_3]\left(-\frac{1}{\tau}\right) = \sum_{\mathcal{L}_k} \left[F^{\mathcal{L}_1,\mathcal{L}_2,\overline{\mathcal{L}_1}}_{\mathcal{L}_2}\right]_{\mathcal{L}_3 \mathcal{L}_k} Z[\mathcal{L}_2,\overline{\mathcal{L}_1},\mathcal{L}_k](\tau) ~.
\end{equation}
States in a given defect Hilbert space $\scriptH_{\scriptL_a}$ will organize into certain irreducible representations of the symmetries commuting with $\scriptL_a$. The subspace $\scriptH_{\scriptL_a}^\mu$ of $\scriptH_{\scriptL_a}$ containing all the states transforming under a particular irrep $\mu$ is called a topological sector. Notice that states in the same topological sector $\scriptH_{\scriptL_a}^\mu$ have the same spin mod $1$.

\,

Let's now consider some simple examples of the fusion category symmetries. For ordinary symmetry $\mathbb{G}$, the corresponding simple TDLs are denoted as $\CL_g, g\in \mathbb{G}$. The juxtaposition of two such TDLs satisfies the group multiplication rule, $\CL_g \times \CL_h = \CL_{gh}$. The TDL corresponding to the identity $\dsi$ in the group $\mathbb{G}$ is the identity line which we will also denote as $\dsi \equiv \scriptL_\dsi$. Notice that all the simple TDLs $\scriptL_g$'s are \textit{invertible}, in the sense that there exists $\scriptL_{g^{-1}}$ such that $\scriptL_g \times \scriptL_{g^{-1}} = \dsi$. The equivalence class of the $F$-symbols satisfying the pentagon equations is not unique, and is classified by the group cohomology classes $[\omega]$ in $H^3(\doubleG,U(1))$ via $F^{g_1, g_2, g_3}_{g_1 g_2 g_3} = \omega(g_1, g_2, g_3)$. Physically, $[\omega]$ is known as the 't Hooft anomaly of the symmetry $\doubleG$. We denote the fusion category with a given $\doubleG$ and $[\omega]\in H^3(\doubleG,U(1))$ as $\VEC_{\doubleG}^{\omega}$.

Another class of interesting fusion category is the Tambara-Yamagami category which involves a self-duality line $\scriptN$ of Abelian group symmetry $\doubleA$ with no 't Hooft anomaly \cite{tambara1998tensor,tambara2000representations,Thorngren:2019iar,Thorngren:2021yso}. The fusion rules involve the duality line $\scriptN$ are
\begin{equation}
    a \times \scriptN = \scriptN \times a ~, \quad \forall a \in \doubleA ~; \quad  \scriptN \times \scriptN = \bigoplus_{a\in \doubleA} a ~.
\end{equation}
From the above fusion rules, one notices that the line $\scriptN$ is  \textit{non-invertible} as it does not admit an inverse. Similar to $\VEC_\doubleG^\omega$, the above fusion rules do not uniquely determine a fusion category, and the equivalent classes of the $F$-symbol solutions are classified by a symmetric non-degenerate bicharacter of $\doubleA$, $\chi:\doubleA \times \doubleA \rightarrow U(1)$ and the Frobenius-Schur indicator $\epsilon\in H^3(\IZ_2 ,U(1))$. Physically, if a theory is self-dual under gauging the Abelian group $\doubleA$, then the self-duality line $\scriptN$ maps the original theory to its gauged version. The choice of $\chi$ specifies how to identify the original symmetry with the dual symmetry, and $\epsilon$ relates to the self-anomaly of the duality line \cite{tambara2000representations,Thorngren:2019iar,Seiberg:2024gek}. 

The \textit{group-theoretical fusion categories} are a subclass of fusion categories with integer Frobenius-Perron dimensions, they
are constructed explicitly from finite group symmetries $\VEC_{\doubleG}^\omega$. They include many physically relevant fusion categories. For symmetry $\doubleG$ with anomaly $[\omega] \in H^3(\doubleG,U(1))$. It is possible to gauge the anomaly-free subgroup $\doubleH$, with $[\omega]\Big|_{\doubleH} = 1 \in H^3(\doubleH, U(1))$. The gauging procedure requires a choice of discrete torsion $[\psi] \in H^2(G,U(1))$. The resulting fusion category is denoted as $\scriptC(\doubleG,\omega,\doubleH,\psi)$ and called a group-theoretical fusion category. For example, if $\doubleH=\{\dsi\}$, $\scriptC(\doubleG,\omega,\doubleZ_1,1)=\VEC_{\doubleG}^\omega$. Although admitting a simple construction, group-theoretical fusion categories may still contain interesting non-invertible symmetries. For instance, gauging the $\doubleZ_2$ symmetry of $S_3$ with $\omega = 1$ leads to the fusion category $\Rep(S_3)$ \cite{Bhardwaj:2017xup} (gauging total $S_3$ also gives $\Rep(S_3)$) and gauging the $\doubleZ_2$ symmetry of $A_4$ with $\omega = 1$ leads to the triality fusion category \cite{Thorngren:2019iar,Thorngren:2021yso}. The $\TY(\doubleA,\chi,\epsilon)$ can be group-theoretical if and only if $\doubleA$ contains a Lagrangian subgroup with respect to $\chi$. For $\doubleA=\IZ_N$, $\TY(\doubleA,\chi,\epsilon)$ is group-theoretical when $N$ is a perfect square \cite{Gelaki:2009blp,Sun:2023xxv}. As another example, $\TY(\IZ_2 \times \IZ_2,\chi_\text{off-diag},+) = \Rep(D_8)$ can be obtained by gauging $D_8$.

\,

Like the ordinary global symmetries, the non-invertible symmetry could have anomaly. Unlike the 't Hooft anomaly which is an obstruction to gauging as well as a symmetric gapped phase with a unique ground state, the anomaly of non-invertible symmetry generally only prohibits a symmetric gapped phase with a unique ground state\footnote{Interestingly, for anomalous fusion categories, there can still be notion of ``gauging'' non-invertible symmetries, see \cite{Choi:2023xjw} for more details.}. For example, the Ising fusion category $\TY(\IZ_2,+)$ is anomalous and obstructs a trivially symmetric gapped phase. With the $\TY(\IZ_2,+)$ symmetric deformation, the Ising CFT is driven to a first-order transition with 3 ground state degeneracies \cite{Seiberg:2024gek}. On the other hand, the $\TY(\IZ_2\times \IZ_2,\chi_\text{off-diag},+)=\Rep(D_8)$ is anomaly-free and admits a symmetric gapped phase with a unique ground state \cite{Sahand2024cluster,Warman:2024lir}. 

For the group-theoretical fusion category $\scriptC(\mathbb{G},\omega;\mathbb{H},\psi)$, its symmetric gapped phases are easier to understand \cite{ostrik2002module}. Any symmetric gapped phases can be acquired by starting with a symmetric gapped phase of $\VEC_\mathbb{G}^\omega$ and gauging the subgroup $\mathbb{H}$ with discrete torsion $\psi$. The symmetric gapped phases of $\VEC_{\mathbb{G}}^\omega$ are labeled by $\scriptM_{(\mathbb{K},[\psi_{\mathbb{K}}])}$, where $\mathbb{K}$ denotes the unbroken subgroup of $\mathbb{G}$ (therefore must be anomaly-free, i.e., $[\omega]\Big|_{\mathbb{K}\times \mathbb{K}\times \mathbb{K}} = 1$) and the ground states (vacua) are labeled by $\mathbb{K}$-coset. On each vacuum, $\mathbb{K}$ realizes the SPT labeled by $[\psi_{\mathbb{K}}]$. Gauging $\mathbb{H}$ with discrete torsion $\psi$ maps this phase to a symmetric gapped phase of $\scriptC(\mathbb{G},\omega;\mathbb{H},\psi)$, and the number of ground states will change and are labeled by the pairs $(\doubleH g \doubleK, \pi_g)$, where $\doubleH g \doubleK$ is a double coset in $\doubleH \textbackslash \doubleG / \doubleK$ with representative $g$, and $\pi_g$ is an irreducible representation of the little group $\doubleH_g = \{(h,k) \in \doubleH \times \doubleK: hgk = g\}$ twisted by certain 2-cocycle $\psi_g$ (see \cite{ostrik2002module} for concrete formula, which we will not use explicitly here).

Requiring there is a unique $(\doubleH g \doubleK, \pi_g)$, one gets the classification of the trivially symmetric gapped phase (mathematically known as fiber functors) of $\scriptC(\doubleG,\omega;\doubleH,\psi)$. First, the uniqueness of $\doubleH g \doubleK$ holds if and only if $\doubleH \doubleK$ contains $\doubleG$. When this does hold, consider the little group $\doubleH_{\dsi} = \doubleH \cap \doubleK$ of the identity $\dsi \in \doubleG$. The irreducible representation $\pi_{\dsi}$ is unique if and only if the 2-cocycle $\psi_\dsi = \frac{\psi|_{\doubleH_{\dsi}}}{\psi_\doubleK|_{\doubleH_{\dsi}}} \in Z^2(\doubleH_{\dsi},U(1))$ is non-degenerate. Therefore, we conclude that the trivially symmetric gapped phases of $\scriptC(\doubleG,\omega;\doubleH,\psi)$ are classified by pairs $(\doubleK, [\psi_{\doubleK}])$ where
\begin{equation}\label{eq:anomaly_free_criteria}
\begin{aligned}
    & \text{1. $\doubleK$ is an anomaly-free subgroup of $\doubleG$ as $\omega|_{\doubleK} = d\psi_{\doubleK}$} ~, \quad \quad \quad \quad \quad \quad \quad \quad \quad \quad \quad \quad \\
    & \text{2. $\doubleH \doubleK$ contains $\doubleG$} ~, \\
    & \text{3. The 2-cocycle $\psi_\dsi = \frac{\psi|_{\doubleH_{\dsi}}}{\psi_\doubleK|_{\doubleH_{\dsi}}}$ is non-degenerate} ~,
\end{aligned}
\end{equation}
The existence of $(\doubleK,[\psi_\doubleK])$ satisfying the above three conditions can be viewed as a criterion for the group-theoretical fusion category $\scriptC(\doubleG,\omega;\doubleH,\psi)$ to be anomaly-free.

For example, the 3 fiber functors of $\Rep(D_8) = \CC(D_8,1,D_8,1)$ are given by $(\langle 1\rangle,1)$, $(\langle s,r^2\rangle,\xi)$ and $(\langle sr,r^2\rangle,\xi)$, where $r,s$ are generators of $D_8 = \langle r,s|r^4 = s^2 = 1, sr = r^{-1}s \rangle $ and $\xi$ is the non-trivial element of $H^2(K,U(1))$ \cite{Diatlyk:2023fwf}. 

Another feature regarding the group-theoretical fusion categories is that their topological sectors can be conveniently computed. This is because discrete gauging does not change each topological sector, but merely organizes them into defect Hilbert spaces differently. Hence, one only needs to compute the topological sectors for $\VEC_G^\omega$, and keep track of how they recombine under discrete gauging. From the bulk point of view, the topological sectors are in 1-to-1 correspondence with the simple anyons in its SymTFT $\scriptZ(\scriptC)$. Thus, the same calculation will also compute the anyon spectrum in the 3d Dijkgraaf-Witten theory with gauge group $\doubleG$ and the twist $\omega$\cite{Coste:2000tq,Hu:2012wx}. We will make use of this later, and we include a summary of the approach in Appendix \ref{app:DW_theory}.

\subsection{Graded fusion categories and $G$-ality defects}\label{sec:review_G_ality}
Graded fusion categories are a generalization of the well-known Tambara-Yamagami fusion categories. Let $G$ be a finite group, then a fusion category $\scriptC$ is $G$-graded if it admits a direct sum decomposition
\begin{equation}
    \scriptC = \bigoplus_{g\in G} \scriptC_g
\end{equation}
such that the tensor product $\otimes$ maps $\scriptC_g \times \scriptC_h$ into $\scriptC_{gh}$ for all $g,h \in G$. In particular, the trivial component $\scriptC_\dsi$ is a tensor subcategory of $\scriptC$ and each $\scriptC_g$ is an invertible $\scriptC_\dsi$-bimodule category. For convenience, we will assume the grading is faithful, which means $\scriptC_g$ is non-empty for all $g \in G$. On the other hand, given the fusion category $\scriptD$, any $G$-graded fusion category $\scriptC$ such that $\scriptC_{\dsi} = \scriptD$ is called a $G$-extension of $\scriptD$.

In this description, TY fusion category is a $\doubleZ_2$-graded fusion category, where $\scriptC_\dsi = \VEC_{\doubleA}$ and $\scriptC_\eta = \{\scriptN\}$ where $\{\dsi, \eta\} \in \doubleZ_2$. Alternatively, we may say that the TY fusion category is a $\doubleZ_2$-extension of the fusion category of the invertible Abelian symmetry $\VEC_{\doubleA}$. Notice that the graded fusion category does not necessarily contain non-invertible symmetries. For instance, we could consider the pointed fusion category $\VEC_{S_3}$ where we take $S_3 = \langle \ra, \rb| \ra^3 = \rb^2 = \dsi, \rb\ra\rb = \ra^2\rangle$. There's a $\doubleZ_2$-grading given by
\begin{equation}
    \scriptC_\dsi = \{\dsi,\ra,\ra^2\} ~, \quad \scriptC_\eta = \{\rb,\ra\rb,\ra^2\rb\} ~.
\end{equation}

Physically, given a fusion category $\scriptC$, we are interested in whether the actual symmetry in a given theory can be larger than $\scriptC$. The $G$-extension of a fusion category $\scriptC$ exactly characterizes the invariance under the additional invertible topological manipulations (which form a group $G$). Notice that here by invertible topological manipulations, we mean auto-equivalences of $\scriptC$\footnote{When $\scriptC = \VEC_{\doubleG}$, an auto-equivalence of $\scriptC$ is given by some combination of the outer automorphism of $\doubleG$ as well as stacking $\doubleG$-SPT.} (which lead to invertible symmetries in the corresponding grading component) as well as discrete gaugings (which lead to non-invertible symmetries in the corresponding grading component). 

The classification of $G$-extension of a given fusion category $\scriptC$ has been worked out in math literature\cite{Etingof:2009yvg} and is related to the classification of symmetry-enriched topological field theory in 2+1d\cite{Barkeshli:2014cna}. The boundary interpretation of the classification result is reviewed in \cite{yichul2024selfdual} and we will not repeat it in detail here. To summarize, the $G$-extension fusion category of a fusion category $\scriptC$ is parameterized by triples $(\rho,[\nu],\alpha)$ up to autoequivalence of $\scriptC$. We will denote such $G$-extension as
\begin{equation}
    \scriptE^{(\rho,[\nu],\alpha)}_{G}\scriptC ~.
\end{equation}
Notice that sometimes when the explicit classification data is unknown, or we simply want to refer to a generic $G$-extension of $\scriptC$, we will suppress the superscript $(\rho,[\nu],\alpha)$.

The first ingredient $\rho: G \rightarrow \operatorname{BrPic}(\scriptC)$ is an injective group homomorphism from $G$ to the so-called Brauer-Picard group $\operatorname{BrPic}(\scriptC)$ whose elements are invertible $\scriptC$-bimodule categories and the group multiplication is the balanced Deligne tensor product of $\scriptC$-bimodule categories. The identity of this multiplication in $\operatorname{BrPic}(\scriptC)$ is $\scriptC$ itself. The choice of $\rho$ determines the simple objects in each non-trivial grading component $(\scriptE_G\scriptC)_g$, as well as the associativity maps ($F$-symbols) describing their fusion with the elements in $\scriptC$. Then, any two non-trivial grading components $(\scriptE_G\scriptC)_g$ and $(\scriptE_G\scriptC)_h$, when viewed as $\scriptC$-bimodule, fuse into another $\scriptC$-bimodule $(\scriptE_G\scriptC)_{gh}$. The second ingredient $[\nu]$ characterizes such fusion by specifying a specific way of identifying $(\scriptE_G\scriptC)_g\boxtimes_{\CC} (\scriptE_G\scriptC)_h$ with $(\scriptE_G\scriptC)_{gh}$ as $\scriptC$-bimodule. Finally, the last data $\alpha \in H^3(G,U(1))$ labels the freedom of attaching an anomalous invertible $G$-symmetry defect $\scriptL_g$ to every simple line in the corresponding grading component $(\scriptE_G \scriptC)_g$. Notice that there are certain obstruction classes $\CO^3_{[\rho]}$ and $\CO^4(\rho,[\nu]) \in H^4(G,U(1))$ which must vanish in order for the corresponding $F$-symbols to satisfy the pentagon equations. Notice that all the data mentioned here has a concrete physical interpretation in terms of the SymTFT, which we will discuss in detail in Section \ref{sec:G_ext_general}: the connection follows from that the Brauer-Picard group $\operatorname{BrPic}(\scriptC)$ is canonically isomorphic to the symmetry group $\EqBr(\mathcal{Z}(\mathcal{C}))$ of the SymTFT $\mathcal{Z}(\mathcal{C})$ mentioned in the introduction \cite{Etingof:2009yvg,2013arXiv1309.5026N}; and the classes $[\nu],\mathcal{O}_{[\rho]}^3$ mentioned above take value in $H^2_\rho(G,\mathbf{A})$ and $H^3_\rho(G,\mathbf{A})$ respectively where $\mathbf{A}$ is the group of Abelian anyons in the SymTFT.

In this paper, we are interested in the case where $\scriptC_g$ contains a single simple object $\CN_g$ for each $g \neq \dsi$ and $\scriptC_\dsi = \VEC_{\doubleA}$ for some Abelian group $\doubleA$ and introduce the notation
\begin{equation}\label{eq:G_ality_ext}
    \underline{\CE}_G^{(\rho,[\nu],\epsilon)} \VEC_{\doubleA} \quad \text{or} \quad \underline{\CE}_G \VEC_{\doubleA}
\end{equation}
to denote this particular type of $G$-extension of $\VEC_{\doubleA}$. Following from the definition that $\otimes$ takes $\scriptC_g \times \scriptC_h$ into $\scriptC_{gh}$, one can derive the following fusion rules \cite{jordan2009classification}
\begin{equation}\label{eq:fusion_Ng_a}
    \CN_g \otimes a = a \otimes \CN_g = \CN_g ~,
\end{equation}
which then implies
\begin{equation}
    \CN_g \otimes \CN_{g^{-1}} = \bigoplus_{a\in \doubleA}a, \quad \CN_g \otimes \CN_h = \sqrt{|A|} \CN_{gh}, \quad g \neq h^{-1} ~.
\end{equation}
We will name the defects $\CN_g$'s as \textit{$G$-ality defects}, generalizing the duality defect in Tambara-Yamagami fusion categories where $G = \doubleZ_2$. Physically, the fusion rules \eqref{eq:fusion_Ng_a} imply that the $\doubleA$-symmetry defects become transparent when moving across the non-invertible symmetries $\mathcal{N}_g$. As a result, $\doubleA$ is effectively gauged in some way as we move across $\mathcal{N}_g$, and the existence of the defect $\mathcal{N}_g$ implies that the theory is invariant under a certain way of gauging $\doubleA$. Notice that here by a certain way of gauging $\doubleA$, we mean a generic topological manipulation which is a composition of stacking $\doubleA$-SPT, gauging $\doubleA$, applying an automorphism of the dual symmetry $\widehat{\doubleA}$ and then stacking $\widehat{\doubleA}$-SPT. We will refer to this as twisted gauging $\doubleA$. Conversely, if the theory is invariant under some twisted gauging of $\doubleA$, then we can consider implementing this topological manipulation on half-space to generate a topological defect which fuses with TDLs in $\doubleA$ like \eqref{eq:fusion_Ng_a}, for more details, see \cite{yichul2024selfdual}. Notice that a generic twisted gauging is not of order-$2$, thus the corresponding defect is not necessarily orientation invariant like the duality line in the $\TY$ fusion categories.

\subsection{Review of the SymTFT}\label{sec:symtft_review}
Symmetries of quantum field theories can be characterized by one higher dimensional bulk with topological quantum field theory, the bulk is dubbed as the symmetry topological field theory or SymTFT in short. The SymTFT was initialized in \cite{Freed:2012bs}, and is subsequently developed in condensed matter and high energy literature \cite{symtft2019XGW,symtft2020XGW2,symtft2021XGW3,symtft2023XGW4,symtft2022XGW5,symtft2021Gaiotto,symtft2021Sakura,moradi2023topoholo,symtft2022Apruzzi,Freed:2022qnc,Kaidi:2022cpf,symtft2022Kulp,symtft2023kaidi2}.

\begin{figure}[!tbp]
	\centering
	\begin{tikzpicture}[scale=0.8]
	\shade[line width=2pt, top color=blue!30, bottom color=blue!5] 
	(0,0) to [out=90, in=-90]  (0,3)
	to [out=0,in=180] (6,3)
	to [out = -90, in =90] (6,0)
	to [out=180, in =0]  (0,0);
	
	\draw[very thick, blue] (-7,0) -- (-7,3);
	\node[below] at (-7,0) {$\scriptZ_{\scriptX}[A]$};
	\draw[thick, snake it, <->] (-1.7,1.5) -- (-5, 1.5);
	
	\draw[thick] (0,0) -- (0,3);
	\draw[thick] (6,0) -- (6,3);
	\node at (2.5,1.5) {SymTFT};
	\node[below] at (0,0) {$\langle \scriptX|$};
	\node[below] at (6,0) {$|D(A)\rangle $}; 
	
	\end{tikzpicture}
    \caption{Schematic picture of the SymTFT.}
    \label{fig:symTFT}
\end{figure}

More concretely, the theory $\scriptX$ in $d$ spacetime dimension with higher fusion category $\scriptC$ can be ``blown up'' into a slab in $d+1$ spacetime dimension with the SymTFT living inside it as shown in Figure \ref{fig:symTFT}. When $d=2$, as considered in this paper, the SymTFT is given by the Turaev-Viro theory of $\scriptC$ whose topological line operators are given by the Drinfeld center of $\scriptC$, i.e. $\scriptZ(\scriptC)$. The right boundary of SymTFT has the topological Dirichlet boundary condition where the symmetry $\scriptC$ is realized, while the left boundary is the physical boundary with the degrees of freedom in the theory $\scriptX$ and thus is generically non-topological. The original theory $\scriptX$ can be reproduced by shrinking the slab. Higher categorical generalization of the above construction is needed when $d>2$.

One advantage of this construction is that theory-independent consequences of symmetries and anomalies are separated from the specific choice of QFTs or lattice Hamiltonians. Topological manipulations of the theory $\scriptX$, such as gauging the global symmetry, can be obtained by simply changing the topological boundary condition and the anomaly is reflected in the absence of certain topological boundary conditions.
For example, let us consider a theory $\scriptX$ with non-anomalous symmetry $G$, and the SymTFT is Dijkgraaf-Witten theory, with the right topological boundary condition being the Dirichlet boundary condition. If we change the right topological boundary condition to a Neumann boundary condition, then the global symmetry $G$ is gauged \cite{Kaidi:2022cpf,symtft2023kaidi2}.

To be more specific, consider the 2d theory $\scriptX$ with anomaly-free $\doubleZ_N$ 0-form symmetry, the SymTFT is given by the $\doubleZ_N$ gauge theory in 3d,
\begin{equation}
    S=\frac{2\pi}{N} \int_{\Sigma_2  \times I_{[0,1]}} a \cup \delta \ha ~,
\end{equation}
where $a,\ha\in H^1(\Sigma_2  \times I_{[0,1]},\doubleZ_N)$ are the dynamical $\doubleZ_N$ valued 1-cochains, and the $\delta$ is the coboundary operator. The base manifold $\Sigma_2  \times I_{[0,1]}$ is the slab in 3d as in Figure \ref{fig:A_gauge_theory}, and the interval $I_{[0,1]}$ has the coordinate $x$. The simple boundary condition on each boundary is represented as a state in the Hilbert space of the SymTFT quantized on $\Sigma_2$, and the latter is spanned by $|A\rangle$ where $A \in H^1(\Sigma_2, \mathbb{Z}_N)$. The left boundary $x=0$ is the physical boundary specified by the boundary state
\begin{equation}
    \bra{\scriptX} = \sum_{a\in H^1(\Sigma_2 |_0,\doubleZ_N)} \bra{a} Z_\scriptX[a] ~,
\end{equation}
where $Z_{\scriptX}[a]$ is the partition function of $\scriptX$ coupled to the flat $\doubleZ_N$ gauge field $a$ on $\Sigma_2$, while the right boundary is the topological boundary,
\begin{align}
    & \ket{D(A)} = \sum_{a\in H^1(\Sigma_2|_1,\doubleZ_N)} \delta(a-A)\ket{a} = \ket{A} ~, \\
    &\ket{N(A)} =\frac{1}{\abs{H^0(\Sigma_2|_1,\doubleZ_N)}} \sum_{a\in H^1(\Sigma_2|_1,\doubleZ_N)} e^{\frac{2\pi \ii}{N} \int_{\Sigma_2|_1} a\cup A}\ket{a} ~,
\end{align}
where $\ket{D(A)}$ ($\ket{(N(A)}$) is the Dirichlet (Neumann) boundary condition. We always choose the left boundary to be the physical boundary and the right boundary to be the topological boundary, and omit the coordinate in the following expressions.
\begin{figure}[!tbp]
	\centering
	\begin{tikzpicture}[scale=0.8]
	\shade[line width=2pt, top color=blue!30, bottom color=blue!5] 
	(0,0) to [out=90, in=-90]  (0,3)
	to [out=0,in=180] (6,3)
	to [out = -90, in =90] (6,0)
	to [out=180, in =0]  (0,0);
	
	\draw[very thick, blue] (-7,0) -- (-7,3);
	\node[below] at (-7,0) {$\scriptZ_{\scriptX}[A]$};
	\draw[thick, snake it, <->] (-1.7,1.5) -- (-5, 1.5);
	
	\draw[thick] (0,0) -- (0,3);
	\draw[thick] (6,0) -- (6,3);
	\node at (2.5,1.5) {$\doubleZ_N$ gauge theory};
	\node[below] at (0,0) {$\langle \scriptX|$};
	\node[below] at (6,0) {$|D(A)\rangle $}; 
	
	\end{tikzpicture}
 
 	\begin{tikzpicture}[scale=0.8]
	\shade[line width=2pt, top color=blue!30, bottom color=blue!5] 
	(0,0) to [out=90, in=-90]  (0,3)
	to [out=0,in=180] (6,3)
	to [out = -90, in =90] (6,0)
	to [out=180, in =0]  (0,0);
	
	\draw[very thick, blue] (-7,0) -- (-7,3);
	\node[below] at (-7,0) {$\scriptZ_{\scriptX/\doubleZ_N}[A]$};
	\draw[thick, snake it, <->] (-1.7,1.5) -- (-5, 1.5);
	
	\draw[thick] (0,0) -- (0,3);
	\draw[thick] (6,0) -- (6,3);
	\node at (2.5,1.5) {$\doubleZ_N$ gauge theory};
	\node[below] at (0,0) {$\langle \scriptX|$};
	\node[below] at (6,0) {$|N(A)\rangle $}; 
	
	\end{tikzpicture}
    \caption{SymTFT for $\doubleZ_N$ 0-form symmetry.}
    \label{fig:A_gauge_theory}
\end{figure}

One can check, when shrinking the slab with the Dirichlet boundary condition, we get,
\begin{equation}
    \bra{\scriptX} \ket{D(A)} = \sum_{a,a'\in H^1(\Sigma_2,\doubleZ_N)} Z_\scriptX [\Sigma_2,a] \delta(a'-A)\bra{a}\ket{a'} =Z_\scriptX[\Sigma_2,A] ~,
\end{equation}
where $\bra{a} \ket{a'}=\delta(a-a')$. Similarly, when choosing the Neumann boundary condition, the partition function becomes,
\begin{equation}
    \bra{\scriptX}\ket{N(A)} = \frac{1}{\abs{H^0(\Sigma_2,\doubleZ_N)}} \sum_{a\in H^1(\Sigma_2,\doubleZ_N)} e^{\frac{2\pi \ii}{N} \int_{\Sigma_2 }a\cup A}Z_\scriptX[\Sigma_2,a] = Z_{\scriptX/\doubleZ_N}[\Sigma_2,A] ~.
\end{equation}
Therefore, the Neumann boundary condition corresponds to gauging the global symmetry $\doubleZ_N$ of the original theory $\scriptX$.

General topological manipulations of the physical theory $\scriptX$ are in one-to-one correspondence with the topological boundary conditions of the SymTFT. Given this, consider an invertible $0$-form symmetry operator $U$ in the bulk. It will generically transform the Dirichlet boundary to some other gapped boundaries, thus implementing some invertible topological manipulation on the boundary theory $\scriptX$. More concretely, the symmetry TFT allows us to express the partition function $\scriptZ_{\scriptX}[A]$ as
\begin{equation}
    \scriptZ_{\scriptX}[A] = \langle \scriptX | A \rangle ~,
\end{equation}
and the topological manipulation implemented by $U$ on $\scriptZ_{\scriptX}[A]$ is given by
\begin{equation}
    U(\scriptZ_{\scriptX}[A]) = \langle \scriptX | U |A\rangle ~,
\end{equation}
and is depicted in Figure \ref{fig:bdy_bulk_anyonp}. To perform a concrete calculation, one can make use of the result shown in \cite{shuheng:conddefect,Fuchs:2012dt} that one can always construct anyon permutation symmetry $U$ as a condensation defect, and write $U$ in terms of a mesh of anyon operators. We will demonstrate this in detail in Section \ref{sec:trans_part}.

\begin{figure}[!tbp]
	\centering
	\begin{tikzpicture}[scale=0.8]
	\shade[line width=2pt, top color=blue!30, bottom color=blue!5] 
	(0,0) to [out=90, in=-90]  (0,3)
	to [out=0,in=180] (6,3)
	to [out = -90, in =90] (6,0)
	to [out=180, in =0]  (0,0);
	
	\draw[very thick, red] (-7,0) -- (-7,3);
	\node[below] at (-7,0) {$U(\scriptZ_{\scriptX}[A]) \equiv \langle \scriptX| U |A\rangle$};
	\draw[thick, snake it, <->] (-1.7,1.5) -- (-5, 1.5);
	
	\draw[thick] (0,0) -- (0,3);
	\draw[thick] (6,0) -- (6,3);
	\node at (2.5,1.5) {$\doubleA$-gauge theory};
	\node[below] at (0,0) {$\langle \scriptX|$};
	\node[below] at (6,0) {$|A\rangle $}; 
    \draw[very thick, red] (5,0) -- (5,3);
    \node[below, red] at (5,0) {$U$};
	
	\end{tikzpicture}
    \caption{The action of the anyon permutation symmetry on $\scriptZ_{\scriptX}[A]$.}
    \label{fig:bdy_bulk_anyonp}
\end{figure}

In general, the topological surface operator admits a boundary, that is, it can end on some topological line operator. The surface operator together with its boundary is known as the twist defect. Consider the configuration as depicted in Figure \ref{fig:bdy_bulk_tdefect}, where a twist defect $\scriptS$ is placed along the boundary direction. Upon shrinking the slab, the twist defect then becomes an interface between the theory $\scriptX$ coupled to the background field $A$ and the theory $\scriptS(\scriptX)$ coupled to some background field $\scriptS(A)$.  

\begin{figure}[H]
	\centering
	\begin{tikzpicture}[scale=0.8]
	\shade[line width=2pt, top color=blue!30, bottom color=blue!5] 
	(0,0) to [out=90, in=-90]  (0,3)
	to [out=0,in=180] (6,3)
	to [out = -90, in =90] (6,0)
	to [out=180, in =0]  (0,0);
	
	\draw[very thick, red] (-7,0) -- (-7,1.5);
    \draw[very thick, blue] (-7,1.5) -- (-7,3);
	\node[below] at (-7,0) {$\scriptS(\scriptZ_{\scriptX}[A]) \equiv \scriptZ_{\scriptS(\scriptX)}[\scriptS(A)]$};
    \node[above] at (-7,3) {$\scriptZ_{\scriptX}[A]$};
    \node[left] at (-7,1.5) {\color{red} $\scriptL_\scriptS$};
    \filldraw [red] (-7,1.5) circle (2pt);
	\draw[thick, snake it, <->] (-1.7,1.5) -- (-5, 1.5);
	
	\draw[thick] (0,0) -- (0,3);
	\draw[thick] (6,0) -- (6,3);
	\node at (3,2.5) {$\doubleA$-gauge theory};
    \node[right,red] at (3,1.5) {$\scriptS$};
	\node[below] at (0,0) {$\langle \scriptX|$};
	\node[below] at (6,0) {$|A\rangle $}; 

    \filldraw[red] (3,1.5) circle (2pt);
    \draw[very thick, red] (3,1.5) -- (3,0);
	
	\end{tikzpicture}
    \caption{Twist defects and the boundary topological interfaces. }
    \label{fig:bdy_bulk_tdefect}
\end{figure}

If it turns out that the theory $\scriptX$ is the same as the theory $\scriptS(\scriptX)$, then the topological interface $\scriptS$ becomes a topological defect for the theory $\scriptX$. It is then natural to ask if the $\scriptS$ is invertible or not. To answer this question, notice that in this setup, $\scriptS$ transformation of $\scriptX$ is generically a composition of discrete gauging, stacking with SPT phases, and automorphism of the symmetry group $\doubleA$. If there's a discrete gauging involved, then the line operator $\scriptL_a$ gauged by $\scriptS$-operation would be absorbed by the defect $\scriptL_\scriptS$. This implies the following fusion rule:
\begin{equation}
    \scriptL_a \times \scriptL_\scriptS = \scriptL_\scriptS ~, \quad \forall a \in \mathbb{A} ~.
\end{equation}
This fusion rule would imply that the $\scriptL_{\scriptS}$ is non-invertible: assuming it is invertible, multiplying $\mathcal{L}_{\mathcal{S}}^{-1}$ on both sides would imply $\mathcal{L}_a = \dsi$ for any $a \in \mathbb{A}$ which is a contradiction. On the other hand, if $\scriptS$ is simply stacking a SPT phase or an automorphism of $\doubleA$, it is clear that the corresponding symmetry defect $\scriptL_{\scriptS}$ is invertible. 

\subsection{$G$-ality defects of $\VEC_{\doubleA}$ from SymTFT $\mathcal{Z}(\VEC_{\doubleA})$}\label{sec:G_ext_general}
In this subsection, we review generically how to describe $G$-ality defects of $\VEC_{\doubleA}$ for a generic Abelian group $\doubleA$ in SymTFT following \cite{Etingof:2009yvg,jordan2009classification}. In summary, to specify a $\underline{\scriptE}_{G}\VEC_{\doubleA}$, one needs to choose a \textit{$\beta$-invertible} $G^{(0)}$ symmetries in the bulk, a choice of the symmetry fractionalization class $[\nu] \in H^2_{\rho}(G,\doubleA \times \widehat{\doubleA})$, as well as an FS indicator $\alpha \in H^3(G,U(1))$. The obstruction to symmetry fractionalization always vanishes for any $G$'s, and for $G = \doubleZ_N, S_3$, the 't Hooft anomaly also vanishes trivially as $H^4(G,U(1)) = \doubleZ_1$ for these $G$'s.

\

We begin by reviewing how to interpret the classification data $(\rho,[\nu],\alpha)$ and the obstruction classes $\mathcal{O}^3_\rho, \mathcal{O}^4(\rho,[\nu])$ in terms of the SymTFT $\scriptZ(\scriptC)$ of the fusion category symmetry $\scriptC$. Briefly, specifying a $G$-extension of the fusion category symmetry $\scriptC$ is the same as specifying the gauging of a $0$-form symmetry $G$ in the SymTFT $\scriptZ(\scriptC)$. $\mathcal{O}^3_\rho, \mathcal{O}^4(\rho,[\nu])$ measures the obstruction to gauging $G$, while $(\rho,[\nu],\alpha)$ specifies inequivalent ways of gauging $G$ symmetry.

More concretely, any invertible topological manipulation corresponds to an invertible $0$-form symmetry of the SymTFT $\scriptZ(\scriptC)$, which forms a group that we denote as $\EqBr(\scriptZ(\scriptC))$. The first piece of data $\rho:G \rightarrow \operatorname{BrPic}(\scriptC)$ can be interpreted as choosing a $G$ subgroup of the bulk symmetry $\EqBr(\scriptZ(\scriptC))$\footnote{Notice that here we are interested in the case in which $G$ implements a non-trivial topological manipulation on the boundary, therefore, $\rho$ is injective.}, thanks to the isomorphism between $\operatorname{BrPic}(\scriptC)$ and $\EqBr(\scriptZ(\scriptC))$ \cite{Etingof:2009yvg}. We will denote the concrete choice of $G$-subgroup of $\EqBr(\scriptZ(\scriptC))$ as $G^{(0)}$ to distinguish it from the abstract group $G$. In the 3d SymTFT, there are also invertible $1$-form symmetries $\mathbf{A}$ generated by all the Abelian anyons in $\scriptZ(\scriptC)$. To consistently gauge the $G^{(0)}$-symmetry, we must be able to specify a choice of symmetry fractionalization class $[\nu] \in H^2_{\rho}(G,\mathbf{A})$, and the corresponding obstruction class $\scriptO^3_{\rho} \in H^3_\rho(G,\mathbf{A})$ must vanish. Provided $\scriptO^3_\rho$ is trivial and a choice of $[\nu] \in H^2_{\rho}(G,\mathbf{A})$ is made, the 't Hooft anomaly $\scriptO^4(\rho,[\nu]) \in H^4(G,U(1))$ is unambiguously determined. When $\scriptO^4(\rho,[\nu])$ vanishes, we can proceed to gauge $G$ after making a choice of the discrete torsion $\alpha \in H^3(G,U(1))$. Notice that the data $(\rho,[\nu],\alpha)$ is considered up to conjugation by elements in some bulk symmetry group $I_{\scriptC}$ for the purpose of classification. $I_{\scriptC}$ is defined as follows. The gapped boundary condition realizing $\scriptC$ symmetry corresponds to a particular Lagrangian algebra $\scriptL_{\scriptC}$. Consider the subgroup $I_{\scriptC}\subset \EqBr(\scriptZ(\scriptC))$ whose elements leave $\scriptL_{\scriptC}$ invariant; the bulk symmetry in $I_{\scriptC}$ implements autoequivalence of the symmetry $\scriptC$ on the boundary, and thus is considered as a gauge transformation to classify $\underline{\scriptE}_G\scriptC$ extensions.

\,

Now, we specialize to the case where $\scriptC = \VEC_{\doubleA}$. As pointed out in \cite{Etingof:2009yvg}, $\EqBr(\scriptZ(\VEC_{\doubleA}))$ is the split orthogonal group $O(\doubleA \oplus \widehat{\doubleA})$. We then proceed to pick a $G$-subgroup $G^{(0)}$ of $\EqBr(\scriptZ(\VEC_{\doubleA}))$. As pointed out previously, not every $G^{(0)}$ will lead to $G$-ality defects on the boundary; furthermore, two $G$-subgroups $G^{(0)}$ and $\widetilde{G}^{(0)}$ lead to equivalent $G$-ality defects if they are related by conjugating by elements in $I_{\doubleA}$. We will discuss this in detail shortly. Once $G^{(0)}$ is picked, then one must proceed to check if there is an obstruction to the symmetry fractionalization. Luckily, \cite{Cordova:2018cvg} shows that for any Abelian TQFT, the obstruction to symmetry fractionalization always vanishes. Hence, we can directly proceed to choose a symmetry fractionalization class $[\nu] \in H^2_\rho(G,\mathbf{A})$ where $\mathbf{A} = \doubleA \times \widehat{\doubleA}$. After that, one must check the 't Hooft anomaly $\scriptO^4(G^{(0)},[\nu]) \in H^4(G^{(0)},U(1))$. For the $G$'s we are interested in this work, that is, $G = \doubleZ_N, S_3$, $H^4(G,U(1))$ is trivial, which means there is no 't Hooft anomaly for any $0$-form symmetry $G$ in 3d in the first place. Thus, we can proceed to the final step where we choose a discrete torsion $\alpha \in H^3(G,U(1))$.

In summary, for $G = \doubleZ_N$ or $S_3$ and $\scriptC = \VEC_{\doubleA}$, the $G$-ality defects are specified by the triple $(G^{(0)},[\nu],\alpha)$, and there is no non-trivial obstruction class. Furthermore, for many choices of $G$ and $\doubleA$, it can be shown that the $H^2_\rho(G,\mathbf{A})$ is trivial, hence there is a unique trivial choice of the symmetry fractionalization class. The well-known TY fusion category is an example of this, and hence is classified by the symmetric bilinear character $\chi$ (which specifies the $\doubleZ_2$ subgroup of $O(\doubleA \oplus \widehat{\doubleA})$) and the discrete torsion $\epsilon$ (which specifies the FS indicator). 

\,

Let us now discuss which choice of $G^{(0)}$ will lead to $G$-ality defects following \cite{jordan2009classification}. The $3d$ SymTFT $\scriptZ(\VEC_{\doubleA})$ of a $0$-form Abelian symmetry $\doubleA$ is alternatively known as $\doubleA$-gauge theory with no DW twist. The simple anyons in the $3d$ SymTFT of $0$-form Abelian symmetry $\doubleA$ is parameterized by $L_{(a,\hat{a})}$ where $(a,\hat{a}) \in \doubleA\oplus \widehat{\doubleA}$, where $\widehat{\doubleA} = \Hom(\doubleA,U(1))$. We can think of $\doubleA$ contains all the pure magnetic lines (the discrete flux labeled by the conjugacy classes, which are single group elements, in Abelian group $\doubleA$), while the $\widehat{\doubleA}$ contains all the pure electric lines (the Wilson lines of the gauge group $\doubleA$ labeled by the irreps in $\widehat{\doubleA}$). Then, following \cite{jordan2009classification}, an anyon permutation symmetry can be parameterized as
\begin{equation}\label{eq:Uoper}
    U = \begin{pmatrix} \alpha & \beta \\ \gamma & \delta \end{pmatrix},
\end{equation}
where $\alpha: \doubleA \rightarrow \doubleA, \beta: \widehat{\doubleA} \rightarrow \doubleA, \gamma: \doubleA \rightarrow \widehat{\doubleA}, \delta: \widehat{\doubleA} \rightarrow \widehat{\doubleA}$\footnote{Notice that here we abuse the notation slightly to use $\alpha:\doubleA\rightarrow \doubleA$ to denote certain component of the bulk symmetry $U$, and this has nothing to do with the FS indicator mentioned previously and the reader should be able to distinguish the two based on context.}. Notice that $\alpha, \beta, \gamma,\delta$ must satisfy some relation to preserve the self-statistics and the mutual braiding. 

The global symmetries $U$'s in the SymTFT correspond to the invertible topological manipulations that one can perform on the QFT $\scriptX$ with $\doubleA$ symmetries. A generic invertible topological manipulation is a composition of acting with an automorphism of the symmetry group of the boundary theory, stacking SPTs, and discrete gauging. Notice that the first two basic operations do not change the theory $\scriptX$, while the last one generically maps $\scriptX$ to a different theory. We now discuss how to interpret the $U$'s symmetry in terms of these operations in detail. 

Recall that our convention is that the Dirichlet boundary condition corresponds to condensing the Lagrangian algebra generated by the pure electric lines $\widehat{\doubleA}$. Whether the boundary theory $\scriptX$ is mapped to a different theory depends on whether $U$ preserves the pure electric Lagrangian algebra $\scriptL_e = \bigoplus_{\hat{a} \in \widehat{\doubleA}} \scriptL_{(0,\hat{a})}$. From \eqref{eq:Uoper}, it is straightforward to see that this is controlled by the component $\beta \subset U$, and $U$ preserves $\scriptL_e$ if and only if $\beta = 0$. Collectively, we denote the subgroup formed by the $U$'s with $\beta = 0$ as $I_{\doubleA}$. Since $U \in I_{\doubleA}$ does not change the theory $\scriptX$, it then corresponds to an autoequivalence of fusion category symmetry $\VEC_{\doubleA}$ on the boundary, which consists of automorphisms of the group $\doubleA$ and stacking SPT phases. As pointed out in \cite{Bhardwaj:2017xup, Fuchs:2014ema}, the autoequivalences of the fusion category symmetry $\VEC_{\doubleA}$ are parameterized by $H^2(\doubleA,U(1)) \rtimes \Aut(\doubleA)$, where $H^2(\doubleA,U(1))$ parameterizes the possible SPT phases one can stack on the boundary and $\Aut(\doubleA)$ is the group of automorphisms of the Abelian group $\doubleA$. \footnote{The group $\Aut(\VEC_G^\omega)$ of autoequivalences for generic $\VEC_G^\omega$ is characterized by the following exact sequence
\begin{equation}
    0 \rightarrow H^2(G,U(1)) \rightarrow \Aut(\VEC_G^\omega) \rightarrow Stab(\omega) \rightarrow 0
\end{equation}
where $Stab(\omega)$ is the subgroup of $\Aut(G)$ which leaves the cohomological class of $\omega$ unchanged. When $\omega$ is trivial, $Stab(\omega) = \Aut(G)$ and we recover the previous result.} 

From the bulk point of view, given an $\alpha \in \Aut(\doubleA)$, naturally one can construct a bulk $0$-form anyon permutation symmetry which does not mix electric lines with magnetic lines at all. It acts on the pure electric line according to $\alpha$, while the action on the pure magnetic line can be uniquely determined by requiring the action preserving the braiding and the self-statistic, which is given by \footnote{Here, $\alpha^*:\widehat{\doubleA}\rightarrow \widehat{\doubleA}$ is defined as $\alpha^*(\hat{a})(b) := \hat{a}(\alpha(b))$. Then $(\alpha^*)^{-1}(\hat{a})(b) = \hat{a}(\alpha^{-1}(b))$, which then implies $(\alpha^*)^{-1}(\hat{a}) (\alpha(b)) = \hat{a}(b)$ and the self-statistic and mutual braidings are preserved.}
\begin{equation}
    U_\alpha = \begin{pmatrix} \alpha & 0 \\ 0 & (\alpha^*)^{-1} \end{pmatrix} ~.
\end{equation}

Given an $[\varphi] \in H^2(\doubleA, U(1))$, to construct the corresponding bulk symmetry, we first notice that element in $H^2(\doubleA, U(1))$ is in bijection with the anti-symmetric bilinear form valued in $U(1)$ on $\doubleA$. And the latter is given by $\xi_\varphi(a_1, a_2) := \frac{\varphi(a_1, a_2)}{\varphi(a_2,a_1)}$ independent of the choice of the representative $\varphi$ in $[\varphi] \in H^2(\doubleA,U(1))$. Since $\xi_{\varphi}:\doubleA \times \doubleA \rightarrow U(1)$, we can alternatively view it as $\xi_{\varphi}:\doubleA\rightarrow \widehat{\doubleA}$, therefore it can be viewed as a map from magnetic electric line to pure electric line. And the corresponding anyon permutation matrix is 
\begin{equation}\label{eq:stacking_SPT}
    U_\varphi = \begin{pmatrix} id & 0 \\ \xi_{\varphi} & id \end{pmatrix} ~.
\end{equation}
Namely, this symmetry acts trivially on all the pure electric lines, but when it acts on a pure magnetic line $L_{(a,0)}$, it will attach an additional pure electric charge $\xi_\varphi(a) \in \widehat{\doubleA}$ to the original magnetic line. The transformation preserves the spin and the braiding as $\xi_\varphi$ is anti-symmetric.

On the other hand, if $\beta$ is invertible, then the symmetry $U$ will map the electric Lagrangian algebra to a magnetic Lagrangian algebra, therefore the new boundary condition corresponds to the topological manipulation where the boundary symmetry $\doubleA$ is completely gauged. Hence, the topological defect line $\mathcal{L}_{\mathcal{S}}$ corresponding to the twist defect $\mathcal{S}$ of the symmetry $U$ will have the fusion rule
\begin{equation}
    \scriptL_a \times \scriptL_\scriptS = \scriptL_\scriptS ~, \quad \forall a \in \doubleA ~.
\end{equation}
We will refer the $U$ satisfies this as $U$ with invertible $\beta$-component, or even more briefly, $\beta$-invertible $U$. In order for a $G^{(0)}$ anyon permutation symmetry group to lead to a $G$-ality type extension, every non-trivial element in $G$ must be $\beta$-invertible. We will call these $G^{(0)}$ as $\beta$-invertible $G^{(0)}$. Given a specific choice of $\beta$-invertible $G^{(0)} \subset \EqBr(\mathcal{Z}(\VEC_{\doubleA}))$, one naturally has a group homomorphism $\rho:G \rightarrow \EqBr(\mathcal{Z}(\VEC_{\doubleA}))$ in the classification data: denote the group elements in $G^{(0)}$ as $U_g$'s where $g \in G$, then $\rho(g) = U_g \in \EqBr(\mathcal{Z}(\VEC_{\doubleA}))$. This provides us a useful tool to search for $G$-ality defects in general. 

Given two $\beta$-invertible $G$-subgroups $G^{(0)}$ and $\widetilde{G}^{(0)}$ of $O(\doubleA \oplus \widehat{\doubleA})$, they may lead to identical $G$-ality extension if they are related by conjugation of an element in $I_{\doubleA}$\footnote{Notice that for any $\beta$-invertible $U$, the $\widetilde{U}$ acquired by conjugating $U$ with an element in $I_{\doubleA}$ will also be $\beta$-invertible.}. From the boundary point of view, $I_{\doubleA}$ parameterizes the gauge transformations of the $F$-symbols of $\underline{\scriptE}_G \VEC_{\doubleA}$ as they leave $F$-symbols of $\VEC_{\doubleA}$ invariant. The bulk symmetry parameterized by $\alpha \in \Aut(\doubleA)$ will change the $F$-symbols of $\underline{\scriptE}_G\VEC_{\doubleA}$ by changing $a \in \doubleA$ to $\alpha(a) \in \doubleA$, e.g.
\begin{equation}
    \alpha: F^{a_1 a_2 a_3} \mapsto F^{\alpha(a_1) \alpha(a_2) \alpha(a_3)} ~, \quad F^{a_1 a_2 \scriptN_g}_{\scriptN_g} \mapsto F^{\alpha(a_1) \alpha(a_2) \scriptN_g}_{\scriptN_g} ~, \quad  etc.
\end{equation}
Notice that since we have chosen $F^{a_1 a_2 a_3} = 1$, this transformation will leave $F^{a_1 a_2 a_3}$ invariant. The bulk symmetry parameterized by $[\varphi] \in H^2(\doubleA,U(1))$ will add the phase factors $\varphi(a_1,a_2)$ to all the junctions $a_1 \times a_2 \rightarrow a_1 a_2$, that is,
\begin{equation}
    \begin{tikzpicture}[scale=0.8,baseline={([yshift=-.5ex]current bounding box.center)},vertex/.style={anchor=base,
    circle,fill=black!25,minimum size=18pt,inner sep=2pt},scale=0.50]
    \draw[thick, black] (-2,-2) -- (0,0);
    \draw[thick, black] (+2,-2) -- (0,0);
    \draw[thick, black] (0,0) -- (0,2);
    \draw[thick, black, -stealth] (-2,-2) -- (-1,-1);
    \draw[thick, black, -stealth] (+2,-2) -- (1,-1);
    \draw[thick, black, -stealth] (0,0) -- (0,1);
    \node[black, below] at (-2,-2) {$a_1$};
    \node[black, below] at (2,-2) {$a_2$};
    \node[black, above] at (0,2) {$a_1 a_2$};
\end{tikzpicture} \mapsto \varphi(a_1,a_2) \begin{tikzpicture}[scale=0.8,baseline={([yshift=-.5ex]current bounding box.center)},vertex/.style={anchor=base,
    circle,fill=black!25,minimum size=18pt,inner sep=2pt},scale=0.50]
    \draw[thick, black] (-2,-2) -- (0,0);
    \draw[thick, black] (+2,-2) -- (0,0);
    \draw[thick, black] (0,0) -- (0,2);
    \draw[thick, black, -stealth] (-2,-2) -- (-1,-1);
    \draw[thick, black, -stealth] (+2,-2) -- (1,-1);
    \draw[thick, black, -stealth] (0,0) -- (0,1);
    \node[black, below] at (-2,-2) {$a_1$};
    \node[black, below] at (2,-2) {$a_2$};
    \node[black, above] at (0,2) {$a_1 a_2$};
\end{tikzpicture} ~.
\end{equation} 
This transformation acts non-trivially on the following $F$-symbols (see e.g. \cite{Barkeshli:2014cna}):
\begin{equation}
\begin{aligned}
    & F^{a b \scriptN_g}_{\scriptN_g} \mapsto \frac{1}{\varphi(a, b)} F^{ab \scriptN_g}_{\scriptN_g} ~, \quad F^{\scriptN_g ab}_{\scriptN_g} \mapsto \varphi(a,b) F^{\scriptN_g ab}_{\scriptN_g} ~, \quad \\
    & F^{\scriptN_g \scriptN_h a}_{b} \mapsto \frac{1}{\varphi(ba^{-1},a)} F^{\scriptN_g \scriptN_h a}_{b} ~, \quad F^{a \scriptN_g \scriptN_h}_{b} \mapsto \varphi(a,a^{-1}b) F^{a \scriptN_g \scriptN_h}_{b} ~.
\end{aligned}
\end{equation}
Both types of transformations on the $F$-symbols are considered as gauge transformations to classify inequivalent fusion categories.

But notice that these transformations in $I_{\doubleA}$ will change the twisted partition function $\mathcal{Z}_{\scriptX}[A]$ of a QFT $\scriptX$ coupled to the $\doubleA$-background gauge field $A$. The symmetry parameterized by $(\varphi,\alpha) \in H^2(\doubleA,U(1)) \rtimes \Aut(\doubleA) = I_{\doubleA}$ will transform $\mathcal{Z}_{\scriptX}[A]$ as
\begin{equation}
    \mathcal{Z}_{\scriptX}[A] \mapsto Z_{\scriptX}'[A] = Z_{SPT,\varphi}[A] Z_{\scriptX}[\alpha(A)] ~.
\end{equation}
Notice that $Z_{\scriptX}'[A]$ relates to $Z_{\scriptX}[A]$ by first applying an automorphism $\alpha$ of $\doubleA$ to the background fields and then stack an $\doubleA$-SPT $Z_{SPT,\varphi}[A]$ specified by $[\varphi]\in H^2(\doubleA,U(1))$. $Z'_{\scriptX}[A]$ is also a good twisted partition function of the theory $\scriptX$. This phenomenon should be interpreted as the twisted partition function of $\scriptX$ is ambiguous up to the relabeling of the background fields and adding counter terms. Suppose $Z_{\scriptX}[A]$ is invariant under the topological manipulations $\mathcal{G}$ corresponding to the bulk symmetry $G^{(0)}$; if $G^{(0)}$ is not invariant under conjugating by the element $U_{(\phi, \alpha)} \in I_{\mathbb{A}}$, i.e. $G^{(0)} \neq \widetilde{G}^{(0)} \equiv U_{(\phi,\alpha)}G^{(0)} U^{-1}_{(\phi,\alpha)}$, then the other twisted partition function $\mathcal{Z}_{\scriptX}'[A]$ will not be invariant under the topological manipulation given by $G^{(0)}$, but will be invariant under a different group of topological manipulations specified by $\widetilde{G}^{(0)}$. 

Practically, this means if one wants to check if a theory $\scriptX$ could admit a $G$-ality defect corresponding to $G^{(0)}$ up to conjugation by an element in $I_\doubleA$, one should fix the set of topological manipulations given by $G^{(0)}$ and test using every possible twisted partition function $Z_{\scriptX}[A]$ (which is related to each other by the action of $I_\doubleA$). Alternatively, one can consider fixing a choice of the twisted partition function $Z_{\scriptX}[A]$, and test it with every group of the form $U_{(\phi,\alpha)}G^{(0)} U^{-1}_{(\phi,\alpha)}$ where $U_{(\phi,\alpha)} \in I_{\doubleA}$. We will demonstrate this subtlety in concrete examples in Section \ref{sec:trans_part}.

\,

Using the gauge transformation $U_\varphi \in H^2(\doubleA,U(1)) \subset I_\doubleA$, the $\delta$-component of any $\beta$-invertible $U$ can be set to $0$, and after that it takes the following form:
\begin{equation}\label{eq:gf_generator}
    U = \begin{pmatrix} \alpha & \gamma^{*-1} \\ \gamma & 0 \end{pmatrix} ~,
\end{equation}
where $\gamma^* \alpha$ is anti-symmetric\footnote{Here, $\gamma:\doubleA\rightarrow \widehat{\doubleA}$ can be alternatively viewed as $\gamma: \doubleA\times \doubleA\rightarrow U(1)$ via $\gamma(a,b) = \gamma(a)(b)$. Then, $\gamma^*:\doubleA\rightarrow \widehat{\doubleA}$ is defined as $\gamma^*(a)(b) = \gamma(b,a)$.}. To see this, starting with a generic $\beta$-invertible $U$ given in \eqref{eq:Uoper} and considering the gauge transformation of conjugating with $U_\varphi \in I_{\doubleA}$:
\begin{equation}
    U_\varphi U U_\varphi^{-1} = \begin{pmatrix} \alpha - \beta \xi_{\varphi} & \beta \\ \xi_{\varphi} \alpha + \gamma - \xi_\varphi \beta\xi_\varphi - \delta \xi_\varphi & \xi_\varphi \beta + \delta \end{pmatrix} ~.
\end{equation}
Since $\beta$ is invertible, one can always choose $\xi_\varphi = - \beta^{-1}\delta$ to set the $\delta$-component to be $0$ in $U_\varphi U U_\varphi^{-1}$. Notice that in general this does not completely fix the gauge transformation via conjugating using elements in $I_{\doubleA}$ (as we will demonstrate in the concrete example in Section \ref{sec:trans_part}). Notice that this particular form of $U$ maps the pure electric Lagrangian algebra to the pure magnetic Lagrangian algebra; therefore, its boundary operation corresponds to $\doubleA$ gauging with trivial discrete torsion. But because a magnetic line is mapped to a dyonic line, hence a SPT is stacked after gauging the $\doubleA$-symmetry \footnote{Alternatively, one can consider using $U_\varphi$ to set the $\alpha$-component to be $0$ instead, which on the boundary it corresponds to first stack SPT and then gauge.}. Notice that generically it is not possible to fix every non-trivial element in $\beta$-invertible $G^{(0)}$ in the form of \eqref{eq:gf_generator}. As a simple demonstration of this, generically $U^2$ will not take the form \eqref{eq:gf_generator} even if $U$ does. Thus, the boundary topological manipulation implemented by $U^2$ will be stacking SPT, gauging, and stacking SPT again. In other words, for generic $G$, there is no simple choice of $\beta$-invertible $G^{(0)}$ in a given equivalence class such that the corresponding boundary manipulations consist of only gauging and stacking SPTs--a generic element in $G^{(0)}$ will always implement stacking SPT, gauging, and then stacking another SPT.

Let's consider $G = \doubleZ_p$. Then classifying $\beta$-invertible $\doubleZ_p^{(0)}$ subgroup of the bulk symmetries is the same as classifying its generator $U$ such that $U^p = 1$ as well as each $U^m$ ($m =1,\cdots,p-1$) is $\beta$-invertible. In the special case where $p = 2$, we recover the familiar case of the TY fusion category. In this case, the condition $U^2 = 1$ precisely implies that $\alpha = 0$ and $\gamma = \gamma^*$. This $\gamma$ is precisely the data of a non-degenerate symmetric bicharacter appearing in the classification of TY fusion categories. Furthermore, using the symmetric non-degenerate character $\gamma$, one can identify $\doubleA \oplus \widehat{\doubleA}$ as $\doubleZ_2 \otimes \doubleA$, and then by Shapiro's lemma, $H^n_{\rho_\gamma}(\doubleZ_2, \doubleA\oplus \widehat{\doubleA}) \simeq H^n(\doubleZ_2, \doubleZ_2 \otimes \doubleA) = \doubleZ_1$ identically for $n > 0$ \cite{Etingof:2009yvg}. Therefore, the obstruction to the fractionalization of the $\doubleZ_2^{em,\gamma}$ vanishes identically and there is always a unique choice of the symmetry fractionalization class. Hence, we reach the familiar conclusion that the TY fusion category is classified by a symmetric non-degenerate bicharacter and the FS indicator.

We want to emphasize again that in the case where $G$ is not generated by a unique element, then to search for $\beta$-invertible $G^{(0)}$ subgroup, one cannot assume every generator is of the form \eqref{eq:gf_generator}. This will become important later when we study the $S_3$-ality extension in Section \ref{sec:S3_ality}. 

\

After specifying $G^{(0)}$ which leads to $G$-ality defect, one must then choose a symmetry fractionalization class $H^2(G^{(0)},\mathbf{A})$. Just as the TY fusion categories where $G = \doubleZ_2$, for $G$ and $\doubleA$ such that their orders are coprime, i.e. $\gcd(|G|,|\doubleA|) = 1$, $H^*_\rho(G, \doubleA \oplus \widehat{\doubleA})$ are also trivial identically (which follows from Zassenhaus theorem instead), therefore the choice of symmetry fractionalization class is unique. This means the corresponding $G$-ality fusion category is classified by the choice of $\beta$-invertible $G$ symmetries (up to the equivalence relation discussed above) together with an FS indicator in $H^3(G,U(1))$. But in general, there could be multiple choices of the symmetry fractionalization classes leading to distinct $G$-ality defects. In this work, we will provide two inequivalent classes of the $p$-ality defects acquired from group-theoretical constructions, and show that they correspond to distinct choices of the symmetry fractionalization classes and have different physical implications in Section \ref{Sec:p_ality}. 

\

Once a $\underline{\scriptE}_G^{(\rho,[\nu],\alpha)}\VEC_{\doubleA}$ is specified via $(\rho,[\nu],\alpha)$, one can ask whether $\underline{\scriptE}_G^{(\rho,[\nu],\alpha)}\VEC_{\doubleA}$ is anomalous, namely if it admits a trivially symmetric gapped phase. A necessary condition is that there exists a trivially gapped phase with $\doubleA$-symmetry (that is, a $\doubleA$-SPT) invariant under the topological manipulations specified by $G^{(0)}$. In the bulk, this is equivalent to the existence of a $G^{(0)}$-stable magnetic Lagrangian algebra. For any $\underline{\scriptE}_G^{(\rho,[\nu],\alpha)}\VEC_{\doubleA}$ passes this check, it must be group-theoretical (see \cite{Gelaki:2009blp,Sun:2023xxv} for more details). By finding out its group-theoretical form explicitly, we can use the criterion in \eqref{eq:anomaly_free_criteria} to determine the anomaly of the $\underline{\scriptE}_G^{(\rho,[\nu],\alpha)}\VEC_{\doubleA}$.\footnote{Notice that a similar approach is described in \cite{Antinucci:2023ezl} from the equivalent characterization of the anomaly via gauging the entire $\underline{\scriptE}_G^{(\rho,[\nu],\alpha)}\VEC_{\doubleA}$. There, one considers formulating gauging $\underline{\scriptE}_G^{(\rho,[\nu],\alpha)}\VEC_{\doubleA}$ as a sequential gauging involves two steps, where the first step involves gauging some $\doubleB \subset \doubleA$ to turn $\underline{\scriptE}_G^{(\rho,[\nu],\alpha)}\VEC_{\doubleA}$ into invertible symmetries $(\doubleA/\doubleB \times \widehat{\doubleB})\rtimes_\rho G$ with certain anomaly, and the second step is to gauge the remaining symmetries (roughly speaking $\doubleA/\doubleB \rtimes_\rho G$). The anomaly of the group-theoretical $\underline{\scriptE}_G^{(\rho,[\nu],\alpha)}\VEC_{\doubleA}$ is characterized by the obstruction to performing the second step.}

\section{Triality defects under twisted gauging $\mathbb{Z}_N \times \mathbb{Z}_N$ ($\gcd(N,3) = 1$)}\label{sec:triality}

In this section, we consider the triality defects under twisted gauging $\doubleA = \mathbb{Z}_N \times \mathbb{Z}_N$ where $N$ is coprime with $3$ as an example, and demonstrate how various tools can be combined to study the triality defects. Specifically, we are interested in the fusion categories with simple objects given by the invertible symmetries $\doubleZ_N \times \doubleZ_N$ and the triality defect $\mathcal{N}$ with its orientation reversal $\overline{\mathcal{N}}$, and the fusion rules are given by
\begin{equation}
\begin{aligned}
    & a \times \scriptN = \scriptN \times a = \scriptN ~, \quad a \times \overline{\scriptN} = \overline{\scriptN} \times a = \overline{\scriptN} ~, \quad \forall a \in \doubleZ_N\times \doubleZ_N ~, \\
    & \scriptN \times \scriptN = N \overline{\scriptN} ~, \quad \overline{\scriptN} \times \overline{\scriptN} = N \scriptN ~, \quad \scriptN \times \overline{\scriptN} = \sum_{a\in \mathbb{Z}_N\times \doubleZ_N} a ~.
\end{aligned}
\end{equation}
These have been considered in \cite{jordan2009classification} where the fusion categories are classified. The classification is similar to the TY fusion categories, which is specified by $(\mathbf{T}, \alpha)$. Here, $\mathbf{T}$ labels the choice of (the generator of) the bulk $\beta$-invertible $\doubleZ_3$ symmetry, and $\alpha \in \doubleZ_3$ is the choice of the FS indicator. Given the prime decomposition $\displaystyle N = \prod_{i=1}^n p_i^{r_i}$ (where $p_i \neq 3$), there are two inequivalent choices of the $\doubleZ_3$ symmetries for each prime factor, and in total there are $2^n$ inequivalent choices of the bulk $\doubleZ_3$ symmetries. 

We begin by explicitly constructing the corresponding bulk $\doubleZ_3$ symmetries after setting up the SymTFT convention in Section \ref{sec:triality_setup}, and we construct the corresponding condensation defects for those symmetries. Then, using the condensation defects we derive the corresponding $\doubleZ_N \times \doubleZ_N$ twisted gauging in Section \ref{sec:trans_part}. Then, we determine the anomaly of the triality defect using the group-theoretical technique in Section \ref{sec:pr_anomaly}. Finally, we consider the example of $\Spin(8)_1$ diagonal WZW model and use the result in \ref{sec:trans_part} to conclude that this CFT admits the non-group-theoretical triality defect under twisted gauging $\doubleZ_2 \times \doubleZ_2$.

\subsection{Triality defects from $\mathbb{Z}_3$-symmetry in the SymTFT}\label{sec:triality_setup}

In this subsection, we summarize the classification of the triality defects for gauging $\mathbb{Z}_N \times \mathbb{Z}_N$ symmetry with $N$ coprime with $3$ from the SymTFT acquired in \cite{jordan2009classification} following the generic discussion in Section \ref{sec:G_ext_general}, and recast the result in a physics-friendly language. 

For $\mathbb{A} = \doubleZ_N \times \doubleZ_N$ where $\gcd(N,3) = 1$, first consider the prime decomposition of $N$
\begin{equation}\label{eq:prime_decom}
    N = \prod_{i = 1}^n p_i^{r_i}
\end{equation}
where $p_i$ are distinct prime numbers and $r_i \in \doubleZ_{>0}$. Then, then invertible symmetry can be written as $\displaystyle \mathbb{A} = \prod_{i=1}^n \doubleZ_{p_i^{r_i}} \times \doubleZ_{p_i^{r_i}}$. There are $3 \cdot 2^n$ inequivalent $\underline{\mathcal{E}}_{\doubleZ_3} \VEC_{\doubleZ_N \times \doubleZ_N}$ where each factor $\doubleZ_{p_i^{r_i}} \times \doubleZ_{p_i^{r_i}}$ contains two choices and the overall factor $3$ encodes the choice of FS indicator $H^3(\doubleZ_3, U(1)) \simeq \doubleZ_3$.

We now describe how to understand the classification data from the $3d$ SymTFT. The $3d$ SymTFT for $\VEC_{\doubleZ_N \times \doubleZ_N}$ $0$-form symmetry in $2d$ contains $N^4$ anyons generated by two bosonic $\doubleZ_N$ pure electric lines $e_1$ and $e_2$ and two bosonic $\doubleZ_N$ pure magnetic lines $m_1$ and $m_2$ with the following non-trivial mutual braiding:
\begin{equation}
    B(e_i, m_j) = e^{\frac{2\pi i}{N}\delta_{i,j}} ~.
\end{equation}
A generic simple anyon takes the form $m_1^{a_1} m_2^{a_2} e_1^{\hat{a}_1} e_2^{\hat{a}_2}$, where $a_i, \hat{a}_i \in \doubleZ_N$, and we denote it as a column vector
\begin{equation}\label{eq:anyon_charge}
    \mathbf{a} = \begin{pmatrix} a_1 & a_2 & \hat{a}_1 & \hat{a}_2 \end{pmatrix}^T ~.
\end{equation}
The topological spin and the braiding are given by 
\begin{equation}
    \theta_{\mathbf{a}} = \exp\left(\frac{\pi i}{N} \mathbf{a}^T K \mathbf{a} \right) ~, \quad 
    B_{\mathbf{a}, \mathbf{b}} = \exp\left(\frac{2 \pi i}{N} \mathbf{b}^T K \mathbf{a} \right) ~,
\end{equation}
where the matrix $K$ is given by \footnote{Notice that the matrix $K$ is not the $K$-matrices in the Abelian Chern-Simons theory literature.} 
\begin{equation}
    K = \begin{pmatrix} 0 & 0 & 1 & 0 \\ 0 & 0 & 0 & 1 \\ 1 & 0 & 0 & 0 \\ 0 & 1 & 0 & 0 \end{pmatrix} ~.
\end{equation}
The anyon permutation symmetry can be parameterized as a $4\times 4$ matrix $U$ such that
\begin{equation}
    U^T K U = K \mod N ~.
\end{equation}

To construct the corresponding $\doubleZ_3$ symmetry in the bulk symmetry TFT, we can proceed as follows. As in \eqref{eq:prime_decom}, we can rewrite the symmetry group as 
\begin{equation}
    (\doubleZ_N)^2 = \prod_{i=1}^{m}(\doubleZ_{p_i^{r_i}})^2,
\end{equation}
then the bulk symmetry TFT can also be written as a trivial tensor product of $(\doubleZ_{p_i^{r_i}})^2$-gauge theories. For each $(\doubleZ_{p_i^{r_i}})^2$-gauge theory, there are two inequivalent choices of $\beta$-invertible $\doubleZ_3$ anyon permutation symmetries. And to construct a $\beta$-invertible $\doubleZ_3$ anyon permutation symmetries in the full SymTFT, we only need to consider the tensor product of the $\beta$-invertible $\doubleZ_3$ anyon permutation symmetries in each factor. Therefore, for now, we will restrict ourselves to $N = p^r$ where $p$ is a prime number and $r\in \doubleZ_{>0}$.
The generators of the two inequivalent $\beta$-invertible $\doubleZ_3$ symmetries are given by
\begin{equation}\label{eq:Z3_sym}
    T_1 = \begin{pmatrix} -1 & 0 & 0 & 1 \\ 0 & -1 & -1 & 0 \\ 0 & 1 & 0 & 0 \\ -1 & 0 & 0 & 0  \end{pmatrix}, \quad T_2 = \begin{pmatrix} 0 & 1 & 1 & 0 \\ -1 & 1 & 1 & 1 \\ 1 & -1 & 0 & 0 \\ 0 & 1 & 0 & 0 \end{pmatrix},
\end{equation}
where they act on the simple anyon $m_1^{a_1} m_2^{a_2} e_1^{\hat{a}_1} e_2^{\hat{a}_2}$ via matrix multiplication with the column vector $\mathbf{a}$ in \eqref{eq:anyon_charge}, and we have
\begin{equation}
\begin{aligned}
    & T_1: e_1 \rightarrow m_2^{-1} ~, \quad e_2 \rightarrow m_1 ~,  \quad m_1 \rightarrow e_2^{-1} m_1^{-1} ~, \quad m_2 \rightarrow e_1 m_2^{-1} ~, \\
    & T_2: e_1 \rightarrow m_1 m_2 ~, \quad e_2 \rightarrow m_2 ~, \quad m_1 \rightarrow m_2^{-1} e_1 ~, \quad m_2 \rightarrow e_1^{-1} e_2 m_1 m_2 ~.
\end{aligned}
\end{equation}
Notice that for both $T_i$, one may consider replacing $T_i$ with $(T_i)^2$, however, they will lead to the same triality fusion categories as there exist $V_i \in I_{\doubleZ_N \times \doubleZ_N}$'s such that $(V_i)^T K V_i = K$ and $(V_i)^{-1} (T_i)^2 V_i = T_i$ for both $T_i$'s. 

It is possible to construct condensation defects \cite{shuheng:conddefect} corresponding to the above two symmetries. For $T_1$, notice that anyons built from $e_1 m_2$, $e_2 m_1^{-1}$, $e_2 m_1^2$ and $e_1 m_2^{-2}$ can be absorbed into the condensation defects. With the assumption that $N$ is coprime with $3$, in this case, every anyon can be absorbed, therefore the $S_{T_1}(\Sigma)$ is acquired from condensing the entire anyon spectrum on the surface $\Sigma$. For $T_2$, the condensation defects are acquired from higher gauging $e_1^{-1} m_1 m_2$ and $e_1^{-1} e_2 m_1$ regardless of $N$. This distinction implies that indeed $T_1$ and $T_2$ are not equivalent, as conjugating by another bulk invertible symmetry should not change the number of condensed anyons. We find the corresponding condensation defects are given by 
\begin{equation}
\begin{aligned}
    S_{T_1}(\Sigma) &= \frac{1}{|H_1(\Sigma,\doubleZ_N)|^2} \sum_{\gamma_i \in H_1(\Sigma,\doubleZ_N)} e^{-\frac{2\pi i x}{N} \langle  \gamma_1  + \gamma_4, \gamma_2 - \gamma_3\rangle}  m_1(\gamma_1) m_2(\gamma_2) e_1(\gamma_3) e_2(\gamma_4)  ~, \\
    S_{T_2}(\Sigma) &= \frac{1}{|H^1(\Sigma,\doubleZ_N)|} \sum_{\gamma_i \in H_1(\Sigma,\doubleZ_N)} m_1(\gamma_1 + \gamma_2) m_2(\gamma_1) e_1(-\gamma_1 - \gamma_2) e_2(\gamma_2) ~,
\end{aligned}
\end{equation}
where $x$ is the integer inverse of the $3$ mod $N$. Using the commutation relation is given by
\begin{equation}
    e_i(\gamma) m_j(\gamma') = e^{-\frac{2\pi i}{N}\langle \gamma,\gamma'\rangle \delta_{ij} }m_j(\gamma')e_i(\gamma) ~,
\end{equation}
where $\langle \gamma,\gamma'\rangle$ denotes the intersection number between the cycles $\gamma$ and $\gamma'$ on $\Sigma$. One can check indeed these symmetries acting as the simple anyons in the desired way. For instance,
\begin{equation}
\begin{aligned}
    & S_{T_1}(\Sigma) m_1(\gamma) \\
    = & \frac{1}{|H_1(\Sigma,\doubleZ_N)|^2} \sum_{\gamma_i \in H_1(\Sigma,\doubleZ_N)} e^{-\frac{2\pi i x}{N} \langle  \gamma_1  + \gamma_4, \gamma_2 - \gamma_3\rangle}  m_1(\gamma_1) m_2(\gamma_2) e_1(\gamma_3) e_2(\gamma_4) m_1(\gamma) \\
    = & \frac{1}{|H_1(\Sigma,\doubleZ_N)|^2} \sum_{\gamma_i \in H_1(\Sigma,\doubleZ_N)} e^{-\frac{2\pi i x}{N} \langle  \gamma_1  + \gamma_4, \gamma_2 - \gamma_3\rangle} e^{-\frac{2\pi i}{N}\langle \gamma_3,\gamma\rangle}  m_1(\gamma_1 + \gamma) m_2(\gamma_2) e_1(\gamma_3) e_2(\gamma_4) \\
    = & \frac{1}{|H_1(\Sigma,\doubleZ_N)|^2} \sum_{\gamma_i \in H_1(\Sigma,\doubleZ_N)} e^{-\frac{2\pi i x}{N} \langle  \gamma_1 - 2\gamma + \gamma_4 -\gamma, \gamma_2 - \gamma_3\rangle} e^{-\frac{2\pi i}{N}\langle \gamma_3,\gamma\rangle}  m_1(\gamma_1 - \gamma) m_2(\gamma_2) e_1(\gamma_3) e_2(\gamma_4-\gamma) \\
    = & \frac{1}{|H_1(\Sigma,\doubleZ_N)|^2} \sum_{\gamma_i \in H_1(\Sigma,\doubleZ_N)} e^{-\frac{2\pi i x}{N} \langle  \gamma_1 + \gamma_4 , \gamma_2 - \gamma_3\rangle} e^{\frac{2\pi i}{N}\langle \gamma,\gamma_2\rangle}  m_1(-\gamma) m_1(\gamma_1) m_2(\gamma_2) e_1(\gamma_3) e_2(-\gamma) e_2(\gamma_4) \\
    = & \frac{1}{|H_1(\Sigma,\doubleZ_N)|^2} \sum_{\gamma_i \in H_1(\Sigma,\doubleZ_N)} e^{-\frac{2\pi i x}{N} \langle  \gamma_1 + \gamma_4 , \gamma_2 - \gamma_3\rangle} e^{\frac{2\pi i}{N}\langle \gamma,\gamma_2\rangle} e^{\frac{2\pi i}{N}\langle -\gamma,\gamma_2\rangle}  e_2(-\gamma) m_1(-\gamma) m_1(\gamma_1) m_2(\gamma_2) e_1(\gamma_3) e_2(\gamma_4) \\
    = & e_2(-\gamma) m_1(-\gamma) S_{T_1}(\Sigma) ~,
\end{aligned}
\end{equation}
where in the third equal sign we make the shift of dummy variables $\gamma_1\rightarrow \gamma_1 - 2\gamma, \gamma_4\rightarrow \gamma_4 - \gamma$.

It is also straightforward to work out the fusion of the condensation defects and verify that indeed they generate $\doubleZ_3$ invertible symmetries
\begin{equation}
\begin{aligned}
     S_{T_1}^2(\Sigma) &= \frac{1}{|H_1(\Sigma,\doubleZ_N)|^2} \sum_{\gamma_i \in H_1(\Sigma,\doubleZ_N)} e^{\frac{2\pi i x}{N} [\langle \gamma_1-2\gamma_4, \gamma_2\rangle +\langle 2\gamma_1-\gamma_4, \gamma_3\rangle]}m_1(\gamma_1) m_2(\gamma_2) e_1(\gamma_3) e_2(\gamma_4) ~, \\ 
     S_{T_2}^2(\Sigma) &= \frac{1}{|H^1(\Sigma,\doubleZ_N)|} \sum_{\gamma_i \in H_1(\Sigma,\doubleZ_N)}e^{\frac{2\pi i}{N} \langle \gamma_1, \gamma_2\rangle} m_1(\gamma_1 +\gamma_2)m_2(\gamma_1) e_1(-\gamma_1 - \gamma_2)e_2(\gamma_2) ~, \\
     S_{T_1}^3(\Sigma) &= \mathds{1} ~,\\
      S_{T_2}^3(\Sigma) &= \mathds{1} ~.
\end{aligned}
\end{equation}
At this stage, it is possible to construct the bulk correspondence of the triality defects as the twist defects of $S_{T_1}$ and $S_{T_2}$. Furthermore, following \cite{Kaidi:2022cpf}, it is possible to compute the corresponding $F$-symbols of the triality fusion categories. However, we will not pursue this direction and leave this for future study. 

\subsection{Transformation on the partition functions}\label{sec:trans_part}
In this subsection, we establish the connection between the bulk symmetry and the transformation on the torus partition functions. Generally speaking, the theory $\scriptX$ would admit a triality defect if it is invariant under some order-$3$ twisted gauging $\mathcal{T}$ of $\doubleZ_N \times \doubleZ_N$. However, it remains to determine the specific fusion category which, for $N = p^r$, involves two pieces of data $(T_i, \epsilon)$, where $T_i$ ($i=1,2$) denotes the choice of bulk $\mathbb{Z}_3$ symmetry generator given in \eqref{eq:Z3_sym} and $\epsilon = e^{\frac{2\pi i}{3}k}$ ($k=0,1,2$) represents a choice of $H^3(\mathbb{Z}_3,U(1)) \simeq \mathbb{Z}_3$. Knowing the corresponding bulk symmetry $T$ of the order-$3$ twisted gauging $\mathcal{T}$ allows us to determine the data $T_i$. 

As a warm-up to set the stage, we first derive the transformation $\mathcal{T}_i$ on the boundary theory induced by two $\doubleZ_3$ symmetries $T_i$ in the bulk. In the SymTFT, one can represent the boundary QFT $\scriptX$ as a vector
\begin{equation}
    \langle \scriptX | = \sum_{a_i \in H^1(\Sigma, \doubleZ_N)} \langle a_1, a_2| \scriptZ_{\scriptX}[a_1,a_2] ~,
\end{equation}
where $Z_{\scriptX}[a_1,a_2]$ is the partition function of $\scriptX$ coupled to background field $a_1$ and $a_2$, and the states $|a_1,a_2\rangle$'s for SymTFT are normalized as
\begin{equation}
    \langle b_1 ,b_2| a_1, a_2\rangle = \delta(a_1 - b_1) \delta(a_2 - b_2) ~.
\end{equation}
The simple anyons act on them as
\begin{equation}
    e_i(\gamma)|a_1,a_2\rangle = e^{\frac{2\pi i}{N}\langle \gamma, [a_i]\rangle }|a_1, a_2\rangle ~, \quad m_i(\gamma)|a_1, a_2\rangle = |a_1 - \delta_{i,1} [\gamma], a_2 - \delta_{i,2}[\gamma]\rangle ~,
\end{equation}
where $[a_i] \in H_1(\Sigma, \doubleZ_N)$ is the Poincare dual of $a_i$ while $[\gamma] \in H^1(\Sigma, \doubleZ_N)$ is the Poincare dual of $\gamma$. In our convention, the Dirichlet boundary conditions realizing $\doubleA = \doubleZ_N \times \doubleZ_N$ symmetries are chosen to diagonalize all the electric lines, therefore
\begin{equation}
    |D(A_1, A_2)\rangle = |A_1, A_2\rangle ~, \quad \scriptZ_{\scriptX}[A_1,A_2] = \langle \scriptX|D(A_1,A_2)\rangle ~.
\end{equation}
The action of the $\doubleZ_3$ symmetry $T_i$ on the partition function can be evaluated using the condensation defects $S_{T_i}(\Sigma)$ as
\begin{equation}
    \scriptT_i: \scriptZ_{\scriptX}[A_1, A_2] \mapsto \langle \scriptX|S_{T_i}(\Sigma)|D(A_1,A_2)\rangle ~.
\end{equation}
And we find
\begin{equation}\label{eq:4.19}
\begin{aligned}
    &\scriptT_1: \scriptZ_{\scriptX}[A_1, A_2] \mapsto \frac{1}{|H^1(\Sigma,\doubleZ_N)|}\sum_{a_i \in H^1(\Sigma,\doubleZ_N)}\scriptZ_{\scriptX}[a_1, a_2]e^{\frac{2\pi i}{N}\int a_1 \cup A_2 - a_2 \cup A_1 + A_1 \cup A_2} ~, \\
    &\scriptT_1^2: \scriptZ_{\scriptX}[A_1, A_2] \mapsto \frac{1}{|H^1(\Sigma,\doubleZ_N)|} \sum_{a_i \in H^1(\Sigma,\doubleZ_N)}\scriptZ_{\scriptX}[a_1, a_2]e^{-\frac{2\pi i}{N}\int a_1 \cup a_2 + a_1 \cup A_2 - a_2 \cup A_1} ~, \\
    &\scriptT_1^3: \scriptZ_{\scriptX}[A_1,A_2] \mapsto \scriptZ_{\scriptX}[A_1, A_2] ~,
\end{aligned}
\end{equation}
and 
\begin{equation}\label{eq:T2_pf}
\begin{aligned}
    & \scriptT_2: \scriptZ_{\scriptX}[A_1, A_2] \mapsto \frac{1}{|H^1(\Sigma,\doubleZ_N)|} \sum_{a_i \in H_1(\Sigma, \doubleZ_N)} \scriptZ_{\scriptX}[a_1, a_2] e^{\frac{2\pi i}{N}\int a_1 \cup (A_1 -A_2) + a_2 \cup A_2 + A_1 \cup A_2} ~, \\
    & \scriptT_2^2: \scriptZ_{\scriptX}[A_1,A_2] \mapsto \frac{1}{|H^1(\Sigma,\doubleZ_N)|} \sum_{a_i \in H_1(\Sigma,\doubleZ_N)} \scriptZ_{\scriptX}[a_1,a_2] e^{-\frac{2\pi i}{N} \int a_1 \cup a_2 - a_1 \cup A_1 + a_2 \cup (A_1 - A_2)} ~, \\
    & \scriptT_2^3: \scriptZ_{\scriptX}[A_1,A_2] \mapsto \scriptZ_{\scriptX}[A_1,A_2] ~.
\end{aligned}
\end{equation}
Indeed, we find the transformation corresponding $\scriptT_i$ is of order $3$ on a generic theory $\scriptX$. In particular, for $N = 2$, the transformation $(\mathcal{T}_1)^2$ is the twisted gauging considered in \cite{Thorngren:2019iar,Thorngren:2021yso}. If a theory $\scriptX$ is invariant under the triality transformation, we require that
\begin{equation}
    \langle \scriptX| S_{T_1}(\Sigma) = \langle \scriptX| ~,
\end{equation}
this is equivalent to for any Dirichlet state $|D(A_1,A_2)\rangle$, we have
\begin{equation}
    \langle \scriptX| S_{T_1}(\Sigma)|D(A_1,A_2)\rangle = \langle \scriptX|D(A_1,A_2)\rangle ~,
\end{equation}
and we find for $S_{T_1}$ and $S_{T_2}$ respectively
\begin{equation}\label{eq:self_triality_part}
\begin{aligned}
    & \scriptZ_{\scriptX}[A_1, A_2] = \frac{1}{|H^1(\Sigma,\doubleZ_N)|}\sum_{a_i \in H^1(\Sigma,\doubleZ_N)}\scriptZ_{\scriptX}[a_1, a_2]e^{\frac{2\pi i}{N}\int a_1 \cup A_2 - a_2 \cup A_1 + A_1 \cup A_2} ~, \\
    &\scriptZ_{\scriptX}[A_1, A_2] = \frac{1}{|H^1(\Sigma,\doubleZ_N)|} \sum_{a_i \in H^1(\Sigma, \doubleZ_N)} \scriptZ_{\scriptX}[a_1, a_2] e^{\frac{2\pi i}{N}\int a_1 \cup (A_1 - A_2) + a_2 \cup A_2 + A_1 \cup A_2} ~.
\end{aligned}
\end{equation}
At this stage, one might think that it is possible to check whether a theory $\scriptX$ admits a triality defect by checking if its partition function satisfies any of the two equations in \eqref{eq:self_triality_part}. We want to emphasize here an important subtlety in the procedure, as mentioned in Section \ref{sec:G_ext_general}. Namely, if we fix the choice of the twisted partition function $Z_{\scriptX}[A_1, A_2]$, then the classification of the triality defects do not uniquely determine the transformation on the partition functions. 

Notice that such a phenomenon already appears in the $\TY$ fusion category. For instance, let's consider $\doubleA = \doubleZ_2 \times \doubleZ_2$, then it is said that there are two inequivalent choices of symmetric, non-degenerate bicharacters, the diagonal one and the off-diagonal one, and in terms of the matrices, we have
\begin{equation}
    D_{\chi_d} = \begin{pmatrix} 0 & 0 & 1 & 0 \\ 0 & 0 & 0 & 1 \\ 1 & 0 & 0 & 0 \\ 0 & 1 & 0 & 0\end{pmatrix} ~, \quad D_{\chi_{od}} = \begin{pmatrix} 0 & 0 & 0 & 1 \\ 0 & 0 & 1 & 0 \\ 0 & 1 & 0 & 0 \\ 1 & 0 & 0 & 0\end{pmatrix} ~.
\end{equation}
Actually, there are really four symmetric, non-degenerate bicharacters, and the additional two correspond to the bulk symmetries given by
\begin{equation}
    D_{\chi_{d}'} = \begin{pmatrix} 0 & 0 & 1 & 1 \\ 0 & 0 & 1 & 0 \\ 0 & 1 & 0 & 0 \\ 1 & 1 & 0 & 0\end{pmatrix} ~, \quad D_{\chi_{d}''} = \begin{pmatrix} 0 & 0 & 0 & 1 \\ 0 & 0 & 1 & 1 \\ 1 & 1 & 0 & 0 \\ 1 & 0 & 0 & 0\end{pmatrix} ~,
\end{equation}
and both of them are related to $D_{\chi_{d}}$ by conjugating with some element in $I_{\doubleZ_2 \times \doubleZ_2}$ respectively. Hence, at the level of classification, they are considered equivalent to $D_{\chi_{d}}$. But $D_{\chi_d}, D_{\chi_d}', D_{\chi_d}''$ will lead to distinct $\doubleZ_2 \times \doubleZ_2$ gauging for the boundary theories:
\begin{equation}
\begin{aligned}
    & \mathcal{D}_{\chi_d}: Z_{\scriptX}[A_1,A_2] \rightarrow \frac{1}{|H^1(\Sigma,\doubleZ_2)|} \sum_{a_1, a_2} Z_{\scriptX}[a_1,a_2] (-1)^{\int a_1 \cup A_1 + a_2 \cup A_2} ~, \\
    & \mathcal{D}_{\chi_d'}: Z_{\scriptX}[A_1,A_2] \rightarrow \frac{1}{|H^1(\Sigma,\doubleZ_2)|} \sum_{a_1, a_2} Z_{\scriptX}[a_1,a_2] (-1)^{\int a_1 \cup A_2 + a_2 \cup (A_1 + A_2)} ~, \\
    & \mathcal{D}_{\chi_d''}: Z_{\scriptX}[A_1,A_2] \rightarrow \frac{1}{|H^1(\Sigma,\doubleZ_2)|} \sum_{a_1, a_2} Z_{\scriptX}[a_1,a_2] (-1)^{\int a_1 \cup (A_1 + A_2) + a_2 \cup A_1} ~.
\end{aligned}
\end{equation}
The theory $\scriptX$ invariant under the $D_{\chi_d}, D_{\chi_d'}, D_{\chi_d''}$ respectively admits $\TY$ fusion category with $F$-symbols that depends on the explicit form of the bicharacters $\chi_d,\chi_d',\chi_d''$, therefore leads to distinct gauging transformation on the twisted partition function. But because the three sets of $F$-symbols relate to each other via automorphism of $\doubleZ_2 \times \doubleZ_2$, they are considered equivalent. \footnote{A non-example is the $\TY(\doubleZ_3, \chi_k, \epsilon)$, where $\chi_k(\eta^a,\eta^b) = e^{\frac{2\pi i}{3} k ab}$. $k = 1,-1$ are two inequivalent bicharacters, therefore lead to inequivalent TY fusion categories. Notice that both bicharacter is invariant under the charge conjugate $\eta^a \rightarrow \eta^{-a}$ as $\chi_k(\eta^a,\eta^b) = \chi_k(\eta^{-a},\eta^{-b})$. This is the only non-trivial automorphism of $\doubleZ_3$, which then implies $\chi_1$ and $\chi_2$ parameterizes different TY fusion categories.}

To illustrate this on the triality defects, let's consider $T_2$ conjugated by the following symmetry $V$:
\begin{equation}
    V = \begin{pmatrix} 1 & 0 & 0 & 0 \\ 0 & 1 & 0 & 0 \\ 0 & -1 & 1 & 0 \\ 1 & 0 & 0 & 1 \end{pmatrix}, \quad T_2' = V^{-1}T_2 V = \begin{pmatrix}
 0 & 0 & 1 & 0 \\
 0 & 0 & 1 & 1 \\
 1 & -1 & 1 & 1 \\
 0 & 1 & -1 & 0 \\
\end{pmatrix}.
\end{equation}
$T_2'$ corresponds to equivalent triality category as $T_2$, and but $T_2'$ action on the partition functions differently. To see this, let's first work out the condensation defect
\begin{equation}
    S_{T_2'}(\Sigma) = \frac{1}{|H^1(\Sigma,\doubleZ_N)|} \sum_{\gamma_i \in H_1(\Sigma,\doubleZ_N)} e^{\frac{2\pi i}{N}\langle \gamma_1,\gamma_2\rangle} m_1(\gamma_1) m_2(\gamma_1 + \gamma_2) e_1(\gamma_2) e_2(-\gamma_1 - \gamma_2) ~,
\end{equation}
and the actions on the partition functions are given by
\begin{equation}
\begin{aligned}
    \scriptT_2' & : \scriptZ_{\scriptX}[A_1,A_2] \mapsto \frac{1}{|H^1(\Sigma,\doubleZ_N)|}  \sum_{a_i \in H^1(\Sigma,\doubleZ_N)}Z_{\scriptX}[a_1,a_2] e^{\frac{2\pi i}{N}\int a_1 \cup a_2 + a_1\cup (A_1 - A_2) + a_2 \cup A_2} ~, \\
    (\scriptT_2')^2 & : \scriptZ_{\scriptX}[A_1,A_2] \mapsto \frac{1}{|H^1(\Sigma,\doubleZ_N)|} \sum_{a_i \in H^1(\Sigma,\doubleZ_N)}Z_{\scriptX}[a_1,a_2] e^{\frac{2\pi i}{N}\int a_1 \cup A_1 + a_2\cup (A_2 - A_1) - A_1 \cup A_2} ~, \\
    (\scriptT_2')^3 & : \scriptZ_{\scriptX}[A_1,A_2] \mapsto \scriptZ_{\scriptX}[A_1,A_2] ~.
\end{aligned}
\end{equation}
Compared with \eqref{eq:T2_pf}, indeed, we see $\scriptT_2'$ or $(\scriptT_2')^2$ acts differently compared with $\scriptT_2$, even though they are equivalent at the level of classification. 

To conclude, if a theory $\scriptX$ is invariant under any order-$3$ operation $\mathcal{T}$ which gauges the $\doubleZ_N \times \doubleZ_N$ symmetry, then by the half-gauging argument, one knows that the theory $\scriptX$ admits a triality defect. Once a concrete $\mathcal{T}$ is known, it is straightforward to find the corresponding $\doubleZ_3$ symmetry generator $T$ in the bulk. Then, there exists $V \in \mathcal{I}_{\doubleZ_N \times \doubleZ_N}$ such that either $V T V^{-1} = T_1$ or $V T V^{-1} = T_2$. The corresponding $T_i$ is the one appearing in the classification data for the corresponding triality defect related to $\mathcal{T}$.

In practice, when computing the bulk symmetry $T$ from a given twisted gauging $\scriptT$, it is convenient to decompose the $\scriptT$ as first stacking $\doubleZ_N \times \doubleZ_N$ SPT, then gauge $\doubleZ_N \times \doubleZ_N$ with no discrete torsion, and then applying automorphism of the dual symmetries and finally stack certain SPTs for the dual symmetries. The symmetry $T$ is then a product of the symmetries corresponding to the above steps, and the latter is easy to write down. Namely, there are 3 types of symmetry in the bulk, 
\begin{equation}\label{eq:ZN_sym_generators}
    s_i = \begin{pmatrix}
        I-J & J \\ J & I-J
    \end{pmatrix},\quad t^n  = \begin{pmatrix}
        I & 0 \\ \begin{pmatrix}
            0 & n \\ -n & 0
        \end{pmatrix} & I
    \end{pmatrix}, \quad r_U = \begin{pmatrix}
      U   &  0 \\ 0 & (U^{-1})^T
    \end{pmatrix}
\end{equation}
where $I$ is the identity matrix, $J_{i,i}=1$ and $0$ otherwise. $s_i$ is the EM duality between the $i$-th $\IZ_N$ and corresponds to gauging the $i$-th $\IZ_N$ symmetry at the boundary. $t^1$ corresponds to stacking an $\IZ_N\times \IZ_N$ SPT. Lastly, $r_U$ corresponds to the automorphism of the boundary $\IZ_N\times \IZ_N$ global symmetry, given by $A\rightarrow U^T A$. For example,
\begin{align}\label{eq:T1T2_decomp}
    T_1 = s_1 s_2 r_{U_1} t^1 ,\quad    T_2 =  s_1 s_2 r_{U_2} t^1,\quad     T_2' = t^1 s_1 s_2 r_{U_2}  
\end{align}
where $U_1 = \left(\begin{smallmatrix}      0 & 1 \\ -1 & 0   \end{smallmatrix}\right)$, and $U_2= \left(\begin{smallmatrix}      1 & -1 \\ 0 & 1     \end{smallmatrix}\right)$.  

On the other hand, in order to rule out the possibility of a theory admitting a triality defect, one may consider fixed the twisted partition function to be some $Z_{\scriptX}[A_1,A_2]$ and check with every order-$3$ twisted gauging of the $\doubleZ_N \times \doubleZ_N$-symmetry will change $Z_{\scriptX}[A_1,A_2]$. Equivalently, one could consider fixed the form of the order-$3$ topological manipulations to be \eqref{eq:self_triality_part}, and check with all possible alternative twisted partition functions of the form $e^{\frac{2\pi i m}{N}\int A_1 \cup A_2 }Z_{\scriptX}[\alpha(A_1),\alpha(A_2)]$ where $m \in \doubleZ_N$ labels the possible 2d $\doubleZ_N \times \doubleZ_N$-SPT and $\alpha$ is an automorphism of $\doubleZ_N \times \doubleZ_N$.

\subsection{Analysis of the anomaly}\label{sec:pr_anomaly}
In this subsection, we demonstrate how to combine the technique in \cite{Zhang:2023wlu} and the group-theoretical fusion category to determine the anomaly of the triality fusion category with $\doubleZ_N \times \doubleZ_N$ for $N$ coprime with $3$. Our main result is that in the prime decomposition of $N$ \footnote{Notice that no $p_i = 3$ as $N$ is coprime with $3$.}
\begin{equation}
    N = \prod_{i=1}^n p_i^{r_i} ~, 
\end{equation}
for the triality fusion category specified by $\left(\{T^{(i)}\}, \alpha \right)$ where $T^{(i)} = T_1, T_2$, it is anomaly-free (that is, admits at least a fiber functor) if and only if the following two conditions are satisfied:
\begin{enumerate}
    \item $p_i = 1 \mod 3$ for all prime factors $p_i$ appearing in the decomposition of $N$;
    \item the FS indicator $\alpha = 1$.
\end{enumerate}
In the following, we will prove the above claim for the special case $N = p^r$ (where $p$ is a prime number other than $3$) to demonstrate our approach. This is a building block for the proof of the result for generic $N$ coprime with $3$, and we will leave the proof of the generic case to the Appendix \ref{app:general_proof}.

For $N = p^r$, the two classes of triality defect correspond to gauging $\doubleZ_3$ symmetry generated by $T_1$ and $T_2$ respectively. Our strategy to determine the anomaly-free triality symmetry is to first rule out a large class of theories by checking if the SymTFT contains a stable magnetic Lagrangian algebra \cite{Zhang:2023wlu}. Only the theories that satisfy this condition could be anomaly-free, and this also implies the symmetry is group-theoretical \cite{Gelaki:2009blp,Sun:2023xxv}. Then, we proceed to write the triality category in the form $\scriptC(\doubleG, \omega, \doubleH,\psi)$ with the help of the stable magnetic Lagrangian algebra we find, and use the criterion \eqref{eq:anomaly_free_criteria} to determine the anomaly by explicitly checking the existence of the fiber functor. Notice that \eqref{eq:anomaly_free_criteria} can also be used to classify all the fiber functors, but for generic $N$ this is rather complicated. We will consider some special cases in the end.

Let's first consider the $T_1$-symmetry. There are only $N$ magnetic Lagrangian algebras in the SymTFT, generated by the following two lines
\begin{equation}
    m_1 e_2^x ~, \quad m_2 e_1^{-x} ~,
\end{equation}
for some $x \in \doubleZ_N$.
Under the action of $T_1$-symmetry, the two generators of the Lagrangian algebra are mapped to
\begin{equation}
    T_1: m_1 e_2^x \mapsto m_1^{x-1} e_2^{-1} ~, \quad m_2 e_1^{-x} \mapsto m_2^{-x+1}e_1 ~.
\end{equation}
To check the Lagrangian algebra is $T_1$-stable, i.e., invariant under $T_1$ action, we only need to check the image of the two generators are spanned by the two generators. Here, this is the case if and only if $m_1 e_2^x$ is generated by $m_1^{x-1} e_2^{-1}$, and $m_2 e_1^{-x}$ is generated by $m_2^{-x+1}e_1$. This is the case if and only if $x$ solves
\begin{equation}
    x(x-1) = -1 \mod p^r \Longleftrightarrow x^2 - x + 1 = 0 \mod p^r ~.
\end{equation}
The equation $x^2 - x + 1 = 0 \mod p^r$ has solution if and only if $p = 1 \mod 3$\footnote{We first notice that $x^2 - x + 1 = 0$ has solutions mod $p^r$ if and only if $x^2 - x + 1 = 0$ has solutions mod $p$. The ``only if'' direction is not hard to see, as if $x^2 - x + 1$ divides $p^r$, then it must divide $p$. To see the ``if'' direction, let's notice that for $p = 2$, this is trivially true because $x^2-x+1 = x(x-1) + 1$ is an odd number, therefore can not equal $0$ mod $2^r$ for any $r$. Then, for odd $p$, one can prove this in a recursive sense by showing if there is a $x_m$ such that $x_m^2 - x_m + 1$ divides $p^m$, then $x_{m+1} = x_m - (x_m^2 - x_m + 1) y_m$ divides $p^{m+1}$, where $y_m$ is a solution to $2 x_m y_m = 1 \mod p^m$. Notice that $y_m$ always exists because $x_m$ is coprime with $p^m$ as $x_m(1 - x_m) = 1 \mod p^m$. Thus, the existence of the solution to $x^2 - x + 1 = 0 \mod p$ would imply the existence of the solution to $x^2 - x + 1 = 0 \mod p^r$ for any $r \in \doubleZ_{>0}$. 

Now, restricting to solve $x^2 - x + 1 = 0 \mod p$ where $p$ is an odd prime. Assume $x$ is a solution, because $(-x)^3 - 1 = -(x+1)(x^2 - x + 1) = 0$, $-x$ generates a non-trivial order-$3$ subgroup of the multiplicative group modulo $p$, $\doubleZ_p^{\times} = \{1,\cdots, p-1\}$. The latter is isomorphic to the cyclic group of order $p-1$. For $p = -1 \mod 3$, $\doubleZ_p^{\times}$ does not contain an order-$3$ subgroup, hence the solution can not exist. When $p = 1 \mod 3$, $\doubleZ_p^\times$ does contain an order-$3$ subgroup, say generated by $a$ where $a^3 = 1 \mod p$ and $a \neq 1$. Then, $-a$ and $a^2$ are two solutions to $x^2 - x + 1 = 0 \mod p$.}. Hence, we immediately conclude that for $p = 2 \mod 3$ in $N = p^r$, the triality from $T_1$ is anomalous. 

To determine the case where $p = 1 \mod 3$, we use the group-theoretical fusion category techniques. Actually, for $T_1$-type of triality defect, it is group-theoretical regardless of $p = \pm 1 \mod 3$. To determine the data $(\doubleG,\omega,\doubleH,\psi)$, we observe that under the conjugation of the bulk symmetry $V_1$, $(V_1)^{-1} T_1 V_1$ takes the form of the automorphism of the boundary $\doubleZ_N \times \doubleZ_N$ symmetry:
\begin{equation}
    V_1 = \begin{pmatrix} 0 & 0 & 1 & 0 \\ 0 & 1 & 0 & 0 \\ 1 & 0 & 0 & 0 \\ 0 & 0 & 0 & 1 \end{pmatrix} ~, \quad (V_1)^{-1} T_1 V_1 = \begin{pmatrix} 0 & 1 & 0 & 0 \\ -1 & -1 & 0 & 0 \\ 0 & 0 & -1 & 1 \\ 0 & 0 & -1 & 0 \end{pmatrix} ~.
\end{equation}
Since $V_1$ is the $\doubleZ^{em}_2$-symmetry for the pair $(e_1, m_1)$, the above relation implies that after gauging $\doubleZ_N$ symmetry, the triality defect on the boundary becomes an invertible $\doubleZ_3$-symmetry acting as outer automorphism of the quantum $\doubleZ_N^a \times \doubleZ_N^b$ symmetry as $a \mapsto a^{-1} b^{-1}, b \mapsto a$.

The same statement can be checked at the level of the partition function of the boundary theory. Given a theory $\scriptX$ invariant under the twisted gauging $\mathcal{T}_1$, 
\begin{equation}\label{eq:T1_invar_part}
    \scriptZ_{\scriptX}[A_1, A_2] = \frac{1}{|H^1(\Sigma,\doubleZ_N)|}\sum_{a_i \in H^1(\Sigma,\doubleZ_N)}\scriptZ_{\scriptX}[a_1, a_2]e^{\frac{2\pi i}{N}\int a_1 \cup A_2 - a_2 \cup A_1 + A_1 \cup A_2} ~.
\end{equation}
Gauging the $\doubleZ_N$ symmetry represented by the background field $A_1$ leads to the theory $\scriptX/\doubleZ_N^{(1)}$, whose partition function is given by
\begin{equation}
    \scriptZ_{\scriptX/\doubleZ_N^{(1)}}[A_1, A_2] = \frac{1}{\sqrt{|H^1(\Sigma,\doubleZ_N)|}} \sum_{b \in H^1(\Sigma,\doubleZ_N)} \scriptZ_{\scriptX}[b, A_2] e^{-\frac{2\pi i}{N} \int b \cup A_1} ~.
\end{equation}
Using \eqref{eq:T1_invar_part}, one can show that
\begin{equation}\label{eq:T1_gauged_Z3_autom}
    \scriptZ_{\scriptX/\doubleZ_N^{(1)}}[A_1, A_2] = \scriptZ_{\scriptX/\doubleZ_N^{(1)}}[-A_1 + A_2, - A_1] ~.
\end{equation}
Reversing this, we conclude that the triality fusion category can be acquired from gauging the $\doubleZ_N^\ra$ subgroup in the symmetry $(\doubleZ_N^\ra \times \doubleZ_N^\rb) \rtimes \doubleZ_3^\rc$ captures the symmetry of $\scriptX/\doubleZ_N^{(1)}$ \eqref{eq:T1_gauged_Z3_autom}, where
\begin{equation}
    (\doubleZ_N^\ra \times \doubleZ_N^\rb) \rtimes \doubleZ_3^{\rc} = \langle \ra,\rb,\rc | \ra^N = \rb^N = \rc^3 = 1, \ra\rb=\rb\ra, \rc\ra\rc^{-1} = \ra^{-1}\rb^{-1}, \rc\rb\rc^{-1} = \ra\rangle ~.
\end{equation}
Different FS indicator $\alpha = e^{\frac{2\pi i \kappa}{3}}$ can be engineered by taking the anomaly
\begin{equation}
    \omega_{\kappa}(\ra^{i_1}\rb^{j_1}\rc^{k_1},\ra^{i_2}\rb^{j_2}\rc^{k_2},\ra^{i_3}\rb^{j_3}\rc^{k_3}) = e^{\frac{2\pi i \kappa}{9}k_1(k_2 + k_3 - [k_2 + k_3]_3)} ~, \quad \kappa = 0,1,2 ~.
\end{equation}
In other words, the triality fusion category from $T_1$ is of the form $\mathcal{C}((\doubleZ_N^\ra \times \doubleZ_N^\rb) \rtimes \doubleZ_3^{\rc}, \omega_{\kappa}, \doubleZ_N^\ra, 1)$. To apply the anomaly-free criterion from group-theoretical fusion category \eqref{eq:anomaly_free_criteria}, one first searches for an anomaly-free subgroup $\doubleK$ of $\doubleG =  (\doubleZ_N^\ra \times \doubleZ_N^\rb) \rtimes \doubleZ_3$, such that $(\doubleZ_N^\ra \times \doubleZ_N^\rb) \rtimes \doubleZ_3$ is contained in $\doubleH \doubleK$. This would imply that for any $\ra^i \rb^j \rc^k \in \doubleG$, we can find a $\ra^l \in \doubleH = \doubleZ_N^a$, such that
\begin{equation}
    a^{i-l} b^j c^k \in \doubleK ~.
\end{equation}

Consider $c \in \doubleG$, for instance, then we must have 
\begin{equation}
    a^l c \in \doubleK, \quad \text{for some} \quad l ~.
\end{equation}
It is straightforward to check that $a^l c$ forms a order-$3$ subgroup $\doubleZ_3^{a^lc} \subset \doubleK$, whose anomaly is captured by $\omega_\kappa$. This means the $T_1$-type triality is anomalous when the FS indicator (which characterizes the self-anomaly $\doubleZ_3^c$) is non-trivial, since $\doubleK$ must be anomaly-free. When $\omega_\kappa = 1$, it is straightforward to find an anomalous free subgroup $\doubleK$ meet all the conditions in the criterion given in \eqref{eq:anomaly_free_criteria} as $\doubleH\cap \doubleK = \{\dsi\}$ for $p = 1 \mod 3$, which is generated by $h_1 = a^{-m}b$ and $c$
\begin{equation}
    \doubleK = \langle h_1, c| c h_1 c^{-1} = (h_1)^m\rangle \subset \doubleG ~,
\end{equation}
where $m$ satisfies $m^2 + m + 1 = 0 \mod N$. $\doubleH\cap \doubleK = \{1\}$ can be seen from the facts that $\doubleG \subset \doubleH \doubleK$ and $\doubleK$ has order $3N$.

Hence, for the triality corresponding to $T_1$ where $N = p^r$, it is anomaly-free if and only if $p = 1 \mod 3$ and the FS indicator $\alpha = 1$.

\

Next, let's analyze triality defects coming from $T_2$. First, we check the existence of the stable magnetic Lagrangian algebras. The $T_2$ action on the generators of the magnetic Lagrangian algebras is given by
\begin{equation}
    T_2: m_1 e_2^x \mapsto m_2^{x-1} e_1 ~, \quad m_2 e_1^{-x} \mapsto (m_1^{-x+1} e_2)(m_2^{-x+1} e_1^{-1}) ~.
\end{equation}
Again, we find that the magnetic Lagrangian algebra is $T_2$ stable, if and only if $x$ solves the following equation
\begin{equation}
    x^2 - x + 1 = 0 \mod p^r ~.
\end{equation}
Again, such $x$ only exists when $p = 1  \mod 3 $. This implies that for $p = 2 \mod 3$, the triality fusion category corresponding to $T_2$ is anomalous. And when $p \mod 3 = 1$, we could write the corresponding triality fusion category in terms of $\scriptC(\doubleG, \omega, \doubleH, \psi)$, which allows us to determine if it is anomalous from the group-theoretical techniques.

To find $\doubleG, \omega, \doubleH, \psi$, let's start with a theory with partition function $\scriptZ_{\scriptX}[A_1,A_2]$ such that 
\begin{equation}\label{eq:T2X_PF}
    \scriptZ_{\scriptX}[A_1,A_2] = \frac{1}{|H^1(\Sigma,\doubleZ_N)|} \sum_{a_i \in H^1(\Sigma,\doubleZ_N)} \scriptZ_{\scriptX}[a_1,a_2] e^{\frac{2\pi i}{N}\int a_1 \cup (A_1 - A_2) + a_2 \cup A_2 + A_1 \cup A_2} ~.
\end{equation}
Consider the following discrete gauging of the theory $\scriptX$ to get a new theory $\widetilde{\scriptX}$, 
\begin{equation}\label{eq:T2groupth_trans}
    \scriptZ_{\widetilde{\scriptX}}[A_1, A_2] = \frac{1}{|H^1(\Sigma,\doubleZ_N)|} \sum_{a_i \in H^1(\Sigma, \doubleZ_N)} \scriptZ_{\scriptX}[a_1, a_2] e^{\frac{2\pi i}{N} \int - x a_1 \cup a_2  + a_1 \cup (A_1 - A_2) - a_2 \cup (x A_1 - (1 - x) A_2) + A_1 \cup A_2} ~,
\end{equation}
where again $x$ satisfies $x^2 - x + 1 = 0 \mod N$. The self-triality of the theory $\scriptX$ then implies the theory $\widetilde{\scriptX}$ satisfies
\begin{equation}
    Z_{\widetilde{\scriptX}}[A_1, A_2] = Z_{\widetilde{\scriptX}}[A_1, -(1-x) A_2] ~.
\end{equation}
Hence, we conclude that if we twisted gauge the $\doubleZ_N \times \doubleZ_N$, the triality defect becomes an invertible symmetry which acts as a $\doubleZ_3$ automorphism on the dual $\doubleZ_N \times \doubleZ_N$-symmetries. The inverse transformation of \eqref{eq:T2groupth_trans} is given by
\begin{equation}\label{eq:T2_trans_inver}
    \scriptZ_{\scriptX}[A_1,A_2] = \frac{1}{|H^1(\Sigma,\doubleZ_N)|} \sum_{b_i \in H^1(\Sigma,\doubleZ_N)} \scriptZ_{\widetilde{\scriptX}}[b_1,b_2] e^{\frac{2\pi i}{N}\int - a_1 \cup a_2 + a_1 \cup (A_1 - x A_2) + a_2 \cup (- A_1 + (1 - x) A_2) + x A_1 \cup A_2} ~.
\end{equation}
From the above, we could determine the $T_2$-type triality fusion category can be written as $\scriptC(\doubleZ_N^\ra \times (\doubleZ_N^\rb \rtimes \doubleZ_3^\rc), \omega_k, \doubleZ_N^\ra \times \doubleZ_N^\rb, \psi)$ where 
\begin{equation}
    \doubleZ_N^{\ra} \times (\doubleZ_N^\rb \rtimes \doubleZ_3^\rc) = \langle \ra,\rb,\rc| \ra^N = \rb^N = \rc^3 = 1, \ra\rb = \rb\ra, \ra\rc = \rc\ra, \rc\rb\rc^{-1} = \rb^{-(1-x)}\rangle ~,
\end{equation}
and
\begin{equation}\label{eq:T2_discrete_torsion}
    \psi(\ra^{i_1}\rb^{j_1},\ra^{i_2}\rb^{j_2}) = e^{-\frac{2\pi i}{N} j_1 i_2} ~.
\end{equation}
Here, the non-trivial $2$-cocycle $\psi$ guarantees the remnant of $\rc$ after gauging $\doubleZ_N^\ra \times \doubleZ_N^\rb$ is a single non-invertible triality defect, rather than $N^2$-invertible symmetry defects.

Notice that naively the transformation \eqref{eq:T2_trans_inver} is more than just gauging $\doubleZ_N^\ra \times \doubleZ_N^{\rb}$ with discrete torsion, it also contains applying the outer automorphism of the dual $\doubleZ_N \times \doubleZ_N$ symmetry then followed by stacking an SPT. While the latter operations will modify the $F$-symbol such that the resulting global triality transformation takes the form of \eqref{eq:T2_pf}, it does not change the equivalence class of the triality fusion category.

For the theory to be anomaly-free, we must be able to find a subgroup $\doubleH_1$ such that $\doubleZ_N^a \times (\doubleZ_N^b \rtimes \doubleZ_3^c) \subset \doubleH \doubleK$ where $\doubleH = \doubleZ_N^a \times \doubleZ_N^b$. This implies $a^i b^j c \in \doubleK$ for some $i,j \in \doubleZ_N$. First, we observe that $b^j c$ is an order-$3$ element. Then taking the $N$-th power of $a^i b^j c$, we see $(b^j c)^N$ is a non-trivial element as $\gcd(N,3) = 1$, hence $\doubleZ_3^{b^j c} \subset \doubleK$. Thus, we immediately see that the FS indicator must be trivial for the $T_2$-type triality to be anomaly-free, since $\doubleK$ must be anomaly-free. And when this is the case, taking $\doubleK = \doubleZ_3^c$, we confirm there exists a fiber functor for this triality fusion category, and the theory is anomaly-free. 

To summarize, for the triality corresponding to $T_1$ and $T_2$, for $N = p^r$, it is anomaly-free if and only if $p = 1 \mod 3$ and the FS indicator $\alpha = 1$. It is rather straightforward to generalize this proof to the case of generic $N$ where $\gcd(N,3) = 1$, which we include in the appendix. 

\

To conclude, we consider the special case where $N = 7$. In this case, both triality defects with $\alpha = 1$ admit fiber functors. Furthermore, one can easily compute all the fiber functors in both cases.

For $T_1$, the group-theoretical construction is given by
\begin{equation}
    \doubleG = (\doubleZ_7^{\ra} \times \doubleZ_7^{\rb}) \rtimes \doubleZ_3 = \langle \ra,\rb,\rc | \ra^7 = \rb^7 = \rc^3 = 1, \ra \rb=\rb\ra, \rc\ra\rc^{-1} = \ra^{-1}\rb^{-1}, \rc\rb\rc^{-1} = \ra\rangle
\end{equation}
while $\doubleH = \doubleZ_7^{\ra}$. For the choice of $\doubleK$, in order for $\doubleH \doubleK$ to contain the entire $\doubleG$, $\doubleK$ can only be $\langle \rc, \ra^{-2}\rb\rangle, \langle \rc, \ra^3 \rb\rangle, \doubleG$. The last option is not possible, as $\doubleH \cap \doubleK \simeq \doubleZ_{7}$ which always has $7$ irreducible representations. Physically, this means if we start with a trivially gapped phase with $\doubleG$ symmetry and gauge $\doubleZ_7^\ra$ symmetry, we will find $7$ ground states degeneracy. For the rest two choices of $\doubleK$, $\doubleK \cap \doubleH = \{\dsi\}$ therefore there is a unique ground state (for any choice of $\psi_1$). However, given the form of $\doubleK$ (which both are equivalent to $\doubleZ_7 \rtimes \doubleZ_3$), there is no non-trivial choice of $\psi_1$. Hence, we conclude that for $T^1$, there are two fiber functors specified by $\doubleK = \langle \rc, \ra^{-2}\rb\rangle, \langle \rc, \ra^3 \rb\rangle$.

The $\doubleG$-symmetric gapped phase corresponding to $\doubleK = \langle \rc, \ra^{-2}b\rangle$ is a partial SSB phase where $\langle \rc, \ra^{-2}\rb\rangle$ remains unbroken. Its twisted torus partition function is given by
\begin{equation}
    \widetilde{Z}_{\langle \rc, \ra^{-2}\rb\rangle}[A,B,C] = 7 \delta(A+2B) ~.
\end{equation}
Gauging $\doubleZ_7^\ra$ symmetry leads to a $\doubleZ_7 \times \doubleZ_7$ SPT admitting the triality defect which arises as the remnant of $\rc$
\begin{equation}
    Z_{\langle \rc, \ra^{-2}\rb\rangle}[A_1,A_2] = \frac{1}{7} \sum_{a\in H^1(T^2,\doubleZ_7)} \widetilde{Z}_{\langle \rc, \ra^{-2}\rb\rangle}[a,A_2] e^{-\frac{2\pi i}{7}\int a\cup A_1} = e^{-\frac{2\pi i}{7} \int 2 A_1\cup A_2} ~.
\end{equation}
Similarly, $\doubleK = \langle \rc, \ra^{3}b\rangle$ describes a SSB phase where $\langle \rc, \ra^{3}\rb\rangle$ remains unbroken. Its twisted torus partition function is given by
\begin{equation}
    \widetilde{Z}_{\langle \rc, \ra^{3}b\rangle}[A,B,C] = 7 \delta(A+4B) ~,
\end{equation}
and gauging $\doubleZ_7^\ra$ symmetry leads to a different $\doubleZ_7 \times \doubleZ_7$ SPT which also admits the triality defect
\begin{equation}
    Z_{\langle \rc, \ra^{3}\rb\rangle}[A_1,A_2] = e^{-\frac{2\pi i}{7} \int 4 A_1 \cup A_2} ~.
\end{equation}

For $T_2$, the group-theoretical construction is given by
\begin{equation}
    \doubleG = \doubleZ_7^\ra \times (\doubleZ_7^\rb \rtimes \doubleZ^\rc_3) = \langle \ra,\rb,\rc| \ra^7 = \rb^7 = \rc^3 = 1, \ra\rb = \rb\ra, \ra\rc = \rc\ra, \rc\rb\rc^{-1} = \rb^2 \rangle ~,
\end{equation}
and $(\doubleH,\psi)$ is given by
\begin{equation}
    \doubleH = \doubleZ_7^\ra \times \doubleZ_7^\rb ~, \quad \psi(\ra^{i_1}\rb^{j_1},\ra^{i_2}\rb^{j_2}) = e^{-\frac{2\pi i}{7} j_1 i_2} ~.
\end{equation}
For the choice of $\doubleK$, any $\doubleK$ contains $\doubleZ_3^\rc$ would satisfy the requirement that $\doubleH \doubleK$ contains $\doubleG$. However, for $\doubleH \cap \doubleK$ to have a unique irrep, we must either choose $\doubleK = \doubleZ_3^\rc$ or $\doubleK = \doubleG$. The first case is obvious, as $\doubleZ_3^\rc \cap \doubleH = \doubleZ_1$. The second case works as $\doubleG \cap \doubleH = \doubleH$, but because $\psi$ is non-degenerate, there is a unique projective irrep of $\doubleH$ twisted by $\psi$, hence also leads to a unique ground state. Notice that for either choice of $\doubleK$, there is no non-trivial choice of $\psi_{\doubleK}$\footnote{Notice that $H^2(\doubleZ_7^\ra \times (\doubleZ_7^\rb \rtimes \doubleZ^\rc_3),U(1)) = \doubleZ_1$. This can be shown by viewing $\doubleZ_7^\ra \times (\doubleZ_7^\rb \rtimes \doubleZ^\rc_3)$ as a product between $\doubleZ_7^\ra$ and $\doubleZ_7^\rb \rtimes \doubleZ^\rc_3$, and apply the LHS spectral sequence.}.

The $\doubleG$-symmetric gapped phase corresponds to $\doubleK = \doubleZ_3^\rc$ is a partial SSB phase where $\doubleZ_3^\rc$ is unbroken. Its twisted torus partition function is simply given by
\begin{equation}
    \widetilde{Z}_{\doubleZ_3^{\rc}}[A,B,C] = 7^2 \delta(A)\delta(B) ~.
\end{equation}
Gauging $\doubleZ_7^\ra \times \doubleZ_7^\rb$ with discrete torsion (follows by a sequence of applying outer automorphisms as well as stacking SPT for the dual symmetry given by \eqref{eq:T2_trans_inver} where we have chosen $x = -2$) leads to a non-trivial SPT under the dual invertible $\doubleZ_7 \times \doubleZ_7$ symmetries
\begin{equation}
    Z_{\doubleZ_3^{\rc}}[A_1,A_2] = e^{-\frac{2\pi i}{7}\int 2 A_1\cup A_2} ~.
\end{equation}
Similarly, $\doubleK = \doubleG$ describes the trivially gapped phase with $\doubleG$-symmetry and its twisted torus partition function is simply given by
\begin{equation}
    \widetilde{Z}_{\doubleG}[A,B,C] = 1 ~.
\end{equation}
The same discrete gauging \eqref{eq:T2_trans_inver} leads to a trivially gapped phase for the triality defect, which itself is a $\doubleZ_7 \times \doubleZ_7$ SPT given by
\begin{equation}
    Z_{\doubleG}[A_1,A_2] = e^{-\frac{2\pi i}{7}\int 4 A_1\cup A_2} ~.
\end{equation}

It is a good place to point out an interesting subtlety. First, exactly which two $\doubleZ_7 \times \doubleZ_7$ SPTs can have $T_2$-type triality defect is not the property of the equivalence class of the fusion category. This is not a surprise because multiplying the phase factor $\phi(a,a')$ (where $\phi \in Z^2(\doubleA,U(1))$) to the fusion junction $a\times a' \rightarrow aa'$ does not change the equivalence class of the $F$-symbols. In particular, it will not change the $F$-symbols of the invertible symmetries, but will change the $F$-symbols (such as $F^{\ra,\ra',\scriptN}_{\scriptN}$) involving non-invertible symmetries. Hence, this map between two equivalent sets of $F$-symbols will induce a non-trivial map on the $\doubleZ_N \times \doubleZ_N$ SPTs when $\phi$ is cohomologically non-trivial. Applying the automorphism of $\doubleA$ on the $F$-symbols will have similar effects. As a concrete example, let's consider strictly follow the group-theoretical construction $\scriptC(\doubleZ_N^{\ra} \times (\doubleZ_N^\rb \rtimes \doubleZ_3^\rc), \omega_k, \doubleZ_N^\ra \times \doubleZ_N^\rb, \psi)$ without following up with applying the automorphism of the dual symmetry and stacking SPT. Then, we would start with the theory $\widetilde{\scriptX}$ with the invertible symmetry
\begin{equation}
    Z_{\widetilde{\scriptX}}[A,B] = Z_{\widetilde{\scriptX}}[A,-3B] ~,
\end{equation}
but now we would define the dual theory $\scriptX'$ as
\begin{equation}
    Z_{\scriptX'}[A_1,A_2] = \frac{1}{7^2} \sum_{a,b} Z_{\widetilde{\scriptX}}[a,b] e^{\frac{2\pi i}{7} \int - a \cup b + a \cup A_1 + b \cup A_2} ~.
\end{equation}
Then, the theory $\scriptX'$ is invariant under an order-$3$ twisted gauging
\begin{equation}
    Z_{\scriptX'}[A_1,A_2] = \frac{1}{7^2} \sum_{a_i} Z_{\scriptX'}[a_1,a_2] e^{\frac{2\pi i}{7} \int a_1 \cup a_2 -2 a_1 \cup A_2 + a_2 \cup A_1 + 2 A_1 \cup A_2} ~,
\end{equation}
therefore admits a triality defect. Even though the concrete form of the twisted gauging is different from the one \eqref{eq:T2_pf} given by $T_2$, the corresponding triality defects are still equivalent as fusion categories. The two $\doubleZ_7 \times \doubleZ_7$ SPTs invariant under the triality transformations are now given by
\begin{equation}
    Z_{\doubleZ_3^\rc}'[A_1,A_2] = 1 ~, \quad Z_{\doubleG}'[A_1,A_2] = e^{\frac{2\pi i}{7}\int A_1 \cup A_2} ~.
\end{equation}

\subsection{Example: Diagonal $\operatorname{Spin}(8)_1$ WZW model}
To conclude this section, we consider a CFT example which realizes the two triality defects corresponding to $T_1$ and $T_2$ under gauging $\mathbb{Z}_2 \times \mathbb{Z}_2$ symmetry. 

To start, let's consider the $\Spin(8)_1$ chiral algebra, which has chiral central charge $c_L = 4$. It has $4$ affine weights denoted as $\wo_0, \wo_1, \wo_2, \wo_3$, and the corresponding primary operators have conformal weights $h_{\wo_0} = 0, h_{\wo_1} = h_{\wo_3} = h_{\wo_4} = \frac{1}{2}$ respectively. Notice that $\wo_1$ corresponds to the vector representation of $\Spin(8)$ and $\wo_3$ and $\wo_4$ corresponds to the two spinor representations of $\Spin(8)$. The corresponding characters are given by
\begin{equation}
    \chi_{\wo_0} = \frac{\theta_3^4 + \theta_4^4}{2 \eta^4}, \quad \chi_{\wo_1} = \frac{\theta_3^4 - \theta_4^4}{2 \eta^4}, \quad \chi_{\wo_3} = \chi_{\wo_4} = \frac{\theta_2^4}{2\eta^4}.
\end{equation}
The triality of $\Spin(8)_1$ which permutes $\chi_{\wo_1}, \chi_{\wo_3}, \chi_{\wo_4}$ is realized by the identity of the Jacobi $\theta$ function
\begin{equation}
    \theta_2^4 = \theta_3^4 - \theta_4^4.
\end{equation}
The $S$ and $T$ matrices characterize the modular transformation property of the above characters are given by
\begin{equation}\label{eq:Spin81ST}
    S = \frac{1}{2} \begin{pmatrix} 1 & 1 & 1 & 1 \\ 1 & 1 & -1 & -1 \\ 1 & -1 & 1 & -1 \\ 1 & -1 & -1 & 1 \end{pmatrix}, \quad T = \begin{pmatrix} e^{-\frac{\pi i}{3}} & 0 & 0 & 0 \\ 0 & e^{\frac{2\pi i}{3}} & 0 & 0 \\ 0 & 0 & e^{\frac{2\pi i}{3}} & 0 \\ 0 & 0 & 0 & e^{\frac{2\pi i}{3}} \end{pmatrix}.
\end{equation}
Alternatively, one may use the modular property of the $\theta$ functions and the $\eta$ function to compute the modular transformation:
\begin{equation}
\begin{aligned}
    & \theta_2(-1/\tau) = \sqrt{-i\tau} \theta_4(\tau), \quad \theta_3(-1/\tau) = \sqrt{-i\tau} \theta_3(\tau), \quad \theta_4(-1/\tau) = \sqrt{-i\tau} \theta_2(\tau), \quad \eta(-1/\tau) = \sqrt{-i\tau} \eta(\tau), \\
    & \theta_2(\tau+1) = e^{i\pi/4} \theta_2(\tau), \quad \theta_3(\tau+1) = \theta_4(\tau), \quad \theta_4(\tau+1) = \theta_3(\tau), \quad \eta(\tau+1) = e^{i\pi/12} \eta(\tau). 
\end{aligned}
\end{equation}

We are interested in the diagonal CFT constructed from $\Spin(8)_1$ chiral algebra, which contains $4$ primary operators $\Spin(8)_1 \times \overline{\Spin(8)_1}$ by combining each chiral primary with its anti-chiral counterpart. The partition function is then given by
\begin{equation}
    Z(\tau) = |\chi_{\wo_0}(\tau)|^2 + |\chi_{\wo_1}(\tau)|^2 + |\chi_{\wo_3}(\tau)|^2 + |\chi_{\wo_4}(\tau)|^2.
\end{equation}
The four Verlinde lines of this theory form the $\mathbb{Z}_2^A \times \mathbb{Z}_2^B$ global symmetry whose actions are
\begin{equation}
\begin{aligned}
    & \mathbb{Z}_2^{A}: (\chi_{\wo_0} \overline{\chi_{\wo_0}}, \chi_{\wo_1} \overline{\chi_{\wo_1}}, \chi_{\wo_3} \overline{\chi_{\wo_3}}, \chi_{\wo_4} \overline{\chi_{\wo_4}}) \mapsto (\chi_{\wo_0} \overline{\chi_{\wo_0}}, \chi_{\wo_1} \overline{\chi_{\wo_1}}, -\chi_{\wo_3} \overline{\chi_{\wo_3}}, -\chi_{\wo_4} \overline{\chi_{\wo_4}}), \\
    & \mathbb{Z}_2^{B}: (\chi_{\wo_0} \overline{\chi_{\wo_0}}, \chi_{\wo_1} \overline{\chi_{\wo_1}}, \chi_{\wo_3} \overline{\chi_{\wo_3}}, \chi_{\wo_4} \overline{\chi_{\wo_4}}) \mapsto (\chi_{\wo_0} \overline{\chi_{\wo_0}}, -\chi_{\wo_1} \overline{\chi_{\wo_1}}, \chi_{\wo_3} \overline{\chi_{\wo_3}}, -\chi_{\wo_4} \overline{\chi_{\wo_4}}). \\
\end{aligned}
\end{equation}

To confirm that this theory realizes both types of triality defects given by the bulk symmetry $T_1$ and $T_2$ explicitly, we only need to check the self-triality at the level of partition functions with coupling to background fields. For simplicity, we use the fact that $\chi_{\wo_1} = \chi_{\wo_3} = \chi_{\wo_4}$ and denote them collectively as $\chi_{\frac{1}{2}}$; we also denote $\chi_{\wo_0} = \chi_0$. Their modular properties are 
\begin{equation}
    \begin{cases} & \chi_0(-1/\tau) = \frac{1}{2} \chi_0(\tau) + \frac{3}{2} \chi_{\frac{1}{2}}(\tau), \\ & \chi_{\frac{1}{2}}(-1/\tau) = \frac{1}{2} \chi_0(\tau) - \frac{1}{2} \chi_{\frac{1}{2}}(\tau) \end{cases}, \quad 
    \begin{cases} & \chi_{0}(\tau+1) = e^{-\frac{\pi i}{3}} \chi_0(\tau), \\ & \chi_{\frac{1}{2}}(\tau+1) = e^{\frac{2\pi i}{3}} \chi_{\frac{1}{2}}(\tau)\end{cases}.
\end{equation}

Then, the twisted partition functions can be solved by generalized modular bootstrap \cite{Thorngren:2021yso}. Alternatively, they can be derived by noticing that the $\Spin(8)_1$ chiral algebra can be realized as the boundary of the bulk $\Spin(8)_1$ MTC. This MTC contains 3 fermions $\psi_i$ where $i = 1,2,3$ mutual semionic statistics, and the fusion rule $\psi_i ^2 =1$, $\psi_1\times \psi_2 \times \psi_3 =1 $. The diagonal $\Spin(8)_1$ CFT can then be realized as the boundary of the bulk $\Spin(8)_1 \times \Spin(8)_1$ MTC; but the latter is nothing but SymTFT of $\mathbb{Z}_2 \times \mathbb{Z}_2$ \cite{Teo:2015xla}, via the identification:
\begin{equation}
    e_1 \equiv \psi_1 \widetilde{\psi}_2, \quad e_2 = \psi_2 \widetilde{\psi}_1, \quad m_1 \equiv \psi_1 \widetilde{\psi}_3, \quad m_2 \equiv \psi_3 \widetilde{\psi}_1, 
\end{equation}
where we use the $\widetilde{\psi}_i$ to denote the fermions in the second copy of $\Spin(8)_1$. The spectrum of (non-)local and their charges can be derived by identifying pure electric lines as charges and pure magnetic lines as symmetry defects. The twisted torus partition functions are given by
\begin{equation}\label{eq:spin81twistedpart}
\begin{aligned}
    & Z^{10} = Z^{01} = Z^{11} = \chi_0 \ochi_0 - \chi_{\frac{1}{2}} \ochi_{\frac{1}{2}}, \\
    & Z_{10} = Z_{01} = Z_{11} = \chi_0 \ochi_{\frac{1}{2}} + \chi_{\frac{1}{2}} \ochi_0 + 2 \chi_{\frac{1}{2}} \ochi_{\frac{1}{2}},  \\
    & Z^{10}_{10} = Z^{01}_{01} = Z^{11}_{11} = -\chi_0 \ochi_{\frac{1}{2}} - \chi_{\frac{1}{2}} \ochi_{0} + 2 \chi_{\frac{1}{2}}\ochi_{\frac{1}{2}}, \\
    & Z^{10}_{01} = Z^{11}_{10} = Z^{01}_{11} = - Z^{01}_{10} = - Z^{11}_{01} = - Z^{10}_{11} = \chi_0 \ochi_{\frac{1}{2}} - \chi_{\frac{1}{2}}\ochi_0,
\end{aligned}
\end{equation}
where we use $\{0,1\}$ to denote $\mathbb{Z}_2$, and by $Z^{ab}_{cd}$, we mean the torus partition function with $(a,b), (c,d) \in\mathbb{Z}_2^A\times \mathbb{Z}_2^B$ defects inserting along the spatial direction and the temporal direction respectively. 

Notice that the twisted partition functions are determined up to stacking a SPT phase as $H^2(\mathbb{Z}_2^A\times \mathbb{Z}_2^B, U(1)) = \mathbb{Z}_2$. This means the twisted partition functions in the last line are determined up to an overall minus sign. Here, we pick the convention as written in \eqref{eq:spin81twistedpart}. In the bulk, this ambiguity amounts to identify $e_2 m_1$ and $e_1 m_2$ as the generators of the $\doubleZ_2 \times \doubleZ_2$ symmetry. It is then straightforward to check that the partition functions satisfy \eqref{eq:4.19} and \eqref{eq:T2_pf}, and therefore admit two types of triality at the same time. Therefore, the $\Spin(8)_1 \times \overline{\Spin(8)_1}$ CFT admits both triality defects. To figure out the FS indicator $\alpha \in \doubleZ_3$, one needs to bootstrap the action of the triality defect and then compute the spins from its defect Hilbert space explicitly, we leave this to future work.

\section{Two classes of $p$-ality defects under twisted gauging $\doubleZ_p\times \doubleZ_p$ for odd prime $p$}\label{Sec:p_ality}
In this section, we present and study two classes of the $p$-ality defects ($p$ is an odd prime number) under twisted $\doubleZ_p \times \doubleZ_p$ gauging acquired from the group-theoretical construction. We then analyze the corresponding data describing these two fusion categories in terms of the SymTFT language. We discover that the two distinct classes of $p$-ality defects correspond to the same $\doubleZ_p$ symmetry in the bulk, but correspond to different symmetry fractionalization classes. As a result, they have different anomaly structures when choosing different FS indicators.

Interestingly, the two $p$-ality defects can be constructed by discrete gauging from the so-called extra special group $\p_\pm^3$, where
\begin{equation}
\begin{aligned}
    & \p_+^3 \equiv \He_p = \langle \ra,\rb,\rc| \ra^p = \rb^p = \rc^p = 1, \rb\rc = \rc\rb, \rc\ra\rc^{-1} = \ra\rb = \rb\ra \rangle ~, \\
    & \p_-^3 = \langle \ra,\rc| \ra^{p^2} = \rc^p = 1, \rc\ra\rc^{-1} = \ra^{p+1} \rangle ~.    
\end{aligned}
\end{equation}
It's not hard to check following \cite{2023triality} that gauging the $\doubleZ_p^c$ in $\p_+^3$ and gauging the $\doubleZ_p^{c}$ in $\p_-^3$ leads to the $p$-ality defects $\scriptN_i$ where $i = 1,\cdots, p-1$ together with a $\doubleZ_p \times \doubleZ_p$ symmetry with the following fusion rule
\begin{equation}
    g \times \scriptN_i =  \scriptN_i \times g = \scriptN_i, \quad \scriptN_i \times \scriptN_j = \begin{cases}
        \sum_{g\in \doubleZ_p\times \doubleZ_p} g, \quad i = -j ~, \\
        p \scriptN_{i+j}, \quad \text{otherwise} ~.
    \end{cases}
\end{equation}
We will denote these two classes of $p$-ality fusion categories as $\scriptP_{\pm,m}$, where $m \in \doubleZ_p$ labels the $\doubleZ_p \in H^3(\doubleZ_p, U(1))$ FS indicator. It is natural to ask if these two classes of $p$-ality defects are equivalent. It is straightforward to compute the spin selection rules for the two classes of the $p$-ality defects following the approach in \cite{2023triality}, and we find
\begin{equation}
    \begin{cases}
        s \in \frac{1}{p}\doubleZ + \frac{m}{p^2} ~, \quad \text{for} \quad \mathcal{P}_{+,m} ~, \\
        s \in \frac{1}{p^2}\doubleZ ~, \quad \text{for} \quad \scriptP_{-,m} ~,
    \end{cases} ~.
\end{equation}
From the above, it is clear that the two $p$-ality defects are inequivalent, as they lead to distinct spin selection rules; but it also raises the following question. There exists a trivial spin $s \in \doubleZ$ for the $\scriptP_{-,m}$ cases regardless of the FS indicator $m$, and this suggests that the corresponding $p$-ality defect could be anomaly-free even when the FS indicator is non-trivial. To verify this, we must use the criterion given in \eqref{eq:anomaly_free_criteria}. For this purpose, it is easier to consider the two $p$-ality defects in different group-theoretical constructions, which we will describe. 

This section is organized as follows. We first study the symmetries $\scriptP_{\pm,m}$ from the boundary and give an alternative construction of $\scriptP_{\pm}^m$ which is used to conveniently determine the anomaly of $\scriptP_{\pm,m}$ using the criterion in \eqref{eq:anomaly_free_criteria}. We then discuss how to study $\scriptP_{\pm,m}$ from the bulk point of view. We first work out the corresponding $\doubleZ_p$ anyon permutation symmetries correspond to $p$-ality defects in the $\doubleZ_p \times \doubleZ_p$-gauge theories. We then use the data from the alternative construction to work out the SymTFTs for $\scriptP_{\pm,m}$ and analyze the topological sectors for $\scriptP_{\pm,m}$. We will also demonstrate the anomaly analysis in the SymTFT. Finally, some details on the group-theoretical analysis using $\p_\pm^3$ can be found in the Appendix \ref{app:ppm3_detail}.

\subsection{Boundary Analysis}
In this subsection, we analyze the $p$-ality defects $\mathcal{P}_{\pm,m}$ from two equivalent group-theoretical constructions. We will first consider the construction from gauging $\doubleZ_p$ subgroup in the extra special group $\p_\pm^3$'s. This presentation of the $p$-ality defect is convenient to compute the $p$-ality transformation on the partition function, and also is convenient to determine the spin selection rule of the $p$-ality. We find the two classes of the $p$-ality defects correspond to the same twisted gauging, and have different spin selection rules. To study the anomaly of the two fusion categories with an arbitrary FS indicator, we introduce alternative group-theoretical presentations to the two classes of $p$-ality fusion categories and determine the anomaly using the group-theoretical technique. 

\subsubsection{$p$-ality Transformation on the Partition Functions and the Spin Selection Rules}\label{sec:p_ality_gt}
We first present the construction of the $p$-ality defects from discrete gauging in the extra special group $\p_\pm^3$. The groups $\p_\pm^3$ are given by
\begin{equation}\label{eq:extra_special_group}
\begin{aligned}
    & \p_+^3 \equiv \He_p = \langle \ra,\rb,\rc| \ra^p = \rb^p = \rc^p = 1, \rb\rc = \rc\rb, \rc\ra\rc^{-1} = \ra\rb = \rb\ra \rangle ~, \\
    & \p_-^3 = \langle \ra,\rc| \ra^{p^2} = \rc^p = 1, \rc\ra\rc^{-1} = \ra^{p+1} \rangle ~.    
\end{aligned}
\end{equation}
In both cases, we consider gauge $\doubleZ_p^c \equiv \langle \rc\rangle$. For $\p_+^3$, the subgroup $\doubleZ_p^{\rb} = \langle \rb \rangle$ commutes with $\rc$ thus survives the gauging. In the gauged theory, the invertible symmetry is $\doubleZ_p^{\rb} \times \doubleZ_p^{\hat{\rc}}$ where $\doubleZ_p^{\hat{\rc}}$ is the dual symmetry of $\doubleZ_p^{\rc}$-gauging. The ${\ra}^k$ symmetry does not commute with $\doubleZ_p^{\rc}$, therefore becomes non-invertible upon gauging $\doubleZ_p^{\rc}$. To see how its remnant acts in the gauged theory, we follow the approach in \cite{Sun:2023xxv} and start with a theory $\widetilde{\scriptX}$ with $\p_+^3$-symmetry. We then couple the theory $\widetilde{\scriptX}$ to the background fields $B, C \in H^1(\Sigma,\doubleZ_p)$ of the $\doubleZ_p^{\rb} \times \doubleZ_p^{\rc}$ subgroup of $\p_+^3$. Then, the global transformation of $\ra$ implies the following relation of the partition functions:
\begin{equation}\label{eq:p_symmetry_part}
    Z_{\widetilde{\scriptX}}[B,C] = Z_{\widetilde{\scriptX}}[B+C,C] ~.
\end{equation}
To see this relation, one can, for instance, consider the twisted partition function $Z_{\widetilde{X}}[B,C]$ on the torus $T^2$ and insert a bubble of $a$ symmetry defect:
\begin{equation}\label{eq:pt_trans}
\begin{aligned}
    Z_{\widetilde{\scriptX}}[B,C]&=\begin{tikzpicture}[baseline={([yshift=-.5ex]current bounding box.center)},vertex/.style={anchor=base,
    circle,fill=black!25,minimum size=18pt,inner sep=2pt},scale=0.5]
    \filldraw[grey] (-2,-2) rectangle ++(4,4);
    \draw[thick, dgrey] (-2,-2) rectangle ++(4,4);
    \draw[thick, black, ->-=.5] (0,-2) -- (0.707,-0.707);
    \draw[thick, black, -<-=.5] (2,0) -- (0.707,-0.707);
    \draw[thick, black, -<-=.5] (0,2) -- (-0.707,0.707);
    \draw[thick, black, ->-=.5] (-2,0) -- (-0.707,0.707);
    \draw[thick, black, ->-=.5] (0.707,-0.707) -- (-0.707,0.707);
    \node[black, below] at (0,-2) {\scriptsize $\rb^{B_x}\rc^{C_x}$};
    \node[black, right] at (2,0) {\scriptsize $\rb^{B_y}\rc^{C_y}$};
\end{tikzpicture}=\begin{tikzpicture}[baseline={([yshift=-.5ex]current bounding box.center)},vertex/.style={anchor=base,
    circle,fill=black!25,minimum size=18pt,inner sep=2pt},scale=0.5]
    \filldraw[grey] (-2,-2) rectangle ++(4,4);
    \draw[thick, dgrey] (-2,-2) rectangle ++(4,4);
    \draw[thick, black, ->-=.5] (0,-2) -- (0.707,-0.707);
    \draw[thick, black, -<-=.5] (2,0) -- (0.707,-0.707);
    \draw[thick, black, -<-=.5] (0,2) -- (-0.707,0.707);
    \draw[thick, black, ->-=.5] (-2,0) -- (-0.707,0.707);
    \draw[thick, black, ->-=.5] (0.707,-0.707) -- (-0.707,0.707);
    \draw[thick, red, ->-=1.0] (-0.7,-0.8) arc (0:360:0.4);
    \node[black, below] at (0,-2) {\scriptsize $\rb^{B_x}\rc^{C_x}$};
    \node[black, right] at (2,0) {\scriptsize $\rb^{B_y}\rc^{C_y}$};
    \node[red, below] at (-0.7,-1.) {\scriptsize $\ra$};
\end{tikzpicture}=\begin{tikzpicture}[baseline={([yshift=-.5ex]current bounding box.center)},vertex/.style={anchor=base,
    circle,fill=black!25,minimum size=18pt,inner sep=2pt},scale=0.5]
    \filldraw[grey] (-2,-2) rectangle ++(4,4);
    \draw[thick, dgrey] (-2,-2) rectangle ++(4,4);
    \draw[thick, black, ->-=.5] (0,-2) -- (0.707,-0.707);
    \draw[thick, black, -<-=.5] (2,0) -- (0.707,-0.707);
    \draw[thick, black, -<-=.5] (0,2) -- (-0.707,0.707);
    \draw[thick, black, ->-=.5] (-2,0) -- (-0.707,0.707);
    \draw[thick, black, ->-=.5] (0.707,-0.707) -- (-0.707,0.707);
    \draw [thick, red, ->-=.5] plot [smooth, tension=0.3] coordinates {(-0.4,-2) (0.3,-0.8) (-0.8,0.3) (-2,-0.4)};
    \draw [thick, red, ->-=.5] plot [smooth, tension=0.3] coordinates {(0.4,2) (-0.3,0.8) (0.8,-0.3) (2,0.4)};
    \draw [thick, red, ->-=.6] plot [smooth, tension=0.4] coordinates {(2,-0.4) (0.9,-1.0) (0.4,-2)};
    \draw [thick, red, ->-=.6] plot [smooth, tension=0.4] coordinates {(-2,0.4) (-0.9,1.0) (-0.4,2)};
    \node[black, below] at (0,-2) {\scriptsize $\rb^{B_x}\rc^{C_x}$};
    \node[black, right] at (2,0) {\scriptsize $\rb^{B_y}\rc^{C_y}$};
    \node[red, below] at (-0.4,-1.) {\scriptsize $\ra$};
    \node[red, below] at (0.99,-1.) {\scriptsize $\ra$};
\end{tikzpicture}\nonumber\\
&=\begin{tikzpicture}[baseline={([yshift=-.5ex]current bounding box.center)},vertex/.style={anchor=base,
    circle,fill=black!25,minimum size=18pt,inner sep=2pt},scale=0.5]
    \filldraw[grey] (-2,-2) rectangle ++(4,4);
    \draw[thick, dgrey] (-2,-2) rectangle ++(4,4);
    \draw[thick, black, ->-=.5] (0,-2) -- (0.707,-0.707);
    \draw[thick, black, -<-=.5] (2,0) -- (0.707,-0.707);
    \draw[thick, black, -<-=.5] (0,2) -- (-0.707,0.707);
    \draw[thick, black, ->-=.5] (-2,0) -- (-0.707,0.707);
    \draw[thick, black, ->-=.5] (0.707,-0.707) -- (-0.707,0.707);
    \node[black, below] at (0,-2) {\scriptsize $\rb^{B_x+C_x}\rc^{C_x}$};
    \node[black, right] at (2,0) {\scriptsize $\rb^{B_y+C_y}\rc^{C_y}$};
\end{tikzpicture}=Z_{\widetilde{\scriptX}}[B+C,C] ~.
\end{aligned}
\end{equation}
Then, we gauge the $\doubleZ_p^{\rc}$ subgroup to get a theory $\scriptX$ with $\doubleZ_p^{\rb} \times \doubleZ_p^{\hat{\rc}}$ where the latter is the dual symmetry of $\doubleZ_p^{\rc}$. The partition function of $\scriptX$ is related to $\widetilde{X}$ via
\begin{equation}
\begin{aligned}
    Z_{\scriptX}[B,C] &= \frac{1}{\sqrt{|H^1(\Sigma,\doubleZ_p)|}} \sum_{c\in H^1(\Sigma,\doubleZ_p)} Z_{\widetilde{\scriptX}}[B,\hat{c}] e^{\frac{2\pi i}{p}\int \hat{c} \cup C} ~, \\
    Z_{\widetilde{\scriptX}}[B,\widehat{C}] &= \frac{1}{\sqrt{|H^1(\Sigma,\doubleZ_p)|}} \sum_{c\in H^1(\Sigma,\doubleZ_p)} Z_{\scriptX}[B,c] e^{\frac{2\pi i}{p}\int c \cup \widehat{C}} ~.
\end{aligned}
\end{equation}
The invertible symmetry transformation imposed by $a$ in the theory $\widetilde{\scriptX}$ then becomes the invariance under the twisted gauging in the theory $\scriptX$:
\begin{equation}\label{eq:p_part_trans}
\begin{aligned}
    Z_{\scriptX}[B,\widehat{C}] &= \frac{1}{\sqrt{|H^1(\Sigma,\doubleZ_p)|}} \sum_{c\in H^1(\Sigma,\doubleZ_p)} Z_{\widetilde{\scriptX}}[B,c] e^{\frac{2\pi i}{p}\int c \cup \widehat{C}} \\
    &= \frac{1}{\sqrt{|H^1(\Sigma,\doubleZ_p)|}} \sum_{c\in H^1(\Sigma,\doubleZ_p)} Z_{\widetilde{\scriptX}}[B+c,c] e^{\frac{2\pi i}{p}\int c \cup \widehat{C}} \\
    &= \frac{1}{|H^1(\Sigma,\doubleZ_p)|} \sum_{c,\hat{c}\in H^1(\Sigma,\doubleZ_p)} Z_{\scriptX}[B+c, \hat{c}] e^{\frac{2\pi i}{p}\int c\cup \widehat{C} + \hat{c}\cup c} \\
    &= \frac{1}{|H^1(\Sigma,\doubleZ_p)|} \sum_{b,\hat{c}\in H^1(\Sigma,\doubleZ_p)} Z_{\scriptX}[b,\hat{c}] e^{\frac{2\pi i}{p}\int (b - B) \cup (-\hat{c} + \widehat{C})} ~.
\end{aligned}
\end{equation}
That both $\doubleZ_p^{\rb} \times \doubleZ_p^{\hat{\rc}}$ is gauged in the transformation is compatible with the $p$-ality fusion rule \eqref{eq:p_ality_frule}. The global transformation of the remnant of the $a$-defect implements the $\doubleZ_p \times \doubleZ_p$-gauging implies that it is a $p$-ality defect in the gauged theory.

It is straightforward to repeat the analysis for $\p_-^3$. Again, we start with a theory $\widetilde{\scriptX}$ with symmetry $\p_-^3$ and couple it to background fields $B,C \in H^1(\Sigma,\doubleZ_p)$ for the $\doubleZ_p$ generated by $\ra^p$ and $\rc$ respectively. The global transformation by $\ra$ on the twisted partition function $Z_{\widetilde{\scriptX}}[B,C]$ then leads to the same relation as 
\begin{equation}
    Z_{\widetilde{\scriptX}}[B,C] = Z_{\widetilde{\scriptX}}[B+C,C] ~.
\end{equation}
Again, gauging the $\doubleZ_p^\rc$ symmetry leads to a theory $\scriptX$ with $\doubleZ_p^{\rb} \times \doubleZ_p^{\hat{\rc}}$ invertible symmetries. Then the remnant of the $\ra$-defect in the theory $\scriptX$ leads to the same $p$-ality transformation as the $\p_+^3$ case
\begin{equation}
    Z_{\scriptX}[B,\widehat{C}] = \frac{1}{|H^1(\Sigma,\doubleZ_p)|} \sum_{b,\hat{c}\in H^1(\Sigma,\doubleZ_p)} Z_{\scriptX}[b,\hat{c}] e^{\frac{2\pi i}{p}\int (b - B) \cup (-\hat{c} + \widehat{C})} ~.
\end{equation}
This implies that in the $\p_-^3$ case, the remnant of $\ra$-symmetry also becomes a $p$-ality defect in the gauged theory.

To confirm this from the fusion rule and to derive the spin selection of the $p$-ality defects, one can study these two categories using the language of algebraic objects and bimodule objects (e.g. see \cite{yichul2024selfdual}), where the details of the calculation can be found in the Appendix \ref{app:ppm3_detail}. The subgroup $\doubleZ_p^c$ we gauged corresponds to an algebraic object $\displaystyle \mathcal{A}_\pm = \bigoplus_{n = 0}^{p-1} \rc^n$, while the simple objects in the dual fusion symmetry category are described by indecomposable bimodule objects of $\mathcal{A}_\pm$. In both cases, we find the dual symmetry category, which we denote as $\mathcal{P}_\pm$ respectively, admitting $\doubleZ_p \times \doubleZ_p$ invertible symmetries together with $p$-ality defects $\scriptN_k$ ($k = 1,\cdots, p-1$):
\begin{equation} \label{eq:p_ality_frule}
    g \times \scriptN_i =  \scriptN_i \times g = \scriptN_i, \quad \scriptN_i \times \scriptN_j = \begin{cases}
        \displaystyle \sum_{g\in \doubleZ_p\times \doubleZ_p} g, \quad i = -j \mod p ~, \\
        p \scriptN_{i+j}, \quad \text{otherwise} ~.
    \end{cases}
\end{equation}

For $\mathcal{P}_+$, the algebraic object is given by $\displaystyle \mathcal{A}_+ = \bigoplus_{n = 0}^{p-1} c^n$. There are $p^2$ invertible bimodule objects of the algebra $\mathcal{A}_+$ which we denote as $\displaystyle \mathcal{M}_+^{(i,j)} = \bigoplus_{n=0}^{p-1} \rb^i \rc^n$, and the $\mathcal{M}_+^{(i,j)}$ is distinguished from $\mathcal{M}_+^{(i,j')}$ via different action of the $\mathcal{A}_+$. $\mathcal{M}_+^{(1,0)}$ generates the $\doubleZ_p^b$ symmetry while $\mathcal{M}_{+}^{(0,1)}$ generates the dual symmetry $\doubleZ_p^{\hat{\rc}}$. There are $(p-1)$ non-invertible bimodule objects which we denote as $\displaystyle \mathcal{M}_+^k = \bigoplus_{i,j=0}^{p-1} \rb^i \rc^j \ra^k$ which correspond to the $p$-ality defects $\scriptN_k$ in \eqref{eq:p_ality_frule}.

For $\mathcal{P}_-$, the algebraic object is again given by $\displaystyle \mathcal{A}_- = \bigoplus_{n = 0}^{p-1} \rc^n$. There are $p^2$ invertible bimodule objects of the algebra $\mathcal{A}_-$ which we denote as $\displaystyle \mathcal{M}_-^{(i,j)} = \bigoplus_{n=0}^{p-1} \ra^{ip} \rc^n$, and the $\mathcal{M}_-^{(i,j)}$ is distinguished from $\mathcal{M}_-^{(i,j')}$ via different action of the $\mathcal{A}_-$. Notice that the subgroup $\doubleZ_p = \langle \ra^p\rangle \subset \p_-^3$ commutes with $\doubleZ_p^{\rc}$, therefore survives the gauging and is generated by $\mathcal{M}_-^{(1,0)}$. $\mathcal{M}_{-}^{(0,1)}$ generates the dual symmetry $\doubleZ_p^{\hat{\rc}}$. There are $(p-1)$ non-invertible bimodule objects which we denote as $\displaystyle \mathcal{M}_-^k = \bigoplus_{i,j=0}^{p-1} \rc^j \ra^{k + p i}$ which correspond to the $p$-ality defects $\scriptN_k$ in \eqref{eq:p_ality_frule}.

The details of the fusion and split junctions characterizing the left and right action of the algebraic object $\mathcal{A}_\pm$ on the bimodule objects $\mathcal{M}_\pm$'s can be found in the Appendix \ref{app:ppm3_detail}. Knowing this allows us to relate the defect Hilbert space of the $p$-ality defects in the theory $\scriptX$ with the defect Hilbert space of the corresponding invertible symmetries in the theory $\widetilde{\scriptX}$, therefore to derive the spin selection rules from the latter. 

Let's first consider $\mathcal{P}_+$. In this case, the defect Hilbert space $\mathcal{H}^{\scriptX}_{\scriptN_1}$ of the $p$-ality defect $\scriptN_1$ in the theory $\scriptX$ can be captured by the following twisted partition function 
\begin{equation}\label{eq:pp3_twisted_part}
    Z_{\scriptX}[\scriptN_1,\dsi, \scriptN_1] = \Tr_{\mathcal{H}^{\scriptX}_{\scriptN_1}} q^{L_0 - c/24} \overline{q}^{\overline{L}_0 - c/24} = \begin{tikzpicture}[baseline={([yshift=+.5ex]current bounding box.center)},vertex/.style={anchor=base,
    circle,fill=black!25,minimum size=18pt,inner sep=2pt},scale=0.5]
    \filldraw[grey] (-2,-2) rectangle ++(4,4);
    \draw[thick, dgrey] (-2,-2) -- (-2,+2);
    \draw[thick, dgrey] (-2,-2) -- (+2,-2);
    \draw[thick, dgrey] (+2,+2) -- (+2,-2);
    \draw[thick, dgrey] (+2,+2) -- (-2,+2);
    \draw[thick, dgreen, -stealth] (0,-2) -- (0.354,-1.354);
    \draw[thick, dgreen] (0,-2) -- (0.707,-0.707);
    \draw[thick, dgreen, -stealth] (-0.707,0.707) -- (-0.354,1.354);
    \draw[thick, dgreen] (0,2) -- (-0.707,0.707);
    \draw[thick, dgreen, -stealth] (0.707,-0.707) -- (0,0);
    \draw[thick, dgreen] (0.707,-0.707) -- (-0.707,0.707);
    \node[dgreen, below] at (0,-2) {\scriptsize $\scriptN_1$};
    \node[black] at (-1.5,1.5) {\footnotesize $\scriptX$};
\end{tikzpicture} ~,
\end{equation}
where in the diagrammatic representation of the twisted torus partition function, the $\scriptX$ on the top-left corner indicates this is for the theory $\scriptX$. This partition function $Z_{\scriptX}[\CN_1,\dsi,\CN_1]$ can be computed using the twisted torus partition function of the theory $\widetilde{\scriptX}$ with $\p^3_+$-symmetry by expanding the following diagram of algebras and bimodules:
\begin{equation}
    \begin{tikzpicture}[baseline={([yshift=+.5ex]current bounding box.center)},vertex/.style={anchor=base,
    circle,fill=black!25,minimum size=18pt,inner sep=2pt},scale=0.5]
    \filldraw[grey] (-2,-2) rectangle ++(4,4);
    \draw[thick, dgrey] (-2,-2) -- (-2,+2);
    \draw[thick, dgrey] (-2,-2) -- (+2,-2);
    \draw[thick, dgrey] (+2,+2) -- (+2,-2);
    \draw[thick, dgrey] (+2,+2) -- (-2,+2);
    \draw[thick, dgreen, -stealth] (0,-2) -- (0.354,-1.354);
    \draw[thick, dgreen] (0,-2) -- (0.707,-0.707);
    \draw[thick, dgreen, -stealth] (-0.707,0.707) -- (-0.354,1.354);
    \draw[thick, dgreen] (0,2) -- (-0.707,0.707);
    \draw[thick, dgreen, -stealth] (0.707,-0.707) -- (0,0);
    \draw[thick, dgreen] (0.707,-0.707) -- (-0.707,0.707);
    \node[dgreen, below] at (0,-2) {\scriptsize $\scriptN_1$};
    \node[black] at (-1.5,1.5) {\footnotesize $\scriptX$};
\end{tikzpicture} = \begin{tikzpicture}[baseline={([yshift=+.5ex]current bounding box.center)},vertex/.style={anchor=base,
    circle,fill=black!25,minimum size=18pt,inner sep=2pt},scale=0.5]
    \filldraw[grey] (-2,-2) rectangle ++(4,4);
    \draw[thick, dgrey] (-2,-2) -- (-2,+2);
    \draw[thick, dgrey] (-2,-2) -- (+2,-2);
    \draw[thick, dgrey] (+2,+2) -- (+2,-2);
    \draw[thick, dgrey] (+2,+2) -- (-2,+2);
    \draw[thick, dgreen, -stealth] (0,-2) -- (0.354,-1.354);
    \draw[thick, dgreen] (0,-2) -- (0.707,-0.707);
    \draw[thick, red, -stealth] (2,0) -- (1.354,-0.354);
    \draw[thick, red] (2,0) -- (0.707,-0.707);
    \draw[thick, dgreen, -stealth] (-0.707,0.707) -- (-0.354,1.354);
    \draw[thick, dgreen] (0,2) -- (-0.707,0.707);
    \draw[thick, red, -stealth] (-0.707,0.707) -- (-1.354,0.354);
    \draw[thick, red] (-2,0) -- (-0.707,0.707);
    \draw[thick, dgreen, -stealth] (0.707,-0.707) -- (0,0);
    \draw[thick, dgreen] (0.707,-0.707) -- (-0.707,0.707);
    \filldraw[black] (0.707,-0.707) circle (2pt);
    \filldraw[black] (-0.707,0.707) circle (2pt);
    \node[dgreen, below] at (0,-2) {\scriptsize $\mathcal{M}_+^1$};
    \node[black] at (-1.5,1.5) {\footnotesize $\widetilde{\scriptX}$};
    \node[red, right] at (2,0) {\scriptsize $\mathcal{A}_+$};
    \node[black, below] at (0.9,-0.607) {\scriptsize $\rho^{1}_+$};
    \node[black, below] at (-0.707,0.707) {\scriptsize $\lambda_+^{1,\vee}$};
\end{tikzpicture} = \frac{1}{p} \sum_{i,j = 0}^{p-1} \begin{tikzpicture}[baseline={([yshift=+.5ex]current bounding box.center)},vertex/.style={anchor=base,
    circle,fill=black!25,minimum size=18pt,inner sep=2pt},scale=0.5]
    \filldraw[grey] (-2,-2) rectangle ++(4,4);
    \draw[thick, dgrey] (-2,-2) -- (-2,+2);
    \draw[thick, dgrey] (-2,-2) -- (+2,-2);
    \draw[thick, dgrey] (+2,+2) -- (+2,-2);
    \draw[thick, dgrey] (+2,+2) -- (-2,+2);
    \draw[thick, black, -stealth] (0,-2) -- (0.354,-1.354);
    \draw[thick, black] (0,-2) -- (0.707,-0.707);
    \draw[thick, black, -stealth] (-0.707,0.707) -- (-0.354,1.354);
    \draw[thick, black] (0,2) -- (-0.707,0.707);
    \draw[thick, black, -stealth] (0.707,-0.707) -- (0,0);
    \draw[thick, black] (0.707,-0.707) -- (-0.707,0.707);
    \node[black, below] at (0,-2) {\scriptsize $\rb^i \rc^j \ra$};
    \node[black] at (-1.5,1.5) {\footnotesize $\widetilde{\scriptX}$};
\end{tikzpicture} = \sum_{i=0}^{p-1} \begin{tikzpicture}[baseline={([yshift=+.5ex]current bounding box.center)},vertex/.style={anchor=base,
    circle,fill=black!25,minimum size=18pt,inner sep=2pt},scale=0.5]
    \filldraw[grey] (-2,-2) rectangle ++(4,4);
    \draw[thick, dgrey] (-2,-2) -- (-2,+2);
    \draw[thick, dgrey] (-2,-2) -- (+2,-2);
    \draw[thick, dgrey] (+2,+2) -- (+2,-2);
    \draw[thick, dgrey] (+2,+2) -- (-2,+2);
    \draw[thick, black, -stealth] (0,-2) -- (0.354,-1.354);
    \draw[thick, black] (0,-2) -- (0.707,-0.707);
    \draw[thick, black, -stealth] (-0.707,0.707) -- (-0.354,1.354);
    \draw[thick, black] (0,2) -- (-0.707,0.707);
    \draw[thick, black, -stealth] (0.707,-0.707) -- (0,0);
    \draw[thick, black] (0.707,-0.707) -- (-0.707,0.707);
    \node[black, below] at (0,-2) {\scriptsize $\rc^i \ra$};
    \node[black] at (-1.5,1.5) {\footnotesize $\widetilde{\scriptX}$};
\end{tikzpicture} ~,
\end{equation}
where in the last step we used the twisted partition function is invariant under conjugation. Therefore, we establish that
\begin{equation}
    Z_{\scriptX}[\scriptN_1, \dsi, \scriptN_1] = \sum_{i=0}^{p-1} Z_{\widetilde{X}}[\rc^i \ra, \dsi, \rc^i \ra] \Longleftrightarrow \mathcal{H}^{\scriptX}_{\scriptN_1} \simeq \bigoplus_{i=0}^{p-1} \mathcal{H}^{\widetilde{X}}_{\rc^i \ra} ~,
\end{equation}
therefore the spin selection rules for the defect Hilbert space $\mathcal{H}^{\scriptX}_{\scriptN_1}$ are determined by the spin selection rules for the defect Hilbert space $\mathcal{H}^{\widetilde{X}}_{\rc^i \ra}$ of the invertible symmetries. Since the symmetry $\p_+^3$ is anomaly-free, the spins of the states in $\mathcal{H}^{\widetilde{X}}_{\rc^i \ra}$ is given by the character $\chi_r(\rc^i a)$ where $r$ is an irrep of the centralizer subgroup $C_{p^3_+}(\rc^i \ra) = \langle \rb, \rc^i \ra\rangle \simeq \doubleZ_p \times \doubleZ_p$ of $\rc^i \ra$. We then conclude that the spin $s$ of the states in $\mathcal{H}^{\scriptX}_{\scriptN_1}$ must take value in $\frac{1}{p}\doubleZ$. Furthermore, different FS indicator can be realized by stacking a theory $\mathcal{T}$ with anomalous $\doubleZ_p$ symmetry generated by $\eta$ with 't Hooft anomaly $m \in H^3(\doubleZ_p, U(1)) \simeq \doubleZ_p$, and consider the new $p$-ality symmetry $\scriptN_k' = \scriptN_k \eta^k$. The spin selection rules for anomalous $\doubleZ_p$ (which is given by $s \in \frac{1}{p}\doubleZ + \frac{m}{p^2}$) would then modifies the spin selection rules for the $\CN_1'$ line, and we find
\begin{equation}\label{eq:spin_pp}
    s \in \frac{1}{p} \doubleZ + \frac{m}{p^2} ~, \quad \text{for} \quad \mathcal{P}_{+,m} ~,
\end{equation}
where we use superscript $m = \{0,\cdots, p-1\}$ in $\mathcal{P}_+^m$ to denote the FS indicator.

Next, let's consider $\mathcal{P}_-$. The twisted partition function which computes the trace over the defect Hilbert space $\mathcal{H}^{\scriptX}_{\CN_1}$ is now related to the twisted partition function of the theory $\widetilde{X}$ via
\begin{equation}\label{eq:pm3_twisted_part}
    \begin{tikzpicture}[baseline={([yshift=+.5ex]current bounding box.center)},vertex/.style={anchor=base,
    circle,fill=black!25,minimum size=18pt,inner sep=2pt},scale=0.5]
    \filldraw[grey] (-2,-2) rectangle ++(4,4);
    \draw[thick, dgrey] (-2,-2) -- (-2,+2);
    \draw[thick, dgrey] (-2,-2) -- (+2,-2);
    \draw[thick, dgrey] (+2,+2) -- (+2,-2);
    \draw[thick, dgrey] (+2,+2) -- (-2,+2);
    \draw[thick, dgreen, -stealth] (0,-2) -- (0.354,-1.354);
    \draw[thick, dgreen] (0,-2) -- (0.707,-0.707);
    \draw[thick, dgreen, -stealth] (-0.707,0.707) -- (-0.354,1.354);
    \draw[thick, dgreen] (0,2) -- (-0.707,0.707);
    \draw[thick, dgreen, -stealth] (0.707,-0.707) -- (0,0);
    \draw[thick, dgreen] (0.707,-0.707) -- (-0.707,0.707);
    \node[dgreen, below] at (0,-2) {\scriptsize $\scriptN_1$};
    \node[black] at (-1.5,1.5) {\footnotesize $\scriptX$};
\end{tikzpicture} = \begin{tikzpicture}[baseline={([yshift=+.5ex]current bounding box.center)},vertex/.style={anchor=base,
    circle,fill=black!25,minimum size=18pt,inner sep=2pt},scale=0.5]
    \filldraw[grey] (-2,-2) rectangle ++(4,4);
    \draw[thick, dgrey] (-2,-2) -- (-2,+2);
    \draw[thick, dgrey] (-2,-2) -- (+2,-2);
    \draw[thick, dgrey] (+2,+2) -- (+2,-2);
    \draw[thick, dgrey] (+2,+2) -- (-2,+2);
    \draw[thick, dgreen, -stealth] (0,-2) -- (0.354,-1.354);
    \draw[thick, dgreen] (0,-2) -- (0.707,-0.707);
    \draw[thick, red, -stealth] (2,0) -- (1.354,-0.354);
    \draw[thick, red] (2,0) -- (0.707,-0.707);
    \draw[thick, dgreen, -stealth] (-0.707,0.707) -- (-0.354,1.354);
    \draw[thick, dgreen] (0,2) -- (-0.707,0.707);
    \draw[thick, red, -stealth] (-0.707,0.707) -- (-1.354,0.354);
    \draw[thick, red] (-2,0) -- (-0.707,0.707);
    \draw[thick, dgreen, -stealth] (0.707,-0.707) -- (0,0);
    \draw[thick, dgreen] (0.707,-0.707) -- (-0.707,0.707);
    \filldraw[black] (0.707,-0.707) circle (2pt);
    \filldraw[black] (-0.707,0.707) circle (2pt);
    \node[dgreen, below] at (0,-2) {\scriptsize $\mathcal{M}_-^1$};
    \node[black] at (-1.5,1.5) {\footnotesize $\widetilde{\scriptX}$};
    \node[red, right] at (2,0) {\scriptsize $\mathcal{A}_-$};
    \node[black, below] at (0.9,-0.607) {\scriptsize $\rho^{1}_-$};
    \node[black, below] at (-0.707,0.707) {\scriptsize $\lambda_-^{1,\vee}$};
\end{tikzpicture} = \frac{1}{p} \sum_{i,j = 0}^{p-1} \begin{tikzpicture}[baseline={([yshift=+.5ex]current bounding box.center)},vertex/.style={anchor=base,
    circle,fill=black!25,minimum size=18pt,inner sep=2pt},scale=0.5]
    \filldraw[grey] (-2,-2) rectangle ++(4,4);
    \draw[thick, dgrey] (-2,-2) -- (-2,+2);
    \draw[thick, dgrey] (-2,-2) -- (+2,-2);
    \draw[thick, dgrey] (+2,+2) -- (+2,-2);
    \draw[thick, dgrey] (+2,+2) -- (-2,+2);
    \draw[thick, black, -stealth] (0,-2) -- (0.354,-1.354);
    \draw[thick, black] (0,-2) -- (0.707,-0.707);
    \draw[thick, black, -stealth] (-0.707,0.707) -- (-0.354,1.354);
    \draw[thick, black] (0,2) -- (-0.707,0.707);
    \draw[thick, black, -stealth] (0.707,-0.707) -- (0,0);
    \draw[thick, black] (0.707,-0.707) -- (-0.707,0.707);
    \node[black, below] at (0,-2) {\scriptsize $\rc^j \ra^{1+p i}$};
    \node[black] at (-1.5,1.5) {\footnotesize $\widetilde{\scriptX}$};
\end{tikzpicture} = \sum_{i=0}^{p-1} \begin{tikzpicture}[baseline={([yshift=+.5ex]current bounding box.center)},vertex/.style={anchor=base,
    circle,fill=black!25,minimum size=18pt,inner sep=2pt},scale=0.5]
    \filldraw[grey] (-2,-2) rectangle ++(4,4);
    \draw[thick, dgrey] (-2,-2) -- (-2,+2);
    \draw[thick, dgrey] (-2,-2) -- (+2,-2);
    \draw[thick, dgrey] (+2,+2) -- (+2,-2);
    \draw[thick, dgrey] (+2,+2) -- (-2,+2);
    \draw[thick, black, -stealth] (0,-2) -- (0.354,-1.354);
    \draw[thick, black] (0,-2) -- (0.707,-0.707);
    \draw[thick, black, -stealth] (-0.707,0.707) -- (-0.354,1.354);
    \draw[thick, black] (0,2) -- (-0.707,0.707);
    \draw[thick, black, -stealth] (0.707,-0.707) -- (0,0);
    \draw[thick, black] (0.707,-0.707) -- (-0.707,0.707);
    \node[black, below] at (0,-2) {\scriptsize $\rc^i \ra$};
    \node[black] at (-1.5,1.5) {\footnotesize $\widetilde{\scriptX}$};
\end{tikzpicture} ~,
\end{equation}
which allows us to establish
\begin{equation}
    \mathcal{Z}_{\scriptX}[\scriptN_1, \dsi, \scriptN_1] = \sum_{i=0}^{p-1} Z_{\widetilde{\scriptX}}[\rc^i\ra,\dsi,\rc^i\ra] \Longleftrightarrow \mathcal{H}^{\scriptX}_{\scriptN_1} = \bigoplus_{i=0}^{p-1} \mathcal{H}^{\widetilde{\scriptX}}_{\rc^i \ra} ~.
\end{equation}
The crucial difference is that the defect Hilbert space $\mathcal{H}_{\rc^i \ra}^{\widetilde{\scriptX}}$ contains states with spin $s \in \frac{1}{p^2}\doubleZ$, because the centralizer subgroup $C_{\p_-^3}(\rc^i \ra) = \langle \rc^i \ra\rangle \simeq \doubleZ_{p^2}$, hence the spin selection rule for the $\scriptN_1$ defect is given by $s \in \frac{1}{p^2}\doubleZ$. If we realize different FS indicators by stacking anomalous $\doubleZ_p$ invertible symmetry line $\eta$, then the $m/p^2$ shift will not change the spin selection rule for the $\scriptN_1' = \scriptN_1 \eta$. Hence, we conclude that 
\begin{equation}\label{eq:spin_pm}
    s \in \frac{1}{p^2} \doubleZ ~, \quad \text{for} \quad \mathcal{P}_{-,m} ~.
\end{equation}
The distinction in the spin selection rules \eqref{eq:spin_pp} and \eqref{eq:spin_pm} implies that the two $p$-ality fusion categories $\mathcal{P}_{\pm,m}$ are inequivalent. Furthermore, since they correspond to the twisted gauging, the corresponding bulk $\doubleZ_p$ anyon permutation symmetry in the $\doubleZ_p\times \doubleZ_p$-SymTFT must be identical. The fact that they lead to different fusion categories indicates that they must differ by the choice of symmetry fractionalization class, and we will explore this in more detail later. The other thing one can learn from the spin selection rules is that for $\mathcal{P}_{-,m}$ may be anomaly-free for any FS indicator, as the defect Hilbert space $\mathcal{H}^{\scriptX}_{\CN_1}$ may contain integer spins for any $m$. To check whether this is the case, we now introduce an alternative group-theoretical construction to the $\mathcal{P}_{\pm,m}$ fusion categories. 

\subsubsection{Alternative group-theoretical Constructions and the Anomaly of the $p$-ality defects}
We now present an alternative group-theoretical construction of the $\mathcal{P}_{\pm,m}$ defects. The idea is the groups $\p_\pm^3$ themselves can be constructed as discrete gauging some $\doubleZ_p$ subgroup of the $\doubleZ_p^3$ with suitable anomalies $\omega_\pm$ respectively, therefore the category $\mathcal{P}_{\pm,m}$ can be constructed as from discrete gauging subgroups in $\VEC_{\doubleZ_p^3}^{\omega_{\pm,m}}$. The advantage of this construction is that the anomaly structure $\doubleZ_p^3$ is well-studied, therefore, it is easier to match the choice of the FS indicator with the choice of the anomaly $\omega_{\pm,m}$ of $\doubleZ_p^3$. We then use these group-theoretical presentations to determine the anomaly of the two classes of $p$-ality fusion categories.

As pointed out in \cite{Tachikawa:2017gyf}, consider the following group extension where both $\widehat{\doubleA}$ and $\doubleG$ are Abelian, 
\begin{equation}
    0 \rightarrow \widehat{\doubleA} \rightarrow \Gamma \rightarrow \doubleG \rightarrow 0 ~,
\end{equation}
where the group extension is characterized by a class $\psi \in H^2(\doubleG,\widehat{\doubleA})$. For our purpose, $G$ acts trivially on $\widehat{\doubleA}$. Then the element in $\Gamma$ is parameterized by $(\hat{a},g)$ and the multiplication rule is given by
\begin{equation}
    (\hat{a},g)\times (\hat{a}',g') = (\hat{a}\hat{a}'\psi(g,g'), gg') ~.
\end{equation}
Gauging the global symmetry $\widehat{\doubleA}$, leads to the group $\doubleG \times \widehat{\doubleA}$ with the mixed anomaly given by
\begin{equation}
    \omega((g_1, a_1),(g_2,a_2),(g_3,a_3)) = \psi(g_2,g_3)(a_1) ~.
\end{equation}

Let's first consider $\p_+^3$. The group fits into the following exact sequence
\begin{equation}
    0 \rightarrow \doubleZ_p^{\hat{\rb}} \rightarrow \p_+^3 \rightarrow \doubleZ_p^{\ra} \times \doubleZ_p^{\rc} \rightarrow 0
\end{equation}
with the $\psi_+ \in H^2(\doubleZ_p^{\ra} \times \doubleZ_p^{\rc}, \doubleZ_p^{\hat{\rb}}$) given by
\begin{equation}
    \psi_+(\ra^i \rc^j, \ra^{i'}\rc^{j'}) = \hat{\rb}^{ji'}
\end{equation}
where $\hat{\rb}(\rb) = e^{\frac{2\pi i}{p}}$. Notice that here we have renamed the generator $\rb$ in the \eqref{eq:extra_special_group} as $\hat{\rb}$. Using the above result, we learn that gauging $\doubleZ_p^{\hat{\rb}}$ in $\p_+^3$ leads to the symmetry $\doubleZ_p^{\ra} \times \doubleZ_p^{\rb} \times \doubleZ_p^{\rc}$ with the class III mixed 't Hooft anomaly
\begin{equation}
    \omega_{+}(\ra^{i_1}\rb^{j_1}\rc^{k_1},\ra^{i_2}\rb^{j_2}\rc^{k_2},\ra^{i_3}\rb^{j_3}\rc^{k_3}) = e^{\frac{2\pi i}{p} j_1 k_2 i_3} ~.
\end{equation}
But since $\doubleZ_p^{\rb}$ is the dual symmetry of the $\doubleZ_p^{\hat{\rb}}$, gauging it will give back the symmetry $\p_+^3$. Furthermore, combining this with the fact that gauging $\doubleZ_p^{\rc}$ in $\p_+^3$ leads to the category $\scriptP_+$, gauging $\doubleZ_p^{\rb} \times \doubleZ_p^{\rc}$ with trivial discrete torsion in $\VEC_{\doubleZ_p^3}^{\omega_+}$ will also lead to $\scriptP_+$ category:
\begin{equation}
\begin{tikzcd}
    \VEC_{\doubleZ_p^\ra \times \doubleZ_p^\rb \times \doubleZ_p^\rc}^{\omega_+} \arrow[rd, "\text{ gauge } \doubleZ_p^\rb \times \doubleZ_p^\rc", swap]\arrow[r, "\quad \text{ gauge } \doubleZ_p^\rb \quad"] & [4em] \p_+^3 \arrow[d, "\text{ gauge } \doubleZ_p^\rc"] \\ [6ex]
    & [4em] \mathcal{P}_+ \\
\end{tikzcd} ~.
\end{equation}
Furthermore, since the remnant of $\ra$ becomes $p$-ality defects, different FS indicator $m \in H^3(\doubleZ_p, U(1)) \simeq \doubleZ_p$ can be realized as starting with the corresponding type I 't Hooft anomaly of the $\doubleZ_p^{\ra}$ in $\doubleZ_p^{\ra} \times \doubleZ_p^\rb \times \doubleZ_p^\rc$. Hence, we conclude that
\begin{equation}
    \mathcal{P}_{+,m} = \mathcal{C}(\doubleZ_p^\ra \times \doubleZ_p^\rb \times \doubleZ_p^\rc, \omega_{+,m}, \doubleZ_p^\rb \times \doubleZ_p^\rc, 1) ~,
\end{equation}
where
\begin{equation}\label{eq:omegap}
    \omega_{+,m}(\ra^{i_1}\rb^{j_1}\rc^{k_1},\ra^{i_2}\rb^{j_2}\rc^{k_2},\ra^{i_3}\rb^{j_3}\rc^{k_3}) = e^{\frac{2\pi i}{p} j_1 k_2 i_3 + \frac{2\pi i m }{p^2}i_1 (i_2 + i_3 - [i_2 + i_3]_p)} ~.
\end{equation}

Then, let's consider $\p_-^3$. Similarly, the group fits into the following exact sequence
\begin{equation}
    0 \rightarrow \doubleZ_p^{\hat{\rb}} \rightarrow \p_-^3 \rightarrow \doubleZ_p^{\ra} \times \doubleZ_p^{\rc} \rightarrow 0 ~,
\end{equation}
where the class $\psi_- \in H^2(\doubleZ_p^{\ra} \times \doubleZ_p^{\rc},\doubleZ_p^{\hat{\rb}})$ is given by
\begin{equation}
    \psi_-(\ra^i \rc^j, \ra^{i'} \rc^{j'}) = \hat{\rb}^{ji' + \frac{1}{p}(i + i' - [i+i']_p)} ~.
\end{equation}
Under this extension, $\doubleZ_p^{\ra}$ and $\doubleZ_p^{\hat{\rb}}$ combine non-trivially into $\doubleZ_{p^2}$ generated by $\ra$. Repeating the same analysis, we find
\begin{equation}
    \mathcal{P}_{-,m} = \mathcal{C}(\doubleZ_p^\ra \times \doubleZ_p^\rb \times \doubleZ_p^\rc, \omega_{-,m}, \doubleZ_p^\rb \times \doubleZ_p^\rc, 1) ~,
\end{equation}
where \footnote{Notice that one can consider 
\begin{equation}
    \omega(\ra^{i_1}\rb^{j_1}\rc^{k_1},\ra^{i_2}\rb^{j_2}\rc^{k_2},\ra^{i_3}\rb^{j_3}\rc^{k_3}) = e^{\frac{2\pi i}{p} j_1 k_2 i_3 + \frac{2\pi i l}{p^2} j_1 (i_2 + i_3 - [i_2 + i_3]_p) + \frac{2\pi i m }{p^2}i_1 (i_2 + i_3 - [i_2 + i_3]_p)}
\end{equation}
for generic $l,m \in \doubleZ_p$. But if $l\neq 0$, since $\gcd(l,p) = 1$, one can consider a redefinition of group generator $(j_i,k_i) \rightarrow (l^{-1} j_i, l k_i)$ where $l^{-1}$ is the mod-$p$ inverse of $l$  to set $l = 1$. Hence, it will not lead to anything qualitatively new and we will only consider the case where $l = 1$ here. Similarly, one can also consider turning on mixed 't Hooft anomaly between $\doubleZ_p^\ra$ and $\doubleZ_p^\rc$, but this is also equivalent to the current case up to a relabeling of the $\doubleZ_p^\rb \times \doubleZ_p^\rc$ group elements.
}
\begin{equation}\label{eq:omegam}
    \omega_{-,m}(\ra^{i_1}\rb^{j_1}\rc^{k_1},\ra^{i_2}\rb^{j_2}\rc^{k_2},\ra^{i_3}\rb^{j_3}\rc^{k_3}) = e^{\frac{2\pi i}{p} j_1 k_2 i_3 + \frac{2\pi i}{p^2} j_1 (i_2 + i_3 - [i_2 + i_3]_p) + \frac{2\pi i m }{p^2}i_1 (i_2 + i_3 - [i_2 + i_3]_p)} ~.
\end{equation}

\

It is then straightforward to use the criterion \eqref{eq:anomaly_free_criteria} to determine the anomaly of the category $\mathcal{P}_{\pm,m}$. Let's start with $\mathcal{P}_{+,m} = \mathcal{C}(\doubleZ_p^\ra \times \doubleZ_p^\rb \times \doubleZ_p^\rc, \omega_{+,m}, \doubleZ_p^\rb \times \doubleZ_p^\rc, 1)$. When the FS indicator is trivial, i.e., $m = 0$, choosing $(\doubleK,\psi_{\doubleK}) = (\doubleZ_p^\ra, 1)$ would meet the criterion, therefore $\mathcal{P}_+ \equiv \mathcal{P}_{+,0}$ is anomaly-free. On the other hand, when the FS indicator is non-trivial, no choice of $(\doubleK,\psi_{\doubleK})$ can meet the criterion. To see this, notice that if $\doubleH \doubleK$ contains $\doubleZ_p^\ra \times \doubleZ_p^\rb \times \doubleZ_p^\rc$, $\doubleK$ must contain an element of the form $\ra \rb^x \rc^y$ for some $x,y \in \doubleZ_p^2$. However, the subgroup generated by $\ra \rb^x \rc^y$ is always anomalous when $m \neq 0$, as the first class III anomaly always trivializes when restricting to a $\doubleZ_p$ subgroup as $p > 2$, so that it can not soak up the type I anomaly due to the non-trivial FS indicator. Notice that this is consistent with the spin selection rule where $\mathcal{P}_{+,m}$ can not contain integer spins in the defect Hilbert of the $p$-ality line when $m \neq 0$.

For the $p$-ality fusion category $\scriptP_{-,m} = \mathcal{C}(\doubleZ_p^\ra \times \doubleZ_p^\rb \times \doubleZ_p^\rc, \omega_{-,m}, \doubleZ_p^\rb \times \doubleZ_p^\rc, 1)$, however, this is not the case. Looking at the anomaly $\omega_{-,m}$, we can simply take $\doubleH_1 = \langle \ra \rb^{-m}\rangle$ such that the class II mixed anomaly between $\doubleZ_p^\ra$ and $\doubleZ_p^\rb$ is used to cancel the class I self-anomaly of $\doubleZ_p^\ra$ in the subgroup $\langle \ra \rb^{-m} \rangle$. It is straightforward to see that $\doubleK = \langle \ra\rb^{-m} \rangle$ satisfies the conditions in \eqref{eq:anomaly_free_criteria} for all $m$, hence $\scriptP_{-,m}$'s are all anomaly-free, regardless of $m$.

Before moving on to the SymTFT analysis of the fusion category $\mathcal{P}_{\pm,m}$, we want to quickly point out what happens if one takes $p = 2$. From the $\p_\pm^3$ construction point of view, $\mathbf{2}^3_+$ is isomorphic to the group $\mathbf{2}^3_-$ by identifying $\rc_+ \equiv \rc_-$, $\rc_+ \ra_+ = \ra_-$ in \eqref{eq:extra_special_group} and hence $\mathcal{P}_{\pm,m}$ are the same fusion categories. We can also see this in the construction using $\doubleZ_2^\ra \times \doubleZ_2^\rb \times \doubleZ_2^\rc$, where the two anomalies $\omega_{\pm,m}$ can be rewritten as
\begin{equation}\label{eq:p=2_omega}
\begin{aligned}
    & \omega_{+,m}(\ra^{i_1}\rb^{j_1}\rc^{k_1},\ra^{i_2}\rb^{j_2}\rc^{k_2},\ra^{i_3}\rb^{j_3}\rc^{k_3}) = (-1)^{j_1 k_2 i_3 + m i_1 i_2 i_3} ~, \\
    & \omega_{-,m}(\ra^{i_1}\rb^{j_1}\rc^{k_1},\ra^{i_2}\rb^{j_2}\rc^{k_2},\ra^{i_3}\rb^{j_3}\rc^{k_3}) = (-1)^{j_1 k_2 i_3 + j_1 i_2 i_3 + m i_1 i_2 i_3} = (-1)^{j_1 (k_2 + i_2) i_3 + m i_1 i_2 i_3} ~.
\end{aligned}
\end{equation}
This implies that $\VEC_{\doubleZ_2^\ra \times \doubleZ_2^\rb \times \doubleZ_2^\rc}^{\omega_{+,m}}$ is isomorphic to $\VEC_{\doubleZ_2^\ra \times \doubleZ_2^\rb \times \doubleZ_2^\rc}^{\omega_{-,m}}$ under the identification $\ra_+ \equiv \ra_-, \rb_+ \equiv \rb_-, \rc_+ \equiv \rc_- \ra_-$. Therefore, gauging $\doubleZ_2^\rb \times \doubleZ_2^\rc$ in both cases leads to the equivalent fusion category, known as $\Rep(D_8)$ and $\Rep(Q_8)$ for $m = 0,1$ respectively. The two fusion categories are anomaly-free can be checked by choosing $(\doubleK,\psi_{\doubleK}) = (\langle \ra\rangle,1)$ and $(\doubleK,\psi_{\doubleK}) = (\langle \ra\rb\rc\rangle, 1)$ respectively, due to the special form of the anomaly \eqref{eq:p=2_omega} when $p = 2$.

\subsection{SymTFT Analysis}\label{sec:palityDW}
In this subsection, we study the two classes of $p$-ality defects from the SymTFT. We first start with $\doubleZ_p \times \doubleZ_p$-SymTFT, and the $p$-ality defects correspond to some $\doubleZ_p$ symmetry in the bulk. We will show here that they correspond to the same $\doubleZ_p$ symmetry, therefore their distinction must correspond to different choices of symmetry fractionalization class. Next, since the $\mathcal{P}_\pm^m$ admits the group-theoretical construction $\mathcal{C}(\doubleZ_p^\ra \times \doubleZ_p^\rb \times \doubleZ_p^\rc, \omega_{\pm,m}, \doubleZ_p^b \times \doubleZ_p^\rc, 1)$, their SymTFT is given by the DW theory with gauge group $\doubleZ_p^\ra \times \doubleZ_p^\rb \times \doubleZ_p^\rc$ with twist $\omega_{\pm,m}$. The latter is well-studied and we use it to explicitly confirm our results on the anomaly of the $\mathcal{P}_{\pm,m}$. We also compute the topological sectors from the SymTFT and the result matches the spin selection rules derived above in \eqref{eq:spin_pp} and \eqref{eq:spin_pm}.

\subsubsection{$\doubleZ_p$-symmetry in the $\doubleZ_p \times \doubleZ_p$-SymTFT}
We first present how to identify the $\doubleZ_p$ global symmetries correspond to the $\mathcal{P}_{\pm,m}$ in the $(\doubleZ_p \times \doubleZ_p)$-SymTFT. This follows from the transformation on the partition function \eqref{eq:p_part_trans} and the generator $P$ of the bulk $\doubleZ_p$-symmetry is given by
\begin{equation}
    P = \begin{pmatrix} 1 & 0 & 0 & 1 \\ 0 & 1 & -1 & 0 \\ 0 & 0 & 1 & 0 \\ 0 & 0 & 0 & 1 \end{pmatrix} ~,
\end{equation}
where we parameterize the bulk anyon as $(\rb,\rc,\hat{\rb},\hat{\rc})^T$ and the gapped boundary corresponds to condensing all the pure electric anyons $(0,0,\hat{\rb},\hat{\rc})^T$. The corresponding condensation defect is
\begin{equation}
    S_P(\Sigma) = \frac{1}{|H^1(\Sigma,\doubleZ_p)|} \sum_{\gamma_i} e^{-\frac{2\pi i}{p}\langle \gamma_1, \gamma_2\rangle} m_{\mathrm{b}}(\gamma_1) m_{\mathrm{c}}(\gamma_2) ~.
\end{equation}
Gauging $\doubleZ_p^\rc$ symmetry on the boundary is implemented by the $\doubleZ_2^{em,\mathrm{c}}$, and the symmetry $P_{\mathrm{c}}$ in this frame (where the bottom two components of the anyons are condensed on the boundary) can be acquired by conjugating $P$ with the generator of the $\doubleZ_2^{em,\mathrm{c}}$:
\begin{equation}\label{eq:bulk_Zp_generator}
    P_{\mathrm{c}} = \begin{pmatrix} 1 & 0 & 0 & 0 \\ 
    0 & 0 & 0 & 1 \\
    0 & 0 & 1 & 0 \\
    0 & 1 & 0 & 0 \\
    \end{pmatrix} \begin{pmatrix} 1 & 0 & 0 & 1 \\ 0 & 1 & -1 & 0 \\ 0 & 0 & 1 & 0 \\ 0 & 0 & 0 & 1 \end{pmatrix} \begin{pmatrix} 1 & 0 & 0 & 0 \\ 
    0 & 0 & 0 & 1 \\
    0 & 0 & 1 & 0 \\
    0 & 1 & 0 & 0 \\
    \end{pmatrix} = \begin{pmatrix} 1 & 1 & 0 & 0 \\ 0 & 1 & 0 & 0 \\ 0 & 0 & 1 & 0 \\ 0 & 0 & -1 & 1 \end{pmatrix} ~, 
\end{equation}
which indeed corresponds to outer automorphism of the boundary $\doubleZ_p \times \doubleZ_p$ symmetry \eqref{eq:p_symmetry_part}.

Gauging $\doubleZ_p^\rb\times \doubleZ_p^\rc$ symmetry on the boundary is implemented by the $\doubleZ_2^{em,\mathrm{b},\mathrm{c}}$, and similarly the symmetry $P_{\mathrm{b},\mathrm{c}}$ in this frame is given by 
\begin{equation}
    P_{\mathrm{b},\mathrm{c}} = 
    \begin{pmatrix} 0 & 0 & 1 & 0 \\ 
    0 & 0 & 0 & 1 \\
    1 & 0 & 0 & 0 \\
    0 & 1 & 0 & 0 \\
    \end{pmatrix} \begin{pmatrix} 1 & 0 & 0 & 1 \\ 0 & 1 & -1 & 0 \\ 0 & 0 & 1 & 0 \\ 0 & 0 & 0 & 1 \end{pmatrix} \begin{pmatrix} 0 & 0 & 1 & 0 \\ 
    0 & 0 & 0 & 1 \\
    1 & 0 & 0 & 0 \\
    0 & 1 & 0 & 0 \\
    \end{pmatrix} = \begin{pmatrix}
        1 & 0 & 0 & 0 \\
        0 & 1 & 0 & 0 \\
        0 & 1 & 1 & 0 \\
        -1 & 0 & 0 & 1 \\
    \end{pmatrix} ~.
\end{equation}
Comparing with \eqref{eq:stacking_SPT}, we see that $P_{\mathrm{b},\mathrm{c}}$ corresponds to the invertible $\doubleZ_p^{\mathrm{a}}$ symmetry commute with dual $\doubleZ_p \times \doubleZ_p$ symmetry, and its global action stacks a non-trivial $\doubleZ_p \times \doubleZ_p$ SPT. This is consistent with the type III anomaly that appeared in \eqref{eq:omegam} and \eqref{eq:omegap}.

The fact that the two classes of $p$-ality defects correspond to the same $\doubleZ_p$ symmetry in bulk, and yet are inequivalent implies that their distinctions correspond to different choices of the symmetry fractionalization class. 

\

The group-theoretical construction $\scriptC(\doubleZ_p^\ra \times \doubleZ_p^\rb \times \doubleZ_p^\rc, \omega_{\pm,m}, \doubleZ_p^b\times \doubleZ_p^\rc, 1)$ of $\mathcal{P}_{\pm,m}$ leads to interesting insights on the SymTFT side. The SymTFT $\scriptZ(\mathcal{P}_{\pm,m}) = \scriptZ(\VEC_{\doubleZ_p^\ra\times \doubleZ_p^\rb \times \doubleZ_p^\rc}^{\omega_{\pm,m}})$ can be constructed as gauging suitable $\doubleZ_p$ symmetry (which corresponds to the $p$-ality operation) in the SymTFT $\scriptZ(\VEC_{\doubleZ_p^\rb \times \doubleZ_p^\rc})$ of the $\doubleZ_p^\rb \times \doubleZ_p^\rc$ symmetry. Therefore, the twist $\omega_{\pm,m}$ must correspond to data in the gauging of the $\doubleZ_p$ symmetry. 

To motivate our proposal for the match of the data, view $\doubleZ_p^\ra \times \doubleZ_p^\rb \times \doubleZ_p^\rc$ as the extension of $\doubleZ_p^\rb \times \doubleZ_p^\rc$ by $\doubleZ_p^\ra$
\begin{equation}
    0 \rightarrow (\doubleA \equiv \doubleZ_p^\rb \times \doubleZ_p^\rc) \rightarrow \doubleZ_p^a \times \doubleZ_p^b \times \doubleZ_p^c \rightarrow (G \equiv \doubleZ_p^a) \rightarrow 0 ~,
\end{equation}
then as pointed out in \cite{Tachikawa:2017gyf}, the LHS spectral sequence separates the possible twists $H^3(\doubleZ_p^\ra \times \doubleZ_p^\rb \times \doubleZ_p^\rc, U(1))$ into different layers $H^p(\doubleZ_p^\ra, H^q(\doubleZ_p^\rb\times \doubleZ_p^\rc,U(1)))$ where $p+q = 3$. $H^0(G,H^3(\doubleA,U(1))) \simeq H^3(\doubleA, U(1))$ represents the twist of $\doubleZ_p^\rb \times \doubleZ_p^\rc$ gauge theory, which in our case takes trivial element by input (as we want $\doubleZ_p^\rb \times \doubleZ_p^\rc$ to be gaugable in the first place). 

The next layer $H^1(G, H^2(\doubleA,U(1))) \simeq \doubleZ_p$ correspond to the type III twist involving all three $\doubleZ_p$ groups, it appears in both $\omega_{\pm,m}$. A representative of a class in $H^1(G, H^2(\doubleA,U(1)))$ maps the generator of $G = \doubleZ_p^\ra$ to an SPT of $H^2(\doubleA,U(1))$, thus naturally identifies with the choices of the bulk symmetry of the form \eqref{eq:stacking_SPT}, which acts trivially on the electric lines $\widehat{\doubleA}$. Concretely, it is specified by the matrix of the form
\begin{equation}
    \begin{pmatrix}
        1 & 0 & 0 & 0 \\
        0 & 1 & 0 & 0 \\
        0 & n & 1 & 0 \\
        -n & 0 & 0 & 1 \\
    \end{pmatrix} ~,
\end{equation}
where $\omega_{\pm,m}$ corresponds to the choice where $n = 1$. This is consistent with the analysis above. 

Then, the layer $H^2(G,H^1(\doubleA,U(1))) = H^2(G,\widehat{\doubleA}) = \doubleZ_p \times \doubleZ_p$ corresponds to the type II twist between $\doubleZ_p^\ra$ and $\doubleZ_p^\rb$ or $\doubleZ_p^\rc$ respectively. $\omega_{+,m}$ does not contain a class in this layer, while $\omega_{-,m}$ does contain a class in this layer. Its form $H^2(G,\widehat{\doubleA})$ suggests that we may interpret it as a symmetry fractionalization class $H^2_{\rho}(G,\doubleA \times \widehat{\doubleA})$ where $\rho$ denotes the choice of the bulk symmetry $G$ specified by $H^1(G,H^2(\doubleA,U(1)))$. The fact that $\rho$ always acts trivially on the electric line $\widehat{\doubleA}$ implies that a class $[\nu]$ in $H^2(G,\widehat{\doubleA})$ can be viewed as a class $[\tilde{\nu}]$ in $H^2_\rho(G,\doubleA\times\widehat{\doubleA})$. Indeed, this lift of $[\nu]$ into $[\tilde{\nu}]$ is a bijection between $H^2(G,\widehat{\doubleA})$ and $H^2_\rho(G,\doubleA \times \widehat{\doubleA})$. To see this explicitly, first notice that $H^2(\doubleZ_p, \doubleZ_p\times \doubleZ_p) \simeq \doubleZ_p \times \doubleZ_p$, and $H^2_\rho(\doubleZ_p, (\doubleZ_p)^4)$ with the $\rho$ specified by the $P$ in \eqref{eq:bulk_Zp_generator} can be computed based on the following well-known result (see e.g. \cite{brown2012cohomology})\footnote{Notice that when $p = 2$, $H^2_\rho(\doubleZ_2, (\doubleZ_2)^4)$ is trivial, consistent with the result from Shapiro's lemma mentioned previously.}
\begin{equation}
    H^2_\rho(\doubleZ_p, (\doubleZ_p)^4)) \simeq \frac{\operatorname{Ker}(\sum_{i=0}^{p-1} P^i)}{\operatorname{Im}(1-P)} \simeq \doubleZ_p \times \doubleZ_p ~.
\end{equation}
Thus, to show the bijective nature of the lift, we only need to show it is injective, namely, any non-trivial class $[\nu]$ is lifted to a non-trivial symmetry fractionalization class $[\tilde{\nu}]$, which can be done explicitly\footnote{To see this, let's assume $[\nu] \in H^2(\doubleZ_p^\ra, \doubleZ_p^{\hat{\rb}}\times \doubleZ_p^{\hat{\rc}})$ is a non-trivial class described by $\nu(i,j) = \begin{pmatrix} \hat{b}_{i,j} \\ \hat{c}_{i,j} \end{pmatrix}$. Then, lifted it to $H^2_\rho(G,\doubleA\times \widehat{\doubleA})$, we can represent it as $\tilde{\nu}(i,j) = \begin{pmatrix} 0 \\ 0 \\ \hat{b}_{i,j} \\ \hat{c}_{i,j} \end{pmatrix}$ where $i,j \in \doubleZ_p$ denotes the element in $\doubleZ_p^a$ in an additive convention. Assuming there exists $\mu \in C^1(\doubleZ_p^\ra, \doubleZ_p^\rb \times \doubleZ_p^\rc \times \doubleZ_p^{\hat{\rb}} \times \doubleZ_p^{\hat{\rc}})$ where $\mu(i) = \begin{pmatrix} b_i' \\ c_i' \\ \hat{b}_i' \\ \hat{c}_i' \end{pmatrix}$ such that $(d_\rho \mu)(i,j) = \tilde{\nu}(i,j)$, we find 
\begin{equation}
    \begin{pmatrix} 0 \\ 0 \\ \hat{b}_{i,j} \\ \hat{c}_{i,j} \end{pmatrix} = \begin{pmatrix} b'_j - b'_{i+j} + b'_{i} \\ c'_j - c'_{i+j} + c'_i \\  \hat{b}'_j - \hat{b}'_{i+j} + \hat{b}'_i - i n c'_j \\ \hat{c}'_j - \hat{c}'_{i+j} + \hat{c}'_i + i n b'_j  \end{pmatrix} \mod p ~.
\end{equation} 
The first two equations imply that $b'_j = k_b j$ and $c'_j = k_c j$ where $k_b,k_c \in \doubleZ_p$. Then for $p > 2$ (therefore $p$ is odd),
\begin{equation}
\begin{aligned}
    & \hat{b}'_j - \hat{b}'_{i+j} + \hat{b}'_i + i n c'_j = \left(\hat{b}'_j + \frac{p+1}{2} n k_c j^2\right) - \left(\hat{b}'_{i+j} + \frac{p+1}{2} n k_c (i+j)^2\right) + \left(\hat{b}'_i + \frac{p+1}{2} n k_c i^2 \right) ~, \\
    & \hat{c}'_j - \hat{c}'_{i+j} + \hat{c}'_i - i n b'_j = \left(\hat{b}'_j - \frac{p+1}{2} n k_b j^2\right) - \left(\hat{b}'_{i+j} - \frac{p+1}{2} n k_b (i+j)^2\right) + \left(\hat{b}'_i - \frac{p+1}{2} n k_b i^2 \right) ~,
\end{aligned}
\end{equation}
implies that the existence of $\mu$ would contradicts the assumption that $\nu$ represents a non-trivial class in $H^2(\doubleZ_p^\ra, \doubleZ_p^{\hat{\rb}} \times \doubleZ_p^{\hat{\rc}})$. Hence, we conclude that $[\tilde{\nu}]$ can be interpreted as a non-trivial symmetry fractionalization class.

On the other hand, when $p = 2$ and $n = 1$, since the non-trivial $[\nu] \in H^2(\doubleZ_2^\ra, \doubleZ_2^{\hat{\rb}} \times \doubleZ_2^{\hat{\rc}})$ can be represented as $\nu(i,j) = \begin{pmatrix} l_b ij \\ l_c ij \end{pmatrix}$ for $(l_b,l_c) \in \doubleZ_2 \times \doubleZ_2$, it is possible to choose $\mu(i) = \begin{pmatrix} l_c i \\ l_b i \\ 0 \\ 0\end{pmatrix}$ to trivialize the lift $\tilde{\nu}$. This is consistent with the fact that $\VEC_{\doubleZ_2^\ra \times \doubleZ_2^\rb \times \doubleZ_2^\rc}^{\omega_{+,m}}$ is equivalent to $\VEC_{\doubleZ_2^\ra \times \doubleZ_2^\rb \times \doubleZ_2^\rc}^{\omega_{-,m}}$, and the fact that in the TY fusion category, and there is no non-trivial choice of the symmetry fractionalization data in the classification of TY fusion category.}. Then naturally, gauging the symmetry $G$ with the corresponding choice of the symmetry fractionalization classes in $[\tilde{\nu}] \in H^2_\rho(G,\doubleA\times \widehat{\doubleA})$ leads to the DW theory with the twist containing $[\nu]$.

Finally, $H^3(G,H^0(\doubleA,U(1))) \simeq H^3(G, U(1)) = \doubleZ_p$ corresponds to the type I twist of the $\doubleZ_p^\ra$, and in the bulk it is interpreted as the choice of the discrete torsion. Different choices are encoded in $m \in \doubleZ_p$ in both classes $\omega_{\pm,m}$.

To conclude, we note that a similar anomaly structure also appears for the duality defect. Namely, one can compare the anomaly structure for $\TY(\doubleZ_2 \times \doubleZ_2, \chi_{\text{diag}}, \epsilon)$ and $\TY(\doubleZ_2 \times \doubleZ_2, \chi_{\text{off-diag}}, \epsilon)$. It is well-known that for the off-diagonal bicharacter, regardless of the FS indicator $\epsilon = \pm 1$, the fusion category is anomaly-free; while for the diagonal bicharacter, only the one with $\epsilon = +1$ is anomaly-free. However, this distinction is caused by different choices of the $\doubleZ_{2}^{em}$-symmetry in the SymTFT; while for the $\mathcal{P}_{\pm,m}$ categories, the distinction results from choosing different symmetry fractionalization classes for the same $\doubleZ_p$ symmetry in the SymTFT.

\subsubsection{SymTFT for the $p$-ality defects}\label{sec:p_ality_SymTFT}
We now study the SymTFT $\scriptZ(\mathcal{P}_{\pm,m})$ directly using the group-theoretical construction $\mathcal{C}(\doubleZ_p^\mathrm{a}\times \doubleZ_p^\mathrm{b}\times \doubleZ_p^\mathrm{c}, \omega_{\pm,m}, \doubleZ_p^{\mathrm{b}}\times \doubleZ_p^{\mathrm{c}},1)$, which implies that $\scriptZ(\mathcal{P}_{\pm,m})$ is the DW theory with the gauge group $\doubleZ_p^\mathrm{a}\times \doubleZ_p^\mathrm{b}\times \doubleZ_p^\mathrm{c}$ and the twist $\omega_{\pm,m}$. We can work out the spectrum of their SymTFT using the technique in \cite{Hu:2012wx,Coste:2000tq}. The detailed calculation of the spectrum of the SymTFT is included in the Appendix \ref{app:DW_theory}. Here, we will only list the details of the SymTFT and discuss the physical implications.

As reviewed in the Appendix \ref{app:DW_theory}, the simple anyons in DW theory $G$ are labeled by a conjugacy class $[\ra]$ of $G$ and an irreducible (projective) representation $\pi_{[\ra]}$ of the centralizer subgroup of $\ra$ twisted by the effective 2-cocycle $\beta_\ra$. In the case here, since $G = \doubleZ_p^\ra \times \doubleZ_p^\rb \times \doubleZ_p^\rc$ is Abelian, each element in $G$ forms its own conjugacy class, hence we will denote the simple anyons as $\mathbf{a}_{\rg,\pi_\rg}$'s. For $\DW(\doubleZ_p^\ra \times \doubleZ_p^\rb \times \doubleZ_p^\rc, \omega_{+,m})$, the simple anyons $\mathbf{a}_{\rg,\pi_\rg}$ are
\begin{equation}
\begin{aligned}
    & \mathbf{a}_{(0,0,0),\vec{u}} ~, \quad \theta_{(0,0,0),\vec{u}} = 1 ~, \quad d_{(0,0,0),\vec{u}} = 1 ~, \quad \vec{u} \in \doubleZ_p^3 ~, \\
    & \mathbf{a}_{(i,j,k),u} ~, \quad \theta_{(i,j,k),u} = e^{\frac{2\pi i}{p} u - \frac{2\pi i}{p}\frac{p+1}{2}ijk + \frac{2\pi i}{p^2} mi^2} ~, \quad d_{(i,j,k),u} = p ~, \quad (i,j,k) \neq (0,0,0) ~, u \in \doubleZ_p ~.
\end{aligned}
\end{equation}
Similarly, the simple anyons $\mathbf{b}_{\rg,\pi_\rg}$ in the $\DW(\doubleZ_p^\ra \times \doubleZ_p^\rb \times \doubleZ_p^\rc, \omega_{-,m})$ are given by
\begin{equation}
\begin{aligned}
    & \mathbf{b}_{(0,0,0),\vec{u}} ~, \quad \theta_{(0,0,0),\vec{u}} = 1 ~, \quad d_{(0,0,0),\vec{u}} = 1 ~, \quad \vec{u} \in \doubleZ_p^3 ~, \\
    & \mathbf{b}_{(i,j,k),u} ~, \quad \theta_{(i,j,k),u} = e^{\frac{2\pi i}{p} u - \frac{2\pi i}{p}\frac{p+1}{2} ijk  + \frac{2\pi i}{p^2} mi^2 + \frac{2\pi i}{p^2} ij} ~, \quad d_{(i,j,k),u} = p ~, \quad (i,j,k) \neq (0,0,0) ~, u \in \doubleZ_p ~.
\end{aligned}
\end{equation}

\subsubsection*{Anomaly from the SymTFT}
Let's first study the anomaly of $\scriptP_{+,m}$. The gapped boundary determining the symmetry corresponds to the electric Lagrangian algebra $\scriptL_{\operatorname{e}}$, and it can be constructed from the subgroup $\doubleZ_p^b \times \doubleZ_p^c$ with the trivial 2-cocycle:
\begin{equation}
    \scriptL_{\operatorname{e}} = \left(\bigoplus_{u \in \doubleZ_p} \mathbf{a}_{(0,0,0),(u,0,0)}\right)\bigoplus \left(\bigoplus_{(j,k)\in \doubleZ_p^2/\{\vec{0}\}} \mathbf{a}_{(0,j,k),0}\right).
\end{equation}
The existence of the trivially gapped phase is equivalent to the existence of the magnetic Lagrangian algebra, which must contain the flux $(i,0,0)$. In general, it only has to contain an anyon of the form $\mathbf{a}_{(i,j,k),u}$ for every $i\neq 0$. But this is only possible when the FS indicator $m = 0$, as when $m \neq 0$, every simple anyon of the form $\mathbf{a}_{(i,j,k),u}$ where $i\neq 0$ has fractional spin. Hence, we conclude that when $m = 0$, the fusion category is anomaly-free and the magnetic Lagrangian algebra given by
\begin{equation}
    \scriptL_{\operatorname{m}} = \left(\bigoplus_{i=1}^{p}\mathbf{a}_{(i,0,0),0}\right)\bigoplus\left(\bigoplus_{(u_2,u_3)\in \doubleZ_p\times \doubleZ_p} \mathbf{a}_{(0,0,0),(0,u_2,u_3)}\right) ~.
\end{equation}
And there's no trivially symmetric gapped phase when $m \neq 0$.

Then, let's study the anomaly of $\scriptP_{-,m}$. Similarly, the electric Lagrangian algebra is constructed as
\begin{equation}
    \scriptL_{\operatorname{e}} = \left(\bigoplus_{u \in \doubleZ_p} \mathbf{b}_{(0,0,0),(u,0,0)}\right)\bigoplus \left(\bigoplus_{(j,k)\in \doubleZ_p^2/\{\vec{0}\}} \mathbf{b}_{(0,j,k),0}\right).
\end{equation}
In this case, however, because the additional term in the spin $\theta_{(i,j,k),u}$, for every FS indicator $m \in \doubleZ_p$, it is possible to construct the magnetic Lagrangian algebra $\scriptL_{\operatorname{m}}$ given by
\begin{equation}
    \scriptL_{\operatorname{m}} = \left(\bigoplus_{i = 1}^p \mathbf{b}_{(i,-mi,0),0}\right)\bigoplus\left(\bigoplus_{(u_2,u_3)\in \doubleZ_p\times \doubleZ_p} \mathbf{b}_{(0,0,0),(m u_2,u_2,u_3)}\right). 
\end{equation}
Hence, we conclude that $\scriptP_{-,m}$ is anomaly-free regardless of the FS indicator.

\subsubsection*{Topological Sectors}
It is interesting to find out the topological sectors of the $\scriptP_{\pm,m}$ symmetry from their SymTFT. Notice that the boundary TDLs are characterized by the $\scriptL_{\operatorname{e}}$-modules $\ell$'s, and each simple anyon in a given $\ell$ represents a topological sector in the defect Hilbert space $\scriptH_{\ell}$. We expect there are $p^2$ modules corresponding to the $\doubleZ_p \times \doubleZ_p$ invertible symmetries on the boundary, and $p-1$ modules realizing the $p$-ality defects on the boundary. 

For the $\DW(\doubleZ_p^a \times \doubleZ_p^b \times \doubleZ_p^c, \omega_{+,m})$, the $p^2$ modules corresponding to invertible symmetries can be constructed by fusing the simple anyon $\mathbf{a}_{(0,0,0),(0,u_2,u_3)}$ with $\scriptL_{\operatorname{e}}$, and the $p-1$ modules corresponding to non-invertible symmetries can be constructed by fusing the simple anyon $\mathbf{a}_{(i,0,0),0}$ with $\scriptL_{\operatorname{e}}$ where $i = 1, \cdots, p-1$. The fusion can be computed using the fusion coefficients via the Verlinde formula and the $S$-matrix listed in the Appendix \ref{app:DW_theory}. We list the $\scriptL_{\operatorname{e}}$-module as follows:
\begin{equation}
\begin{aligned}
    & \ell_{(a,b)} = \left(\bigoplus_{u\in\doubleZ_p} \mathbf{a}_{(0,0,0),(u,a,b)}\right) \bigoplus \left(\bigoplus_{(j,k)\in\doubleZ_p^2/\{\vec{0}\}}\mathbf{a}_{(0,j,k),ja+kb}\right), \quad (a,b) \in \doubleZ_p\times \doubleZ_p ~, \\
    & \ell_{\scriptN_i} = \left(\bigoplus_{u \in \doubleZ_p} \mathbf{a}_{(i,0,0),iu}\right)\bigoplus \left(\bigoplus_{\substack{u\in \doubleZ_p \\ (j,k) \in \doubleZ_p^2/\{\vec{0}\}}} \mathbf{a}_{(i,j,k),u}\right), \quad i = 1,\cdots,p-1 ~. \quad 
\end{aligned}
\end{equation}
We list the topological sectors and their spin selection rules in the Table \ref{tab:top_sec_omegapm}:
\begin{table}[h]
    \centering
    \begin{tabular}{|c|c|}
        \hline
        Topological Sector $\scriptH_\ell^a$ & spin selection rules  \\
        \hline
        $\scriptH_{(a,b)}^{\vec{0},(u,a,b)}$ & $s\in \doubleZ $  \\
        \hline
        $\scriptH_{(a,b)}^{(0,j,k),ja+kb}, \quad (j,k) \neq (0,0)$ & $s\in \doubleZ + \frac{ja+kb}{p}$  \\
        \hline
        $\scriptH_{\scriptN_i}^{(i,0,0),iu}, \quad i \neq 0$ & $s \in \doubleZ + \frac{iu}{p} + \frac{mi^2}{p^2}$ \\
        \hline
        $\scriptH_{\scriptN_i}^{(i,j,k),u}, \quad (j,k) \neq 0$ & $s\in \doubleZ + \frac{u}{p} - \frac{p+1}{2}\frac{ijk}{p} + \frac{mi^2}{p^2}$ \\
        \hline
    \end{tabular}
    \caption{Topological Sectors for $\scriptP_+^m$.}
    \label{tab:top_sec_omegapm}
\end{table}

The calculation for $\DW(\doubleZ_p^a \times \doubleZ_p^b \times \doubleZ_p^c, \omega_{-,m})$ can be done identically. The module categories are given by
\begin{equation}
\begin{aligned}
    & \ell_{(a,b)} = \left(\bigoplus_{u\in\doubleZ_p} \mathbf{b}_{(0,0,0),(u,a,b)}\right) \bigoplus \left(\bigoplus_{(j,k)\in\doubleZ_p^2/\{\vec{0}\}}\mathbf{b}_{(0,j,k),ja+kb}\right), \quad (a,b) \in \doubleZ_p\times \doubleZ_p, \\
    & \ell_{\scriptN_i} = \left(\bigoplus_{u \in \doubleZ_p} \mathbf{b}_{(i,0,0),iu}\right)\bigoplus \left(\bigoplus_{\substack{u\in \doubleZ_p \\ (j,k) \in \doubleZ_p^2/\{\vec{0}\}}} \mathbf{b}_{(i,j,k),u}\right), \quad i = 1,\cdots,p-1. \quad 
\end{aligned}
\end{equation}
And we list the result in the Table \ref{tab:top_sec_omegamm}:
\begin{table}[h]
    \centering
    \begin{tabular}{|c|c|}
        \hline
        Topological Sector $\scriptH_\ell^a$ & spin selection rules  \\
        \hline
        $\scriptH_{(a,b)}^{\vec{0},(u,a,b)}$ & $s\in \doubleZ $  \\
        \hline
        $\scriptH_{(a,b)}^{(0,j,k),ja+kb}, \quad (j,k) \neq (0,0)$ & $s\in \doubleZ + \frac{ja+kb}{p}$  \\
        \hline
        $\scriptH_{\scriptN_i}^{(i,0,0),iu}, \quad i \neq 0$ & $s \in \doubleZ + \frac{iu}{p} + \frac{mi^2}{p^2}$ \\
        \hline
        $\scriptH_{\scriptN_i}^{(i,j,k),u}, \quad (j,k) \neq 0$ & $s\in \doubleZ + \frac{u}{p} - \frac{p+1}{2} \frac{ijk}{p} + \frac{(m i+j)i}{p^2}$ \\
        \hline
    \end{tabular}
    \caption{Topological Sectors for $\scriptP_-^m$.}
    \label{tab:top_sec_omegamm}
\end{table}
Notice that the above result is consistent with the spin selection rules derived in \eqref{eq:spin_pp} and \eqref{eq:spin_pm}.

\section{Constructing $S_3$-ality under twisted gauging $\doubleZ_N \times \doubleZ_N$}\label{sec:S3_ality}
In this section, we study the non-trivial combination of duality and triality lines such that they form a non-Abelian $S_3$-ality extension. In the notation introduced earlier, we are interested in an $S_3$-extension of the fusion category $\VEC_{\doubleA}$, $\underline{\CE}_{S_3}\VEC_{\doubleA}$. This fusion category contains invertible simple objects labeled by the Abelian group $\doubleA$, and non-invertible objects $\mathcal{N}_g$ where $g \in S_3-\{\dsi\}$. They satisfy the fusion rule
\begin{equation}\label{eq:S3_fusion_rule}
    a \times \mathcal{N}_g = \mathcal{N}_g \times a = \mathcal{N}_g ~, \quad  \mathcal{N}_g \times \mathcal{N}_{g'} = \begin{cases} & \sum_{a\in \doubleA} a ~, \quad \text{if} \quad gg' = \dsi ~, \\ & \sqrt{|\doubleA|} \scriptN_{gg'} ~,  \quad \text{otherwise} ~. \end{cases} 
\end{equation}

In this section, we will consider the case where $\doubleA = \doubleZ_N\times \doubleZ_N$. We will first search for the $S_3$-ality defects from the SymTFT. The first step is to find an $S_3$ subgroup of the anyon permutation symmetries such that every non-trivial group element has invertible $\beta$-component. We numerically search for such $S_3$'s for $N < 20$, where we find any $N$ containing a factor of $2$ or $3$ does not admit the desired $S_3$, therefore $\underline{\CE}_{S_3}\VEC_{\doubleZ_N\times \doubleZ_N}$ does not exist for such $N$. We prove this for the case where $N$ contains a factor of $2$, and conjecture this also holds when $N$ contains a factor of $3$ based on the numerical evidence. 

Assuming this indeed holds, then since $|S_3| = 6$ is coprime with $|\mathbf{A}| = N^4$ (where $N$ does not contain a factor of $2$ or $3$), $H^2(S_3,\mathbf{A})$ is trivial and there is a unique choice of the symmetry fractionalization. Then, $\underline{\CE}_{S_3}\VEC_{\doubleZ_N\times \doubleZ_N}$, when it exists, is classified by a choice of inequivalent bulk $S_3$ symmetries and a FS indicator $\alpha \in H^3(S_3,U(1)) \simeq \mathbb{Z}_6$. We then find the numbers of inequivalent $\underline{\CE}_{S_3}\VEC_{\doubleZ_N\times \doubleZ_N}$ for $N<20$ by explicit calculation, where the smallest $N$ such that $\underline{\CE}_{S_3}\VEC_{\doubleZ_N\times \doubleZ_N}$ exists is $N = 5$. The smallest $N$ such that some $\underline{\CE}_{S_3}\VEC_{\doubleZ_N\times \doubleZ_N}$ is group-theoretical is $N = 11$. For every $N<20$, none of the $S_3$-graded fusion categories we find admits a stable magnetic Lagrangian algebra in the bulk; therefore, they are all anomalous.

The section is organized as follows. In Section \ref{sec:S3_ality_SymTFT}, we describe our results on searching $S_3$-ality defects utilizing the SymTFT. In Section \ref{sec:S3_ality_gpt}, we provide a group-theoretical construction of the $S_3$-ality defects, which works for any $N$ coprime with $2,3,5,7$. This is consistent with the fact that $N = 11$ is the minimal $N$ can admit group-theoretical $S_3$-ality defects.
 
\subsection{Searching for $S_3$-ality defects from SymTFT}\label{sec:S3_ality_SymTFT}
From the generic discussion, we know that the first step is to find an $S_3$ subgroup of the bulk anyon permutation symmetry group $O(\doubleA \oplus \widehat{\doubleA})$, such that each non-trivial group element has invertible $\beta$. In this case, the $S_3$ is generated by an order-$3$ symmetry, which we denote as $T$, together with an order-$2$ symmetry $D$, satisfying the relation
\begin{equation}
    T^3 = 1 ~, \quad D^2 = 1 ~, \quad DTD = T^2 ~.
\end{equation}
Notice that in this case, one can no longer assume $T$ and $D$ both have the canonical form \eqref{eq:gf_generator}, since starting with generic $T$ and $D$, the conjugation by elements in $I_{\scriptA}$ can be used to fix one of them to be \eqref{eq:gf_generator}. For instance, when $N$ is coprime with $3$, we could set $T$ to be of $T_1,T_2$ given in \eqref{eq:Z3_sym}, and let the order-$2$ symmetry generator $D$ be generic. On the other hand, since any order-$2$ duality symmetry generator can be put into the form
\begin{equation}
    D_\chi = \begin{pmatrix} 0 & \chi \\ \chi^{-1} & 0 \end{pmatrix} ~,
\end{equation}
where $\chi$ is a $2\times 2$ $\doubleZ_N$-valued symmetric matrix, by conjugating with the matrices $t^n$ given in \eqref{eq:ZN_sym_generators}. This means a generic order-$2$ symmetry with invertible $\beta$-component in the bulk takes the form
\begin{equation}
    D = t^n D_\chi t^{-n} ~, \quad n = 0,\cdots, N-1 ~.
\end{equation}
Then, for any choice of $T$, we consider the $D$'s satisfies the following conditions
\begin{itemize}
    \item $T^2D = DT$,
    \item $TD$ and $T^2D$ both have invertible $\beta$-components.
\end{itemize}
To find inequivalent bulk $S_3$ symmetries, we begin by noticing that fixing $T = T_i$ does not fully fix the gauge freedom, generically there are still residual gauge symmetries captured by the subgroup $I_i \subset I_{\doubleZ_N \times \doubleZ_N}$ leaving the $\doubleZ_3 \subset S_3$ invariant. Notice that this subgroup $I_i$ is given by
\begin{equation}
    I_i = \{\phi \in I_{\doubleZ_N \times \doubleZ_N}: \phi T_i \phi^{-1} = T_i \quad or \quad \phi T_i \phi^{-1} = (T_i)^2 \} ~,
\end{equation}
as the relabeling of group elements in $\doubleZ_3$ via automorphism is also part of the redundancy. Then, any $D$ and $D'$ satisfying
\begin{equation}
    D = \phi T_i^n D' \phi^{-1} \quad
\end{equation}
for some $n \in \doubleZ_3$ and $\phi \in I_i$ are considered as equivalent.

Using this, we compute inequivalent choices of $S_3$ subgroups in the $\doubleZ_N \times \doubleZ_N$ numerically for $N < 20$ and $N = 3^3 = 27$, and summarized the result in Table \ref{tab:S3_class} and additional details on the concrete choices of the $S_3$ symmetries can be found in Appendix \ref{app:S3_class}

\begin{table}[h]
    \centering
    \begin{tabular}{|c|c|c|c|}
        \hline
        $N$ in $\doubleZ_N \times \doubleZ_N$ & \# of $T$'s & \# of $S_3$'s & \# of group-theoretical $S_3$'s \\
        \hline
        $2$ & 2 & 0 & 0 \\ 
        \hline
        $3$ & 3 & 0 & 0 \\
        \hline
        $4$ & 2 & 0 & 0 \\
        \hline
        {\color{red}5} & {\color{red}2} & {\color{red}3} & {\color{red}0} \\
        \hline
        $6$ & 6 & 0 & 0 \\
        \hline
        {\color{red}7} & {\color{red}2} & {\color{red}3} & {\color{red}0} \\ 
        \hline
        $8$ & 2 & 0 & 0 \\
        \hline
        $9$ & 7 & 0 & 0 \\
        \hline
        $10$ & 4 & 0 & 0 \\
        \hline
        {\color{red}11} & {\color{red}2} & {\color{red}6} & {\color{red}1} \\
        \hline
        $12$ & 6 & 0 & 0 \\
        \hline
        {\color{red}13} & {\color{red}2} & {\color{red}6} & {\color{red}1} \\
        \hline
        $14$ & 4 & 0 & 0 \\
        \hline
        $15$ & 6 & 0 & 0 \\
        \hline
        $16$ & 2 & 0 & 0 \\
        \hline
        {\color{red}$17$} & {\color{red}2} & {\color{red}9} & {\color{red}1} \\
        \hline
        $18$ & 14 & 0 & 0 \\
        \hline
        {\color{red}$19$} & {\color{red}2} & {\color{red}9} & {\color{red}1} \\
        \hline
        $\cdots$ & $\cdots$ & $\cdots$ & $\cdots$ \\
        \hline
        $27$ & 7 & 0 & 0 \\
        \hline
    \end{tabular}
    \caption{The classification of inequivalent choices of $S_3$ subgroup for $\doubleZ_N \times \doubleZ_N$ SymTFT. Notice that for $N$ coprime with $3$, the number of inequivalent $\doubleZ_3$ generators is given in Section \ref{sec:triality_setup}. For $N = 3, 9, 27$, we compute the inequivalent $\doubleZ_3\subset S_3$ generator $T$'s explicitly. }
    \label{tab:S3_class}
\end{table}

Observing the result, we find that any $N$ not coprime with $2$ or $3$ does not admit $S_3$-ality defects. By the same argument in the classification of the triality defects, given the prime decomposition of $N$,
\begin{equation}
    N = \prod_{i=1}^n p_i^{r_i} ~, \quad \doubleZ_N \times \doubleZ_N = \prod_{i=1}^n \doubleZ_{p_i^{r_i}} \times \doubleZ_{p_i^{r_i}} ~,
\end{equation}
specifying the $S_3$-symmetry in $\doubleZ_N \times \doubleZ_N$ SymTFT is equivalent to specifying an $\beta$-invertible $S_3$-symmetry for each $\doubleZ_{p_i^{r_i}} \times \doubleZ_{p_i^{r_i}}$ SymTFT. If no choice of $S_3$ exists in any prime factor, then there is no choice of $S_3$ for the full $\doubleZ_N \times \doubleZ_N$ SymTFT. Thus, to prove that there is no $S_3$-ality for $\doubleZ_N \times \doubleZ_N$ where $N$ is not coprime with $2$ or $3$, we only need to prove that there is no choice of $S_3$ subgroups for $N = 2^r, 3^r$. 

For $N = 2^r$, it is not hard to prove this explicitly, since there are only two inequivalent choices of $\doubleZ_3\subset S_3$ generator $T = T_1,T_2$ given in \eqref{eq:Z3_sym}. To proceed, first notice that a generic order-$2$ generator $D$ can be parameterized explicitly as
\begin{equation}
    D = \left(
\begin{array}{cccc}
 n b & -na  & a & b \\
 n c & -nb & b & c \\
 (n^2+p)c & -(n^2+p)b & nb & nc \\
 -(n^2+p)b& (n^2+p)a& -na  & -nb  \\
\end{array}
\right) ~, \quad n = 0,\cdots, N-1 ~, \quad p(ac-b^2) = 1 \mod N ~.
\end{equation}
It is straightforward to check for any choices of $T = T_1$ or $T_2$, $TD$ and $T^2D$ can not simultaneously have invertible $\beta$-component, as
\begin{equation}
    \det(\beta(TD)) = (n+1)^2 (ac - b^2) ~, \quad \det(\beta(T^2D)) = n^2 (ac - b^2) ~.
\end{equation}
In order for $\beta$ to be invertible, its determinant must be coprime with $N = 2^r$, that is, be an odd number. However, there is no way that $(n+1)^2$ and $n^2$ can be odd simultaneously; hence we conclude that there cannot be a valid choice of $S_3$ when $N = 2^r$.

For $N = 3^r$, however, at the moment we do not know the numbers and the generic forms of inequivalent $\doubleZ_3 \subset S_3$ generator $T$'s for generic $r$, we therefore conjecture that this is true based on the explicit result with $r = 1,2,3$.

With this, it is interesting to notice that in the case where the choice of the $\beta$-invertible $S_3$ symmetry exists, $|G|$ is coprime with $|\mathbf{A}|$, since $|\mathbf{A}|$ is coprime with $2,3$. Hence, the corresponding $H^2_\rho(S_3,\mathbf{A})$ is universally trivial. Furthermore, since the possible obstruction $H^4(S_3,U(1)) = \doubleZ_1$ is also always trivial, $\underline{\scriptE}_{S_3}\VEC_{\doubleZ_N \times \doubleZ_N}$ are expected to be classified by the choice of inequivalent $\beta$-invertible $S_3$ symmetries as well as the FS indicator $\alpha \in H^3(S_3, U(1)) \simeq \mathbb{Z}_6$.

\

The minimal $N$ with $S_3$-ality is $N = 5$. In this case, we find $3 \cdot 6$ inequivalent fusion categories $\underline{\scriptE}_{S_3}\VEC_{\doubleZ_5 \times \doubleZ_5}$. The first factor $3$ corresponds to $3$ inequivalent choices of the $S_3$ subgroups, while the second factor $6$ is from the choices of the FS indicator. Regarding the inequivalent choice of the $S_3$ symmetries, when choosing $T = T_1$, there is a unique inequivalent choice of the order-$2$ generators
\begin{equation}
    D_{1,1} = \left(
\begin{array}{cccc}
 0 & 3 & 1 & 0 \\
 4 & 0 & 0 & 2 \\
 4 & 0 & 0 & 4 \\
 0 & 2 & 3 & 0 \\
\end{array}
\right) ~.
\end{equation}
For $T = T_2$, there are two inequivalent choices of the order-$2$ generators
\begin{equation}
D_{2,1} = \left(
\begin{array}{cccc}
 4 & 3 & 1 & 2 \\
 2 & 1 & 2 & 1 \\
 2 & 1 & 4 & 2 \\
 1 & 2 & 3 & 1 \\
\end{array}
\right) ~, \quad 
D_{2,2} = \left(
\begin{array}{cccc}
 2 & 1 & 2 & 1 \\
 3 & 3 & 1 & 4 \\
 3 & 3 & 2 & 3 \\
 3 & 4 & 1 & 3 \\
\end{array}
\right) ~.
\end{equation}
Given the concrete form of the $S_3$'s, one can explicitly check if they admit stable Lagrangian algebras. The $\doubleZ_5 \times \doubleZ_5$-gauge theory admits $12$ Lagrangian algebras with generators $(e_1,e_2)$, $(e_2,m_1)$, $(e_1 e_2^k, m_1^{-k} m_2)$, $(m_1 e_2^k, m_2 e_1^{-k})$ where $k = 0,\cdots,4$. It is straightforward to check that none of the $S_3$'s admit stable Lagrangian algebras, and therefore are all non-group-theoretical. This means all $\underline{\scriptE}_{S_3}\VEC_{\doubleZ_5 \times \doubleZ_5}$ fusion categories are anomalous and do not admit trivial symmetric gapped phases.

As we increase $N$, we find the minimal $N$ such that the $S_3$-ality can be group-theoretical is $N = 11$. We further check if these group-theoretical $\underline{\scriptE}_{S_3} \VEC_{\doubleZ_N\times \doubleZ_N}$ admits a self-$S_3$-ality SPT phase by checking if the bulk $S_3$ symmetry admits a stable magnetic Lagrangian algebra. However, for all the examples we considered where $N < 20$, none of them admit stable magnetic Lagrangian algebra; therefore, all $\underline{\scriptE}_{S_3} \VEC_{\doubleZ_N\times \doubleZ_N}$'s for $N < 20$ are anomalous.

\subsection{A group-theoretical construction of $S_3$-ality defect}\label{sec:S3_ality_gpt}
As shown in Table \ref{tab:S3_class}, the minimal $N$ with group-theoretical $S_3$-ality is $N = 11$. Below, we present an example of the $S_3$-ality defects from group-theoretical construction. This construction works for generic $N$ coprime with $2,3,5,7$, and $N = 11$ is the smallest valid $N$. 

We start with the $S_3$ subgroup of $I_{\doubleZ_N \times \doubleZ_N}$ is given by
\begin{equation}
    \widetilde{D} = \begin{pmatrix} 0 & -1 & 0 & 0 \\ -1 & 0 & 0 & 0 \\ 0 & 0 & 0 & -1 \\ 0 & 0 & -1 & 0  \end{pmatrix} ~, \quad  \widetilde{T} = \begin{pmatrix} 0 & -1 & 0 & 0 \\ 1 & -1 & 0 & 0 \\ 0 & 0 & -1 & -1 \\ 0 & 0 & 1 & 0  \end{pmatrix} ~.
\end{equation}
From the boundary point of view, $\widetilde{D}$ and $\widetilde{T}$ generates the following $S_3$ automorphism of the boundary symmetry $\doubleZ_{N}^{\mathrm{a}} \times \doubleZ_{N}^{\mathrm{b}}$:
\begin{equation}
    \widetilde{D}: \ra \mapsto \rb^{-1} ~, \, \rb \mapsto \ra^{-1} ~, \quad \widetilde{T}: \ra \mapsto \rb ~, \, \rb \mapsto \ra^{-1} \rb^{-1} ~.
\end{equation}
The extended boundary symmetry is described as $G = (\doubleZ_{N} \times \doubleZ_{N}) \rtimes S_3$ with no 't Hooft anomaly, and denote the generator of $S_3$ as $\rt$ and $\rd$ (where $\rt^3 = \rd^2 = 1$), the semi-direct product is characterized by
\begin{equation}\label{eq:S3_ext}
    \rd \ra \rd = \rb^{-1} ~, \quad \rd \rb \rd = \ra^{-1} ~, \quad \rt \ra \rt^{-1} = \rb ~, \quad \rt \rb \rt^{-1} = \ra^{-1} \rb^{-1} ~.
\end{equation}
Under the conjugation of the following bulk symmetry $\phi$:
\begin{equation}
    \phi = \begin{pmatrix} 0 & 0 & 1 & 0 \\ 0 & 1 & 0 & 0 \\1 & 0 & 0 & 0 \\0 & 0 & 0 & 1 \\  \end{pmatrix}\begin{pmatrix} 1 & 0 & 0 & 0 \\ -3 & 1 & 0 & 0 \\0 & 0 & 1 & 3 \\0 & 0 & 0 & 1 \end{pmatrix} = \begin{pmatrix} 0 & 0 & 1 & 3 \\ -3 & 1 & 0 & 0 \\ 1 & 0 & 0 & 0 \\ 0 & 0 & 0 & 1 \end{pmatrix} ~,
\end{equation}
the generators $\widetilde{D}$ and $\widetilde{T}$ become
\begin{equation}
    D = \phi \widetilde{D} \phi^{-1} = \begin{pmatrix} -3 & 0 & 0 & 8 \\ 0 & 3 & 8 & 0 \\ 0 & -1 & -3 & 0 \\ -1 & 0 & 0 & 3 \end{pmatrix} ~, \quad T = \phi \widetilde{T} \phi^{-1} = \begin{pmatrix} 2 & 0 & 0 & -7 \\ 0 & 2 & 7 & 0 \\ 0 & -1 & -3 & 0 \\ 1 & 0 & 0 & -3 \end{pmatrix} ~.
\end{equation}
The other non-trivial group elements are given by
\begin{equation}
    T^2 = \begin{pmatrix} -3 & 0 & 0 & 7 \\ 0 & -3 & -7 & 0 \\  0 & 1 & 2 & 0 \\ -1 & 0 & 0 & 2 \end{pmatrix} ~, \quad TD = \begin{pmatrix} 1 & 0 & 0 & -5 \\ 0 & -1 & -5 & 0 \\ 0 & 0 & 1 & 0 \\ 0 & 0 & 0 & -1 \end{pmatrix} ~, \quad DT = \begin{pmatrix} 2 & 0 & 0 & -3 \\ 0 & -2 & -3 & 0 \\ 0 &1 & 2 & 0 \\ 1 & 0 & 0 & 2\end{pmatrix} ~.
\end{equation}
The $\beta$ components of all the non-trivial group elements are invertible as long as $N$ is coprime with $2,3,5,7$.

One can also check this from the boundary point of view. For simplicity, we only demonstrate with $N = 11$, as generic results will depend on the mod $N$ inverse of several numbers and therefore not particularly illuminating. Start with a theory $\widetilde{\scriptX}$ with $G = (\doubleZ_{11} \times \doubleZ_{11}) \rtimes S_3$ and couple the theory to $\doubleZ_{11}^{\ra} \times \doubleZ_{11}^{\rb}$ background fields. Following a similar derivation as in \eqref{eq:pt_trans}, the twisted partition functions satisfy
\begin{equation}\label{eq:S3_gt_inv_trans}
    Z_{\widetilde{\scriptX}}[A,B] = Z_{\widetilde{\scriptX}}[-B,-A] ~, \quad Z_{\widetilde{\scriptX}}[A,B] = Z_{\widetilde{\scriptX}}[-B,A-B] ~.
\end{equation}
The gauging implemented by the above $\phi$ defines a new theory $\scriptX$ related to $\widetilde{\scriptX}$ via
\begin{equation}
\begin{aligned}
    Z_{\scriptX}[A,B] &= \frac{1}{\sqrt{|H^1(\Sigma,\doubleZ_{11})|}} \sum_{a} Z_{\widetilde{\scriptX}}[a,B + 3a] e^{\frac{2\pi i}{11}\int a\cup A} ~, \\
    Z_{\widetilde{\scriptX}}[A,B] &= \frac{1}{\sqrt{|H^1(\Sigma,\doubleZ_{11})|}} \sum_{a} Z_{\scriptX}[a, B - 3A] e^{\frac{2\pi i}{11}\int a\cup A} ~.
\end{aligned}
\end{equation}
It is then straightforward to check \eqref{eq:S3_gt_inv_trans} imply the following properties on the twisted partition function $Z_{\scriptX}[A,B]$:
\begin{equation}
\begin{aligned}
    Z_{\scriptX}[A,B] &= \frac{1}{|H^1(\Sigma,\doubleZ_{11})|} \sum_{a,b} Z_{\scriptX}[a,b] e^{\frac{2\pi i}{11}\int a\cup b - a\cup (4B) + b\cup (7A) - A\cup B}~, \\
    Z_{\scriptX}[A,B] &= \frac{1}{|H^1(\Sigma,\doubleZ_{11})|} \sum_{a,b} Z_{\scriptX}[a,b] e^{\frac{2\pi i}{11}\int -2 a\cup b + a\cup (3B) + b\cup (-3A) - 6 A\cup B} ~.
\end{aligned}
\end{equation}
\subsection*{Group-Theoretical analysis}
We now check our proposal with the group-theoretical analysis of the fusion category $\scriptC((\doubleZ_{N}^\ra\times \doubleZ_{N}^\rb)\rtimes S_3, \omega, \doubleZ_{N}^{\ra\rb^3},1)$ (where $N$ is coprime with $2,3,5,7$) along the line of algebraic objects and bimodule objects as in Section \ref{sec:p_ality_gt}. Here, $\omega \in H^3(S_3,U(1))$ parameterizes the 't Hooft anomaly of $S_3$, which corresponds to different choices of the FS indicator.

The algebraic object $\scriptA$ is the group algebra of the $\doubleZ_{N}$ subgroup generated by $\rc\equiv \ra\rb^3$. The corresponding algebra object $\displaystyle \mathcal{A}=\bigoplus_{n\in \doubleZ_N} {\rc}^n$ with the fusion and split junction given by:
\begin{equation}
\begin{tikzpicture}[scale=0.50,baseline = {(0,0)}]
    \draw[thick, red, ->-=.5] (-1.7,-1) -- (0,0);
    \draw[thick, red, -<-=.5] (0,0) -- (1.7,-1);
    \draw[thick, red, -<-=.5] (0,2) -- (0,0);
    \node[red, above] at (0,2) {\scriptsize$\mathcal{A}$};
    \node[red, below] at (-1.7,-1) {\scriptsize$\mathcal{A}$};
    \node[red, below] at (1.7,-1) {\scriptsize$\mathcal{A}$};
    \node[red, right] at (+0,0.2) {\scriptsize$\mu$};
    \filldraw[red] (0,0) circle (2pt);
\end{tikzpicture} = \frac{1}{\sqrt{N}}\sum_{m,n\in \doubleZ_{N}} \begin{tikzpicture}[scale=0.50,baseline = 0.]
    \draw[thick, black, ->-=.5] (-1.7,-1) -- (0,0);
    \draw[thick, black, -<-=.5] (0,0) -- (1.7,-1);
    \draw[thick, black, -<-=.5] (0,2) -- (0,0);
    \node[black, above] at (0,2) {\scriptsize$ \rc^{m+n}$};
    \node[black, below] at (-1.7,-1) {\scriptsize$ \rc^m$};
    \node[black, below] at (1.7,-1) {\scriptsize$ \rc^n$};
\end{tikzpicture} ~, \quad \begin{tikzpicture}[scale=0.50,baseline = {(0,-0.5)}]
    \draw[thick, red, -<-=.5] (-1.7,1) -- (0,0);
    \draw[thick, red, ->-=.5] (0,0) -- (1.7,1);
    \draw[thick, red, ->-=.5] (0,-2) -- (0,0);
    \node[red, below] at (0,-2) {\scriptsize$\mathcal{A}$};
    \node[red, above] at (-1.7,1) {\scriptsize$\mathcal{A}$};
    \node[red, above] at (1.7,1) {\scriptsize$\mathcal{A}$};
    \node[red, right] at (+0,-0.2) {\scriptsize$\mu^\vee$};
    \filldraw[red] (0,0) circle (2pt);
\end{tikzpicture} = \frac{1}{\sqrt{N}} \sum_{m,n \in \doubleZ_{N}} \begin{tikzpicture}[scale=0.50,baseline = {(0,-0.5)}]
    \draw[thick, black, -<-=.5] (-1.7,1) -- (0,0);
    \draw[thick, black, ->-=.5] (0,0) -- (1.7,1);
    \draw[thick, black, ->-=.5] (0,-2) -- (0,0);
    \node[black, below] at (0,-2) {\scriptsize$ \rc^{m+n}$};
    \node[black, above] at (-1.7,1) {\scriptsize$ \rc^m$};
    \node[black, above] at (1.7,1) {\scriptsize$ \rc^n$};
\end{tikzpicture} ~.
\end{equation}
The dual symmetries are characterized by indecomposable bimodule objects of $\scriptA$. Among them, there are invertible ones which generates the $\doubleZ_{N} \times \doubleZ_{N}$ symmetries, and we denote them as $(\scriptM^{(i,j)},\lambda^{(i,j)},\rho^{(i,j)})$ where $\displaystyle \scriptM^{(i,j)} = \bigoplus_{n\in \doubleZ_N} \ra^i \rc^n$ with left and right action:
\begin{equation}
\begin{tikzpicture}[baseline={([yshift=-.5ex]current bounding box.center)},vertex/.style={anchor=base,
    circle,fill=black!25,minimum size=18pt,inner sep=2pt},scale=0.75]
\draw[red, thick, ->-=.5] (-1.5,-1) -- (-0.5,0);
\draw[dgreen, thick, ->-=.5] (-0.5,0) -- (0.5,1.0);
\draw[dgreen, thick, ->-=.5] (0.5,-1.0) -- (-0.5,0);

\node[red, below] at (-1.5,-1.5+0.5) {\footnotesize $\mathcal{A}$};
\node[dgreen, above] at (0.5,0.5+0.5) {\footnotesize $\mathcal{M}^{(i,j)}$};
\node[dgreen, below] at (0.5,-1.5+0.5) {\footnotesize $\mathcal{M}^{(i,j)}$};

\filldraw[dgreen] (-0.5,-0.5+0.5) circle (1.5pt);

\node[dgreen, left] at (-0.5,-0.5+0.5) {\footnotesize $\lambda^{(i,j)}$};
\end{tikzpicture} = \frac{1}{\sqrt{N}} \sum_{m,n \in \doubleZ_{N}} \begin{tikzpicture}[baseline={([yshift=-.5ex]current bounding box.center)},vertex/.style={anchor=base,
    circle,fill=black!25,minimum size=18pt,inner sep=2pt},scale=0.75]
\draw[black, thick, ->-=.5] (-1.5,-1) -- (-0.5,0);
\draw[black, thick, ->-=.5] (-0.5,0) -- (0.5,1.0);
\draw[black, thick, ->-=.5] (0.5,-1.0) -- (-0.5,0);

\node[black, below] at (-1.5,-1.5+0.5) {\footnotesize $\rc^m$};
\node[black, above] at (0.5,0.5+0.5) {\footnotesize $\ra^i \rc^{m+n}$};
\node[black, below] at (0.5,-1.5+0.5) {\footnotesize $\ra^i \rc^{n}$};

\end{tikzpicture} ~, \quad \begin{tikzpicture}[baseline={([yshift=-.5ex]current bounding box.center)},vertex/.style={anchor=base,
    circle,fill=black!25,minimum size=18pt,inner sep=2pt},scale=0.75]
\draw[dgreen, thick, ->-=.5] (-1.5,-1) -- (-0.5,0);
\draw[dgreen, thick, ->-=.5] (-0.5,0) -- (0.5,1.0);
\draw[red, thick, ->-=.5] (0.5,-1.0) -- (-0.5,0);

\node[dgreen, below] at (-1.5,-1.5+0.5) {\footnotesize $\mathcal{M}^{(i,j)}$};
\node[dgreen, above] at (0.5,0.5+0.5) {\footnotesize $\mathcal{M}^{(i,j)}$};
\node[red, below] at (0.5,-1.5+0.5) {\footnotesize $\mathcal{A}$};

\filldraw[dgreen] (-0.5,-0.5+0.5) circle (1.5pt);

\node[dgreen, left] at (-0.5,-0.5+0.5) {\footnotesize $\rho^{(i,j)}$};
\end{tikzpicture} = \frac{1}{\sqrt{N}}\sum_{m,n\in \doubleZ_{N}} e^{\frac{2\pi i}{N}jn} \begin{tikzpicture}[baseline={([yshift=-.5ex]current bounding box.center)},vertex/.style={anchor=base,
    circle,fill=black!25,minimum size=18pt,inner sep=2pt},scale=0.75]
\draw[black, thick, ->-=.5] (-1.5,-1) -- (-0.5,0);
\draw[black, thick, ->-=.5] (-0.5,0) -- (0.5,1.0);
\draw[black, thick, ->-=.5] (0.5,-1.0) -- (-0.5,0);

\node[black, below] at (-1.5,-1.5+0.5) {\footnotesize $\ra^i \rc^m$};
\node[black, above] at (0.5,0.5+0.5) {\footnotesize $\ra^i \rc^{m+n}$};
\node[black, below] at (0.5,-1.5+0.5) {\footnotesize $\rc^n$};

\end{tikzpicture} ~.
\end{equation}
Furthermore, There are 5 non-invertible bimodule objects $(\scriptM^{g},\lambda^{g},\rho^{g})$ labeled by $g \in S_3 - \{\dsi\} $ in the orbifold theory \footnote{Here, the condition that $N$ is coprime with $2,3,5,7$ enters. Without this, $\mathcal{M}^g$ will generically split into a direct sum of smaller simple bimodule objects.}: 
\begin{equation}
    \mathcal{M}^g = \bigoplus_{m,n \in \doubleZ_N} g \ra^m \rb^n ~, \quad g = \rt,\rt^2,\rd,\rd \rt, \rt \rd ~,
\end{equation}
and left and right action $(\lambda^g,\rho^g)$ are trivial (meaning every junction coefficient is $1$ up to normalization). The split junction from the left and right is also trivial. For instance, for $g = \rt$, we have 
\begin{equation}
\begin{tikzpicture}[baseline={([yshift=-.5ex]current bounding box.center)},vertex/.style={anchor=base,
    circle,fill=black!25,minimum size=18pt,inner sep=2pt},scale=0.7]
\draw[red, thick, ->-=.5] (-1.5,-1) -- (-0.5,0);
\draw[dgreen, thick, ->-=.5] (-0.5,0) -- (0.5,1.0);
\draw[dgreen, thick, ->-=.5] (0.5,-1.0) -- (-0.5,0);

\node[red, below] at (-1.5,-1.5+0.5) {\footnotesize $\mathcal{A}$};
\node[dgreen, above] at (0.5,0.5+0.5) {\footnotesize $\mathcal{M}^{\rt}$};
\node[dgreen, below] at (0.5,-1.5+0.5) {\footnotesize $\mathcal{M}^{\rt}$};

\filldraw[dgreen] (-0.5,-0.5+0.5) circle (1.5pt);

\node[dgreen, left] at (-0.5,-0.5+0.5) {\footnotesize $\lambda^{\rt}$};
\end{tikzpicture} = \frac{1}{\sqrt{N}} \sum_{m,n,l} \begin{tikzpicture}[baseline={([yshift=-.5ex]current bounding box.center)},vertex/.style={anchor=base,
    circle,fill=black!25,minimum size=18pt,inner sep=2pt},scale=0.75]
\draw[black, thick, ->-=.5] (-1.5,-1) -- (-0.5,0);
\draw[black, thick, ->-=.5] (-0.5,0) -- (0.5,1.0);
\draw[black, thick, ->-=.5] (0.5,-1.0) -- (-0.5,0);

\node[black, below] at (-1.5,-1.5+0.5) {\footnotesize $\rc^{l}$};
\node[black, above] at (0.5,0.5+0.5) {\footnotesize $\rt\ra^{m + 2l}\rb^{n - l}$};
\node[black, below] at (1.0,-1.5+0.5) {\footnotesize $\rt\ra^{m}\rb^{n}$};

\end{tikzpicture} ~,
\begin{tikzpicture}[baseline={([yshift=-.5ex]current bounding box.center)},vertex/.style={anchor=base,
    circle,fill=black!25,minimum size=18pt,inner sep=2pt},scale=0.7]
\draw[dgreen, thick, ->-=.5] (-1.5,-1) -- (-0.5,0);
\draw[dgreen, thick, ->-=.5] (-0.5,0) -- (0.5,1.0);
\draw[red, thick, ->-=.5] (0.5,-1.0) -- (-0.5,0);

\node[dgreen, below] at (-1.5,-1.5+0.5) {\footnotesize $\mathcal{M}^{\rt}$};
\node[dgreen, above] at (0.5,0.5+0.5) {\footnotesize $\mathcal{M}^{t}$};
\node[red, below] at (0.5,-1.5+0.5) {\footnotesize $\mathcal{A}$};

\filldraw[dgreen] (-0.5,-0.5+0.5) circle (1.5pt);

\node[dgreen, left] at (-0.5,-0.5+0.5) {\footnotesize $\rho^{t}$};
\end{tikzpicture} = \frac{1}{\sqrt{N}}\sum_{m,n,l} \begin{tikzpicture}[baseline={([yshift=-.5ex]current bounding box.center)},vertex/.style={anchor=base,
    circle,fill=black!25,minimum size=18pt,inner sep=2pt},scale=0.75]
\draw[black, thick, ->-=.5] (-1.5,-1) -- (-0.5,0);
\draw[black, thick, ->-=.5] (-0.5,0) -- (0.5,1.0);
\draw[black, thick, ->-=.5] (0.5,-1.0) -- (-0.5,0);

\node[black, below] at (-1.5,-1.5+0.5) {\footnotesize $\rt\ra^{m}\rb^{n}$};
\node[black, above] at (0.3,0.5+0.5) {\footnotesize $\rt\ra^{m+l}\rb^{n+3l}$};
\node[black, below] at (0.5,-1.5+0.5) {\footnotesize $\rc^{l}$};

\end{tikzpicture} ~.
\end{equation}
Denote the simple TDL corresponding to $\mathcal{M}^g$ as $\scriptN_g$, it is straightforward to check that one has the fusion rule given in \eqref{eq:S3_fusion_rule}. 

Notice that it is then possible to build some basic twisted partition functions capturing the action of $S_3$-ality defects on the Hilbert space, or the partition functions over the defect Hilbert space of the $S_3$-ality defects in the orbifold theory, following the approach in Section \ref{sec:p_ality_gt}. We will not dive into the details here. One interesting result is that one can match the $S_3$-ality symmetry preserving operator in the orbifold theory with the local operator in the original theory, which we summarized in Table \ref{tab:S3_charge}. The upshot is any $S_3$-ality preserving local operator $\scriptO$ in the dual theory must be a $(\doubleZ_N \times \doubleZ_N) \rtimes S_3$ invariant operator in the original theory.

\begin{table}[H]
    \centering
    \begin{tabular}{c|c}
        irrep under $S_3$ &  charges under the $S_3$-ality defects \\
        \hline
         $\mathbf{1}$ & $\mathcal{T} \mathcal{O} = \mathcal{D} \mathcal{O} = N \mathcal{O}$ \\
        \hline
         $\mathbf{1}_s$ & $\mathcal{T} \mathcal{O} = N \mathcal{O} ~, \quad \mathcal{D} \mathcal{O} = - N \mathcal{O}$ \\
        \hline
         $\mathbf{2}$ & $\mathcal{T} \mathcal{O}_i = \omega^i N \mathcal{O}_i ~, \quad \mathcal{D} \mathcal{O}_i = N (\sigma_x)_{ij} \mathcal{O}_j ~, \quad i = 1,2$ \\
    \end{tabular}
    \caption{Here, $\omega = e^{\frac{2\pi i}{3}}$. The operator $\scriptO$ considered here is neutral under the dual $\doubleZ_{N} \times \doubleZ_N$ symmetry, otherwise $\scriptT \scriptO = \scriptD \scriptO =0$ from the fusion rule. As a result, it is an $\doubleZ_N \times \doubleZ_N$ neutral operator in the original theory. For $\mathbf{1}$, the operator $\mathcal{O}$ preserves the full $S_3$-ality defect. For $\mathbf{1}_s$, $\mathcal{O}$ only preserves the triality defect $\mathcal{T}$. For $\mathbf{2}$, the linear combination $\mathcal{O}_1 + \mathcal{O}_2$ will preserve the duality defect $\mathcal{D}$ but breaks $\mathcal{T}$.}
    \label{tab:S3_charge}
\end{table}

To conclude, this group-theoretical construction enables the identification of $S_3$-ality defects in concrete 2d CFTs. For instance, one can consider embedding $(\doubleZ_{11} \times \doubleZ_{11})\rtimes S_3$ into $(U(1) \times U(1))\rtimes S_3$, and look for 2d CFT with such symmetry. A simple example is to consider two compact bosons $\phi^i \simeq \phi^i + 2\pi$ with the simple action
\begin{equation}\label{eq:c2_action}
    S = \frac{R^2}{4\pi}\int d\phi^1 \wedge *d\phi^1 - d\phi^1 \wedge *d\phi^2 + d\phi^2\wedge *d\phi^2 ~.
\end{equation}
This action admits the $S_3$-symmetry, whose generators $\rt,\rd$ act as
\begin{equation}
    \rt: \begin{pmatrix} \phi^1 \\ \phi^2 \end{pmatrix} \mapsto \begin{pmatrix} 0 & -1 \\ 1 & -1 \end{pmatrix} \begin{pmatrix} \phi^1 \\ \phi^2 \end{pmatrix} ~, \quad \rd: \begin{pmatrix} \phi^1 \\ \phi^2 \end{pmatrix} \mapsto \begin{pmatrix} 0 & -1 \\ -1 & 0 \end{pmatrix} \begin{pmatrix} \phi^1 \\ \phi^2 \end{pmatrix} ~.
\end{equation}
Let's consider coupling the theory to two $U(1)$ background fields $A,B$ and consider the corresponding partition functions:
\begin{equation}
    Z[A,B] = \int [\scriptD \phi^1][\scriptD \phi^2] e^{-\frac{R^2}{4\pi}\int (d\phi^1 + A) \wedge *(d\phi^1 + A) - (d\phi^1 + A) \wedge *(d\phi^2 + B) + (d\phi^2 + B) \wedge (*d\phi^2 +B)} ~,
\end{equation}
and the $S_3$-symmetry then implies the following properties of $Z[A,B]$
\begin{equation}
    Z[A,B] = Z[-B,A-B]~, \quad Z[A,B] = Z[-B,-A] ~,
\end{equation}
which matches the form \eqref{eq:S3_gt_inv_trans}. Furthermore, there is no 't Hooft anomaly for the $U(1)^2$ symmetry, therefore we can gauge any $\doubleZ_N \times \doubleZ_N$ subgroup of $U(1) \times U(1)$. For any $N$ coprime with $2,3,5,7$, this will turn the $S_3$-symmetry into $S_3$-ality. As pointed out previously, any $S_3$-ality preserving operator in the dual theory must come from the singlet under the entire $(\doubleZ_N \times \doubleZ_N)\rtimes S_3$ in the original theory. It is straightforward to see that no $S_3$-ality preserving relevant deformation exists, while an $S_3$-ality preserving marginal deformation does. This marginal direction is parameterized by the overall radius $R^2$ in the action. 

\section*{Acknowledgements}
We are grateful to Ken Intriligator, Theo Jacobson, John McGreevy, Aiden Sheckler, Yi-Zhuang You for interesting discussions. We thank Yi-Zhuang You for collaboration on the companion paper regarding the lattice realization \cite{Lu:2024ytl}. D.C.L. is supported by the National Science Foundation (NSF) Grant No. DMR-2238360. Z.S. is partially supported by the US Department of Energy (DOE) under cooperative research agreement DE-SC0009919, Simons Foundation award No. 568420 (K.I.), and the Simons Collaboration on Global Categorical Symmetries. Z.Z. is supported in part by the Simons Collaboration on Global Categorical Symmetries, and by Simons Foundation award 568420.

\appendix

\section{Proof of the results on the anomaly for $\doubleZ_N \times \doubleZ_N$ where $\gcd(N,3) = 1$}\label{app:general_proof}
In this appendix, we provide details on deriving the anomaly result for generic $N$ where $\gcd(N,3) = 1$. Consider the prime decomposition of $\displaystyle N = \prod_{i=1}^n p_i^{r_i}$, we can write the bulk $\doubleZ_N \times \doubleZ_N$ gauge theory as a product of $\doubleZ_{p_i^{r_i}}\times \doubleZ_{p_i^{r_i}}$ gauge theories. For each $\doubleZ_{p_i^{r_i}}\times \doubleZ_{p_i^{r_i}}$ gauge theory, there is a choice of inequivalent $\doubleZ_3$ anyon permutation symmetry, labeled by $T_1$ and $T_2$. The triality in $\doubleZ_N \times \doubleZ_N$ gauge theory then corresponds to the diagonal $\doubleZ_3$-anyon permutation symmetry $\displaystyle \mathbf{T} = \bigotimes_{i=1}^n T^{(i)}$ (where $T^{(i)} = T_1$ or $T_2$) and a choice of FS indicator $\alpha \in \doubleZ_3$. 

Given a choice of $\mathbf{T}$, the existence of the stable magnetic Lagrangian algebra can be checked at the level of each prime factor. To see this, notice that the magnetic Lagrangian algebra is classified by $\psi \in H^2(\doubleZ_N \times \doubleZ_N,U(1))$, or equivalently by the anti-symmetric bilinear form $\xi_\psi(g,h) = \frac{\psi(g,h)}{\psi(h,g)}$. By the prime decomposition, we can write $\doubleZ_N \times \doubleZ_N$ as
\begin{equation}
    \doubleZ_N \times \doubleZ_N = \prod_{i=1}^n (\doubleZ_{p_i^{r_i}} \times \doubleZ_{p_i^{r_i}}) ~.
\end{equation}
The $H^2(\doubleZ_N \times \doubleZ_N,U(1))$ also decomposes in the similar fashion:
\begin{equation}
    \doubleZ_N = H^2(\doubleZ_N \times \doubleZ_N,U(1)) = \prod_{i = 1}^n H^2(\doubleZ_{p_i^{r_i}} \times \doubleZ_{p_i^{r_i}}, U(1)) = \prod_{i=1}^n \doubleZ_{p_i^{r_i}} ~.
\end{equation}
More concretely, one can show that the anti-symmetric bilinear form $\xi$ factorizes in a way that 
\begin{equation}
    \xi(h_i, h_j) = 1 ~, \quad \forall \, h_i \in \doubleZ_{p_i^{r_i}} \times \doubleZ_{p_i^{r_i}} ~, \quad h_j \in \doubleZ_{p_j^{r_j}} \times \doubleZ_{p_j^{r_j}} ~, \quad i \neq j ~.
\end{equation}
Let $m_i$ be the order of $h_i$ and $m_j$ be the order of $h_j$, and clearly $m_i$ is coprime with $m_j$. Then,
\begin{equation}
    [\xi(h_i,h_j)]^{m_i} = \xi(1, h_j) = 1 ~, \quad [\xi_H(h_i,h_j)]^{m_j} = \xi(h_i,1) = 1 ~, 
\end{equation}
and the two equations on $\xi(h_i,h_j)$ then imply $\xi(h_i,h_j) = 1$ as $\gcd(m_i,m_j) = 1$.

This factorization of the anti-symmetric bilinear form $\xi$ implies that any magnetic Lagrangian algebra of the $\doubleZ_N \times \doubleZ_N$ gauge theory can be constructed as the product of the magnetic Lagrangian algebras of each $\doubleZ_{p_i^{r_i}} \times \doubleZ_{p_j^{r_j}}$ gauge theory. This guarantees that we can check the stableness condition at the level of each factor.

Hence, combining our previous result for a specific $p^r$ in Section \ref{sec:pr_anomaly}, we immediately conclude that if $N$ contains a prime factor $p_i \neq 1 \mod 3$, then the corresponding triality fusion category is anomalous. 

For the $N$ such that each of its prime factors $p_i$ satisfies $p_i = 1 \mod 3$, the triality fusion category is group-theoretical and can be constructed from the gauging $\displaystyle \mathbb{H} = \prod_{i=1}^{n} \mathbb{H}^{(i)}$ in the finite group $\mathbb{G}$ where
\begin{equation}
    \mathbb{G} = \left(\prod_{i=1}^n \doubleZ_{p_i^{r_i}}^{\ra_i} \times \doubleZ_{p_i^{r_i}}^{\rb_i} \right)\rtimes \doubleZ_3^{\rc} ~,
\end{equation}
where $\ra_i, \rb_i, \rc$ denotes the generators of the corresponding group and the automorphism $\rc$ acts as
\begin{equation}
    \begin{cases}
        \rc \rb_i \rc^{-1} = (\ra_i \rb_i)^{-1} ~, \quad  \rc \rb_i \rc^{-1} = \ra_i ~, \quad \text{if} \quad T^{(i)} = T_1 ~, \\
        \rc \ra_i \rc^{-1} = \ra_i ~, \quad \rc \rb_i \rc^{-1} = \rb_i^{k_i - 1} ~, \quad \text{where} \quad (k_i)^2 + k_i + 1 = 0 \mod p_i^{r_i} ~, \quad \text{if}  \quad T^{(i)} = T_2 ~. 
    \end{cases}
\end{equation}
And the group $\displaystyle \mathbb{H} = \prod_{i=1}^n \mathbb{H}^{(i)}$ we gauge is specified by its component
\begin{equation}
    \mathbb{H}^{(i)} = \begin{cases}
        \doubleZ_{p_i^{r_i}}^{\ra_i} ~, \quad \text{if} \quad T^{(i)} = T_1 ~, \\
        \doubleZ_{p_i^{r_i}}^{\ra_i} \times \doubleZ_{p_i^{r_i}}^{\rb_i} ~, \quad \text{if} \quad T^{(i)} = T_2 ~.
    \end{cases}
\end{equation}
The discrete torsion is given by
\begin{equation}
    \psi_{\mathbb{H}} = \prod_{i \, s.t. \, T^{(i)} = T_2 } \psi_i ~,
\end{equation}
where $\psi_i \in H^2(\doubleZ_{p_i^{r_i}}\times \doubleZ_{p_i^{r_i}}, U(1))$ given by \eqref{eq:T2_discrete_torsion}. Finally, different choices of the FS indicator are realized by choosing different self-anomalies of the $\doubleZ_3^{\rc}$ subgroup.

To see if the triality specified by $(\mathbf{T}, \alpha)$ is anomaly-free, we again use the criterion \eqref{eq:anomaly_free_criteria}. First, we need to find $\mathbb{K} \subset \mathbb{G}$ such that $\mathbb{G}$ is contained in $\mathbb{H}\mathbb{K}$. This means $\mathbb{K}$ exists a $\doubleZ_3$ subgroup, whose generator is a product of $\rc$ with $\ra_i$'s and $b_j$'s. Then, $\mathbb{K}$ is anomalous unless the $\doubleZ_3^{\rc}$ is anomaly-free, which implies the FS indicator is trivial. When the FS indicator is indeed trivial, we could then choose the subgroup $\mathbb{K}$ to be generated by
\begin{equation}
    \rc ~,  \quad (\ra_i)^{-k_i} \rb_i ~, \quad \text{for the $i$ such that $T^{(i)} = T_1$} ~. 
\end{equation}
It's straightforward to check that $\mathbb{H} \cap \mathbb{K} = \{\dsi\}$, and hence the symmetry is anomaly-free when the FS indicator is trivial. Thus, we reach the conclusion that the triality fusion category $\underline{\scriptE}_{\doubleZ_3}^{(\mathbf{T},\alpha)}\VEC_{\doubleZ_N \times \doubleZ_N}$ is anomaly-free if and only if
\begin{enumerate}
    \item In the prime decomposition \eqref{eq:prime_decomp_N}, every prime factor $p_i = 1 \mod 3$.
    \item The FS indicator $\alpha$ is trivial.
\end{enumerate}

\section{Details on the $\p_\pm^3$ construction of the $p$-ality fusion category}\label{app:ppm3_detail}
In this section, we provide some details on the $\p_\pm^3$ group-theoretical construction of the $p$-ality fusion categories in terms of algebraic objects and the bimodule objects. Notice that the same procedure can be applied directly to work out the group-theoretical construction of the $S_3$-ality defect constructed in Section \ref{sec:S3_ality} as well.

Generically speaking, after picking a gaugable algebra $\scriptA$, a bimodule object $(\mathcal{M},\lambda,\rho)$ of $\scriptA$ is described by a (generically non-simple) object $\scriptM$ of the fusion category, $\lambda \in \Hom(\mathcal{A}\otimes\mathcal{M},\mathcal{M})$ and $\rho \in \Hom(\mathcal{M}\otimes \mathcal{A},\mathcal{M})$ are fusion junctions describing the left and right action of $\mathcal{A}$ on $\mathcal{M}$ respectively, which satisfy the following relations:
\begin{equation}\label{eq:bimodule_condition}
\begin{tikzpicture}[baseline={([yshift=-.5ex]current bounding box.center)},vertex/.style={anchor=base,
    circle,fill=black!25,minimum size=18pt,inner sep=2pt},scale=0.4]
\draw[red, thick, ->-=.5] (-1.5,-1.5) -- (-0.5,-0.5);
\draw[red, thick, ->-=.5] (-0.5,-0.5) -- (0.5,0.5);
\draw[dgreen, thick, ->-=.5] (0.5,0.5) -- (1.5,1.5);
\draw[red, thick, ->-=.5] (0.5,-1.5) -- (-0.5,-0.5);
\draw[dgreen, thick, ->-=.5] (2.5,-1.5) -- (0.5,0.5);

\filldraw[red] (-0.5,-0.5) circle (3pt);
\filldraw[dgreen] (0.5,0.5) circle (3pt);

\node[red, below] at (0.3,0.3) {\scriptsize $\mathcal{A}$};

\node[red, below] at (-1.5,-1.5) {\scriptsize $\mathcal{A}$};
\node[dgreen, above] at (1.5,1.5) {\scriptsize $\mathcal{M}$};
\node[dgreen, below] at (2.5,-1.5) {\scriptsize $\mathcal{M}$};
\node[red, below] at (0.5,-1.5) {\scriptsize $\mathcal{A}$};
\node[red, left] at (-0.3,-0.3) {\scriptsize $\mu$};
\node[dgreen, left] at (0.7,0.7) {\scriptsize $\lambda$};
\end{tikzpicture} = \begin{tikzpicture}[baseline={([yshift=-.5ex]current bounding box.center)},vertex/.style={anchor=base,
    circle,fill=black!25,minimum size=18pt,inner sep=2pt},scale=0.4]
\draw[red, thick, ->-=.5] (-1.5,-1.5) -- (0.5,0.5);
\draw[dgreen, thick, ->-=.5] (0.5,0.5) -- (1.5,1.5);
\draw[red, thick, ->-=.5] (0.5,-1.5) -- (1.5,-0.5);
\draw[dgreen, thick, ->-=.5] (2.5,-1.5) -- (1.5,-0.5);
\draw[dgreen, thick, ->-=.5] (1.5,-0.5) -- (0.5,0.5);

\filldraw[dgreen] (1.5,-0.5) circle (3pt);
\filldraw[dgreen] (0.5,0.5) circle (3pt);

\node[dgreen, below] at (0.6,0.3) {\scriptsize $\mathcal{M}$};

\node[red, below] at (-1.5,-1.5) {\scriptsize $\mathcal{A}$};
\node[dgreen, above] at (1.5,1.5) {\scriptsize $\mathcal{M}$};
\node[dgreen, below] at (2.5,-1.5) {\scriptsize $\mathcal{M}$};
\node[red, below] at (0.5,-1.5) {\scriptsize $\mathcal{A}$};
\node[dgreen, right] at (1.3,-0.3) {\scriptsize $\lambda$};
\node[dgreen, left] at (0.7,0.7) {\scriptsize $\lambda$};
\end{tikzpicture}, \begin{tikzpicture}[baseline={([yshift=-.5ex]current bounding box.center)},vertex/.style={anchor=base,
    circle,fill=black!25,minimum size=18pt,inner sep=2pt},scale=0.4]
\draw[dgreen, thick, ->-=.5] (-1.5,-1.5) -- (-0.5,-0.5);
\draw[dgreen, thick, ->-=.5] (-0.5,-0.5) -- (0.5,0.5);
\draw[dgreen, thick, ->-=.5] (0.5,0.5) -- (1.5,1.5);
\draw[red, thick, ->-=.5] (0.5,-1.5) -- (-0.5,-0.5);
\draw[red, thick, ->-=.5] (2.5,-1.5) -- (0.5,0.5);

\filldraw[dgreen] (-0.5,-0.5) circle (3pt);
\filldraw[dgreen] (0.5,0.5) circle (3pt);

\node[dgreen, below] at (0.3,0.3) {\scriptsize $\mathcal{M}$};

\node[dgreen, below] at (-1.5,-1.5) {\scriptsize $\mathcal{M}$};
\node[dgreen, above] at (1.5,1.5) {\scriptsize $\mathcal{M}$};
\node[red, below] at (2.5,-1.5) {\scriptsize $\mathcal{A}$};
\node[red, below] at (0.5,-1.5) {\scriptsize $\mathcal{A}$};
\node[dgreen, left] at (-0.3,-0.3) {\scriptsize $\rho$};
\node[dgreen, left] at (0.7,0.7) {\scriptsize $\rho$};
\end{tikzpicture} = \begin{tikzpicture}[baseline={([yshift=-.5ex]current bounding box.center)},vertex/.style={anchor=base,
    circle,fill=black!25,minimum size=18pt,inner sep=2pt},scale=0.4]
\draw[dgreen, thick, ->-=.5] (-1.5,-1.5) -- (0.5,0.5);
\draw[dgreen, thick, ->-=.5] (0.5,0.5) -- (1.5,1.5);
\draw[red, thick, ->-=.5] (0.5,-1.5) -- (1.5,-0.5);
\draw[red, thick, ->-=.5] (2.5,-1.5) -- (1.5,-0.5);
\draw[red, thick, ->-=.5] (1.5,-0.5) -- (0.5,0.5);

\filldraw[red] (1.5,-0.5) circle (3pt);
\filldraw[dgreen] (0.5,0.5) circle (3pt);

\node[red, below] at (0.6,0.3) {\scriptsize $\mathcal{A}$};

\node[dgreen, below] at (-1.5,-1.5) {\scriptsize $\mathcal{M}$};
\node[dgreen, above] at (1.5,1.5) {\scriptsize $\mathcal{M}$};
\node[red, below] at (2.5,-1.5) {\scriptsize $\mathcal{A}$};
\node[red, below] at (0.5,-1.5) {\scriptsize $\mathcal{A}$};
\node[red, right] at (1.3,-0.3) {\scriptsize $\mu$};
\node[dgreen, left] at (0.7,0.7) {\scriptsize $\rho$};
\end{tikzpicture}, \begin{tikzpicture}[baseline={([yshift=-.5ex]current bounding box.center)},vertex/.style={anchor=base,
    circle,fill=black!25,minimum size=18pt,inner sep=2pt},scale=0.4]
\draw[red, thick, ->-=.5] (-1.5,-1.5) -- (-0.5,-0.5);
\draw[dgreen, thick, ->-=.5] (-0.5,-0.5) -- (0.5,0.5);
\draw[dgreen, thick, ->-=.5] (0.5,0.5) -- (1.5,1.5);
\draw[dgreen, thick, ->-=.5] (0.5,-1.5) -- (-0.5,-0.5);
\draw[red, thick, ->-=.5] (2.5,-1.5) -- (0.5,0.5);

\filldraw[dgreen] (-0.5,-0.5) circle (3pt);
\filldraw[dgreen] (0.5,0.5) circle (3pt);

\node[dgreen, below] at (0.3,0.3) {\scriptsize $\mathcal{M}$};

\node[red, below] at (-1.5,-1.5) {\scriptsize $\mathcal{A}$};
\node[dgreen, above] at (1.5,1.5) {\scriptsize $\mathcal{M}$};
\node[red, below] at (2.5,-1.5) {\scriptsize $\mathcal{A}$};
\node[dgreen, below] at (0.5,-1.5) {\scriptsize $\mathcal{M}$};
\node[dgreen, left] at (-0.3,-0.3) {\scriptsize $\lambda$};
\node[dgreen, left] at (0.7,0.7) {\scriptsize $\rho$};
\end{tikzpicture} = \begin{tikzpicture}[baseline={([yshift=-.5ex]current bounding box.center)},vertex/.style={anchor=base,
    circle,fill=black!25,minimum size=18pt,inner sep=2pt},scale=0.4]
\draw[red, thick, -stealth] (-1.5,-1.5) -- (-0.5,-0.5);
\draw[red, thick] (-0.5,-0.5) -- (0.5,0.5);
\draw[dgreen, thick, -stealth] (0.5,0.5) -- (1.0,1.0);
\draw[dgreen, thick] (1.0,1.0) -- (1.5,1.5);

\draw[dgreen, thick, -stealth] (0.5,-1.5) -- (1.0,-1.0);
\draw[dgreen, thick] (1.0,-1.0) -- (1.5,-0.5);

\draw[red, thick, -stealth] (2.5,-1.5) -- (2,-1.0);
\draw[red, thick] (2,-1.0) -- (1.5,-0.5);
\draw[dgreen, thick, -stealth] (1.5,-0.5) -- (1.0,0);
\draw[dgreen, thick] (1.0,0) -- (0.5,0.5);

\filldraw[dgreen] (1.5,-0.5) circle (3pt);
\filldraw[dgreen] (0.5,0.5) circle (3pt);

\node[dgreen, below] at (0.6,0.3) {\scriptsize $\mathcal{M}$};

\node[red, below] at (-1.5,-1.5) {\scriptsize $\mathcal{A}$};
\node[dgreen, above] at (1.5,1.5) {\scriptsize $\mathcal{M}$};
\node[red, below] at (2.5,-1.5) {\scriptsize $\mathcal{A}$};
\node[dgreen, below] at (0.5,-1.5) {\scriptsize $\mathcal{M}$};
\node[dgreen, right] at (1.3,-0.3) {\scriptsize $\rho$};
\node[dgreen, left] at (0.7,0.7) {\scriptsize $\lambda$};
\end{tikzpicture} ~.
\end{equation}
To construct the twisted partition function in the orbifold theory, the splitting versions of the junctions $\lambda^\vee,\rho^\vee$ are required and can be easily determined from $\lambda,\rho$ via the following relation:
\begin{equation}
\begin{tikzpicture}[scale=0.50,baseline = {(0,0)}]
    \draw[thick, dgreen, ->-=.5] (0,-2) -- (0,-0.7);
    \draw[thick, dgreen, ->-=.5] (0,0.7) -- (0,2);
    \draw[thick, red] (0,-0.7) arc (-90:-270:0.7);
    \draw[thick, dgreen] (0,-0.7) arc (-90:90:0.7);
    \draw[thick, red, ->-=1.0] (-0.7,0.0) -- (-0.7, 0.1);
    \draw[thick,dgreen, ->-=1.0] (0.7,0.0) -- (0.7, 0.1);
    \filldraw[dgreen] (0,0.7) circle (3pt);
    \filldraw[dgreen] (0,-0.7) circle (3pt);
    \node[dgreen, below] at (0,-2) {\scriptsize$\mathcal{M}$};
    \node[dgreen, above] at (0,2) {\scriptsize$\mathcal{M}$};
    \node[dgreen, right] at (0.7,0) {\scriptsize$\mathcal{M}$};
    \node[red, left] at (-0.7,0) {\scriptsize$\mathcal{A}$};
    \node[dgreen, right] at (0,-1.0) {\scriptsize $\lambda^\vee$};
    \node[dgreen, right] at (0,+1.0) {\scriptsize $\lambda$};
\end{tikzpicture} = \,\,   \begin{tikzpicture}[scale=0.50,baseline = {(0,0)}]
    \draw[thick, dgreen, ->-=.5] (0,-2) -- (0,2);
    \node[dgreen, right] at (0.,0) {\scriptsize$\mathcal{M}$};
\end{tikzpicture} ~, \quad \begin{tikzpicture}[scale=0.50,baseline = {(0,0)}]
    \draw[thick, dgreen, ->-=.5] (0,-2) -- (0,-0.7);
    \draw[thick, dgreen, ->-=.5] (0,0.7) -- (0,2);
    \draw[thick, dgreen] (0,-0.7) arc (-90:-270:0.7);
    \draw[thick, red] (0,-0.7) arc (-90:90:0.7);
    \draw[thick, dgreen, ->-=1.0] (-0.7,0.0) -- (-0.7, 0.1);
    \draw[thick, red, ->-=1.0] (0.7,0.0) -- (0.7, 0.1);
    \filldraw[dgreen] (0,0.7) circle (3pt);
    \filldraw[dgreen] (0,-0.7) circle (3pt);
    \node[dgreen, below] at (0,-2) {\scriptsize$\mathcal{M}$};
    \node[dgreen, above] at (0,2) {\scriptsize$\mathcal{M}$};
    \node[red, right] at (0.7,0) {\scriptsize$\mathcal{A}$};
    \node[dgreen, left] at (-0.7,0) {\scriptsize$\mathcal{M}$};
    \node[dgreen, right] at (0,-1.0) {\scriptsize $\rho^\vee$};
    \node[dgreen, right] at (0,+1.0) {\scriptsize $\rho$};
\end{tikzpicture} = \,\,   \begin{tikzpicture}[scale=0.50,baseline = {(0,0)}]
    \draw[thick, dgreen, ->-=.5] (0,-2) -- (0,2);
    \node[dgreen, right] at (0.,0) {\scriptsize$\mathcal{M}$};
\end{tikzpicture} ~.
\end{equation}

To construct the explicit indecomposable bimodule objects, it is useful to keep in mind the classification of indecomposable bimodule objects derived in \cite{ostrik2002module}. In the group-theoretical fusion category $\scriptC(\doubleG,1;\doubleH,\psi)$, that is, the 't Hooft anomaly of $\doubleG$ is trivial and we are gauging the subgroup $\doubleH$ with the discrete torsion $\psi$, the indecomposable bimodule objects are labeled by $(\doubleH g \doubleH, \pi_g)$, where $\doubleH g \doubleH$ is a double coset of the group $G$ generated by the element $g \in \doubleG$, and $\pi_g$ is an irreducible representation of the little group $\doubleH_g := \{(h_L,h_R)\in \doubleH\times \doubleH: h_L g h_R = g\} \simeq \doubleH \cap g\doubleH g^{-1}$ twisted by $\psi_g(h_1, h_2) = \frac{\psi(h_1,h_2)}{\psi(g^{-1}h_1 g, g^{-1}h_2g)}$ where $h_1,h_2 \in \doubleH \cap g\doubleH g^{-1}$. Furthermore, the object $\scriptM$ of the corresponding bimodule is simply the sum of the $\dim(\pi_g)$ copies of the elements in the double coset. The explicit junction coefficients describing the left/right action of the algebra on the bimodules are subjected to gauge transformation, therefore, it is convenient to construct inequivalent solutions directly rather than solving the equations starting with explicit gauge fixing, since we already know how many inequivalent solutions there are from the classification.

For $\p_+^3$, to gauge $\doubleZ_p^{\rc}$ is implemented by the algebraic object $\displaystyle \mathcal{A}_+ = \bigoplus_{n = 0}^{p-1} \rc^n$, with the fusion junction $\mu_+$ and the split junction $\mu_+^\vee$ given by
\begin{equation}
\begin{tikzpicture}[scale=0.50,baseline = {(0,0)}]
    \draw[thick, red, ->-=.5] (-1.7,-1) -- (0,0);
    \draw[thick, red, -<-=.5] (0,0) -- (1.7,-1);
    \draw[thick, red, -<-=.5] (0,2) -- (0,0);
    \node[red, above] at (0,2) {\scriptsize$\mathcal{A}_+$};
    \node[red, below] at (-1.7,-1) {\scriptsize$\mathcal{A}_+$};
    \node[red, below] at (1.7,-1) {\scriptsize$\mathcal{A}_+$};
    \node[red, right] at (+0,0.2) {\scriptsize$\mu_+$};
    \filldraw[red] (0,0) circle (2pt);
\end{tikzpicture} = \frac{1}{\sqrt{p}}\sum_{m,n\in \doubleZ_p} \begin{tikzpicture}[scale=0.50,baseline = 0.]
    \draw[thick, black, ->-=.5] (-1.7,-1) -- (0,0);
    \draw[thick, black, -<-=.5] (0,0) -- (1.7,-1);
    \draw[thick, black, -<-=.5] (0,2) -- (0,0);
    \node[black, above] at (0,2) {\scriptsize$\rc^{m+n}$};
    \node[black, below] at (-1.7,-1) {\scriptsize$\rc^m$};
    \node[black, below] at (1.7,-1) {\scriptsize$\rc^n$};
\end{tikzpicture} ~, \quad \begin{tikzpicture}[scale=0.50,baseline = {(0,-0.5)}]
    \draw[thick, red, -<-=.5] (-1.7,1) -- (0,0);
    \draw[thick, red, ->-=.5] (0,0) -- (1.7,1);
    \draw[thick, red, ->-=.5] (0,-2) -- (0,0);
    \node[red, below] at (0,-2) {\scriptsize$\mathcal{A}_+$};
    \node[red, above] at (-1.7,1) {\scriptsize$\mathcal{A}_+$};
    \node[red, above] at (1.7,1) {\scriptsize$\mathcal{A}_+$};
    \node[red, right] at (+0,-0.2) {\scriptsize$\mu^\vee_+$};
    \filldraw[red] (0,0) circle (2pt);
\end{tikzpicture} = \frac{1}{\sqrt{p}} \sum_{m,n \in \doubleZ_p} \begin{tikzpicture}[scale=0.50,baseline = {(0,-0.5)}]
    \draw[thick, black, -<-=.5] (-1.7,1) -- (0,0);
    \draw[thick, black, ->-=.5] (0,0) -- (1.7,1);
    \draw[thick, black, ->-=.5] (0,-2) -- (0,0);
    \node[black, below] at (0,-2) {\scriptsize$\rc^{m+n}$};
    \node[black, above] at (-1.7,1) {\scriptsize$\rc^m$};
    \node[black, above] at (1.7,1) {\scriptsize$\rc^n$};
\end{tikzpicture} ~.
\end{equation}
There are $p^2$ invertible indecomposable bimodule objects $(\scriptM^{(i,j)}_+,\lambda_{+}^{(i,j)},\rho_{+}^{(i,j)})$ where $\displaystyle \scriptM^{(i,j)}_+ = \bigoplus_{n=0}^{p-1} \rb^i \rc^n$ \footnote{In this case, $\scriptM_+^{(i,j)}$ corresponds to the double coset $\doubleZ_p^\rc \rb^i \doubleZ_p^\rc$, $j$ corresponds to the label of the irreducible representation of the little group $H_{\rb} \simeq \doubleZ_p$.} and $\lambda_+^{(i,j)}, \rho_{+}^{(i,j)}$ is the fusion junction between the algebraic object $\mathcal{A}_+$ and $\scriptM^{(i,j)}_+$ from the left and the right respectively:
\begin{equation}
\begin{tikzpicture}[baseline={([yshift=-.5ex]current bounding box.center)},vertex/.style={anchor=base,
    circle,fill=black!25,minimum size=18pt,inner sep=2pt},scale=0.75]
\draw[red, thick, ->-=.5] (-1.5,-1) -- (-0.5,0);
\draw[dgreen, thick, ->-=.5] (-0.5,0) -- (0.5,1.0);
\draw[dgreen, thick, ->-=.5] (0.5,-1.0) -- (-0.5,0);

\node[red, below] at (-1.5,-1.5+0.5) {\footnotesize $\mathcal{A}_+$};
\node[dgreen, above] at (0.5,0.5+0.5) {\footnotesize $\mathcal{M}_+^{(i,j)}$};
\node[dgreen, below] at (0.5,-1.5+0.5) {\footnotesize $\mathcal{M}_+^{(i,j)}$};

\filldraw[dgreen] (-0.5,-0.5+0.5) circle (1.5pt);

\node[dgreen, left] at (-0.5,-0.5+0.5) {\footnotesize $\lambda_+^{(i,j)}$};
\end{tikzpicture} = \frac{1}{\sqrt{p}} \sum_{m,n \in \doubleZ_p} \begin{tikzpicture}[baseline={([yshift=-.5ex]current bounding box.center)},vertex/.style={anchor=base,
    circle,fill=black!25,minimum size=18pt,inner sep=2pt},scale=0.75]
\draw[black, thick, ->-=.5] (-1.5,-1) -- (-0.5,0);
\draw[black, thick, ->-=.5] (-0.5,0) -- (0.5,1.0);
\draw[black, thick, ->-=.5] (0.5,-1.0) -- (-0.5,0);

\node[black, below] at (-1.5,-1.5+0.5) {\footnotesize $\rc^m$};
\node[black, above] at (0.5,0.5+0.5) {\footnotesize $\rb^i \rc^{m+n}$};
\node[black, below] at (0.5,-1.5+0.5) {\footnotesize $\rb^i \rc^{n}$};

\end{tikzpicture} ~, \quad \begin{tikzpicture}[baseline={([yshift=-.5ex]current bounding box.center)},vertex/.style={anchor=base,
    circle,fill=black!25,minimum size=18pt,inner sep=2pt},scale=0.75]
\draw[dgreen, thick, ->-=.5] (-1.5,-1) -- (-0.5,0);
\draw[dgreen, thick, ->-=.5] (-0.5,0) -- (0.5,1.0);
\draw[red, thick, ->-=.5] (0.5,-1.0) -- (-0.5,0);

\node[dgreen, below] at (-1.5,-1.5+0.5) {\footnotesize $\mathcal{M}_+^{(i,j)}$};
\node[dgreen, above] at (0.5,0.5+0.5) {\footnotesize $\mathcal{M}_+^{(i,j)}$};
\node[red, below] at (0.5,-1.5+0.5) {\footnotesize $\mathcal{A}_+$};

\filldraw[dgreen] (-0.5,-0.5+0.5) circle (1.5pt);

\node[dgreen, left] at (-0.5,-0.5+0.5) {\footnotesize $\rho_+^{(i,j)}$};
\end{tikzpicture} = \frac{1}{\sqrt{p}}\sum_{m,n\in \doubleZ_p} \omega^{jn} \begin{tikzpicture}[baseline={([yshift=-.5ex]current bounding box.center)},vertex/.style={anchor=base,
    circle,fill=black!25,minimum size=18pt,inner sep=2pt},scale=0.75]
\draw[black, thick, ->-=.5] (-1.5,-1) -- (-0.5,0);
\draw[black, thick, ->-=.5] (-0.5,0) -- (0.5,1.0);
\draw[black, thick, ->-=.5] (0.5,-1.0) -- (-0.5,0);

\node[black, below] at (-1.5,-1.5+0.5) {\footnotesize $\rb^i \rc^m$};
\node[black, above] at (0.5,0.5+0.5) {\footnotesize $\rb^i \rc^{m+n}$};
\node[black, below] at (0.5,-1.5+0.5) {\footnotesize $\rc^n$};

\end{tikzpicture} ~,
\end{equation}
where $\omega$ is the $p$-th root of unity. The split junction can be constructed accordingly, which is given by
\begin{equation}
\begin{tikzpicture}[baseline={([yshift=-.5ex]current bounding box.center)},vertex/.style={anchor=base,
    circle,fill=black!25,minimum size=18pt,inner sep=2pt},scale=0.75]
\draw[red, thick, -<-=.5] (-1.5,1.0) -- (-0.5,0);
\draw[dgreen, thick, ->-=.5] (-0.5,0) -- (0.5,1.0);
\draw[dgreen, thick, ->-=.5] (0.5,-1.0) -- (-0.5,0);

\node[red, above] at (-1.5,1.0) {\footnotesize $\mathcal{A}_+$};
\node[dgreen, above] at (0.5,0.5+0.5) {\footnotesize $\mathcal{M}_+^{(i,j)}$};
\node[dgreen, below] at (0.5,-1.5+0.5) {\footnotesize $\mathcal{M}_+^{(i,j)}$};

\filldraw[dgreen] (-0.5,-0.5+0.5) circle (1.5pt);

\node[dgreen, left] at (-0.5,-0.5+0.3) {\footnotesize $\lambda_+^{(i,j),\vee}$};
\end{tikzpicture} = \frac{1}{\sqrt{p}} \sum_{m,n \in \doubleZ_p} \begin{tikzpicture}[baseline={([yshift=-.5ex]current bounding box.center)},vertex/.style={anchor=base,
    circle,fill=black!25,minimum size=18pt,inner sep=2pt},scale=0.75]
\draw[black, thick, -<-=.5] (-1.5,1) -- (-0.5,0);
\draw[black, thick, ->-=.5] (-0.5,0) -- (0.5,1.0);
\draw[black, thick, ->-=.5] (0.5,-1.0) -- (-0.5,0);

\node[black, above] at (-1.5,1) {\footnotesize $\rc^{m}$};
\node[black, above] at (0.5,0.5+0.5) {\footnotesize $\rb^i \rc^{n}$};
\node[black, below] at (0.5,-1.5+0.5) {\footnotesize $\rb^i \rc^{m+n}$};

\end{tikzpicture} ~, \quad \begin{tikzpicture}[baseline={([yshift=-.5ex]current bounding box.center)},vertex/.style={anchor=base,
    circle,fill=black!25,minimum size=18pt,inner sep=2pt},scale=0.75]
\draw[dgreen, thick, -<-=.5] (-1.5,1.0) -- (-0.5,0);
\draw[red, thick, ->-=.5] (-0.5,0) -- (0.5,1.0);
\draw[dgreen, thick, ->-=.5] (0.5,-1.0) -- (-0.5,0);

\node[dgreen, above] at (-1.5,1.0) {\footnotesize $\mathcal{M}_+^{(i,j)}$};
\node[red, above] at (0.5,0.5+0.5) {\footnotesize $\mathcal{A}_+$};
\node[dgreen, below] at (0.5,-1.5+0.5) {\footnotesize $\mathcal{M}_+^{(i,j)}$};

\filldraw[dgreen] (-0.5,-0.5+0.5) circle (1.5pt);

\node[dgreen, left] at (-0.5,-0.5+0.3) {\footnotesize $\rho_+^{(i,j),\vee}$};
\end{tikzpicture} = \frac{1}{\sqrt{p}} \sum_{m,n \in \doubleZ_p} \omega^{-jn} \begin{tikzpicture}[baseline={([yshift=-.5ex]current bounding box.center)},vertex/.style={anchor=base,
    circle,fill=black!25,minimum size=18pt,inner sep=2pt},scale=0.75]
\draw[black, thick, -<-=.5] (-1.5,1) -- (-0.5,0);
\draw[black, thick, ->-=.5] (-0.5,0) -- (0.5,1.0);
\draw[black, thick, ->-=.5] (0.5,-1.0) -- (-0.5,0);

\node[black, above] at (-1.5,1) {\footnotesize $\rb^i \rc^{m}$};
\node[black, above] at (0.5,0.5+0.5) {\footnotesize $\rc^{n}$};
\node[black, below] at (0.5,-1.5+0.5) {\footnotesize $\rb^i \rc^{m+n}$};

\end{tikzpicture} ~.
\end{equation}
The fusion rules between $\scriptM^{(i,j)}_+$'s are given by
\begin{equation}
    \scriptM^{(i,j)}_+ \times \scriptM^{(i',j')}_+ = \scriptM^{(i+i',j+j')}_+ ~.
\end{equation}
Here, $\scriptM^{(1,0)}_+$ generates the $\doubleZ_p^{\rb}$ symmetry while $\scriptM^{(0,1)}_+$ generates the dual $\doubleZ_p^{\hat{\rc}}$ symmetry of gauging $\doubleZ_p^{\rc}$.

There are $(p-1)$ non-invertible bimodule objects $(\scriptM_+^k,\lambda_+^k, \rho_+^k)$ where $\displaystyle \scriptM_+^k = \bigoplus_{i,j=0}^{p-1} \rb^i \rc^j \ra^k$, and the junctions $\lambda_+^k,\rho_+^k$ are trivial in the sense that the junction coefficients are given by $1$ up to some overall normalization
\begin{equation}
\begin{tikzpicture}[baseline={([yshift=-.5ex]current bounding box.center)},vertex/.style={anchor=base,
    circle,fill=black!25,minimum size=18pt,inner sep=2pt},scale=0.75]
\draw[red, thick, ->-=.5] (-1.5,-1) -- (-0.5,0);
\draw[dgreen, thick, ->-=.5] (-0.5,0) -- (0.5,1.0);
\draw[dgreen, thick, ->-=.5] (0.5,-1.0) -- (-0.5,0);

\node[red, below] at (-1.5,-1.5+0.5) {\footnotesize $\mathcal{A}_+$};
\node[dgreen, above] at (0.5,0.5+0.5) {\footnotesize $\mathcal{M}_+^{k}$};
\node[dgreen, below] at (0.5,-1.5+0.5) {\footnotesize $\mathcal{M}_+^{k}$};

\filldraw[dgreen] (-0.5,-0.5+0.5) circle (1.5pt);

\node[dgreen, left] at (-0.5,-0.5+0.5) {\footnotesize $\lambda_+^{k}$};
\end{tikzpicture} = \frac{1}{\sqrt{p}} \sum_{i,j,m} \begin{tikzpicture}[baseline={([yshift=-.5ex]current bounding box.center)},vertex/.style={anchor=base,
    circle,fill=black!25,minimum size=18pt,inner sep=2pt},scale=0.75]
\draw[black, thick, ->-=.5] (-1.5,-1) -- (-0.5,0);
\draw[black, thick, ->-=.5] (-0.5,0) -- (0.5,1.0);
\draw[black, thick, ->-=.5] (0.5,-1.0) -- (-0.5,0);

\node[black, below] at (-1.5,-1.5+0.5) {\footnotesize $\rc^m$};
\node[black, above] at (0.5,0.5+0.5) {\footnotesize $\rb^i \rc^{j+m} \ra^k$};
\node[black, below] at (0.5,-1.5+0.5) {\footnotesize $\rb^i \rc^{j} \ra^k$};

\end{tikzpicture} ~, \quad \begin{tikzpicture}[baseline={([yshift=-.5ex]current bounding box.center)},vertex/.style={anchor=base,
    circle,fill=black!25,minimum size=18pt,inner sep=2pt},scale=0.75]
\draw[dgreen, thick, ->-=.5] (-1.5,-1) -- (-0.5,0);
\draw[dgreen, thick, ->-=.5] (-0.5,0) -- (0.5,1.0);
\draw[red, thick, ->-=.5] (0.5,-1.0) -- (-0.5,0);

\node[dgreen, below] at (-1.5,-1.5+0.5) {\footnotesize $\mathcal{M}_+^{k}$};
\node[dgreen, above] at (0.5,0.5+0.5) {\footnotesize $\mathcal{M}_+^{k}$};
\node[red, below] at (0.5,-1.5+0.5) {\footnotesize $\mathcal{A}_+$};

\filldraw[dgreen] (-0.5,-0.5+0.5) circle (1.5pt);

\node[dgreen, left] at (-0.5,-0.5+0.5) {\footnotesize $\rho_+^{k}$};
\end{tikzpicture} = \frac{1}{\sqrt{p}}\sum_{i,j,m}  \begin{tikzpicture}[baseline={([yshift=-.5ex]current bounding box.center)},vertex/.style={anchor=base,
    circle,fill=black!25,minimum size=18pt,inner sep=2pt},scale=0.75]
\draw[black, thick, ->-=.5] (-1.5,-1) -- (-0.5,0);
\draw[black, thick, ->-=.5] (-0.5,0) -- (0.5,1.0);
\draw[black, thick, ->-=.5] (0.5,-1.0) -- (-0.5,0);

\node[black, below] at (-1.5,-1.5+0.5) {\footnotesize $\rb^i \rc^j \ra^k$};
\node[black, above] at (0.5,0.5+0.5) {\footnotesize $\rb^{i-km} \rc^{j+km} \ra^k$};
\node[black, below] at (0.5,-1.5+0.5) {\footnotesize $\rc^m$};

\end{tikzpicture} ~,
\end{equation}
and the split junctions are also trivial:
\begin{equation}
\begin{tikzpicture}[baseline={([yshift=-.5ex]current bounding box.center)},vertex/.style={anchor=base,
    circle,fill=black!25,minimum size=18pt,inner sep=2pt},scale=0.75]
\draw[red, thick, -<-=.5] (-1.5,1.0) -- (-0.5,0);
\draw[dgreen, thick, ->-=.5] (-0.5,0) -- (0.5,1.0);
\draw[dgreen, thick, ->-=.5] (0.5,-1.0) -- (-0.5,0);

\node[red, above] at (-1.5,1.0) {\footnotesize $\mathcal{A}_+$};
\node[dgreen, above] at (0.5,0.5+0.5) {\footnotesize $\mathcal{M}_+^{k}$};
\node[dgreen, below] at (0.5,-1.5+0.5) {\footnotesize $\mathcal{M}_+^{k}$};

\filldraw[dgreen] (-0.5,-0.5+0.5) circle (1.5pt);

\node[dgreen, left] at (-0.5,-0.5+0.3) {\footnotesize $\lambda_+^{k,\vee}$};
\end{tikzpicture} = \frac{1}{\sqrt{p}} \sum_{i,j,m} \begin{tikzpicture}[baseline={([yshift=-.5ex]current bounding box.center)},vertex/.style={anchor=base,
    circle,fill=black!25,minimum size=18pt,inner sep=2pt},scale=0.75]
\draw[black, thick, -<-=.5] (-1.5,1) -- (-0.5,0);
\draw[black, thick, ->-=.5] (-0.5,0) -- (0.5,1.0);
\draw[black, thick, ->-=.5] (0.5,-1.0) -- (-0.5,0);

\node[black, above] at (-1.5,1) {\footnotesize $\rc^{m}$};
\node[black, above] at (0.5,0.5+0.5) {\footnotesize $\rb^i \rc^{j-m} \ra^k$};
\node[black, below] at (0.5,-1.5+0.5) {\footnotesize $\rb^i \rc^{j} \ra^k$};

\end{tikzpicture} ~, \quad \begin{tikzpicture}[baseline={([yshift=-.5ex]current bounding box.center)},vertex/.style={anchor=base,
    circle,fill=black!25,minimum size=18pt,inner sep=2pt},scale=0.75]
\draw[dgreen, thick, -<-=.5] (-1.5,1.0) -- (-0.5,0);
\draw[red, thick, ->-=.5] (-0.5,0) -- (0.5,1.0);
\draw[dgreen, thick, ->-=.5] (0.5,-1.0) -- (-0.5,0);

\node[dgreen, above] at (-1.5,1.0) {\footnotesize $\mathcal{M}_+^{k}$};
\node[red, above] at (0.5,0.5+0.5) {\footnotesize $\mathcal{A}_+$};
\node[dgreen, below] at (0.5,-1.5+0.5) {\footnotesize $\mathcal{M}_+^{k}$};

\filldraw[dgreen] (-0.5,-0.5+0.5) circle (1.5pt);

\node[dgreen, left] at (-0.5,-0.5+0.3) {\footnotesize $\rho_+^{k,\vee}$};
\end{tikzpicture} = \frac{1}{\sqrt{p}} \sum_{i,j,m} \begin{tikzpicture}[baseline={([yshift=-.5ex]current bounding box.center)},vertex/.style={anchor=base,
    circle,fill=black!25,minimum size=18pt,inner sep=2pt},scale=0.75]
\draw[black, thick, -<-=.5] (-1.5,1) -- (-0.5,0);
\draw[black, thick, ->-=.5] (-0.5,0) -- (0.5,1.0);
\draw[black, thick, ->-=.5] (0.5,-1.0) -- (-0.5,0);

\node[black, above] at (-1.5,1) {\footnotesize $\rb^{i+km}\rc^{j-km} \ra^k$};
\node[black, above] at (0.5,0.5+0.5) {\footnotesize $\rc^{m}$};
\node[black, below] at (0.5,-1.5+0.5) {\footnotesize $\rb^i \rc^{j} \ra^k$};

\end{tikzpicture} ~.
\end{equation}
Notice that this is the case where it is convenient to keep in mind the classification result mentioned above. For a given object, $\mathcal{M}_+^k$, the little group of the corresponding double coset $\doubleZ_p^\rc a^k \doubleZ_p^\rc$ is trivial; therefore, the solution of the junction coefficients must be unique (up to gauge transformation). And obviously, the trivial junction is a solution, hence we are done.

The analysis for the case of $\p_-^3$ can be done analogously. The algebraic object $\mathcal{A}_- = \displaystyle \bigoplus_{n=0}^{p-1} c^n$ with the fusion $\mu$ and split junction $\mu^\vee$ given by
\begin{equation}
\begin{tikzpicture}[scale=0.50,baseline = {(0,0)}]
    \draw[thick, red, ->-=.5] (-1.7,-1) -- (0,0);
    \draw[thick, red, -<-=.5] (0,0) -- (1.7,-1);
    \draw[thick, red, -<-=.5] (0,2) -- (0,0);
    \node[red, above] at (0,2) {\scriptsize$\mathcal{A}_-$};
    \node[red, below] at (-1.7,-1) {\scriptsize$\mathcal{A}_-$};
    \node[red, below] at (1.7,-1) {\scriptsize$\mathcal{A}_-$};
    \node[red, right] at (+0,0.2) {\scriptsize$\mu_-$};
    \filldraw[red] (0,0) circle (2pt);
\end{tikzpicture} = \frac{1}{\sqrt{p}}\sum_{m,n\in \doubleZ_p} \begin{tikzpicture}[scale=0.50,baseline = 0.]
    \draw[thick, black, ->-=.5] (-1.7,-1) -- (0,0);
    \draw[thick, black, -<-=.5] (0,0) -- (1.7,-1);
    \draw[thick, black, -<-=.5] (0,2) -- (0,0);
    \node[black, above] at (0,2) {\scriptsize$\rc^{m+n}$};
    \node[black, below] at (-1.7,-1) {\scriptsize$\rc^m$};
    \node[black, below] at (1.7,-1) {\scriptsize$\rc^n$};
\end{tikzpicture} ~, \quad \begin{tikzpicture}[scale=0.50,baseline = {(0,-0.5)}]
    \draw[thick, red, -<-=.5] (-1.7,1) -- (0,0);
    \draw[thick, red, ->-=.5] (0,0) -- (1.7,1);
    \draw[thick, red, ->-=.5] (0,-2) -- (0,0);
    \node[red, below] at (0,-2) {\scriptsize$\mathcal{A}_-$};
    \node[red, above] at (-1.7,1) {\scriptsize$\mathcal{A}_-$};
    \node[red, above] at (1.7,1) {\scriptsize$\mathcal{A}_-$};
    \node[red, right] at (+0,-0.2) {\scriptsize$\mu^\vee_-$};
    \filldraw[red] (0,0) circle (2pt);
\end{tikzpicture} = \frac{1}{\sqrt{p}} \sum_{m,n \in \doubleZ_p} \begin{tikzpicture}[scale=0.50,baseline = {(0,-0.5)}]
    \draw[thick, black, -<-=.5] (-1.7,1) -- (0,0);
    \draw[thick, black, ->-=.5] (0,0) -- (1.7,1);
    \draw[thick, black, ->-=.5] (0,-2) -- (0,0);
    \node[black, below] at (0,-2) {\scriptsize$\rc^{m+n}$};
    \node[black, above] at (-1.7,1) {\scriptsize$\rc^m$};
    \node[black, above] at (1.7,1) {\scriptsize$\rc^n$};
\end{tikzpicture} ~.
\end{equation}
The $p^2$ invertible indecomposable bimodule objects $(\scriptM^{(i,j)}_-,\lambda_{-}^{(i,j)},\rho_{-}^{(i,j)})$ where $\displaystyle \scriptM^{(i,j)}_- = \bigoplus_{n=0}^{p-1} \ra^{i p} \rc^n$ and $\lambda_-^{(i,j)}, \rho_-^{(i,j)}$ is the fusion junction between the algebraic object $\mathcal{A}_-$ and $\scriptM^{(i,j)}_-$ from the left and the right respectively:
\begin{equation}
\begin{tikzpicture}[baseline={([yshift=-.5ex]current bounding box.center)},vertex/.style={anchor=base,
    circle,fill=black!25,minimum size=18pt,inner sep=2pt},scale=0.75]
\draw[red, thick, ->-=.5] (-1.5,-1) -- (-0.5,0);
\draw[dgreen, thick, ->-=.5] (-0.5,0) -- (0.5,1.0);
\draw[dgreen, thick, ->-=.5] (0.5,-1.0) -- (-0.5,0);

\node[red, below] at (-1.5,-1.5+0.5) {\footnotesize $\mathcal{A}_-$};
\node[dgreen, above] at (0.5,0.5+0.5) {\footnotesize $\mathcal{M}_-^{(i,j)}$};
\node[dgreen, below] at (0.5,-1.5+0.5) {\footnotesize $\mathcal{M}_-^{(i,j)}$};

\filldraw[dgreen] (-0.5,-0.5+0.5) circle (1.5pt);

\node[dgreen, left] at (-0.5,-0.5+0.5) {\footnotesize $\lambda_-^{(i,j)}$};
\end{tikzpicture} = \frac{1}{\sqrt{p}} \sum_{i,m,n} \begin{tikzpicture}[baseline={([yshift=-.5ex]current bounding box.center)},vertex/.style={anchor=base,
    circle,fill=black!25,minimum size=18pt,inner sep=2pt},scale=0.75]
\draw[black, thick, ->-=.5] (-1.5,-1) -- (-0.5,0);
\draw[black, thick, ->-=.5] (-0.5,0) -- (0.5,1.0);
\draw[black, thick, ->-=.5] (0.5,-1.0) -- (-0.5,0);

\node[black, below] at (-1.5,-1.5+0.5) {\footnotesize $\rc^m$};
\node[black, above] at (0.5,0.5+0.5) {\footnotesize $\ra^{ip} \rc^{m+n}$};
\node[black, below] at (0.5,-1.5+0.5) {\footnotesize $\ra^{ip} \rc^{n}$};

\end{tikzpicture} ~, \quad \begin{tikzpicture}[baseline={([yshift=-.5ex]current bounding box.center)},vertex/.style={anchor=base,
    circle,fill=black!25,minimum size=18pt,inner sep=2pt},scale=0.75]
\draw[dgreen, thick, ->-=.5] (-1.5,-1) -- (-0.5,0);
\draw[dgreen, thick, ->-=.5] (-0.5,0) -- (0.5,1.0);
\draw[red, thick, ->-=.5] (0.5,-1.0) -- (-0.5,0);

\node[dgreen, below] at (-1.5,-1.5+0.5) {\footnotesize $\mathcal{M}_-^{(i,j)}$};
\node[dgreen, above] at (0.5,0.5+0.5) {\footnotesize $\mathcal{M}_-^{(i,j)}$};
\node[red, below] at (0.5,-1.5+0.5) {\footnotesize $\mathcal{A}_-$};

\filldraw[dgreen] (-0.5,-0.5+0.5) circle (1.5pt);

\node[dgreen, left] at (-0.5,-0.5+0.5) {\footnotesize $\rho_+^{(i,j)}$};
\end{tikzpicture} = \frac{1}{\sqrt{p}}\sum_{i,m,n} \omega^{jn} \begin{tikzpicture}[baseline={([yshift=-.5ex]current bounding box.center)},vertex/.style={anchor=base,
    circle,fill=black!25,minimum size=18pt,inner sep=2pt},scale=0.75]
\draw[black, thick, ->-=.5] (-1.5,-1) -- (-0.5,0);
\draw[black, thick, ->-=.5] (-0.5,0) -- (0.5,1.0);
\draw[black, thick, ->-=.5] (0.5,-1.0) -- (-0.5,0);

\node[black, below] at (-1.5,-1.5+0.5) {\footnotesize $\ra^{ip} \rc^m$};
\node[black, above] at (0.5,0.5+0.5) {\footnotesize $\ra^{ip} \rc^{m+n}$};
\node[black, below] at (0.5,-1.5+0.5) {\footnotesize $\rc^n$};

\end{tikzpicture} ~,
\end{equation}
and the splitting junctions are given by
\begin{equation}
\begin{tikzpicture}[baseline={([yshift=-.5ex]current bounding box.center)},vertex/.style={anchor=base,
    circle,fill=black!25,minimum size=18pt,inner sep=2pt},scale=0.75]
\draw[red, thick, -<-=.5] (-1.5,1.0) -- (-0.5,0);
\draw[dgreen, thick, ->-=.5] (-0.5,0) -- (0.5,1.0);
\draw[dgreen, thick, ->-=.5] (0.5,-1.0) -- (-0.5,0);

\node[red, above] at (-1.5,1.0) {\footnotesize $\mathcal{A}_-$};
\node[dgreen, above] at (0.5,0.5+0.5) {\footnotesize $\mathcal{M}_-^{(i,j)}$};
\node[dgreen, below] at (0.5,-1.5+0.5) {\footnotesize $\mathcal{M}_-^{(i,j)}$};

\filldraw[dgreen] (-0.5,-0.5+0.5) circle (1.5pt);

\node[dgreen, left] at (-0.5,-0.5+0.3) {\footnotesize $\lambda_-^{(i,j),\vee}$};
\end{tikzpicture} = \frac{1}{\sqrt{p}} \sum_{i,m,n} \begin{tikzpicture}[baseline={([yshift=-.5ex]current bounding box.center)},vertex/.style={anchor=base,
    circle,fill=black!25,minimum size=18pt,inner sep=2pt},scale=0.75]
\draw[black, thick, -<-=.5] (-1.5,1) -- (-0.5,0);
\draw[black, thick, ->-=.5] (-0.5,0) -- (0.5,1.0);
\draw[black, thick, ->-=.5] (0.5,-1.0) -- (-0.5,0);

\node[black, above] at (-1.5,1) {\footnotesize $\rc^{m}$};
\node[black, above] at (0.5,0.5+0.5) {\footnotesize $\ra^{ip} \rc^{n}$};
\node[black, below] at (0.5,-1.5+0.5) {\footnotesize $\ra^{ip} \rc^{m+n}$};

\end{tikzpicture} ~, \quad \begin{tikzpicture}[baseline={([yshift=-.5ex]current bounding box.center)},vertex/.style={anchor=base,
    circle,fill=black!25,minimum size=18pt,inner sep=2pt},scale=0.75]
\draw[dgreen, thick, -<-=.5] (-1.5,1.0) -- (-0.5,0);
\draw[red, thick, ->-=.5] (-0.5,0) -- (0.5,1.0);
\draw[dgreen, thick, ->-=.5] (0.5,-1.0) -- (-0.5,0);

\node[dgreen, above] at (-1.5,1.0) {\footnotesize $\mathcal{M}_-^{(i,j)}$};
\node[red, above] at (0.5,0.5+0.5) {\footnotesize $\mathcal{A}_-$};
\node[dgreen, below] at (0.5,-1.5+0.5) {\footnotesize $\mathcal{M}_-^{(i,j)}$};

\filldraw[dgreen] (-0.5,-0.5+0.5) circle (1.5pt);

\node[dgreen, left] at (-0.5,-0.5+0.3) {\footnotesize $\rho_-^{(i,j),\vee}$};
\end{tikzpicture} = \frac{1}{\sqrt{p}} \sum_{i,m,n} \omega^{-jn} \begin{tikzpicture}[baseline={([yshift=-.5ex]current bounding box.center)},vertex/.style={anchor=base,
    circle,fill=black!25,minimum size=18pt,inner sep=2pt},scale=0.75]
\draw[black, thick, -<-=.5] (-1.5,1) -- (-0.5,0);
\draw[black, thick, ->-=.5] (-0.5,0) -- (0.5,1.0);
\draw[black, thick, ->-=.5] (0.5,-1.0) -- (-0.5,0);

\node[black, above] at (-1.5,1) {\footnotesize $\ra^{ip} \rc^{m}$};
\node[black, above] at (0.5,0.5+0.5) {\footnotesize $\rc^{n}$};
\node[black, below] at (0.5,-1.5+0.5) {\footnotesize $\ra^{ip} \rc^{m+n}$};

\end{tikzpicture} ~.
\end{equation}
Then, there are $(p-1)$ non-invertible bimodule objects $(\scriptM_-^k,\lambda_-^k,\rho_-^k)$ where $\displaystyle \scriptM_-^k = \bigoplus_{i,j=0}^{p-1} \rc^{i} \ra^{k+jp}$ with the fusion junctions
\begin{equation}
\begin{tikzpicture}[baseline={([yshift=-.5ex]current bounding box.center)},vertex/.style={anchor=base,
    circle,fill=black!25,minimum size=18pt,inner sep=2pt},scale=0.75]
\draw[red, thick, ->-=.5] (-1.5,-1) -- (-0.5,0);
\draw[dgreen, thick, ->-=.5] (-0.5,0) -- (0.5,1.0);
\draw[dgreen, thick, ->-=.5] (0.5,-1.0) -- (-0.5,0);

\node[red, below] at (-1.5,-1.5+0.5) {\footnotesize $\mathcal{A}_-$};
\node[dgreen, above] at (0.5,0.5+0.5) {\footnotesize $\mathcal{M}_-^{k}$};
\node[dgreen, below] at (0.5,-1.5+0.5) {\footnotesize $\mathcal{M}_-^{k}$};

\filldraw[dgreen] (-0.5,-0.5+0.5) circle (1.5pt);

\node[dgreen, left] at (-0.5,-0.5+0.5) {\footnotesize $\lambda_-^{k}$};
\end{tikzpicture} = \frac{1}{\sqrt{p}} \sum_{i,j,m} \begin{tikzpicture}[baseline={([yshift=-.5ex]current bounding box.center)},vertex/.style={anchor=base,
    circle,fill=black!25,minimum size=18pt,inner sep=2pt},scale=0.75]
\draw[black, thick, ->-=.5] (-1.5,-1) -- (-0.5,0);
\draw[black, thick, ->-=.5] (-0.5,0) -- (0.5,1.0);
\draw[black, thick, ->-=.5] (0.5,-1.0) -- (-0.5,0);

\node[black, below] at (-1.5,-1.5+0.5) {\footnotesize $\rc^m$};
\node[black, above] at (0.5,0.5+0.5) {\footnotesize $\rc^{i+m}\ra^{k+jp}$};
\node[black, below] at (0.5,-1.5+0.5) {\footnotesize $\rc^{i} \ra^{k+jp}$};

\end{tikzpicture} ~, \quad \begin{tikzpicture}[baseline={([yshift=-.5ex]current bounding box.center)},vertex/.style={anchor=base,
    circle,fill=black!25,minimum size=18pt,inner sep=2pt},scale=0.75]
\draw[dgreen, thick, ->-=.5] (-1.5,-1) -- (-0.5,0);
\draw[dgreen, thick, ->-=.5] (-0.5,0) -- (0.5,1.0);
\draw[red, thick, ->-=.5] (0.5,-1.0) -- (-0.5,0);

\node[dgreen, below] at (-1.5,-1.5+0.5) {\footnotesize $\mathcal{M}_-^{k}$};
\node[dgreen, above] at (0.5,0.5+0.5) {\footnotesize $\mathcal{M}_-^{k}$};
\node[red, below] at (0.5,-1.5+0.5) {\footnotesize $\mathcal{A}_-$};

\filldraw[dgreen] (-0.5,-0.5+0.5) circle (1.5pt);

\node[dgreen, left] at (-0.5,-0.5+0.5) {\footnotesize $\rho_-^{k}$};
\end{tikzpicture} = \frac{1}{\sqrt{p}}\sum_{i,j,m}  \begin{tikzpicture}[baseline={([yshift=-.5ex]current bounding box.center)},vertex/.style={anchor=base,
    circle,fill=black!25,minimum size=18pt,inner sep=2pt},scale=0.75]
\draw[black, thick, ->-=.5] (-1.5,-1) -- (-0.5,0);
\draw[black, thick, ->-=.5] (-0.5,0) -- (0.5,1.0);
\draw[black, thick, ->-=.5] (0.5,-1.0) -- (-0.5,0);

\node[black, below] at (-1.5,-1.5+0.5) {\footnotesize $\rc^{i}\ra^{k+jp}$};
\node[black, above] at (0.5,0.5+0.5) {\footnotesize $\rc^{i+m} \ra^{k+(j-km)p}$};
\node[black, below] at (0.5,-1.5+0.5) {\footnotesize $\rc^m$};

\end{tikzpicture} ~,
\end{equation}
and the splitting junctions
\begin{equation}
\begin{tikzpicture}[baseline={([yshift=-.5ex]current bounding box.center)},vertex/.style={anchor=base,
    circle,fill=black!25,minimum size=18pt,inner sep=2pt},scale=0.75]
\draw[red, thick, -<-=.5] (-1.5,1.0) -- (-0.5,0);
\draw[dgreen, thick, ->-=.5] (-0.5,0) -- (0.5,1.0);
\draw[dgreen, thick, ->-=.5] (0.5,-1.0) -- (-0.5,0);

\node[red, above] at (-1.5,1.0) {\footnotesize $\mathcal{A}_-$};
\node[dgreen, above] at (0.5,0.5+0.5) {\footnotesize $\mathcal{M}_-^{k}$};
\node[dgreen, below] at (0.5,-1.5+0.5) {\footnotesize $\mathcal{M}_-^{k}$};

\filldraw[dgreen] (-0.5,-0.5+0.5) circle (1.5pt);

\node[dgreen, left] at (-0.5,-0.5+0.3) {\footnotesize $\lambda_-^{k,\vee}$};
\end{tikzpicture} = \frac{1}{\sqrt{p}} \sum_{i,j,m} \begin{tikzpicture}[baseline={([yshift=-.5ex]current bounding box.center)},vertex/.style={anchor=base,
    circle,fill=black!25,minimum size=18pt,inner sep=2pt},scale=0.75]
\draw[black, thick, -<-=.5] (-1.5,1) -- (-0.5,0);
\draw[black, thick, ->-=.5] (-0.5,0) -- (0.5,1.0);
\draw[black, thick, ->-=.5] (0.5,-1.0) -- (-0.5,0);

\node[black, above] at (-1.5,1) {\footnotesize $\rc^{m}$};
\node[black, above] at (0.5,0.5+0.5) {\footnotesize $\rc^{i-m} \ra^{k+jp}$};
\node[black, below] at (0.5,-1.5+0.5) {\footnotesize $\rc^{i} \ra^{k+jp}$};

\end{tikzpicture} ~, \quad \begin{tikzpicture}[baseline={([yshift=-.5ex]current bounding box.center)},vertex/.style={anchor=base,
    circle,fill=black!25,minimum size=18pt,inner sep=2pt},scale=0.75]
\draw[dgreen, thick, -<-=.5] (-1.5,1.0) -- (-0.5,0);
\draw[red, thick, ->-=.5] (-0.5,0) -- (0.5,1.0);
\draw[dgreen, thick, ->-=.5] (0.5,-1.0) -- (-0.5,0);

\node[dgreen, above] at (-1.5,1.0) {\footnotesize $\mathcal{M}_-^{k}$};
\node[red, above] at (0.5,0.5+0.5) {\footnotesize $\mathcal{A}_-$};
\node[dgreen, below] at (0.5,-1.5+0.5) {\footnotesize $\mathcal{M}_-^{k}$};

\filldraw[dgreen] (-0.5,-0.5+0.5) circle (1.5pt);

\node[dgreen, left] at (-0.5,-0.5+0.3) {\footnotesize $\rho_-^{k,\vee}$};
\end{tikzpicture} = \frac{1}{\sqrt{p}} \sum_{i,j,m} \begin{tikzpicture}[baseline={([yshift=-.5ex]current bounding box.center)},vertex/.style={anchor=base,
    circle,fill=black!25,minimum size=18pt,inner sep=2pt},scale=0.75]
\draw[black, thick, -<-=.5] (-1.5,1) -- (-0.5,0);
\draw[black, thick, ->-=.5] (-0.5,0) -- (0.5,1.0);
\draw[black, thick, ->-=.5] (0.5,-1.0) -- (-0.5,0);

\node[black, above] at (-1.5,1) {\footnotesize $\rc^{i-m}\ra^{k + (j + km)p}$};
\node[black, above] at (0.5,0.5+0.5) {\footnotesize $\quad \rc^{m}$};
\node[black, below] at (0.5,-1.5+0.5) {\footnotesize $\rc^{i} \ra^{k+jp}$};

\end{tikzpicture} ~.
\end{equation}
These junctions are then used to compute the twisted partition functions in the gauged theory in \eqref{eq:pp3_twisted_part} and \eqref{eq:pm3_twisted_part}.

\section{Review of the $3d$ DW Theory and the Spectrum of the SymTFTs of the $p$-ality defects}\label{app:DW_theory}
In this appendix, we will briefly review the spectrum of line operators in 3d Dijkgraaf-Witten(DW) theory $\DW(G,\omega)$ with gauge group $G$ and twist $[\omega]\in H^3(G,U(1))$. Notice that the same spectrum also labels the topological sectors of the $2d$ CFT with global symmetry $G$ with 't Hooft anomaly $[\omega]\in H^3(G,U(1))$. We will provide explicit examples of $\DW(\doubleZ_p\times\doubleZ_p\times\doubleZ_p,\omega_{\pm,m})$ utilized in Section \ref{sec:p_ality_SymTFT}. More details can be found in the literature\cite{deWildPropitius:1995cf,Coste:2000tq,Hu:2012wx}. Notice that in this appendix, we will use Latin letters to denote both group elements in multiplicative and additive conventions, and the readers should be able to distinguish the two cases based on context.

The input for the construction of DW theory is a choice of a finite $G$ and a twist $[\omega] \in H^3(G,U(1))$. For all $a,g,h \in G$ we can construct auxiliary quantities: 
\begin{equation}\label{eq:effective_2_cocyle}
    \beta_a(h,g)=\omega(a,h,g)\omega(h,h^{-1}ah,g)^{-1}\omega(h,g,(hg)^{-1}ahg) ~.
\end{equation}
It is straightforward to check that $\beta_a$'s are normalized twisted cocycles on $G$:
\begin{equation}
    \beta_a(x,y)\beta_a(xy,z)=\beta_a(x,yz)\beta_{x^{-1}ax}(y,z), \quad \forall x,y,z\in G ~.
\end{equation}
They become normalized 2-cocycles when $x,y,z \in C_G{(a)}$, the centralizer subgroup of $a$, and define projective representations of $C_G(a)$. 

Line operators are labeled by a pair $(a, \mu)$ where $a$ is a representative of a conjugacy class of $G$ and $\mu$ is an irreducible $\beta_a$-representation of $C_G(a)$. The spin of line operator labeled by $(a,\mu)$ is:
\begin{equation}\label{eq:top_spin}
    \theta_{a,\mu}=\frac{\Tilde{\chi}^a_{\mu}(a)}{\text{dim}_{\mu}} ~,
\end{equation}
where $\Tilde{\chi}^a_{\mu}$ is the projective character of $\beta_a$-representation $\mu$. It is possible to show that the spin is independent of the representative of the conjugacy class. $S$-matrix and $T$-matrix can be found to be:
\begin{equation}\label{eq:modularmat}
    \begin{aligned}
        S_{(a,\mu),(b,\nu)}=&\frac{1}{|G|} \sum_{g\in C_a,h\in C_b,hg=gh} {{\Tilde{\chi}^{g *}}_{\mu}}(h) {{\Tilde{\chi}^{h*}}_{\nu}}(g) ~, \\
        T_{(a,\mu),(b,\nu)}=&\frac{\Tilde{\chi}^a_{\mu}(a)}{\text{dim}_{\mu}}\delta_{a,b}\delta_{\mu,\nu} ~,
    \end{aligned}
\end{equation}
where $C_a = \{g^{-1}a g| g\in G \}$ denotes the conjugacy class of $a$. 

Notice that the same data also parameterizes the topological sectors of a $2d$ CFT with finite group $G$ and anomaly $[\omega] \in H^3(G,U(1))$. Consider the defect Hilbert space $\mathcal{H}_{a}$ where $a \in G$. First, notice that since the twisted partition function characterized the trace over $\mathcal{H}_{a}$ is invariant under conjugation $a \mapsto gag^{-1}$, $\mathcal{H}_{a}$ is isomorphic to $\mathcal{H}_{gag^{-1}}$. Then, the TDL in $G$ which can act on $\mathcal{H}_{a}$ must commute with $a$, hence $\mathcal{H}_a$ admits the symmetry group $C_G(a)$. Due to the 't Hooft anomaly $\omega$, $C_G(a)$ generically will act projectively with the $2$-cocycle $\beta_a$ given in \eqref{eq:effective_2_cocyle}; and $\mathcal{H}_a$ further decomposes into a direct sum of $\mathcal{H}_{(a,\mu)}$, where $\mathcal{H}_{(a,\mu)}$ contains all the states transformed in the (projective) irreducible representation $\mu$ of $C_G(a)$. States in the same topological sector $\mathcal{H}_{(a,\mu)}$ has the same spin $s$ mod $\doubleZ$, specified by $e^{2\pi i s} = \theta_{a,\mu}$ given in \eqref{eq:top_spin}. And the modular properties of the torus partition functions computing the trace over $\mathcal{H}_{(a,\mu)}$ are characterized by the $S$ and $T$ matrices given in \eqref{eq:modularmat}.

Now we consider the case used in Section \ref{sec:p_ality_SymTFT}, where $G=\doubleZ_p^a\times\doubleZ_p^b\times\doubleZ_p^c$ and $p$ is an odd prime number. A general element of $G$ can be labeled by $a^{i}b^{j}c^k$ where $i,j,k=0,1,...,p-1$. Since this group is Abelian, each element also labels a conjugacy class of $G$. The cohomology group $H^3(\doubleZ^3_p,U(1))=\doubleZ^7_p$ has seven generators which can be further divided into three types. The first type involves one copy of $\doubleZ_p$ subgroup:
\begin{equation}
    \begin{aligned}
        \omega^{1}_{\text{\uppercase\expandafter{\romannumeral1}}}(a^{i_1}b^{j_1}c^{k_1},a^{i_2}b^{j_2}c^{k_2},a^{i_3}b^{j_3}c^{k_3})=&e^{\frac{2\pi i}{p^2}i_1(i_2+i_3-[ i_2+i_3]_p)} ~, \\
     \omega^{2}_{\text{\uppercase\expandafter{\romannumeral1}}}(a^{i_1}b^{j_1}c^{k_1},a^{i_2}b^{j_2}c^{k_2},a^{i_3}b^{j_3}c^{k_3})=&e^{\frac{2\pi i}{p^2}j_1(j_2+j_3-[ j_2+j_3]_p)} ~,\\
     \omega^{3}_{\text{\uppercase\expandafter{\romannumeral1}}}(a^{i_1}b^{j_1}c^{k_1},a^{i_2}b^{j_2}c^{k_2},a^{i_3}b^{j_3}c^{k_3})=&e^{\frac{2\pi i}{p^2}k_1(k_2+k_3-[ k_2+k_3]_p)} ~. \\
    \end{aligned}
\end{equation}
The second type involves two copies of $\doubleZ_p$ subgroups and contains 3 independent generators:
\begin{equation}
    \begin{aligned}
          \omega^{1}_{\text{\uppercase\expandafter{\romannumeral2}}}(a^{i_1}b^{j_1}c^{k_1},a^{i_2}b^{j_2}c^{k_2},a^{i_3}b^{j_3}c^{k_3})=&e^{ \frac{2\pi i}{p^2}i_1(j_2+j_3-[ j_2+j_3]_p)} ~, \\        
          \omega^{2}_{\text{\uppercase\expandafter{\romannumeral2}}}(a^{i_1}b^{j_1}c^{k_1},a^{i_2}b^{j_2}c^{k_2},a^{i_3}b^{j_3}c^{k_3})=&e^{ \frac{2\pi i}{p^2}i_1(k_2+k_3-[ k_2+k_3]_p)} ~, \\            
          \omega^{3}_{\text{\uppercase\expandafter{\romannumeral2}}}(a^{i_1}b^{j_1}c^{k_1},a^{i_2}b^{j_2}c^{k_2},a^{i_3}b^{j_3}c^{k_3})=&e^{ \frac{2\pi i}{p^2}j_1(k_2+k_3-[ k_2+k_3]_p)} ~.\\
    \end{aligned}
\end{equation}
The last type involves the whole group and contains one independent generator:
\begin{equation}
    \begin{aligned}
          \omega_{\text{\uppercase\expandafter{\romannumeral3}}}(a^{i_1}b^{j_1}c^{k_1},a^{i_2}b^{j_2}c^{k_2},a^{i_3}b^{j_3}c^{k_3})=&e^{ \frac{2\pi i}{p}j_1k_2i_3} ~.\\            
    \end{aligned}
\end{equation}
The $\beta_a(x,y)$ of type $\text{\uppercase\expandafter{\romannumeral1}}$ and $\text{\uppercase\expandafter{\romannumeral2}}$ is cohomologically trivial, meaning that it can be expressed as $\beta_a(x,y) =\epsilon_a(x)\epsilon_a(y)\epsilon_a(xy)^{-1}$, where $\epsilon_a: C_G(a)\rightarrow U(1)$ is a 1-cochain. Hence, the projective character is related to the irreducible character in the untwisted theory by $\Tilde{\chi}^a_\mu = \epsilon_a \chi^a_\mu$. The modular matrices and other data can then be analyzed similarly to the untwisted theory \cite{Dijkgraaf:1989DW,Dijkgraaf:1989CFT}. 

On the other hand, when the type III twist is involved, some $\beta_a$ will no longer be cohomologically trivial. We now analyze the two cases with $\omega_{\pm,m}$ considered in Section \ref{sec:p_ality_SymTFT} in detail. Let's start with the 3-cocycle $\omega_{+,m}$,
\begin{equation}
    \omega_{+,m}(a^{i_1}b^{j_1}c^{k_1},a^{i_2}b^{j_2}c^{k_2},a^{i_3}b^{j_3}c^{k_3}) = e^{\frac{2\pi i}{p} j_1 k_2 i_3 + \frac{2\pi i m }{p^2}i_1 (i_2 + i_3 - [i_2 + i_3]_p)} ~.
        \label{eq:omegam+}
\end{equation}
Given an element $g_1 = a^{i_1}b^{j_1}c^{k_1}$, the effective 2-cocycle is given by
\begin{equation}\label{eq:omegap2cocycle}
    \beta_{(i_1,j_1,k_1)}(g_2,g_3) = e^{\frac{2\pi i}{p} \left(j_1 k_2 i_3 + j_2 k_3 i_1 -j_2 k_1 i_3\right) + \frac{2\pi i m}{p^2}  i_1 (i_2 + i_3 - [i_2 + i_3]_p)} \equiv \beta^c_{(i_1,j_1,k_1)}(g_2,g_3) \beta^0_{(i_1)}(g_2,g_3) ~.
\end{equation}
where $\beta^c_{(i_1,j_1,k_1)}(g_2,g_3)= e^{\frac{2\pi i}{p} \left(j_1 k_2 i_3 + j_2 k_3 i_1 -j_2 k_1 i_3\right)}$ is cohomologically nontrivial, while $\beta^0_{(i_1)}(g_2,g_3)=e^{\frac{2\pi i m}{p^2}  i_1 (i_2 + i_3 - [i_2 + i_3]_p)}$, which arises from the type I twist of the first $\doubleZ_p$, is cohomologically trivial
\begin{equation}
    \beta^0_{(i_1)}(g_2,g_3)=(\delta \epsilon_{(i_1)})(g_2,g_3)=\epsilon_{(i_1)}(g_2)\epsilon_{(i_1)}(g_3)\epsilon_{(i_1)}(g_2 g_3)^{-1} ~,
\end{equation}
where $\epsilon_{(i_1)}(g_2) = e^{\frac{2\pi i m}{p^2}i_1 i_2}$.
To study the structure of the projective representation, we first find the center subgroup of $\doubleZ_p^a \times \doubleZ_p^b \times \doubleZ_p^c$ in the projective representation, that is, we want to find out all $g \in G$ such that $\beta_{(i_1,j_1,k_1)}(g,h) = \beta_{(i_1,j_1,k_1)}(h,g)$ for any $h \in G$. Such $g$ is known as a $\beta$-regular element\cite{Coste:2000tq}. Denote this subgroup as $C_{(i_1, j_1,k_1)}$ and it is apparently a normal subgroup of $G$. The $2$-cocycle can be reduced to a $2$-cocycle of the quotient group $G/C_{(i_1,j_1,k_1)}$, and there's a unique projective representation of the dimension $|G/C_{(i_1,j_1,k_1)}|$ because there's no central element in the quotient group. Then any irreducible projective representation of $G$ can be constructed from an irreducible representation of $C_{(i_1,j_1,k_1)}$ and the unique projective irrep of $G/C_{(i_1,j_1,k_1)}$.

To see this in action, let's try to solve the center elements for $\beta_{(i_1,j_1,k_1)}$. Notice that $g$ is a central element in $G$, if and only if it commutes with the three generators $a,b,c \in G$. Hence, we want to solve for $(i_2, j_2, k_2)$ such that
\begin{equation}\label{eq:betareg}
    j_1 k_2 - j_2 k_1 = k_2 i_1 - k_1 i_2 = i_1 j_2 - j_1 i_2 = 0 \mod p ~. 
\end{equation}
Using the fact that $p$ is prime, it is not hard to show the above equations have $p$ solutions if and only if $a^{i_1}b^{j_1}c^{k_1} \neq 1$, and have $p^3$ solutions otherwise. Then, we know $C_{(i_1,j_1,k_1)} \simeq \doubleZ_p$ when $(i_1,j_1,k_1) \neq (0,0,0)$ and $C_{(i_1, j_1,k_1)} = \doubleZ_p^3$ when $(i_1,j_1,k_1) = (0,0,0)$. Hence, we conclude that there are $p^3$ Abelian anyons corresponding to the Wilson line of the discrete gauge group, and for each non-trivial flux $[a^{i_1} b^{j_1} c^{k_1}]$, there are $p$ lines with quantum dimension $p$. The explicit representation can be constructed using the Clock and Shift matrices. 

As an example, let's consider the case where $i_1 = j_1 = 0$ and $k_1 \neq 0$. In this case, the central elements are parameterized by $(0,0,k_2)$, which forms the $\doubleZ_p^c$ subgroup, and the multiplication of the projective representation takes the form
\begin{equation}
    U_{(0,0,k_1)}(g_2=a^{i_2}b^{j_2}c^{k_2}) U_{(0,0,k_1)}(g_3=a^{i_3}b^{j_3}c^{k_3}) = e^{-\frac{2\pi i k_1}{p} j_2 i_3} U_{(0,0,k_1)}(g_2 g_3) ~.
\end{equation}
Notice that the $\doubleZ_p^c$ completely decouples in the 2-cocycle, the irreducible projective representation can be constructed by stacking a projective representation of $\doubleZ_p^a \times \doubleZ_p^b$ with an ordinary irreducible representation of $\doubleZ_p^c$. The irrep can be constructed explicitly using the $p\times p$ Clock and Shift matrix:
\begin{equation}\label{eq:csmat}
    S = \begin{pmatrix} 0 & 0 & 0 & \cdots & 0 & 1 \\ 1 & 0 & 0 & \cdots & 0 & 0 \\ 0 & 1 & 0 & \cdots & 0 & 0 \\ 0 & 0 & 1 & \cdots & 0 & 0 \\ \vdots & \vdots & \vdots & \ddots & \vdots & \vdots \\ 0 & 0 & 0 & \cdots & 1 & 0  \end{pmatrix}, \quad C = \begin{pmatrix} 1 & 0 & 0 & \cdots & 0 \\ 0 & e^{\frac{2\pi i}{p}} & 0 & \cdots & 0 \\ 0 & 0 & e^{\frac{4\pi i}{p}} & \cdots & 0 \\ \vdots & \vdots & \vdots & \ddots & \vdots \\ 0 & 0 & 0 & \cdots & e^{\frac{2\pi i (p-1)}{p}}\end{pmatrix},
\end{equation}
which satisfies the commutation relation:
\begin{equation}
    SC = e^{-\frac{2\pi i}{p}}CS.
\end{equation}
And the irreducible projective representations $U_{(0,0,k_1),u}(g)$ where $u = 0,\cdots,p-1$ are given by

\begin{equation}
    U_{(0,0,k_1),u}(g=a^ib^jc^k) = e^{\frac{2\pi i}{p} \frac{uk}{k_1}}  C^{k_1i} S^j 
\end{equation}
where $\frac{1}{k_1}$ is the integer inverse of $k_1 \mod p$. We will use the convention that $\frac{1}{x}$ is the inverse of  $x \mod p$ whenever $x \neq np$.
By similar methods, one can show that projective representations $U_{(0,j_1,k_1),u}(g)$ and $U_{(i_1,j_1,k_1),u}(g)$  where $u=0,...,p-1$ are given by:
\begin{equation}
\begin{aligned}
            U_{(0,j_1,k_1),u}(g=a^ib^jc^k) &= e^{\frac{2\pi i}{p}  \frac{uj}{j_1}}  C^{j_1i} S^{-k+\frac{jk_1}{j_1}}, \quad j_1\neq 0 \\
            U_{(0,j_1,k_1),u}(g_2)U_{(0,j_1,k_1),u}(g_3)&=e^{\frac{2\pi i}{p}(j_1k_2i_3 -j_2k_1i_3)}U_{(0,j_1,k_1),u}(g_2g_3)\\
            U_{(i_1,j_1,k_1),u}(g=a^ib^jc^k) &= e^{\frac{2\pi i}{p} ( \frac{ui}{i_1} +\frac{j_1k_1i^2}{2i_1} -j_1ik)} e^{\frac{2\pi i}{p^2}mi_1i} C^{-k+\frac{ik_1}{i_1}} S^{i_1j-j_1i}, \quad i_1\neq 0 \\
            U_{(i_1,j_1,k_1),u}(g_2)U_{(i_1,j_1,k_1),u}(g_3)&=e^{\frac{2\pi i}{p}(j_1k_2i_3+j_2k_3i_1 -j_2k_1i_3)+\frac{2\pi i m}{p^2}i_1(i_2+i_3-[i_2+i_3]_p)}U_{(i_1,j_1,k_1),u}(g_2g_3)
\end{aligned}
\end{equation}
where we have again used the convention that $\frac{1}{x}$ is the inverse of $x\mod  p$ when $x\neq np$.

Now, we are ready to compute the corresponding character. Notice that the trace of $C^i S^j$ is $0$ unless $i =  j = 0$. We then find:
\begin{equation}
    \chi_{(0,0,k_1),u}(g) = \Tr(U_{(0,0,k_1),u}(g)) = e^{\frac{2\pi i}{p} \frac{u k}{k_1}} \delta_{i,0} \delta_{j,0}.
\end{equation}
This is the general feature of this calculation--the character for the projective representation will be non-zero only for the $\beta_{g_1}$-regular elements. Taking this into account, the rest of the results are worked and summarized in the Table \ref{tab:omegap_spectrum}.

\newgeometry{margin=2cm} 
\begin{landscape}

\begin{table}[]
\begin{tabular}{c||c|c|c|c}
flux $g_1$  & 
         $(0,0,0)$  &  
         $(0,0,k_1), \quad k_1 \neq 0 $  & 
         $(0,j_1,k_1), \quad j_1 \neq 0$  & 
         $(i_1,j_1,k_1), \quad i_1 \neq 0$  \\ \hline  $\beta_{g_1}$-regular subgroup $R_{g_1}$  &  $\doubleZ_p^a\times \doubleZ_p^b \times \doubleZ_p^c$  &  $\{(0,0,k)|k\in\doubleZ_p\}$  &  $\{j (0,1,\frac{k_1}{j_1}): j\in \doubleZ_p\}$  &  $\{i (1,\frac{j_1}{i_1}, \frac{k_1}{i_1}):i\in \doubleZ_p\}$  \\ \hline  projective irreps $U$ on $R_{g_1}$  &  $U_{(0,0,0),\vec{u}}(g) = e^{\frac{2\pi i}{p}(u_1 i + u_2 j + u_3 k)}$ &  $U_{(0,0,k_1),u}(g) = e^{\frac{2\pi i}{p}  \frac{u k}{k_1}} I $  &  $U_{(0,j_1,k_1),u}(g) = e^{\frac{2\pi i}{p}\frac{uj}{j_1}} I $  &  \makecell[l]{$U_{(i_1,j_1,k_1),u}(g)  = $ \\ $e^{\frac{2\pi i}{p}[ \frac{u i}{i_1} -\frac{j_1k_1i^2}{2i_1}]+ \frac{2\pi i}{p^2} m i_1 i} I$}  \\ \hline  
         projective character $\chi$  &  \makecell[l]{$\chi_{(0,0,0),\vec{u}}(g) = $\\$ e^{\frac{2\pi i}{p}(u_1 i + u_2 j + u_3 k)}$}  &  \makecell[l]{$\chi_{(0,0,k_1),u}(g) = $\\$p e^{\frac{2\pi i}{p} \frac{u k}{k_1}} \delta_{i,0}\delta_{j,0}$}  &  \makecell[l]{$\chi_{(0,j_1,k_1),u}(g) =$\\$ p e^{\frac{2\pi i}{p}\frac{uj}{j_1}} \delta_{i,0}\delta_{k,\frac{k_1}{j_1}j} $}  &  \makecell[l]{$\chi_{(i_1,j_1,k_1),u}(g) = $\\ $pe^{\frac{2\pi i}{p}[ \frac{u i}{i_1} -\frac{j_1k_1i^2}{2i_1} ]+ \frac{2\pi i}{p^2} m i_1 i}$\\ $\delta_{j,\frac{j_1}{i_1}i}\delta_{k,\frac{k_1}{i_1}i}$}  \\ \hline  topological spin $\theta$   &  $\theta_{(0,0,0),\vec{u}} = 1$  &  $\theta_{(0,0,k),u} = e^{\frac{2\pi i}{p} u}$  &  $\theta_{(0,j_1,k_1),u} = e^{\frac{2\pi i}{p}u}$  &  \makecell[l]{$\theta_{(i_1,j_1,k_1),u} =$\\ $e^{\frac{2\pi i u}{p} + \frac{2\pi i}{p} \frac{-i_1 j_1 k_1}{2} + \frac{2\pi i m}{p^2} i_1^2} $}  \\ \hline  quantum dimension $d$  &  $d_{(0,0,0),\vec{u}} = 1$  &  $d_{(0,0,k_1),u} = p$  &   $d_{(0,j_1,k_1),u} = p$  &  $d_{(i_1,j_1,k_1),u} = p$
\end{tabular}
\caption{Spectrum of the $\doubleZ_p^a \times \doubleZ_p^b \times \doubleZ_p^c$ gauge theory with the twist $\omega_{+,m}$. $I$ denotes the $p\times p$ identity matrix. g denotes a general element $a^ib^jc^k$. We use the convention that $\frac{1}{x}$ is the inverse of $x\mod  p$ when $x\neq np$.}
\label{tab:omegap_spectrum}
\end{table}

The $S$-matrix for twist \eqref{eq:omegap} is,
\begin{equation}
 S_{(a,u),(b,u')}=  \frac{1}{p} \left(
\begin{array}{cccc}
 \frac{1}{p^2} & \frac{1}{p}{\omega ^{-u_1 i_b -u_2 j_b-u_3 k_b}} & \frac{1}{p}\omega ^{- u_2 j_b  -u_3 k_b} & \frac{1}{p}\omega ^{-u_3 k_b} \\
 \frac{1}{p}\omega ^{- u_1'i_a- u_2'j_a-u_3'k_a } & \begin{matrix}
     \omega^{\frac{1}{2}\frac{j_ak_ai_b^2}{i_a}+\frac{1}{2}\frac{j_bk_bi_a^2}{i_b}-\frac{2 m i_a i_b}{p}-\frac{i_a u'}{i_b}-\frac{u i_b}{i_a}} \times \\  \delta \left(j_b-\frac{j_a i_b}{i_a}\right) \delta \left(k_b-\frac{k_a i_b}{i_a}\right)
 \end{matrix}  & 0 & 0 \\
 \frac{1}{p}\omega ^{-u_2'j_a - u_3'k_a} & 0 & \begin{matrix}
     \omega ^{-\frac{j_a u'}{j_b}-\frac{u j_b}{j_a}}\times\\
     \delta \left(j_a k_b-k_a j_b\right) 
 \end{matrix} & 0 \\
 \frac{1}{p}\omega ^{-u_3'k_a } & 0 & 0 & \omega ^{-\frac{k_a u'}{k_b}-\frac{u k_b}{k_a}} \\
\end{array}
\right) ~,
\end{equation}
where $a=(i_a,j_a,k_a)$, and the same for $b$. The blocks correspond to $(0,0,0),(i,j,k),(0,j,k),(0,0,k)$. The fusion rules are given by the Verlinde's formula. The $T$-matrix is given by,
\begin{equation}
    T_{(a,u),(b,u')}=\delta_{a,b}\delta_{u,u'}\begin{cases}
        1 & a=b=(0,0,0) ~, \\
        \omega^{u-\frac{1}{2}i_a j_a k_a+\frac{m}{p}i_a^2} & \text{otherwise} ~.
    \end{cases}
\end{equation}

\begin{table}[]
    \centering
    \begin{tabular}{c||c|c|c|c}
         flux $g_1$  & 
         $(0,0,0)$  &  
         $(0,0,k_1), \quad k_1 \neq 0 $  & 
         $(0,j_1,k_1), \quad j_1 \neq 0$  & 
         $(i_1,j_1,k_1), \quad i_1 \neq 0$  \\ \hline  $\beta_{g_1}$-regular subgroup $R_{g_1}$  &  $\doubleZ_p^a\times \doubleZ_p^b \times \doubleZ_p^c$  &  $\{(0,0,k)|k\in\doubleZ_p\}$  &  $\{j (0,1,\frac{k_1}{j_1}): j\in \doubleZ_p\}$  &  $\{i (1,\frac{j_1}{i_1}, \frac{k_1}{i_1}):i\in \doubleZ_p\}$  \\ \hline  projective irreps $U$ on $R_{g_1}$  &  $U_{(0,0,0),\vec{u}}(g) = e^{\frac{2\pi i}{p}(u_1 i + u_2 j + u_3 k)}$ &  $U_{(0,0,k_1),u}(g) = e^{\frac{2\pi i}{p}  \frac{u k}{k_1}} I $  &  $U_{(0,j_1,k_1),u}(g) = e^{\frac{2\pi i}{p}\frac{uj}{j_1}} I $  &  \makecell[l]{$U_{(i_1,j_1,k_1),u}(g)  = $ \\ $e^{\frac{2\pi i}{p}\frac{u i}{i_1} - \frac{2\pi i}{p} \frac{1}{2} \frac{j_1 k_1}{i_1} i^2 + \frac{2\pi i}{p^2} m i_1 i +\frac{2\pi i}{p^2} j_1 i} I$}  \\ \hline  
         projective character $\chi$  &  \makecell[l]{$\chi_{(0,0,0),\vec{u}}(g) = $\\$ e^{\frac{2\pi i}{p}(u_1 i + u_2 j + u_3 k)}$}  &  \makecell[l]{$\chi_{(0,0,k_1),u}(g) =$\\$p e^{\frac{2\pi i}{p} \frac{u k}{k_1}} \delta_{i,0}\delta_{j,0}$}  &  \makecell[l]{$\chi_{(0,j_1,k_1),u}(g) =$\\$p e^{\frac{2\pi i}{p}\frac{uj}{j_1}} \delta_{i,0}\delta_{k,\frac{k_1}{j_1}j} $}  &  \makecell[l]{$\chi_{(i_1,j_1,k_1),u}(g) = $\\ $p e^{\frac{2\pi i}{p}\frac{u i}{i_1} - \frac{2\pi i}{p} \frac{1}{2} \frac{j_1 k_1}{i_1} i^2 + \frac{2\pi i}{p^2} m i_1 i  + \frac{2\pi i}{p^2} j_1 i}\delta_{j,\frac{j_1}{i_1}i}\delta_{k,\frac{k_1}{i_1}i}$}  \\ \hline  topological spin $\theta$   &  $\theta_{(0,0,0),\vec{u}} = 1$  &  $\theta_{(0,0,k),u} = e^{\frac{2\pi i}{p} u}$  &  $\theta_{(0,j_1,k_1),u} = e^{\frac{2\pi i}{p}u}$  &  \makecell[l]{$\theta_{(i_1,j_1,k_1),u} =$\\ $e^{\frac{2\pi i u}{p} + \frac{2\pi i}{p}\frac{-1}{2} i_1 j_1 k_1 + \frac{2\pi i m}{p^2} i_1^2 + \frac{2\pi i}{p^2} j_1 i_1} $}  \\ \hline  quantum dimension $d$  &  $d_{(0,0,0),\vec{u}} = 1$  &  $d_{(0,0,k_1),u} = p$  &   $d_{(0,j_1,k_1),u} = p$  &  $d_{(i_1,j_1,k_1),u} = p$
    \end{tabular}
    \caption{Spectrum of the $\doubleZ_p^a \times \doubleZ_p^b \times \doubleZ_p^c$ gauge theory with the twist $\omega_{-,m}$. By $I$, we mean the $p\times p$ identity matrix. We use the convention that $\frac{1}{x}$ is the inverse of $x \mod p$ when $x\neq np$.}
    \label{tab:omegam_spectrum}
\end{table}
We can repeat the above calculation for twist \eqref{eq:omegam}. The $S$-matrix is
\begin{equation}
 S_{(a,u),(b,u')}=  \frac{1}{p} \left(
\begin{array}{cccc}
 \frac{1}{p^2} & \frac{1}{p}{\omega ^{-u_1 i_b -u_2 j_b-u_3 k_b}} & \frac{1}{p}\omega ^{- u_2 j_b  -u_3 k_b} & \frac{1}{p}\omega ^{-u_3 k_b} \\
 \frac{1}{p}\omega ^{- u_1'i_a- u_2'j_a-u_3'k_a } & \begin{matrix}
     \omega ^{\frac{ i_a^2 j_b k_b}{2 i_b}+\frac{ j_a k_a i_b^2}{2 i_a}-\frac{2 m i_a i_b}{p}-\frac{i_a u'}{i_b}-\frac{u i_b}{i_a}-\frac{j_b i_a}{p}-\frac{j_a i_b}{p}}\times\\ \delta \left(j_b-\frac{j_a i_b}{i_a}\right) \delta \left(k_b-\frac{k_a i_b}{i_a}\right)
 \end{matrix}  & 0 & 0 \\
 \frac{1}{p}\omega ^{-u_2'j_a - u_3'k_a} & 0 & \begin{matrix}
     \omega ^{-\frac{j_a u'}{j_b}-\frac{u j_b}{j_a}}\times\\
     \delta \left(j_a k_b-k_a j_b\right) 
 \end{matrix} & 0 \\
 \frac{1}{p}\omega ^{-u_3'k_a } & 0 & 0 & \omega ^{-\frac{k_a u'}{k_b}-\frac{u k_b}{k_a}} \\
\end{array}
\right) ~.
\end{equation}

The $T$-matrix is given by
\begin{equation}
    T_{(a,u),(b,u')}=\delta_{a,b}\delta_{u,u'}\begin{cases}
        1 & a=b=(0,0,0) ~, \\
        \omega^{u-\frac{1}{2}i_a j_a k_a+\frac{m}{p}i_a^2+\frac{j_a i_a}{p}} & \text{otherwise} ~.
    \end{cases}
\end{equation}

And the spectrum of the simple anyons is summarized in Table \ref{tab:top_sec_omegapm}.
\end{landscape}
\restoregeometry
\restoregeometry

\section{Details of inequivalent $\beta$-invertible $S_3$ symmetries in $\doubleZ_N\times \doubleZ_N$-SymTFT}\label{app:S3_class}
In this section, we list the generators of a representative of inequivalent $\beta$-invertible $S_3$ symmetries in $\doubleZ_N\times \doubleZ_N$-SymTFT for $N < 20$. For $N = 3$, while there is no choice of $\beta$-invertible $S_3$ symmetry, we still list representatives of the generators of inequivalent $\beta$-invertible $\doubleZ_3$ symmetries.

\subsection{$N = 3$}
There are three inequivalent $\beta$-invertible $\doubleZ_3$ symmetries; a set of representatives of the generators is given by
\begin{equation}
    T_i = \begin{pmatrix} 1 & 0 & 0 & 1 \\ i & 1 & -1 & i \\ 0 & 0 & 1 & -i \\ 0 & 0 & 0 & 1 \end{pmatrix} ~, \quad i = 0,1,2 ~.
\end{equation}
Notice that $i = 0$ corresponds to the triality defect constructed in Section \ref{Sec:p_ality}. There is no $\beta$-invertible $S_3$ symmetry.

\subsection{$N = 5$}
There are two inequivalent $\beta$-invertible $\doubleZ_3$ given by $T_1, T_2$ in \eqref{eq:Z3_sym}. For $T = T_1$, there is a single inequivalent $\doubleZ_2$ generator with the representative
\begin{equation}
    D_{1,1} = \left(
\begin{array}{cccc}
 0 & 3 & 1 & 0 \\
 4 & 0 & 0 & 2 \\
 4 & 0 & 0 & 4 \\
 0 & 2 & 3 & 0 \\
\end{array}
\right) ~,
\end{equation}
while for $T = T_2$, there are two inequivalent $\doubleZ_2$ generators with representatives
\begin{equation}
    D_{2,1} = \left(
\begin{array}{cccc}
 3 & 3 & 2 & 3 \\
 1 & 2 & 3 & 1 \\
 3 & 1 & 3 & 1 \\
 1 & 1 & 3 & 2 \\
\end{array}
\right) ~, \quad D_{2,2} = \left(
\begin{array}{cccc}
 1 & 1 & 3 & 2 \\
 2 & 4 & 2 & 4 \\
 4 & 3 & 1 & 2 \\
 3 & 3 & 1 & 4 \\
\end{array}
\right) ~.
\end{equation}

None of these $S_3$ symmetries admit stable Lagrangian algebras, therefore are not group-theoretical.

\subsection{$N = 7$}
There are two inequivalent $\beta$-invertible $\doubleZ_3$ given by $T_1, T_2$ in \eqref{eq:Z3_sym}. For $T = T_1$, there is a single inequivalent $\doubleZ_2$ generators with the representative
\begin{equation}
    D_{1,1} = \left(
\begin{array}{cccc}
 0 & 4 & 1 & 0 \\
 3 & 0 & 0 & 1 \\
 3 & 0 & 0 & 3 \\
 0 & 3 & 4 & 0 \\
\end{array}
\right) ~,
\end{equation}
while for $T = T_2$, there are two inequivalent $\doubleZ_2$ generators with representatives
\begin{equation}
    D_{2,1} = \left(
\begin{array}{cccc}
 4 & 4 & 2 & 5 \\
 1 & 3 & 5 & 3 \\
 3 & 2 & 4 & 1 \\
 2 & 2 & 4 & 3 \\
\end{array}
\right) ~, \quad D_{2,2} = \left(
\begin{array}{cccc}
 2 & 2 & 5 & 2 \\
 4 & 5 & 2 & 4 \\
 6 & 4 & 2 & 4 \\
 4 & 4 & 2 & 5 \\
\end{array}
\right) ~.
\end{equation}

\subsection{$N = 11$}
There are two inequivalent $\beta$-invertible $\doubleZ_3$ given by $T_1, T_2$ in \eqref{eq:Z3_sym}. For $T = T_1$, there are two inequivalent $\doubleZ_2$ generators with the representatives
\begin{equation}
    D_{1,1} = \left(
\begin{array}{cccc}
 2 & 0 & 0 & 2 \\
 0 & 9 & 2 & 0 \\
 0 & 4 & 2 & 0 \\
 4 & 0 & 0 & 9 \\
\end{array}
\right) ~, \quad D_{1,2} = \left(
\begin{array}{cccc}
 0 & 5 & 1 & 0 \\
 6 & 0 & 0 & 1 \\
 4 & 0 & 0 & 6 \\
 0 & 4 & 5 & 0 \\
\end{array}
\right) ~.
\end{equation}
while for $T = T_2$, there are four inequivalent $\doubleZ_2$ generators with representatives
\begin{equation}
\begin{aligned}
    & D_{2,1} = \left(
\begin{array}{cccc}
 0 & 9 & 7 & 0 \\
 9 & 0 & 0 & 4 \\
 9 & 0 & 0 & 9 \\
 0 & 2 & 9 & 0 \\
\end{array}
\right) ~, \quad D_{2,2} = \left(
\begin{array}{cccc}
 7 & 5 & 2 & 6 \\
 1 & 4 & 6 & 4 \\
 6 & 2 & 7 & 1 \\
 2 & 3 & 5 & 4 \\
\end{array}
\right) ~, \\
& D_{2,3} = \left(
\begin{array}{cccc}
 5 & 9 & 5 & 7 \\
 3 & 6 & 7 & 2 \\
 6 & 1 & 5 & 3 \\
 1 & 4 & 9 & 6 \\
\end{array}
\right) ~, \quad D_{2,4} = \left(
\begin{array}{cccc}
 7 & 7 & 4 & 7 \\
 3 & 4 & 7 & 3 \\
 5 & 3 & 7 & 3 \\
 3 & 3 & 7 & 4 \\
\end{array}
\right) ~.
\end{aligned}
\end{equation}

Among the six $S_3$'s, only the one generated by $\langle T_1, D_{1,1} \rangle$ admits a stable Lagrangian algebra, therefore is group-theoretical. But it does not admit a stable magnetic Lagrangian algebra, therefore it is anomalous.

\subsection{$N = 13$}
There are two inequivalent $\beta$-invertible $\doubleZ_3$ given by $T_1, T_2$ in \eqref{eq:Z3_sym}. For $T = T_1$, there are two inequivalent $\doubleZ_2$ generators with the representatives
\begin{equation}
    D_{1,1} = \left(
\begin{array}{cccc}
 3 & 0 & 0 & 3 \\
 0 & 10 & 3 & 0 \\
 0 & 6 & 3 & 0 \\
 6 & 0 & 0 & 10 \\
\end{array}
\right) ~, \quad D_{1,2} = \left(
\begin{array}{cccc}
 0 & 6 & 1 & 0 \\
 9 & 0 & 0 & 5 \\
 12 & 0 & 0 & 9 \\
 0 & 5 & 6 & 0 \\
\end{array}
\right) ~.
\end{equation}
while for $T = T_2$, there are four inequivalent $\doubleZ_2$ generators with representatives
\begin{equation}
\begin{aligned}
    & D_{2,1} = \left(
\begin{array}{cccc}
 0 & 10 & 7 & 0 \\
 10 & 0 & 0 & 6 \\
 10 & 0 & 0 & 10 \\
 0 & 3 & 10 & 0 \\
\end{array}
\right) ~, \quad D_{2,2} = \left(
\begin{array}{cccc}
 8 & 6 & 2 & 6 \\
 1 & 5 & 6 & 4 \\
 8 & 1 & 8 & 1 \\
 1 & 4 & 6 & 5 \\
\end{array}
\right) ~, \\
& D_{2,3} = \left(
\begin{array}{cccc}
 6 & 11 & 7 & 8 \\
 4 & 7 & 8 & 1 \\
 8 & 1 & 6 & 4 \\
 1 & 4 & 11 & 7 \\
\end{array}
\right) ~, \quad D_{2,4} = \left(
\begin{array}{cccc}
 8 & 8 & 4 & 9 \\
 3 & 5 & 9 & 5 \\
 5 & 4 & 8 & 3 \\
 4 & 4 & 8 & 5 \\
\end{array}
\right) ~.
\end{aligned}
\end{equation}

Among the six $S_3$'s, only the one generated by $\langle T_1, D_{1,1} \rangle$ admits a stable Lagrangian algebra, therefore is group-theoretical. But it does not admit a stable magnetic Lagrangian algebra, therefore it is anomalous.

\subsection{$N = 17$}
There are two inequivalent $\beta$-invertible $\doubleZ_3$ given by $T_1, T_2$ in \eqref{eq:Z3_sym}. For $T = T_1$, there are two inequivalent $\doubleZ_2$ generators with the representatives
\begin{equation}
    D_{1,1} = \left(
\begin{array}{cccc}
 2 & 0 & 0 & 6 \\
 0 & 15 & 6 & 0 \\
 0 & 8 & 2 & 0 \\
 8 & 0 & 0 & 15 \\
\end{array}
\right) ~, \quad D_{1,2} =\left(
\begin{array}{cccc}
 0 & 6 & 1 & 0 \\
 15 & 0 & 0 & 6 \\
 13 & 0 & 0 & 15 \\
 0 & 5 & 6 & 0 \\
\end{array}
\right) ~, \quad D_{1,3} = \left(
\begin{array}{cccc}
 0 & 9 & 1 & 0 \\
 12 & 0 & 0 & 10 \\
 12 & 0 & 0 & 12 \\
 0 & 8 & 9 & 0 \\
\end{array}
\right) ~.
\end{equation}
while for $T = T_2$, there are four inequivalent $\doubleZ_2$ generators with representatives
\begin{equation}
\begin{aligned}
    & D_{2,1} = \left(
\begin{array}{cccc}
 0 & 11 & 9 & 0 \\
 11 & 0 & 0 & 8 \\
 15 & 0 & 0 & 11 \\
 0 & 2 & 11 & 0 \\
\end{array}
\right) ~, \quad D_{2,2} = \left(
\begin{array}{cccc}
 0 & 8 & 6 & 0 \\
 8 & 0 & 0 & 11 \\
 15 & 0 & 0 & 8 \\
 0 & 2 & 8 & 0 \\
\end{array}
\right) ~, \quad D_{2,3} = \left(
\begin{array}{cccc}
 1 & 8 & 1 & 2 \\
 9 & 16 & 2 & 1 \\
 7 & 3 & 1 & 9 \\
 3 & 7 & 8 & 16 \\
\end{array}
\right) ~, \\
& D_{2,4} = \left(
\begin{array}{cccc}
 7 & 9 & 1 & 3 \\
 16 & 10 & 3 & 2 \\
 16 & 10 & 7 & 16 \\
 10 & 8 & 9 & 10 \\
\end{array}
\right) ~, \quad D_{2,5} = \left(
\begin{array}{cccc}
 1 & 2 & 1 & 8 \\
 3 & 16 & 8 & 7 \\
 7 & 9 & 1 & 3 \\
 9 & 1 & 2 & 16 \\
\end{array}
\right) ~, \quad D_{2,6} = \left(
\begin{array}{cccc}
 7 & 3 & 1 & 9 \\
 10 & 10 & 9 & 8 \\
 16 & 16 & 7 & 10 \\
 16 & 2 & 3 & 10 \\
\end{array}
\right) ~.
\end{aligned}
\end{equation}

Among the nine $S_3$'s, only the one generated by $\langle T_1, D_{1,1} \rangle$ admits a stable Lagrangian algebra, therefore is group-theoretical. But it does not admit a stable magnetic Lagrangian algebra, therefore it is anomalous.

\subsection{$N = 19$}
There are two inequivalent $\beta$-invertible $\doubleZ_3$ given by $T_1, T_2$ in \eqref{eq:Z3_sym}. For $T = T_1$, there are two inequivalent $\doubleZ_2$ generators with the representatives
\begin{equation}
    D_{1,1} = \left(
\begin{array}{cccc}
 2 & 0 & 0 & 12 \\
 0 & 17 & 12 & 0 \\
 0 & 14 & 2 & 0 \\
 14 & 0 & 0 & 17 \\
\end{array}
\right) ~, \quad D_{1,2} = \left(
\begin{array}{cccc}
 0 & 10 & 1 & 0 \\
 7 & 0 & 0 & 5 \\
 7 & 0 & 0 & 7 \\
 0 & 9 & 10 & 0 \\
\end{array}
\right) ~, \quad D_{1,3} = \left(
\begin{array}{cccc}
 0 & 5 & 1 & 0 \\
 12 & 0 & 0 & 9 \\
 17 & 0 & 0 & 12 \\
 0 & 4 & 5 & 0 \\
\end{array}
\right) ~.
\end{equation}
while for $T = T_2$, there are four inequivalent $\doubleZ_2$ generators with representatives
\begin{equation}
\begin{aligned}
    & D_{2,1} = \left(
\begin{array}{cccc}
 0 & 7 & 5 & 0 \\
 7 & 0 & 0 & 14 \\
 17 & 0 & 0 & 7 \\
 0 & 2 & 7 & 0 \\
\end{array}
\right) ~, \quad D_{2,2} = \left(
\begin{array}{cccc}
 0 & 14 & 12 & 0 \\
 14 & 0 & 0 & 7 \\
 17 & 0 & 0 & 14 \\
 0 & 2 & 14 & 0 \\
\end{array}
\right) ~, \quad D_{2,3} = \left(
\begin{array}{cccc}
 11 & 4 & 1 & 2 \\
 15 & 8 & 2 & 1 \\
 3 & 13 & 11 & 15 \\
 13 & 3 & 4 & 8 \\
\end{array}
\right) ~, \\
& D_{2,4} = \left(
\begin{array}{cccc}
 11 & 2 & 1 & 4 \\
 13 & 8 & 4 & 3 \\
 3 & 15 & 11 & 13 \\
 15 & 1 & 2 & 8 \\
\end{array}
\right) ~, \quad D_{2,5} = \left(
\begin{array}{cccc}
 1 & 1 & 16 & 3 \\
 2 & 18 & 3 & 6 \\
 11 & 4 & 1 & 2 \\
 4 & 4 & 1 & 18 \\
\end{array}
\right) ~, \quad D_{2,6} = \left(
\begin{array}{cccc}
 1 & 3 & 16 & 1 \\
 4 & 18 & 1 & 4 \\
 11 & 2 & 1 & 4 \\
 2 & 6 & 3 & 18 \\
\end{array}
\right) ~.
\end{aligned}
\end{equation}

Among the nine $S_3$'s, only the one generated by $\langle T_1, D_{1,1} \rangle$ admits a stable Lagrangian algebra, therefore is group-theoretical. But it does not admit a stable magnetic Lagrangian algebra, therefore it is anomalous.

\bibliographystyle{utphys}
\bibliography{pality}
\end{document}